\documentclass[10pt,draftcls, onecolumn]{IEEEtran}

\usepackage{times}
\usepackage{amsmath,dsfont}
\usepackage{amssymb,amsthm}
\usepackage{epsfig,verbatim}
\usepackage{subfigure}
\usepackage{setspace}
\usepackage{color}
\usepackage{cite}
\usepackage{epstopdf}
\usepackage{graphics}
\usepackage{accents}

\usepackage{enumitem}

\newtheorem{definition}{Definition}
\newtheorem{theorem}{Theorem}
\newtheorem{corollary}{Corollary}
\newtheorem{proposition}{Proposition}
\newtheorem{lemma}{Lemma}
\newtheorem{remark}{Comment}

\newcommand{\E}{\mathds{E}}

\newcommand{\utilde}{\underaccent{\tilde}}
\newcommand{\Cw}[1]{\mathsf{C}_{\bf W}\left[ #1  \right]}
\newcommand{\Cu}[1]{\mathsf{C}_{\bf U}\left[ #1  \right]}
\newcommand{\CCw}[1]{\mathsf{C}'_{\bf W}\!\left( #1 \right)}
\newcommand{\CCu}[1]{\mathsf{C}'_{\bf U}\!\left( #1 \right)}
\newcommand{\CCx}[1]{\mathsf{C}'_{\bf X}\!\left( #1 \right)}
\newcommand{\CCxopt}[1]{\mathsf{C}'_{{\bf X}, opt}\left( #1 \right)}
\newcommand{\CUw}[1]{\mathsf{C}_{\utilde{\bf W}}\left[ #1 \right]}
\newcommand{\CUu}[1]{\mathsf{C}_{\utilde{\bf U}}\left[ #1 \right]}
\newcommand{\CUHw}[1]{\mathsf{C}_{\utilde{\hat{\bf W}}}\left[ #1 \right]}
\newcommand{\CUHu}[1]{\mathsf{C}_{\utilde{\hat{\bf U}}}\left[ #1 \right]}
\newcommand{\CUHx}[1]{\mathsf{C}_{{\hat{\bf X}}}\left[ #1 \right]}
\newcommand{\CUHxOpt}[1]{\mathsf{C}_{{\hat{\bf X}}, opt}\left[ #1 \right]}
\newcommand{\HCw}[1]{\mathsf{H}'_w\!\left( #1 \right)}
\newcommand{\GCw}[1]{\mathsf{G}'_w\!\left( #1 \right)}
\newcommand{\CCG}[1]{C_{s}^{#1-CG}}
\newcommand{\CCGTA}[1]{C_{s,TA}^{#1-CG}}
\newcommand{\CMG}[1]{C_{s}^{#1-MG}}
\newcommand{\Cs}{C_s}
\newcommand{\Lm}{\left\lfloor \frac{n}{2}\right\rfloor}
\newcommand{\Ltm}{\left\lfloor \frac{n-1}{2}\right\rfloor}
\newcommand{\Gm}{\mathsf{G}}
\newcommand{\LmSet}{\mathcal{L}_{CG}}

\newcommand{\MyBeta}[1]{\E\left\{ {{\left\| {\bf X}[i]  \right\|}^2} \right\} \leq P, ~\forall {0\leq i < #1}}

\newcommand{\ScPLCY}{Y_{\rm PLC}}
\newcommand{\ScPLCW}{W_{\rm PLC}}
\newcommand{\ScPLCZ}{Z_{\rm PLC}}
\newcommand{\ScPLCU}{U_{\rm PLC}}
\newcommand{\ScPLCH}{h_{\rm PLC}}
\newcommand{\ScPLCG}{g_{\rm PLC}}
\newcommand{\ScPLCm}{m_{\rm PLC}}
\newcommand{\ScPLCmCh}{m_{ch}}
\newcommand{\ScPLCmN}{m_{noise}}
\newcommand{\ScPLCnt}{n_{\rm PLC}}
\newcommand{\ScPLCnSet}{\mathcal{N}_{\rm PLC}}

\newcommand{\VecPLCY}{{\bf Y}_{\rm PLC}}
\newcommand{\VecPLCW}{{\bf W}_{\rm PLC}}
\newcommand{\VecPLCZ}{{\bf Z}_{\rm PLC}}
\newcommand{\VecPLCU}{{\bf U}_{\rm PLC}}
\newcommand{\VecPLCX}{{\bf X}_{\rm PLC}}
\newcommand{\VecPLCH}{\mathsf{H}_{\rm PLC}}
\newcommand{\VecPLCG}{\mathsf{G}_{\rm PLC}}

\newcommand{\CwPLC}[1]{\mathsf{C}_{\VecPLCW}\left[ #1  \right]}
\newcommand{\CuPLC}[1]{\mathsf{C}_{\VecPLCU}\left[ #1  \right]}
\newcommand{\CCwPLC}[1]{\mathsf{C}'_{\VecPLCW}\!\left( #1 \right)}
\newcommand{\CCuPLC}[1]{\mathsf{C}'_{\VecPLCU}\!\left( #1 \right)}
\newcommand{\CCxPLC}[1]{\mathsf{C}'_{\bf X}\left( #1 \right)}
\newcommand{\CChPLC}[1]{\mathsf{H}'_{\rm PLC}\!\left( #1 \right)}
\newcommand{\CCgPLC}[1]{\mathsf{G}'_{\rm PLC}\!\left( #1 \right)}

\newcommand{\dsN}{\mathds{N}}
\newcommand{\dsR}{\mathds{R}}
\newcommand{\mN}{\mathcal{N}}
\newcommand{\tm}{\tilde{m}}
\newcommand{\tR}{\tilde{R}}
\newcommand{\tb}{\tilde{b}}
\newcommand{\ti}{\tilde{i}}
\newcommand{\mSzro}{\mathcal{S}_0}
\newcommand{\xvec}{\mathbf{x}}

\newif\ifextended
\extendedtrue 

\newif\ifcomments

\long\def\symbolfootnote[#1]#2{\begingroup\def\thefootnote{\fnsymbol{footnote}}\footnote[#1]{#2}\endgroup}

\DeclareMathOperator*{\argmax}{arg\,max}

 \IEEEoverridecommandlockouts
\title{The Secrecy Capacity of Gaussian MIMO Channels with Finite Memory
	\ifextended
	- Full Version
	\fi 
 \thanks{
 \noindent This work was supported by the Ministry of Economy of Israel through the Israeli Smart Grid Consortium. This paper was presented in part at the 2015 IEEE International Symposium on Information Theory. 
}
 \thanks{N. Shlezinger, D. Zahavi, and R. Dabora are with the department of Electrical and Computer
 	Engineering, Ben-Gurion University, Beer-Sheva, Israel (e-mail:	\{nirshl, zahavida\}@post.bgu.ac.il; ron@ee.bgu.ac.il).}
 \thanks{Y. Murin is with the department of Electrical Engineering, Stanford University, Stanford, CA (e-mail: moriny@stanford.edu).  }
}

\vspace{-0.5cm}

\author{
\IEEEauthorblockN{\vspace{-0.2cm} Nir Shlezinger,~\IEEEmembership{Student Member,~IEEE}, Daniel Zahavi,~\IEEEmembership{Student Member,~IEEE}, Yonathan Murin,~\IEEEmembership{Member,~IEEE}, and Ron Dabora,~\IEEEmembership{Senior Member,~IEEE}\\
}

\vspace{-1.0cm}

}

\vspace{-0.75cm}

\begin{document}

\maketitle
\pagestyle{plain}
\thispagestyle{plain}
\begin{abstract}
		In this work we study the secrecy capacity of Gaussian multiple-input multiple-output (MIMO) wiretap channels (WTCs) with a {\em finite memory}, subject to a per-symbol average power constraint on the MIMO channel input.
MIMO channels with finite memory are very common in wireless communications as well as in wireline communications (e.g., in communications over power lines).
		To derive the secrecy capacity of the Gaussian MIMO WTC with finite memory we first construct an asymptotically-equivalent block-memoryless MIMO WTC, which is then transformed into a set of parallel, independent, memoryless MIMO WTCs in the frequency domain. 
				The secrecy capacity of the Gaussian MIMO WTC with finite memory is obtained as the secrecy capacity of the set of parallel independent memoryless MIMO WTCs, and is expressed as a maximization over the input covariance matrices in the frequency domain. 
Lastly, we detail two applications of our result: First, we show that the secrecy capacity of the Gaussian {\em scalar} WTC with {\em finite memory}  can be achieved by waterfilling, and obtain a closed-form expression for this secrecy capacity. Then, we use our result to characterize the secrecy capacity of narrowband powerline channels, thereby resolving one of the major open issues for this channel model.
\end{abstract}

\section{Introduction}
One of the main challenges in the design of communications schemes for shared channels is to reliably transmit information to a destination, while keeping potential eavesdroppers ignorant of the transmitted information. 
The fundamental model for studying secure physical-layer communications over shared mediums is the {\em wiretap channel} (WTC) model \cite{Wyner:75}, which consists of three terminals: A transmitter (Tx), an intended receiver (Rx), and an eavesdropper (Ev).
The secrecy capacity is defined as the maximum information rate for reliable Tx--Rx communications such that the rate of information  leaked to the eavesdropper asymptotically vanishes. 
The initial study of WTCs detailed in \cite{Wyner:75}, considered memoryless WTCs in which the channel inputs and the channel outputs are discrete random variables (RVs) with finite alphabets, and the Tx--Ev channel is a physically degraded version of the Tx--Rx channel.
The general discrete memoryless WTC was studied in \cite{Csiszar:78}, which characterized its secrecy capacity by introducing a virtual channel, also referred to as {\em prefix channel} \cite[Ch. 3.5]{Barros:11}.

Memoryless scalar WTCs with additive white Gaussian noise (AWGN) were first studied in \cite{Hellman:78}, 
which made three important observations: (1) No prefix channel is required, (2) Gaussian codebooks are optimal, and (3) The secrecy capacity is zero when the noise power at the intended receiver is equal to or greater than the noise power at the eavesdropper.
Several works studied the fundamental limits of secure communications over memoryless WTCs with AWGN and multiple antennas at the terminals, referred to as the multiple-input multiple-output (MIMO) WTC:
The work \cite{Ulukus:09} considered the scenario of two antennas at the Tx, two antennas at the Rx, and one antenna at the Ev, where the channel input is subject to a per-codeword average power constraint.
The secrecy capacity of  MIMO WTCs with an arbitrary number of antennas at each node was derived in  \cite{Wornell:10}  subject to a per-codeword average power constraint, and in \cite{Hassibi:11} subject to a per-symbol average power constraint. 
An alternative derivation of the secrecy capacity of MIMO WTCs was carried out in \cite{Shamai:09}, subject to a more general input covariance matrix constraint. In  \cite[Corollary 1]{Shamai:09} it is shown that the secrecy capacity subject to a per-codeword average power constraint on the input can be obtained as a corollary of the main result of \cite{Shamai:09}.
The more general scenario of AWGN MIMO broadcast channels with confidential messages was studied in \cite{Liu:10, Liu:13, Ekrem:12}.
Similarly to the scalar Gaussian case, the secrecy capacity of MIMO WTCs with AWGN is achieved by using a Gaussian codebook without channel prefixing, where the secrecy capacity expression is stated as an optimization over all possible input covariance matrices which satisfy a specified power constraint.
This optimization problem was shown to be non-convex \cite{Wornell:10, Li:13, Fakoorian:15, Loyka:14}, and methods for approaching the maximizing input covariance matrix were proposed in several works. In particular, \cite{Li:13} proposed an algorithm based on alternating optimization for approaching the optimal covariance matrix, \cite{Fakoorian:15} studied the conditions for the covariance matrix to be full rank and characterized the optimal covariance matrix for this case, and in \cite{Loyka:14} rank deficient solutions for the optimal covariance matrix were proposed.
Secrecy in the presence of temporally correlated Gaussian noise was studied in \cite{Fujita:12}, which considered {\em scalar degraded} block-memoryless WTCs with additive colored Gaussian noise. 
Additional scenarios of physical-layer security in modern networks include fading WTCs, studied in \cite{ElGamal:08, Liang:08, Barros:08}, independent parallel channels, studied in \cite{Li:10, Khisti:14}, and an achievable secrecy rate for multi-carrier systems, characterized in \cite{Jorswieck:08} and \cite{Renna:12}. 
The wiretap framework was further extended to multi-user channels in \cite{Liu:08, Khisti:08, Lai:07, Poor:08}
(see also detailed surveys in \cite{Mukherjee:14}, \cite[Ch. 8]{Barros:11}, and \cite[Ch. 22]{ElGamal:10}).
The secrecy capacity of arbitrary wiretap channels was studied in \cite{Bloch:08}, yet, the expression derived in \cite[Thm. 1]{Bloch:08} is rather involved and does not identify the input
distribution which maximizes the secrecy rate.
Finally, we note that a suboptimal precoding scheme for block frequency-selective scalar WTCs with AWGN was proposed in \cite{Kobayashi:08}. 

In this paper we study the secrecy capacity of MIMO Gaussian WTCs with finite memory, i.e., MIMO Gaussian WTCs in which the channel introduces intersymbol interference (ISI) of a finite duration at each receive antenna, and the noise is an additive stationary colored Gaussian process whose temporal correlation has a finite length.
This channel model applies to many communications scenarios, including wireless communications and power line communications. 
However, despite the importance of this model as a fundamental model for secure modern communications, the secrecy capacity of Gaussian {\em finite-memory} MIMO WTCs and also of Gaussian {\em finite-memory} scalar WTCs {\em has not been characterized to date}. 

\smallskip
{\bf {\slshape Main Contributions}:}
 In this paper we derive the secrecy capacity of Gaussian MIMO WTCs with finite memory, subject to a per-MIMO symbol average power constraint on the channel input, where the transmitter knows both the Tx-Rx channel and the Tx-Ev channel. To this aim, we first construct a block-memoryless Gaussian MIMO WTC based on the characteristics of the original finite-memory Gaussian MIMO WTC, and  prove that the two channel models are asymptotically equivalent. Then, we transform the block-memoryless channel into an equivalent set of parallel memoryless Gaussian MIMO WTCs, for which the secrecy capacity has been characterized in \cite{Li:10}.
Our derivation uses concepts from the derivation of the (non-secure) capacity of finite-memory Gaussian point-to-point channels \cite{Massey:88}, multiple-access channels (MACs) \cite{Verdu:89}, and broadcast channels (BCs) \cite{Goldsmith:01}, as well as introduce novel techniques and schemes for the analysis of the information leakage rate at the eavesdropper.
For the special case of the {\em scalar} Gaussian WTC with {\em finite memory}, we show that the secrecy capacity can be obtained via the waterfilling power allocation scheme, and demonstrate the resulting rate via a numerical example.
Finally, we show how our result directly leads to the secrecy capacity of narrowband powerline communications (PLC) channels, which is a major challenge in smart grid communications networks \cite{Tonello:14}. Our results provide insights on the relationship between these seemingly different problems.

The rest of this paper is organized as follows:  Section \ref{sec:Preliminaries} introduces the problem formulation;
Section \ref{sec:SecCap} derives the secrecy capacity for finite-memory Gaussian MIMO WTCs;
Section \ref{sec:Simulations} discusses the results and their application to PLC, and provides a numerical example;
Lastly,  Section \ref{sec:Conclusions} provides some concluding remarks.

\section{Notations and Problem Formulation}
\label{sec:Preliminaries}

\subsection{Notations}
\label{subsec:Pre_Notations}
\vspace{-0.10cm}
We use upper-case letters to denote random variables (RVs), e.g., $X$, and calligraphic letters to denote sets, e.g., $\mathcal{X}$.
We denote column vectors with boldface letters, e.g., ${\bf{X}}$; the $k$-th element of a vector ${\bf{X}}$ ($k \geq 0$) is denoted with $({\bf{X}})_k$. 
Matrices are denoted with Sans-Sarif fonts, e.g., $\mathsf{M}$; the  element at the $k$-th row and the $l$-th column of a matrix $\mathsf{M}$ is denoted by $(\mathsf{M})_{k,l}$.
We use $\mathsf{I}_a$ to denote the $a \times a$ identity matrix,  and $\mathsf{0}_{a\times b}$ to denote the all-zero $a \times b$ matrix.
Hermitian transpose, transpose, trace, complex conjugate, and stochastic expectation are denoted by $(\cdot)^H$, $(\cdot)^T$, ${\rm Tr}(\cdot)$, $(\cdot)^*$, and $\E\{ \cdot \}$, respectively. 
We use $I(X;Y)$ to denote the mutual information between the RVs $X \in \mathcal{X}$ and $Y \in \mathcal{Y}$, $H(X)$ to denote the entropy of a discrete RV $X$, $h(X)$ to denote the differential entropy of a continuous RV $X$, and $p(X)$ to denote the probability density function (PDF) of a continuous RV $X$. 
The symbol $\stackrel{d}{=}$ denotes equality in distribution, and we use $j$  to denote $\sqrt{-1}$; All logarithms are taken to base $2$.
The sets of integers, non-negative integers, real numbers, and complex numbers are denoted by $\mathds{Z}$, $\mathds{N}$, $\mathds{R}$, and $\mathds{C}$, respectively.
We use $((a))_b$ to denote ``$a$ modulo $b$", i.e., writing $c = ((a))_b$ implies that $c$ is satisfies the relationship $a = k \cdot b + c$, where $k \in \mathds{Z}$ and $0 \leq c < b$.
We use  $a^+$ to denote $\max\left\{0,a\right\}$, and $\left|\cdot\right|$ to denote the magnitude when applied to scalars, and the determinant operator when applied to matrices. For 
$x\in \dsR$, $\lfloor x \rfloor$ denotes the largest integer not greater than $x$.
For any sequence, possibly multivariate, ${\bf q}[i]$, $i \in \mathds{Z}$, and for any pair of integers, $a_1$, $a_2$, satisfying
$a_1 < a_2$, we use ${\bf q}_{a_1}^{a_2}$ to denote the column vector obtained by stacking $\left[{\bf q}[a_1]^T, {\bf q}[a_1+1]^T\ldots,{\bf q}[a_2]^T\right]^T$ and define ${\bf q}^{a_2} \equiv {\bf q}_{0}^{a_2}$.
Lastly, we define the discrete Fourier transform (DFT) of a real multivariate sequence  as follows:
For some $n_q \in \mathds{N}$, let  $\left\{{\bf \hat{q}}[k]\right\}_{k=0}^{n-1}$ denote the $n$-point DFT of the multivariate sequence $\left\{{\bf q}[i]\right\}_{i=0}^{n-1}$, ${\bf q}[i] \in \mathds{R}^{n_q}$. The sequence $\left\{{\bf \hat{q}}[k]\right\}_{k=0}^{n-1}$ is computed via
\begin{equation}
\label{eqn:MultiDFT}
{\bf \hat{q}}[k] = \sum\limits_{i=0}^{n-1}{\bf q}[i]e^{-j2\pi \frac{ik}{n}},
\end{equation}
$k \in \{0,1,\ldots,n-1\} \triangleq \mathcal{N}$.
\vspace{-0.2cm}

%
\subsection{Channel Model}
\label{subsec:Pre_Model}
We consider the $n_t \times n_r \times n_e$ MIMO WTC with finite memory. 
Let $m$ be a non-negative integer which denotes the {\em length of the memory of the channel}, and
let ${\bf{W}}[i] \in \mathds{R}^{n_r}$ and ${\bf{U}}[i] \in \mathds{R}^{n_e}$ be two multivariate, zero-mean  stationary real Gaussian processes with autocorrelation functions $\Cw{\tau} \triangleq \E\left\{ {{\bf W} [i + \tau ]{{\left( {{\bf W} [i]} \right)}^T}} \right\}$ and $\Cu{\tau} \triangleq \E\left\{ {{\bf U} [i + \tau ]{{\left( {{\bf U} [i]} \right)}^T}} \right\}$, respectively.
We assume that ${\bf{W}}\left[i_1\right]$ and  ${\bf{U}}\left[i_2\right]$ are uncorrelated $\forall i_1, i_2 \in \mathds{Z}$, and that $\Cw{\tau} = \mathsf{0}_{n_r \times n_r}$ and $\Cu{\tau} = \mathsf{0}_{n_e \times n_e}$ for all $|\tau| > m$. 
We further assume that none of the samples of ${\bf{W}}[i]$ and  ${\bf{U}}[i]$ are deterministically dependent, i.e., there is no index $i_0$ for which either ${\bf{W}}[i_0]$ or  ${\bf{U}}[i_0]$ can be expressed as a linear combination of $\left\{{\bf{W}}[i]\right\}_{i \neq i_0}$ and $\left\{{\bf{U}}[i]\right\}_{i \neq i_0}$, respectively.
Let $\left\{ {\mathsf{H} [\tau ]} \right\}_{\tau  = 0}^{m}$ denote the real $n_r \times n_t$ Tx--Rx channel transfer matrices and $\left\{ {\mathsf{G} [\tau ]} \right\}_{\tau  = 0}^{m}$ denote the real $n_e \times n_t$ Tx--Ev channel transfer matrices. 
The channel transfer matrices, $\left\{ {\mathsf{H} [\tau ]} \right\}_{\tau  = 0}^{m}$ and $\left\{ {\mathsf{G} [\tau ]} \right\}_{\tau  = 0}^{m}$, and the autocorrelation functions of the noises, $\Cw{\tau}$ and $\Cw{\tau}$, $\tau \in \mathds{Z}$, are assumed to be a-priori known at the transmitter. We refer to this assumption as Tx-CSI.
The input-output relationships for the linear time-invariant (LTI) Gaussian MIMO WTC (LGMWTC) are given by
\begin{subequations}
	\label{eqn:RxModel_2}
	\begin{equation}
	{\bf{Y}}[i] = \sum\limits_{\tau = 0}^{m} {{{\mathsf{H}}}[\tau]{{\bf{X}}}[i - \tau]}  + {\bf{W}}[i] \label{eqn:RxModel_2a}
	\end{equation}
	\begin{equation}
	{\bf{Z}}[i] = \sum\limits_{\tau = 0}^{m} {{{\mathsf{G}}}[\tau]{{\bf{X}}}[i - \tau]}  + {\bf{U}}[i], \label{eqn:RxModel_2b}
	\end{equation}
\end{subequations}
$i \in \{0,1,\ldots,l - 1\}$, $l \in \dsN$,
where the channel inputs are subject to a per-MIMO symbol power constraint (hereafter referred to as per-symbol power constraint for brevity)
\begin{equation}
\label{eqn:Constraint1}
\E\left\{ {\left\| {{\bf X} \left[ i \right]} \right\|}^2  \right\} \le P, 
\end{equation}
$i \in \{0,1,\ldots,l - 1\}$. 
We note that restricting the power of the information symbols  at all time instants, rather than over the entire codeword, is very common in the design of practical communications systems, since the dynamic range of practical power amplifiers is limited \cite[Ch. 09]{Cripps:09}, rendering it impossible for transmitters  to "store" power for later channel uses.
This constraint is therefore a natural model for energy-constrained channels \cite[Sec. I-A]{Massey:88}. It should also be noted that similar constraints were used in related works, e.g., the derivation of the secrecy capacity of memoryless MIMO channels in \cite{Hassibi:11}, as well as in the derivation  of some major information theoretic results  including \cite{Massey:88}, \cite[Section VII]{Kramer:05}, and \cite{Weingerten:06}.

In this work we characterize the secrecy capacity of the LGMWTC.

\subsection{Definitions}
\label{subsec:Pre_Definitions}
The framework used in this study is based on the following definitions:

\begin{definition}
\label{def:WTC}
A MIMO WTC with memory, in which the transmitter has $n_t$ antennas, the intended receiver has $n_r$ antennas, and the eavesdropper has $n_e$ antennas, abbreviated as the $n_t \times n_r \times n_e$ MIMO WTC, consists of an input stream ${\bf{X}}[i] \in \mathds{R}^{n_t}$, two output streams  ${\bf{Y}}[i] \in \mathds{R}^{n_r}$ and ${\bf{Z}}[i] \in \mathds{R}^{n_e}$, observed by the intended receiver and by the eavesdropper, respectively, $i \in \mathds{N}$, an initial state ${\bf S}_0 \in \mathcal{S}_0$, and a sequence of transition probabilities $\left\{p\left({\bf{Y}}^{l-1},{\bf{Z}}^{l-1}|{\bf{X}}^{l-1},{\bf S}_0\right)\right\}_{l=1}^{\infty}$.
\end{definition}
In this work we focus on the LGMWTC, which is an instance of the general class of MIMO WTCs with memory defined above. From Def. \ref{def:WTC} it follows that the initial state of the LGMWTC is given by 
\begin{equation*}
{{\bf S} _0} = \left[ \left({\bf X} _{ - m}^{-1}\right)^T,\left({\bf W} _{ - m}^{-1}\right)^T,\left({\bf U} _{ - m}^{-1}\right)^T \right]^T.
\end{equation*}
Note that complex MIMO WTCs with memory can be accommodated by the setup of Def. \ref{def:WTC} by representing all complex vectors using real vectors having twice the number of elements and, representing the complex channel matrices using real matrices having four times the number of elements, corresponding to the real parts and the imaginary parts of the entries, see, e.g., \cite[Sec. I]{Weingerten:06}.

\begin{definition}
\label{def:Code} 
An $\left[R, l \right]$ code with rate $R$ and blocklength $l \in \mathds{N}$ for the WTC consists of: {\em (1)} A source of local randomness at the encoder represented by the RV $D \in \mathcal{D}$ with PDF $p(D)$. {\em (2)} An encoder $e_l$ which maps a message $M$, uniformly distributed over $\mathcal{M} \triangleq \{0,1,\ldots,2^{lR}-1\}$, and a realization of $D$ into a codeword ${\bf{X}}^{l-1}\in \mathcal{X}^l$,
i.e.,
\begin{equation*}
e_l:\mathcal{M} \times \mathcal{D} \mapsto \mathcal{X}^l.
\end{equation*}
{\em (3)}  A decoder $d_l$ which maps the channel output  ${\bf{Y}}^{l-1} \in \mathcal{Y}^l$ into a message $\hat{M} \in \mathcal{M}$. 
i.e.,
\begin{equation*}
d_l: \mathcal{Y}^l \mapsto \mathcal{M}.
\end{equation*}
The source of local randomness $D$ facilitates the random nature of the encoder, and it is emphasized that the realization of $D$ is known only to the encoder.
\end{definition}
Note that we follow the standard setup for channels with memory and let the encoder and decoder operate using only the $l$ symbols corresponding to the currently transmitted codeword \cite{Massey:88, Verdu:89, Goldsmith:01}, \cite[Ch. 5.9]{Gallager:68}, \cite{Dabora:10}. 
The encoder is assumed to be independent of the initial state ${\bf S}_0$. 
%
%
\begin{definition}
\label{def:AvgError} 
The average probability of error of an $[R,l]$ code, when the initial state is ${\bf s}_0$, is defined as:
\vspace{-0.15cm}
\begin{equation*}
\label{eqn:def_avgError}
P_e^l \left( {{\bf s} }_0 \right) = \frac{1}{2^{lR}}\sum\limits_{\tm = 0}^{2^{lR}  - 1} \Pr \left( {\left. {{d_l }\left( {{{{\bf Y} }^{l-1} }} \right) \ne \tm} \right|M = \tm, {{{\bf S} }_0} = {{{\bf s} }_0}} \right).
\end{equation*}
\end{definition}
%
\vspace{-0.25cm}
\begin{definition}
\label{def:SecrecyRate} 
A secrecy rate $R_s$ is achievable for a WTC if for every positive triplet $\epsilon_1,\epsilon_2,\epsilon_3 > 0$, $\exists l_0 >0$ such that $\forall l > l_0$ there exists an $\left[R, l \right]$ code which satisfies:
\begin{subequations}
\label{eqn:def_Rs}
\begin{equation}
\label{eqn:def_Rs1}
\mathop {\sup }\limits_{{{{\bf s} }_0} \in \mathcal{S}_0} P_e^l \left( {{{{\bf s} }_0}} \right) \le {\epsilon _1},
\end{equation}
\begin{equation}
\label{eqn:def_Rs2}
\mathop {\sup }\limits_{{{{\bf s} }_0} \in \mathcal{S}_0} 
\frac{1}{l}I\left( {\left. {M;{{{\bf Z} }^{l-1} }} \right|{{{\bf S} }_0} = {{{\bf s} }_0}} \right) \le {\epsilon _2},
\end{equation}
and
\begin{equation}
\label{eqn:def_Rs3}
R \geq R_s - \epsilon_3.
\end{equation}
\end{subequations}
\end{definition}
Def. \ref{def:SecrecyRate}  extends the definition of codes for memoryless WTCs stated in \cite[Ch. 3.5]{Barros:11}, \cite[Ch. 22.1]{ElGamal:10} to finite-memory WTCs.
The term $\frac{1}{l }I\left( {\left. {M;{{{\bf Z} }^{l-1} }} \right|{{{\bf S} }_0} = {{{\bf s} }_0}} \right)$ represents the maximum achievable information rate at the eavesdropper, while {\em the eavesdropper knows the initial state}. This achievable rate is referred to as the {\em information leakage rate} \cite[Ch. 3.4]{Barros:11}.
%
\begin{definition}
\label{def:SecCapacity}
The {\em secrecy capacity} is defined as the supremum of all achievable secrecy rates.
\end{definition}

\vspace{-0.2cm}
\begin{definition}
\label{def:MemorylessChannel}
A WTC is said to be memoryless if for every non-negative integer $i$ 
\vspace{-0.2cm}
\begin{equation*}
p\left(\left.{\bf{Y}}[i],{\bf{Z}}[i]\right|{\bf{Y}}^{i-1},{\bf{Z}}^{i-1},{\bf{X}}^{i},{\bf S}_0\right) = p\left(\left.{\bf{Y}}[i],{\bf{Z}}[i]\right|{\bf{X}}[i]\right).
\end{equation*}
\end{definition}
Def. \ref{def:MemorylessChannel} corresponds to the general notion of memoryless channels as in, e.g., \cite[Sec. II-A]{Murin:13}. 
Note that if there is no feedback to the transmitter, it follows that
 $p\left({\bf X}[i]\big|{\bf X}^{i-1},{\bf Y}^{i-1},{\bf Z}^{i-1}, {\bf S}_0\right) = p\left({\bf X}[i]\big|{\bf X}^{i-1} \right)$. Then, Def. \ref{def:MemorylessChannel} implies that for every positive integer $l$, 
\begin{equation}
\label{eqn:MemorylessChannel}
p\left({\bf{Y}}^{l-1},{\bf{Z}}^{l-1}|{\bf{X}}^{l-1},{\bf S}_0\right) = \prod\limits_{i=0}^{l-1}p\left({\bf{Y}}[i],{\bf{Z}}[i]\Big|{\bf{X}}[i]\right), 
\end{equation}
which also coincides with the definition of memoryless WTCs stated in \cite[Ch. 3.5]{Barros:11}. 
We henceforth assume that no feedback is present in any of the channels considered.

\vspace{-0.15cm}
\begin{definition}
\label{def:BlockMemorylessChannel}
A WTC is said to be {$n$-block memoryless} if for every positive integer $b$
\begin{equation*}
p\left({\bf{Y}}^{n\cdot b-1},{\bf{Z}}^{n\cdot b-1}|{\bf{X}}^{n\cdot b-1},{\bf S}_0\right) 
  = \prod\limits_{\tb=1}^{b}p\left({\bf{Y}}_{n\cdot (\tb-1)}^{n\cdot \tb-1},{\bf{Z}}_{n\cdot (\tb-1)}^{n\cdot \tb-1}|{\bf{X}}_{n\cdot (\tb-1)}^{n\cdot \tb-1}\right).
\end{equation*}
\end{definition}
Def. \ref{def:BlockMemorylessChannel} corresponds to the definition of $n$-block memoryless BCs stated in \cite[Eq. (8)]{Goldsmith:01}.
Note that codewords of any length can be transmitted over $n$-block memoryless channels, however, when the length of the codeword is an integer multiple of the channel block memory $n$, then the average probability of error is independent of   the initial state ${\bf S}_0$ \cite[Sec. II]{Goldsmith:01}, and similarly, the information leakage rate is also independent of  ${\bf S}_0$. This follows since the outputs of the channels at the receiver and at the eavesdropper corresponding to the transmitted codeword are independent of the initial channel state, by the definition of the channel.

\section{The Secrecy Capacity of the LGMWTC} 
\label{sec:SecCap}
Our main result is the characterization of the secrecy capacity of the LTI Gaussian MIMO WTC with finite memory,
defined in 
Subsection \ref{subsec:Pre_Model}.
This secrecy capacity is stated in the following theorem:
\begin{theorem}
\label{Thm:MainThm2} 
Consider the LGMWTC defined in \eqref{eqn:RxModel_2} subject to the per-symbol power constraint  \eqref{eqn:Constraint1} and with Tx-CSI. 
 Define  $\CCw{\omega} \mspace{-3mu} \triangleq \mspace{-3mu} \sum\limits_{\tau  =  - m }^{m } \mspace{-3mu} \Cw{\tau}e^{ - j \omega \tau}$,  $\CCu{\omega} \mspace{-3mu} \triangleq \mspace{-3mu} \sum\limits_{\tau  =  - m }^{m} \mspace{-3mu} \Cu{\tau}e^{ - j \omega \tau}$,  $\mathsf {H}'(\omega) \mspace{-3mu} \triangleq \mspace{-3mu} \sum\limits_{\tau  =  0}^{m } \mathsf{H}\left[ \tau  \right]e^{ - j \omega \tau}$, and  $\mathsf {G}'(\omega) \mspace{-3mu} \triangleq \mspace{-3mu} \sum\limits_{\tau  =  0}^{m } \mspace{-3mu} \mathsf{G}\left[ \tau  \right]e^{ - j \omega \tau}$.
Let $\mathcal{C}_P$ denote the set of $n_t \times n_t$ positive semi-definite Hermitian matrix functions
$\CCx{\omega}$, defined over the interval $\omega \in \left[0,\pi\right)$, such that
\begin{subequations}
\label{eqn:thm_main2}
\vspace{-0.15cm}
\begin{equation}
\label{eqn:thm_const}
\frac{1}{{\pi }}\int\limits_{\omega = 0}^\pi  {\rm {Tr}}\big( \CCx{\omega} \big)d\omega  \leq P,
\end{equation}
and define  $\psi(\omega)$ as:
\begin{align}
	\psi(\omega) \mspace{-3mu} \triangleq \mspace{-3mu} \frac{\left| {{\mathsf{I}}_{{n_r}}} \mspace{-4mu} + \mspace{-3mu} \mathsf{H}'(\omega)\CCx{\omega}\big(\mathsf {H}'(\omega)\big)^H \big(\CCw{\omega}\big)^{-1} \right|}{\left| {{\mathsf{I}}_{{n_e}}} \mspace{-4mu} + \mspace{-3mu} \mathsf{G}'(\omega)\CCx{\omega}\big(\mathsf {G}'(\omega)\big)^H \big(\CCu{\omega}\big)^{-1} \right|}. \label{eqn:psiomegaDef}
\end{align}
\vspace{-0.15cm}

\noindent Then, the secrecy capacity of the LGMWTC is given by 
\vspace{-0.15cm}
\begin{align}
 \Cs =  \mathop {\max }\limits_{\CCx{\omega} \in \mathcal{C}_P } \frac{1}{2\pi}  \int\limits_{ \omega = 0 }^\pi \log \psi(\omega) d\omega.
\label{eqn:thm_res}
\end{align}
\end{subequations}
\end{theorem}

\smallskip
 In the proof we use elements from the capacity derivation for the finite-memory MAC \cite{Verdu:89} and BC \cite{Goldsmith:01}, as well as
 novel approach and techniques for analyzing the information leakage rate.

{\bf {\slshape Proof Outline}:} 
First, for $n>2m$, we define the $n$-block memoryless circular Gaussian MIMO wiretap channel ($n$-CGMWTC) as follows: 
Let $\utilde{\bf{W}}[i]$ and $\utilde{\bf{U}}[i]$ be zero mean multivariate Gaussian processes, whose autocorrelation functions, denoted $\CUw{\tau}$ and $\CUu{\tau}$, respectively, are defined by 
\vspace{-0.15cm}
\begin{subequations}
\label{eqn:CorrModel1}
\begin{equation}
\label{eqn:CorrModel1a}
\CUw{\tau} \triangleq \Cw{\tau} + \Cw{\tau + n} + \Cw{\tau - n},
\end{equation}
\begin{equation}
\label{eqn:CorrModel1b}
\CUu{\tau} \triangleq \Cu{\tau} + \Cu{\tau + n} + \Cu{\tau - n},
\end{equation}
\end{subequations}
when the noise samples belong to the same $n$-block. 
Noise samples that belong to different $n$-blocks are independent since the channel is $n$-block memoryless.
The outputs of the $n$-CGMWTC over any given $n$-block, i.e., for $i=0,1,\ldots,n-1$, are defined as 
\vspace{-0.15cm}
\begin{subequations}
\label{eqn:CRxModel_2}
\begin{equation}
\utilde{\bf{Y}}[i] = \sum\limits_{\tau = 0}^{m} {{{\mathsf{H}}}[\tau]{{\bf{X}}}\left[\left(\left(i - \tau\right)\right)_n\right]}  + \utilde{\bf{W}}[i] \label{eqn:CRxModel_3a}
\end{equation}
\begin{equation}
\utilde{\bf{Z}}[i] = \sum\limits_{\tau = 0}^{m} {{{\mathsf{G}}}[\tau]{{\bf{X}}}\left[\left(\left(i - \tau\right)\right)_n\right]}  + \utilde{\bf{U}}[i]. \label{eqn:CRxModel_3b}
\end{equation}
\end{subequations}
The $n$-CGMWTC is subject to the same per-symbol average power constraints as the LGMWTC, stated in \eqref{eqn:Constraint1}.
Note that the definition of the $n$-CGMWTC is a natural extension of the definition of the $n$-block memoryless circular Gaussian channel (without secrecy), defined in \cite[Sec. II]{Goldsmith:01}, to secure communications.

The proof now proceeds in the following steps:	
\begin{itemize}	
	 \item In Subsection \ref{subsec:CapStep1}, we prove that the secrecy capacity  of the LGMWTC can be obtained from the secrecy capacity of the $n$-CGMWTC by taking $n \rightarrow \infty$. 
	Note that while the asymptotic relationship between finite-memory channels and their circular block-memoryless counterparts has been used in the (non-secure) capacity analysis of finite-memory Gaussian channels in, e.g.,  \cite{Massey:88}, \cite{Verdu:89}, and \cite{Goldsmith:01}, to the best of our knowledge, this is the first time this approach is applied in the study of the secrecy capacity, and in such scenarios analyzing the {\em information leakage} presents a substantial challenge, as is evident from the analysis in Appendix \ref{app:Proof1}.
	\item Next, in Subsection \ref{subsec:CapStep2}, we derive a closed-form expression for the secrecy capacity of the $n$-CGMWTC for a finite $n$. 
	\item  Lastly, in Subsection \ref{subsec:CapStep3}, we let $n \rightarrow \infty$ and use the capacity expression derived for the $n$-CGMWTC in Subsection \ref{subsec:CapStep2}, to obtain an explicit optimization problem whose maximal solution is the secrecy capacity of the LGMWTC.
\end{itemize}
	%
	%

\vspace{-0.3cm}
\subsection{Equivalence Between the Secrecy Capacity of the LGMWTC and the Asymptotic Secrecy Capacity of the $n$-CGMWTC}
\label{subsec:CapStep1} 

\vspace{-0.1cm}
We now show that the secrecy capacity of the finite-memory LTI Gaussian MIMO WTC can be obtained as the secrecy capacity of the $n$-CGMWTC, by taking $n \rightarrow \infty$.
Letting $\CCG{n}$ denote the secrecy capacity of the $n$-CGMWTC, the result is summarized in the following proposition:

\begin{proposition}
\label{Pro:MainThm1}
The secrecy capacity of the LGMWTC defined in \eqref{eqn:RxModel_2}, subject to the power constraint \eqref{eqn:Constraint1} can be written as
\vspace{-0.15cm}
\begin{equation}
\label{eqn:Thm_main1}
\Cs  = \mathop {\lim }\limits_{n \to \infty } \CCG{n}.
\end{equation}
\end{proposition}
\vspace{-0.15cm}
\begin{IEEEproof}
We provide here an outline of the proof; The detailed proof is provided in Appendix \ref{app:Proof1}. 
First, recall that the $n$-CGMWTC is defined for $n > 2m$. Next, for $n > 2m$ we define the $n$-block memoryless Gaussian MIMO wiretap channel ($n$-MGMWTC) as follows: The $n$-MGMWTC is obtained from the LGMWTC by considering the last $n-m$ vector channel outputs out of each $n$-block at both the eavesdropper and the receiver, i.e., the outputs of the $n$-MGMWTC are defined as the outputs of the LGMWTC for $0\le\left(\left(i\right)\right)_n \geq m$, while for  $\left(\left(i\right)\right)_n < m$ the outputs of the $n$-MGMWTC are not defined, see, e.g., \cite{Kim:08}.
The $n$-MGMWTC is subject to the power constraint \eqref{eqn:Constraint1} on the channel input, similarly to the LGMWTC.
With this definition, we formulate the secrecy capacity of the $n$-MGMWTC in the form of the result of  Csisz{\'a}r and K\"{o}rner \cite[Eq. (11)]{Csiszar:78}.
Note that since the LGMWTC is transformed into the $n$-MGMWTC by setting the first $m$  vector channel outputs out of each $n$-block of channel outputs to be ``undefined" (see also, e.g., \cite[Appendix A]{Goldsmith:01}), then by construction the codeword transmission starts at the beginning of an $n$-block, i.e., an $n$-block begins at time $i=0$.
Next, we show that the secrecy capacity of the LGMWTC can be obtained as the secrecy capacity of the $n$-MGMWTC by taking $n \rightarrow \infty$. For non-secure communications over BCs, it immediately follows that the capacity of the 
$n$-block memoryless Gaussian BC
is not larger than the capacity of the LTI Gaussian BC, as the $n$-block memoryless Gaussian BC is a special case of the LTI Gaussian BC, obtained by letting the intended receiver discard $m$
channel outputs out of every block of $n$ received channel outputs, as	was already shown in \cite[Appendix A]{Goldsmith:01}.
However, in the secure setup, this no longer holds, as the decoder at the eavesdropper {\em cannot be forced to discard $m$ vector channel outputs out of every block of $n$ received channel outputs}, and we conclude that
such an inequality relationship between the secrecy capacities of the $n$-MGMWTC and of the LGMWTC  can be proved only for the asymptotic case $n \rightarrow \infty$. This presents a fundamental difference from non-secure scenarios as will be elaborated in Comment \ref{cmt:Proof1a} in Appendix \ref{app:Proof1}.
Lastly, we show that in the asymptotic regime of $n \rightarrow \infty$, the $n$-MGMWTC and the $n$-CGMWTC have the same secrecy capacity, from which we conclude that $\Cs $ is the secrecy capacity of the $n$-CGMWTC, in the limit $n \rightarrow \infty$. 
\end{IEEEproof}

\vspace{-0.1cm}
\begin{remark}
\label{rem:Thm1_0}
{\em In the proof of Proposition \ref{Pro:MainThm1} in Appendix \ref{app:Proof1}, the secrecy capacity of the $n$-MGMWTC is obtained from the secrecy capacity of a memoryless Gaussian MIMO channel in which the number of antennas is set to be an integer multiple of $n$. Consequently, the {\em computation} of the secrecy capacity of the $n$-MGMWTC for $n \rightarrow \infty$ becomes prohibitive, and the expression for the secrecy capacity of the $n$-MGMWTC provides only little insight on the characterization of the channel inputs that achieve the secrecy capacity. However, the secrecy capacity of the $n$-CGMWTC  for $n \rightarrow \infty$ can be obtained as a maximization problem with a closed-form objective, as will be shown in the sequel.
 For this reason, the $n$-MGMWTC is only an intermediate step, and in order to obtain a useful characterization of $\Cs$
we consider the secrecy capacity of the $n$-CGMWTC.}
\end{remark}

\vspace{-0.1cm}
\begin{remark}
\label{rem:Thm1_1}
{\em As the secrecy capacity of the $n$-MGMWTC is independent of the initial channel state, we conclude that the secrecy capacity of the finite-memory Gaussian MIMO WTC is also independent of the initial state. 
This is intuitive as the finite-memory property of the channel makes the impact of the initial state vanish when considering very large blocklengths.}
\end{remark}

\vspace{-0.1cm}
\begin{remark}
	\label{rem:Thm1_2}
	{\em The coding scheme that achieves the secrecy capacity of the $n$-MGMWTC does not require a prefix channel. In Lemma \ref{lem:Proof_Thm1c}  it is shown that every achievable rate $R_s$ for the LGMWTC can be approached by applying codes  for the $n$-MGMWTC which approach the same $R_s$ after adding a fixed number of zero symbols at the beginning of each codeword.
		Since in Lemma \ref{lem:Proof_Thm1a} it is shown that the coding scheme that achieves the secrecy capacity for the $n$-MGMWTC  does not require a prefix channel,  it  follows from this code construction that the coding scheme that achieves the secrecy capacity of the LGMWTC does not require a prefix channel. This conclusion simplifies the design of secure coding schemes for such channels. }
\end{remark}

\vspace{-0.45cm}
\subsection{Characterizing the Secrecy Capacity of the $n$-CGMWTC}
\label{subsec:CapStep2}

\vspace{-0.05cm}
Next, we derive $\CCG{n}$ for a fixed and finite $n > 2m$. 
The derivation begins with applying the DFT to each $n$-block of the $n$-CGMWTC. 
Let $\left\{\utilde{\bf{\hat{ W}}}[k]\right\}_{k=0}^{n-1}$ and $\left\{\utilde{\bf{\hat{U}}}[k]\right\}_{k=0}^{n-1}$ be the $n$-point DFTs of  $\left\{\utilde{\bf{{W}}}[i]\right\}_{i=0}^{n-1}$ and $\left\{\utilde{\bf{{U}}}[i]\right\}_{i=0}^{n-1}$, respectively, i.e., 
$\utilde{\bf{\hat{ W}}}[k] \triangleq \sum\limits_{i=0}^{n-1}\utilde{\bf{{W}}}[i]e^{-j2\pi \frac{ik}{n}}$ and 
$\utilde{\bf{\hat{ U}}}[k] \triangleq \sum\limits_{i=0}^{n-1}\utilde{\bf{{U}}}[i]e^{-j2\pi \frac{ik}{n}}$.
Let $\CUHw{k}$ and $\CUHu{k}$ denote the covariance matrices of $\utilde{\bf{\hat{ W}}}[k]$ and $\utilde{\bf{\hat{ U}}}[k]$, respectively. 
Define ${\hat{\mathsf{H}}}[k] \triangleq \sum\limits_{\tau=0}^{m}{\mathsf{H}}[\tau]e^{-j2\pi \frac{\tau k}{n}}$ and ${\hat{\mathsf{G}}}[k] \triangleq \sum\limits_{\tau=0}^{m}{\mathsf{G}}[\tau]e^{-j2\pi \frac{\tau k}{n}}$. 
The secrecy capacity of the $n$-CGMWTC for a fixed and finite $n$ is stated in the following proposition:

\begin{proposition}
\label{pro:Expres_Rn}
Let $\mathcal{\hat C}{}_P^n$ denote the collection of $n$-sets of $n_t \times n_t$ positive semi-definite Hermitian matrices $\left\{\CUHx{k}\right\}_{k=0}^{n-1}$, which satisfy $\CUHx{k} = \left(\CUHx{n-k}\right)^*$  for  $\left\lfloor \frac{n}{2}\right\rfloor < k < n$, and
\begin{subequations}
\label{eqn:pro_main2}
\begin{equation}
\label{eqn:pro_main2a}
\sum\limits_{k=0}^{n-1}{\rm {Tr}}\left( \CUHx{k} \right) \leq n^2 P.
\end{equation}

\noindent Further define $\hat{\psi}[k]$ as:
\begin{align}
	\hat{\psi}[k] \mspace{-3mu} \triangleq \mspace{-3mu} \frac{{\left| {{\mathsf{I}}_{{n_r}}} \!+ {\hat{\mathsf{H}}}[k]\CUHx{k}\left({\hat{\mathsf{H}}}[k]\right)^H \left(\CUHw{k}\right)^{-1}\right|}}{{\left| {{\mathsf{I}}_{{n_e}}}\! + {\hat{\mathsf{G}}}[k]\CUHx{k}\left({\hat{\mathsf{G}}}[k]\right)^H \left(\CUHu{k}\right)^{-1}\right|}}. \label{eqn:psihatkDef}
\end{align}

\noindent The secrecy capacity of the $n$-CGMWTC defined in \eqref{eqn:CRxModel_2}, for a fixed and finite $n$, subject to the per-symbol constraint \eqref{eqn:Constraint1} is
\begin{align}
&\CCG{n} = \mathop {\max }\limits_{\left\{\CUHx{k}\right\}_{k=0}^{n-1} \in \mathcal{\hat C}_P^n} \frac{1}{2n} \sum\limits_{k=0}^{n-1} \log \hat{\psi}[k] .
\label{eqn:pro_main2b}
\end{align}
\end{subequations}
\end{proposition}
\begin{IEEEproof}
A detailed proof is provided in Appendix \ref{app:Proof2}, and in the following we present only the outline of the proof:
By applying the multivariate DFT to the channel outputs of the $n$-CGMWTC, we obtain an equivalent set of $n$ MIMO WTCs in the frequency domain, such that each component WTC has no ISI and has additive Gaussian noise. We then show that the noise components in the equivalent set of $n$ MIMO WTCs at different frequency indexes are mutually independent and that each noise component is a circularly symmetric complex normal random process, i.i.d. across different $n$-blocks. Next, {\em relaxing the power constraint to the per $n$-block power constraint}, it follows that the equivalent set of $n$ parallel MIMO WTCs can be analyzed as a set of independent parallel memoryless MIMO WTCs with additive circularly symmetric complex normal noise, i.i.d. in time (here, we refer to the frequency index as ``time").
The secrecy rate of the component MIMO WTCs {\em for a given power allocation}, has already been established in \cite{Wornell:10} and \cite{Hassibi:11}. In Prop. \ref{pro:Expres_Rn}  we state that the secrecy rate of the equivalent set of $n$ MIMO WTCs subject to a given power allocation for each subchannel can be written as the sum of the secrecy rates of the independent subchannels, divided by the number of subchannels. In order to
 arrive at this expression, we use an obvious extension of \cite[Thm. 1]{Li:10} to the memoryless Gaussian MIMO case.
The secrecy capacity for the equivalent set of $n$ parallel MIMO WTCs is obtained by maximizing over all secrecy rates which satisfy the relaxed sum-power constraint, eventually resulting in \eqref{eqn:pro_main2}.
Lastly, we show that the channel input which achieves the secrecy capacity satisfies the per-symbol average power constraint \eqref{eqn:Constraint1}, hence \eqref{eqn:pro_main2} characterizes the secrecy capacity of the $n$-CGMWTC subject to \eqref{eqn:Constraint1}.
\end{IEEEproof}

\vspace{-0.3cm}
\subsection{The Secrecy Capacity of the LGMWTC}
\label{subsec:CapStep3}

\vspace{-0.05cm}
In the final step, we first derive the  asymptotic expression for $\CCG{n}$ in the limit of $n \rightarrow \infty$. Then, we obtain $\Cs$ as the limit $\mathop {\lim }\limits_{n \to \infty } \CCG{n}$.
 The  asymptotic expression for $\mathop {\lim }\limits_{n \to \infty } \CCG{n}$  is stated in the following Proposition:
%
\begin{proposition}
\label{Pro:MainThm2} 
$\mathop {\lim }\limits_{n \to \infty } \CCG{n}$ converges to the expression in \eqref{eqn:thm_main2}. 
\end{proposition}
\begin{IEEEproof}
\ifextended
The detailed proof is provided in Appendix \ref{app:Proof3}. 
The outline of the proof is as follows:  
We begin with the expression for the secrecy capacity of the $n$-CGMWTC with a fixed $n$, $\CCG{n}$, stated in Proposition \ref{pro:Expres_Rn}; We show
\fi 
 that \eqref{eqn:pro_main2b} can be expressed as an average over $n$ samples of a Riemann integrable even function over the range $[0,2\pi)$. Thus, by definition of Riemann integrability \cite[Ch. 6]{Rudin:76}, it follows that for $n \rightarrow \infty$, \eqref{eqn:pro_main2b} converges to \eqref{eqn:thm_res}, and that the energy constraint in \eqref{eqn:pro_main2a} asymptotically coincides with the energy constraint in \eqref{eqn:thm_const}. 
\end{IEEEproof}

From Proposition \ref{Pro:MainThm1} it follows that ${\Cs} = \mathop {\lim }\limits_{n \to \infty } \CCG{n}$.  Therefore, it follows from Proposition \ref{Pro:MainThm2} that $\Cs$ is given by \eqref{eqn:thm_main2}, which completes the proof of Thm. \ref{Thm:MainThm2}.

\vspace{-0.2cm}
\section{Discussion and Numerical Examples}
\vspace{-0.1cm}
\label{sec:Simulations}
In the following we discuss the insights obtained from the results derived above.  
In Subsection \ref{subsec:condition} we present a necessary and sufficient condition for non-zero secrecy capacity;  
Then, in Subsection \ref{subsec:PLC} we present the application of our result to the characterization of the secrecy capacity of narrowband PLC channels;  
Lastly, in Subsection \ref{subsec:ScalarChannels} we show that the secrecy capacity of the {\em scalar} finite-memory LTI Gaussian WTC can be obtained in closed-form, and provide numerical examples.

\subsection{Necessary and Sufficient Condition for $\Cs > 0$}
\label{subsec:condition}
The secrecy capacity expression \eqref{eqn:thm_res} is the solution to a non-convex optimization problem (see, e.g., \cite{Wornell:10} for the memoryless case), which makes it hard to directly develop a practical interpretation. To assist with the understanding of Thm. \ref{Thm:MainThm2} , we now present a necessary and sufficient condition for non-zero secrecy capacity, which follows from Thm. \ref{Thm:MainThm2}.
\begin{proposition}
\label{Pro:Condition_LGMWTC}
Define $\HCw{\omega} \triangleq \left( \CCw{\omega} \right)^{ - \frac{1}{2}}\mathsf{H}'\left( \omega  \right)$ and  $\GCw{\omega} \triangleq \left( \CCu{\omega} \right)^{ - \frac{1}{2}}\mathsf{G}'\left( \omega  \right)$. 
The secrecy capacity of the LGMWTC is strictly positive if and only if $\exists {\Omega}  \subset \left[ {0 ,\pi } \right)$ with a non-zero Lebesgue measure, such that $\forall \omega \in {\Omega}$ 
\begin{equation}
\label{Pro:SuffCond}
\mathop {\sup }\limits_{{\bf{v}}\left( \omega  \right) \in \mathds{C}^{n_t \times 1}} \frac{{\left\| \HCw{\omega}{\bf{v}}\left( \omega  \right) \right\|}}{{\left\| \GCw{\omega}{\bf{v}}\left( \omega  \right) \right\|}} > 1.
\end{equation}
\end{proposition}
\begin{IEEEproof} The proof is similar to that of \cite[Corollary 2]{Wornell:10}, and is provided in Appendix \ref{app:Proof4}.
\end{IEEEproof}

\smallskip
Note that the vector ${\bf{v}}\left( \omega  \right)$ can be considered as a beamforming vector. Therefore, Proposition  \ref{Pro:Condition_LGMWTC} implies that the secrecy capacity is strictly positive only when there exists a continuous set of frequencies for which the sender can beamform the transmitted signal such that the intended receiver observes a higher SNR than the eavesdropper at each frequency in the set of frequencies.

\subsection{Application: The Secrecy Capacity of Narrowband PLC Channels}
\label{subsec:PLC}
An important application of our result is the characterization of the secrecy capacity of {\em scalar} PLC channels in the frequency range of $3-500$ kHz, referred to as narrowband (NB) PLC. 
NB-PLC plays an important role in the realization of smart power grids \cite{Galli:11}, in which secure communications is a critical issue \cite{Galli:11, Farhangi:10, Gungor:11}. 
Despite the importance of secure NB-PLC, to date there is no characterization of the secrecy capacity for this channel, which accounts for its unique characteristics \cite{Tonello:14}:
The NB-PLC channel is a linear channel with additive noise , in which the channel
impulse response (CIR) is commonly modeled as a real  periodically time-varying signal with a finite memory \cite{Corripio:06}, \cite{Evans:12}, while the additive noise is commonly modeled as a real  cyclostationary Gaussian process with a finite correlation length \cite{Evans:12}, \cite{Katayama:06}. 
In the following we fill the knowledge gap of secure communications rates over NB-PLC channels by characterizing the secrecy capacity of the NB-PLC wiretap channel.

Let $\ScPLCW[i]$ and $\ScPLCU[i]$ be zero-mean scalar additive cyclostationary Gaussian noises (ACGNs), each with a period of $t_{noise}$ and a temporal correlation which has a finite-duration, whose length is $\ScPLCmN$.
Let $\ScPLCmCh$ be a non-negative integer representing the length of the memory of the NB-PLC CIR, and let $\left\{ \ScPLCH[i,\tau] \right\}_{\tau = 0}^{\ScPLCmCh}$ and  $\left\{ \ScPLCG[i,\tau] \right\}_{\tau = 0}^{\ScPLCmCh}$ denote the channel coefficients of the Tx-Rx channel and of the Tx-Ev channel, respectively, both with period\footnote{In NB-PLC systems, $t_{ch}$ is equal to the mains period and $t_{noise}$ is equal to half the mains period \cite{Evans:12}. However, our secrecy capacity result applies also to the more general case in which  the periods of the Tx-Rx CIR and of the Tx-Ev  CIR are not identical, by setting $t_{ch}$ to be the least common multiple of these periods. The same applies to the noises and $t_{noise}$, see further explanations in \cite[Footnote 1]{Shlezinger:15}.}  $t_{ch}$, i.e., $\ScPLCH[i,\tau] = \ScPLCH[i+t_{ch},\tau]$ and $\ScPLCG[i,\tau] = \ScPLCG[i+t_{ch},\tau]$, $\forall i, \in \mathds{Z}$, $\forall \tau \in \{0,1,\ldots,\ScPLCmCh-1\}$, and $\ScPLCH[i,\tau] = \ScPLCG[i,\tau] = 0$,   for all integer $\tau < 0$ or $\tau > \ScPLCmCh$.
Let $\ScPLCm = \max \left\{\ScPLCmCh,\ScPLCmN\right\}$.
We use  $X[i]$ to denote the transmitted scalar signal, and  $\ScPLCY[i]$ and $\ScPLCZ[i]$ to denote the channel outputs at the destination and at the eavesdropper, respectively, all at time $i$.
The input-output relationships of the NB-PLC WTC can be written as 
\begin{subequations}
\label{eqn:RxModel_1}
\begin{equation}
\ScPLCY[i] = \sum\limits_{\tau = 0}^{\ScPLCm} {\ScPLCH[i,\tau]X[i - \tau] + \ScPLCW[i]} \label{eqn:RxModel_1a}
\end{equation}
\begin{equation}
\ScPLCZ[i] = \sum\limits_{\tau = 0}^{\ScPLCm} {\ScPLCG[i,\tau]X[i - \tau] + \ScPLCU[i]} \label{eqn:RxModel_1b}.	
\end{equation}
\end{subequations} 
Set $\ScPLCnt$ to be the least common multiple of $t_{ch}$ and $t_{noise}$ which satisfies $\ScPLCnt > \ScPLCm$. 
We assume that the channel input is subject to an average power constraint
\begin{subequations}
\label{eqn:cor_PLCAvgConst}
\begin{equation}
\label{eqn:cor_PLCAvgConsta}
\frac{1}{l}\sum\limits_{k = 0}^{l - 1} \E\left\{{\big| {X\left[ i \right]} \big|}^2\right\} \leq P,
\end{equation}
for any blocklength $l$, and we assume that for all $i \geq 0$ it holds that
\begin{equation}
\label{eqn:cor_PLCAvgConstb}
\frac{1}{{\ScPLCnt}}\sum\limits_{k = 0}^{ \ScPLCnt - 1} \E\left\{{\big| {X\left[ {i \cdot \ScPLCnt + k} \right]} \big|}^2\right\} \leq P,
\end{equation}
\end{subequations}
i.e., over {\em any block of $\ScPLCnt$ symbols}, starting from the first transmitted symbol, the average power is upper bound by $P$.
Applying the decimated components decomposition \cite[Sec. 17.2]{Giannakis:98} to the cyclostationary processes $\ScPLCW[i]$ and $\ScPLCU[i]$, we define the $\ScPLCnt \times 1$ multivariate processes $\VecPLCW\left[\, \tilde i \,\right]$ and $\VecPLCU\left[\, \tilde i \,\right]$, $\tilde{i} \in \mathds{Z}$, whose elements are given by $\left(\VecPLCW\left[\, \tilde i \,\right]\right)_k = \ScPLCW\left[\,\tilde{i}\cdot \ScPLCnt + k\right]$ and $\left(\VecPLCU\left[\, \tilde i \,\right]\right)_k = \ScPLCU\left[\,\tilde{i}\cdot \ScPLCnt + k\right]$, respectively, $k \in \{0,1,\ldots,\ScPLCnt-1\} \triangleq \ScPLCnSet$. 
From \cite[Sec. 17.2]{Giannakis:98} it follows that   $\VecPLCW\left[\, \tilde i \,\right]$ and $\VecPLCU\left[\, \tilde i \,\right]$  are  each a stationary Gaussian process. Let $\CwPLC{\, \tilde{\tau}\,}$ and $\CuPLC{\, \tilde{\tau}\,}$ denote the autocorrelation function of  $\VecPLCW\left[\, \tilde i \,\right]$ and  the autocorrelation function of  $\VecPLCU\left[\, \tilde i \,\right]$, respectively.
Finally, let  $\VecPLCH\left[\, \tilde \tau \,\right]$ and $\VecPLCG\left[\, \tilde \tau\,\right]$, $\tilde{\tau} \in \{0,1\}$, be $\ScPLCnt \times \ScPLCnt$ matrices \textcolor{black}{whose entries at the $k_1$-th row and the $k_2$-th column are given in \eqref{eqn:PLCChDefs}},  $\forall k_1, k_2 \in \ScPLCnSet$.
\begin{subequations}
	\label{eqn:PLCChDefs}
\begin{align}
{\left( {\VecPLCH[0]} \right)_{{k_1},{k_2}}} & = \begin{cases}
																										\ScPLCH\left[ {{k_1},{k_1} - {k_2}} \right], &\qquad\quad  0\le {k_1} - {k_2} \le \ScPLCm \\
																										0, &\qquad\quad \text{otherwise}
																								 \end{cases}, \\
%
%
%
%
{\left( {\VecPLCH[1]} \right)_{{k_1},{k_2}}} & = \begin{cases}
																										\ScPLCH\left[ {{k_1},\ScPLCnt + {k_1} - {k_2}} \right], & 1 \le \ScPLCnt + {k_1} - {k_2} \le \ScPLCm \\
																										0, & \text{otherwise}
																								 \end{cases}, \\
{\left( {\VecPLCG[0]} \right)_{{k_1},{k_2}}} & = \begin{cases}
																										\ScPLCG\left[ {{k_1},{k_1} - {k_2}} \right], &\qquad\quad 0\le {k_1} - {k_2} \le \ScPLCm \\
																										0, &\qquad \quad \text{otherwise}
																								 \end{cases}, \\
{\left( {\VecPLCG[1]} \right)_{{k_1},{k_2}}} & = \begin{cases}
																										\ScPLCG\left[ {{k_1},\ScPLCnt + {k_1} - {k_2}} \right], & 1 \le \ScPLCnt + {k_1} - {k_2} \le \ScPLCm \\
																										0, & \text{otherwise}
																								 \end{cases}.
%
\end{align}
\end{subequations}
The secrecy capacity of the NB-PLC WTC is stated in the following corollary:

\vspace{-0.25cm}

\begin{corollary}
\label{Cor:NBPLC}
Consider the NB-PLC WTC defined in \eqref{eqn:RxModel_1}, subject to the power constraints \eqref{eqn:cor_PLCAvgConst}.
 Define  $\CCwPLC{\omega} \mspace{-3mu} \triangleq \mspace{-3mu} \sum\limits_{\tilde{\tau}  =  - 1 }^{1 } \mspace{-3mu} \CwPLC{\, \tilde{\tau}\,}e^{ - j \omega \tilde{\tau}}$,  $\CCuPLC{\omega}  \triangleq \sum\limits_{\tilde{\tau}  =  - 1 }^{1} \CuPLC{\, \tilde{\tau}\,}e^{ - j \omega \tilde{\tau}}$,  $\CChPLC{\omega} \triangleq \sum\limits_{\tilde{\tau}  =  0}^{1} \VecPLCH \left[ \, \tilde{\tau} \, \right]e^{ - j \omega \tilde{\tau}}$, and   $\CCgPLC{\omega} \triangleq \sum\limits_{\tilde{\tau}  =  0}^{1} \VecPLCG \left[\, \tilde{\tau} \,\right]e^{ - j \omega \tilde{\tau}}$.

\vspace{-0.25cm}

Let $\mathcal{C}_{P}^{\rm PLC}$ denote the set of $\ScPLCnt  \times \ScPLCnt $ positive semi-definite Hermitian matrix functions
$\CCxPLC{\omega}$ defined over the interval $\omega \in \left[0,\pi\right)$, which satisfy 
\begin{subequations}
\label{eqn:cor_PLC2}
\vspace{-0.15cm}
\begin{equation}
\label{eqn:cor_PLCConst}
\frac{1}{{\pi }}\int\limits_{\omega = 0}^\pi  {\rm {Tr}}\big( \CCxPLC{\omega} \big)d\omega  \leq {P}\cdot {\ScPLCnt},
\end{equation}

\vspace{-0.15cm}
\noindent Then, the secrecy capacity of the NB-PLC WTC is given by 
\begin{align}
&C_{s,{\rm PLC}} =  \frac{1}{\ScPLCnt} \mathop {\max }\limits_{\CCxPLC{\omega} \in \mathcal{C}_{P}^{PLC} } \frac{1}{{2\pi }}
\mspace{-5mu} \int\limits_{ \omega = 0 }^\pi   \mspace{-4mu} \log  \psi_{\rm PLC}(\omega) d\omega.
\label{eqn:cor_PLC}
\end{align}
\end{subequations}
\end{corollary}
\begin{IEEEproof}
The proof follows from the representation of NB-PLC channels as Gaussian MIMO channels with finite memory, see, e.g., \cite[Appendix B]{Shlezinger:15}, and is provided in Appendix \ref{app:Proof5}. 
\end{IEEEproof}

\subsection{Scalar Gaussian WTCs with Finite Memory}
\label{subsec:ScalarChannels}
\begin{figure}
    \centering
		 \scalebox{0.35}{\includegraphics{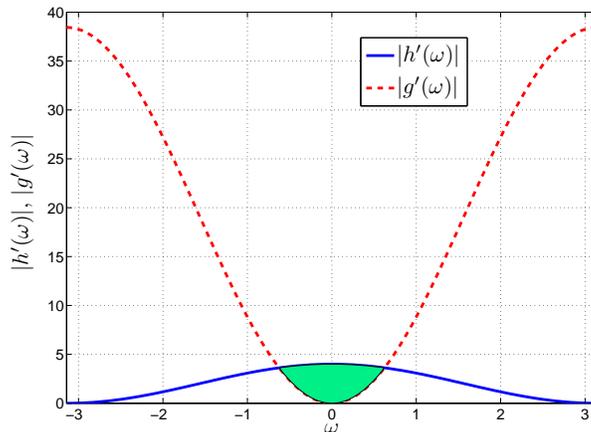}}
    \vspace{-0.8cm}
\caption{The magnitudes of $h'\left(\omega\right)$ and $g'\left(\omega\right)$. The shadowed region corresponds to frequencies in which $|h'\left(\omega\right)|\ge|g'\left(\omega\right)|$.}
\label{fig:NumEx}
\vspace{-0.7cm}
\end{figure}

To analytically evaluate \eqref{eqn:thm_res} it is required to search over all possible input correlation matrix functions in $\mathcal{C}_P$.
However, for the special case of the scalar linear Gaussian WTC (LGWTC), obtained from the general model by setting
$n_t = n_r = n_e = 1$, the secrecy capacity can be obtained explicitly. This result is stated in the following corollary:
\begin{corollary}
\label{Cor:Scalar_LGMWTC}
Consider the scalar LGWTC.
Define the scalar functions $h'\left(\omega\right) \triangleq {\mathsf{H}}'\left(\omega\right)$,  $g'\left(\omega\right) \triangleq {\mathsf{G}}'\left(\omega\right)$,  $c_W'\left(\omega\right) \triangleq \CCw{\omega}$, $c_U'\left(\omega\right) \triangleq \CCu{\omega}$,  $\alpha_r'\left(\omega\right) \triangleq \frac{\left|h'\left(\omega\right)\right|^2}{c_W'\left(\omega\right)}$, and  $\alpha_e'\left(\omega\right) \triangleq \frac{\left|g'\left(\omega\right)\right|^2}{c_U'\left(\omega\right)}$, where the domain for all the functions is $[0, \pi)$. 
The secrecy capacity of the scalar LGWTC is given by
\begin{subequations}
\label{eqn:Cor1_Exam}
\begin{equation}
\label{eqn:Cor1_Exam1}
C_{s,{\rm Scalar}} =  \frac{1}{2\pi}\int\limits_{\omega = 0}^{\pi} \log \left(\frac{ 1 + \alpha_r'\left(\omega\right) c_X'\left(\omega\right)}{ 1 + \alpha_e'\left(\omega\right) c_X'\left(\omega\right)}\right)d\omega,
\end{equation}
where $c_X'\left(\omega\right)$, $\omega \in [0,\pi)$, is obtained as follows: If $\alpha_r'\left(\omega\right) \leq \alpha_e'\left(\omega\right)$ then $c_X'\left(\omega\right) = 0$, otherwise 
\begin{equation}
c_X'\left(\omega\right)= \Bigg(
\sqrt{\left(\frac{\alpha_r'\left(\omega\right)-\alpha_e'\left(\omega\right)}{2\alpha_r'\left(\omega\right)\alpha_e'\left(\omega\right)}\right)^2 + \frac{\alpha_r'\left(\omega\right)-\alpha_e'\left(\omega\right)}{\mu' \cdot  \alpha_r'\left(\omega\right)\alpha_e'\left(\omega\right)}} 
 -\frac{\alpha_r'\left(\omega\right)+\alpha_e'\left(\omega\right)}{2\alpha_r'\left(\omega\right)\alpha_e'\left(\omega\right)}\Bigg)^+,
\label{eqn:Cor1_Exam2}
\end{equation}
\end{subequations}
and $\mu' > 0$ is selected such that $\frac{1}{\pi}\int\limits_{0}^{\pi}c_X'\left(\omega\right)d\omega  = P$.
\end{corollary}
\begin{IEEEproof}
For the scalar $n$-block memoryless circular Gaussian WTC ($n$-CGWTC), 
define the scalar functions $\hat{h}[k] \triangleq \hat{\mathsf{H}}[k]$,  $\hat{g}[k] \triangleq \hat{\mathsf{G}}[k]$,  $c_{\hat{ \utilde{W}}}[k] \triangleq \CUHw{k}$, $c_{\hat{ \utilde{U}}}[k] \triangleq \CUHu{k}$,  $\alpha_r[k] \triangleq \frac{\big|\hat{h}[k]\big|^2}{c_{\hat{ \utilde{W}}}[k]}$, and  $\alpha_e[k] \triangleq \frac{\big|\hat{g}[k]\big|^2}{c_{\hat{\utilde{U}}}[k]}$, $k \in \{0,1,\ldots,n-1\}$.
The secrecy capacity of the scalar $n$-CGWTC is given by the solution of the optimization problem in \eqref{eqn:pro_main2} with the matrices replaced by the corresponding scalar quantities. The resulting expression is \cite[Thm. 1]{Jorswieck:08}, \cite[Thm. 2]{Li:10}:
\begin{subequations}
\label{eqn:pro2_Exam}
\begin{equation}
\label{eqn:pro2_Exam1}
C_{s,{\rm Scalar}}^{n-CG} =  \frac{1}{2n}\sum\limits_{k=0}^{n-1} \log \left(\frac{ 1 + \alpha_r[k] c_{\hat X}[k]}{ 1 + \alpha_e[k] c_{\hat X}[k]}\right),
\end{equation}
where $c_{\hat X}[k]$, $k \in \{0,1,\ldots,n-1\}$, is obtained as follows: If $\alpha_r[k] \leq \alpha_e[k]$ then $c_{\hat X}[k] = 0$, otherwise 
\begin{equation}
c_{\hat X}[k] = \Bigg(
\sqrt{\left(\frac{\alpha_r[k]-\alpha_e[k]}{2\alpha_r[k]\alpha_e[k]}\right)^2 + \frac{\alpha_r[k]-\alpha_e[k]}{\mu \cdot  \alpha_r[k]\alpha_e[k]}}
 -\frac{\alpha_r[k]+\alpha_e[k]}{2\alpha_r[k]\alpha_e[k]}\Bigg)^+,
\label{eqn:pro2_Exam2}
\end{equation}
\end{subequations}
and $\mu > 0$ is selected such that $\sum\limits_{k=0}^{n-1}c_{\hat X}[k]  = n^2 P$.

Now, it follows from Proposition \ref{Pro:MainThm1}  that $C_{s,{\rm Scalar}} \mspace{-4mu} = \mspace{-4mu} \mathop {\lim }\limits_{n \to \infty } C_{s,{\rm Scalar}}^{n-CG}$, thus, the corollary is proved by first showing that in the limit of $n \rightarrow \infty$ the power constraint on $c_{\hat X}[k]$ \eqref{eqn:pro2_Exam2} converges to the power constraint on $c_X'\left(\omega\right)$ in  \eqref{eqn:Cor1_Exam2}, and then showing that \eqref{eqn:pro2_Exam1} can be expressed as an average over $n$ samples of a Riemann integrable even function\footnote{The function is even in the sense that for $1\le k\le\lfloor\frac{n}{2}\rfloor$, then $\xi[k] \triangleq \log \left(\frac{ 1 + \alpha_r[k] c_{\hat X}[k]}{ 1 + \alpha_e[k] c_{\hat X}[k]}\right)$ satisfies $\xi[k] = \xi[n-k]$.} over the range $[0,2\pi)$.
Therefore, from \cite[Ch. 6]{Rudin:76}, it follows for $n \rightarrow \infty$, \eqref{eqn:pro2_Exam1} coincides with \eqref{eqn:Cor1_Exam1}.  As these steps are essentially
the same as the steps in the proof of Proposition \ref{Pro:MainThm2}, they are not repeated here.
\end{IEEEproof}
While Corollary \ref{Cor:Scalar_LGMWTC} applies only to scalar Gaussian WTCs with finite memory, it facilitates a deeper understanding of the main result stated in Thm. \ref{Thm:MainThm2}: Recall that for scalar Gaussian WTCs {\em without memory}, i.e., without ISI and with AWGN, the secrecy capacity is zero if the signal-to-noise ratio (SNR) at the intended receiver (SNR$_r$) is less than or equal to the SNR at the eavesdropper  (SNR$_e$) \cite{Hellman:78}.
In contrast, Corollary \ref{Cor:Scalar_LGMWTC} and Proposition \ref{Pro:Condition_LGMWTC} imply that for scalar Gaussian WTCs {\em with finite memory}, the secrecy capacity is zero if and only if $\alpha_r'\left(\omega\right) \leq \alpha_e'\left(\omega\right)$ for all sets $\Omega \subset [0,\pi)$ of positive Lebesgue measure. This implies that $C_{s,{\rm Scalar}}$ is zero
if and only if the ``SNR density" at the intended receiver, i.e., $\frac{\left|h'\left(\omega\right)\right|^2}{c'_W\left(\omega\right)}$, is less than that at the eavesdropper, i.e., $\frac{\left|g'\left(\omega\right)\right|^2}{c'_U\left(\omega\right)}$, {\em over the entire frequency range}. It thus follows that the finite memory of the channel introduces additional degrees-of-freedom for concealing the information from the eavesdropper.
To demonstrate this, consider the following two-tap channels to the receiver and to the eavesdropper, respectively: $h'\left(\omega\right)=1+e^{-j \omega}$ and $g'\left(\omega\right)=3.1-3.1e^{-j \omega}$. 
Let the noises in both channels be AWGN with unit variance, thus, SNR$_r \approx 3$[dB] while SNR$_e \approx 13$[dB]. 
The magnitude of the frequency response for these two channels is depicted in Fig. \ref{fig:NumEx}.
From the above discussion it follows that the shaded region in which $|h'\left(\omega\right)|\ge|g'\left(\omega\right)|$ facilitates a positive secrecy capacity, and this is achieved
by waterfilling over this region according to \eqref{eqn:Cor1_Exam2}. Therefore, although SNR$_e$ is $10$[dB] higher than SNR$_r$, the secrecy capacity of this channel, derived via \eqref{eqn:Cor1_Exam}, is $0.21$ bits per channel use.

\section{Conclusions}
\label{sec:Conclusions}
In this work we characterized for the first time the secrecy capacity of finite-memory MIMO Gaussian WTCs. The secrecy capacity is derived via the analysis of an equivalent multivariate block-memoryless channel model, and the result is stated as a maximization over the input covariance matrices in the frequency domain. 
Based on the capacity characterization we were able to characterize a necessary and sufficient condition for non-zero secrecy capacity. We also derived the secrecy capacity of narrowband PLC channels, as a  special case of the main result, thereby resolving one of the major open problems for this communications channel.
For the scalar case, we explicitly demonstrated that the frequency selectivity of the channel can be utilized to facilitate secure communications over scenarios in which the SNR at the intended receiver is less than the SNR at the eavesdropper.

\numberwithin{proposition}{section} 
\numberwithin{lemma}{section} 
\numberwithin{corollary}{section} 
\numberwithin{remark}{section} 
\numberwithin{equation}{section}

\begin{appendices}

\section{Proof of Proposition \ref{Pro:MainThm1}}
\label{app:Proof1}
\vspace{-0.1cm}
In order to prove that $\Cs$ is obtained from $\CCG{n}$ by taking $n \rightarrow \infty$, we begin by characterizing the secrecy capacity of the $n$-MGMWTC, which was defined in the proof outline in Subsection \ref{subsec:CapStep1}. This capacity is characterized in Subsection \ref{app:Proof1aa}.
Then, in Subsection \ref{app:Proof1a} we show that $\Cs$ can be obtained from the secrecy capacity of the $n$-MGMWTC by taking $n \rightarrow \infty$. 
Lastly, in Subsection \ref{app:Proof1b} we show that for $n \rightarrow \infty$,  the secrecy capacity of the $n$-MGMWTC is equal to the secrecy capacity of the $n$-CGMWTC. 
Combining these results we obtain \eqref{eqn:Thm_main1}.

Define
\begin{equation}
\label{eqn:Sec_DefQn}
 \CMG{n} \triangleq \frac{1}{n} \mspace{-25mu} \sup_{\substack{p\left({\bf V}^{n-1}, {\bf X} ^{n-1} \right): \\ \MyBeta{n}}} \bigg\{ I\left( {{{{\bf V} }^{n-1}};{\bf Y} _m^{n-1}} \right) 
 - I\left( {{{{\bf V} }^{n-1}};{\bf Z} _m^{n-1}} \right) \bigg\}.
\end{equation}

\subsection{Characterizing the Secrecy Capacity of the $n$-MGMWTC}
\label{app:Proof1aa}
\ifextended
Before obtaining the secrecy capacity of the $n$-MGMWTC, we note that the expression obtained by of  Csisz{\'a}r and K\"{o}rner  \cite[Eq. (11)]{Csiszar:78} can be extended to upper bound the secrecy capacity of $n_t \times n_r \times n_e$  memoryless MIMO channels with AWGN subject to a {per-entry} power constraint.
Specifically, we consider the following MIMO WTC:
\begin{subequations}
\label{eqn:ExtWTC}
\begin{equation}
{\bf{Y}}\left[ i \right] = \mathsf{H}{\bf{X}}\left[ i \right] + {\bf{W}}\left[ i \right]
\label{eqn:ExtWTCa}
\end{equation}
\begin{equation}
{\bf{Z}}\left[ i \right] = \mathsf{G}{\bf{X}}\left[ i \right] + {\bf{U}}\left[ i \right],
\label{eqn:ExtWTCb}
\end{equation}
\end{subequations}
where $n_t = K \cdot n_t'$. 
The noises are AWGNs, hence this WTC is memoryless. 
The channel input whose size is $K\cdot n_t' \times 1$, is treated as a group of $K$ vectors, each of size $n_t' \times 1$, and is subject to an average power constraint $P$. Thus, at time $0 \leq i \leq n-$, the transmitted vector ${\bf X}[i]$ can be written as ${\bf{X}}\left[ i \right] = \left[ {{\bf{X}}_0^T\left[ i \right],{\bf{X}}_1^T\left[ i \right], \ldots ,{\bf{X}}_{K - 1}^T\left[ i \right]} \right]^T$, where each ${\bf X}_k[i]$ is subject to an average power constraint $\E\left\{ {{{\left\| {{{\bf{X}}_k}\left[ i \right]} \right\|}^2}} \right\} \le P$, $0 \leq k \leq K-1$, $0 \leq i \leq n-1$. 
%
\begin{lemma}
\label{lem:Csiszar1}
The secrecy capacity of the WTC \eqref{eqn:ExtWTC} is upper bounded by 
\begin{equation}
\label{eqn:Csiszar1}
{C_s} \le  \mathop {\sup }\limits_{p\left( {\bf{V}}, {\bf{X}} \right):{{\bf{V}} \to {\bf{X}} \to {\bf{Y}},{\bf{Z}}},\left\{ {\E\left\{ {{{\left\| {{{\bf{X}}_k}} \right\|}^2}} \right\} \le P} \right\}_{k=0}^{K-1}} I\left( {{\bf{V}};{\bf{Y}}} \right) - I\left( {{\bf{V}};{\bf{Z}}} \right).
\end{equation}
\end{lemma}
\vspace{-0.2cm}
\begin{proof}[$\mspace{-20mu}$Proof]
Consider $R_s \leq C_s$. As $R_s$ is achievable, then for all ${\epsilon _1},{\epsilon _2},{\epsilon _3} > 0$ and for all sufficiently large $n$ there exists an $\left[R,n\right]$ code which satisfies 
\begin{subequations}
\label{eqn:ProofExtWTC1}
\begin{equation}
\label{eqn:ProofExtWTC1a}
\Pr \left( {M \ne \hat M} \right) \le {\epsilon _1},
\end{equation}
\begin{equation}
\label{eqn:ProofExtWTC1b}
\frac{1}{n}I\left( {M;{{\bf{Z}}^{n - 1}}} \right) \le {\epsilon _2},
\end{equation}
and 
\begin{equation}
\label{eqn:ProofExtWTC1c}
R \ge {R_s} - {\epsilon _3}.
\end{equation}
\end{subequations}
From Fano's inequality it follows that 
\begin{align}
H\left( {M|{{\bf{Y}}^{n - 1}}} \right) 
&\le 1 + \Pr \left( {M \ne \hat M} \right) \cdot nR \notag \\
&\le 1 + {\epsilon _1} \cdot nR.
\label{eqn:ProofExtWTC2}
\end{align}
Therefore, 
\begin{align}
I\left( {M;{{\bf{Y}}^{n - 1}}} \right) - I\left( {M;{{\bf{Z}}^{n - 1}}} \right) 
&= H\left( M \right) - H\left( {M|{{\bf{Y}}^{n - 1}}} \right) - I\left( {M;{{\bf{Z}}^{n - 1}}} \right) \notag \\
&\ge nR - 1 - {\epsilon _1} \cdot nR - {\epsilon _2} \cdot n \notag \\
&= \left( {1 - {\epsilon _1}} \right)nR - 1 - {\epsilon _2} \cdot n \notag \\
&\ge \left( {1 - {\epsilon _1}} \right)\left( {{R_s} - {\epsilon _3}} \right)n - 1 - {\epsilon _2} \cdot n.
\label{eqn:ProofExtWTC3}
\end{align}
It follows from \eqref{eqn:ProofExtWTC3} that 
\begin{equation}
\label{eqn:ProofExtWTC4}
\left( {1 - {\epsilon _1}} \right)\left( {{R_s} - {\epsilon _3}} \right) - \frac{1}{n} - {\epsilon _2} \le \frac{1}{n}\left( {I\left( {M;{{\bf{Y}}^{n - 1}}} \right) - I\left( {M;{{\bf{Z}}^{n - 1}}} \right)} \right).
\end{equation}
From the chain rule for mutual information it follows that 
\begin{equation}
\label{eqn:ProofExtWTC5}
I\left( {M;{{\bf{Y}}^{n - 1}}} \right) - I\left( {M;{{\bf{Z}}^{n - 1}}} \right) = \sum\limits_{i = 0}^{n - 1} {I\left( {\left. {M;{\bf{Y}}\left[ i \right]} \right|{{\bf{Y}}^{i - 1}}} \right)}  - \sum\limits_{i = 0}^{n - 1} I\left( {\left. {M;{\bf{Z}}\left[ i \right]} \right|{\bf{Z}}_{i + 1}^{n - 1}} \right).
\end{equation}
Now, 
\begin{align*}
I\left( {\left. {M;{\bf{Y}}\left[ i \right]} \right|{{\bf{Y}}^{i - 1}}} \right) 
&= h\left( {\left. {{\bf{Y}}\left[ i \right]} \right|{{\bf{Y}}^{i - 1}}} \right) - h\left( {\left. {{\bf{Y}}\left[ i \right]} \right|M,{{\bf{Y}}^{i - 1}}} \right) \notag \\
&\stackrel{(a)}{=} h\left( {\left. {{\bf{Y}}\left[ i \right]} \right|{{\bf{Y}}^{i - 1}}} \right) - h\left( {\left. {{\bf{Y}}\left[ i \right]} \right|M,{{\bf{Y}}^{i - 1}},{\bf{Z}}_{i + 1}^{n - 1}} \right) \notag \\
&\qquad \quad  + h\left( {\left. {{\bf{Y}}\left[ i \right]} \right|M,{{\bf{Y}}^{i - 1}},{\bf{Z}}_{i + 1}^{n - 1}} \right) - h\left( {\left. {{\bf{Y}}\left[ i \right]} \right|M,{{\bf{Y}}^{i - 1}}} \right) \notag \\ 
&\stackrel{(b)}{=} I\left( {\left. {M,{\bf{Z}}_{i + 1}^{n - 1};{\bf{Y}}\left[ i \right]} \right|{{\bf{Y}}^{i - 1}}} \right) - I\left( {\left. {{\bf{Z}}_{i + 1}^{n - 1};{\bf{Y}}\left[ i \right]} \right|{\bf{V}},{{\bf{Y}}^{i - 1}}} \right) \notag \\
&\stackrel{(c)}{=} I\left( {\left. {{\bf{Z}}_{i + 1}^{n - 1};{\bf{Y}}\left[ i \right]} \right|{{\bf{Y}}^{i - 1}}} \right) + I\left( {\left. {M;{\bf{Y}}\left[ i \right]} \right|{{\bf{Y}}^{i - 1}},{\bf{Z}}_{i + 1}^{n - 1}} \right) - I\left( {\left. {{\bf{Z}}_{i + 1}^{n - 1};{\bf{Y}}\left[ i \right]} \right|{\bf{V}},{{\bf{Y}}^{i - 1}}} \right) \notag \\
&\stackrel{(d)}{=} \sum\limits_{j = i + 1}^{n - 1} {I\left( {\left. {{\bf{Z}}\left[ j \right];{\bf{Y}}\left[ i \right]} \right|{{\bf{Y}}^{i - 1}},{\bf{Z}}_{j + 1}^{n - 1}} \right)}  + I\left( {\left. {M;{\bf{Y}}\left[ i \right]} \right|{{\bf{Y}}^{i - 1}},{\bf{Z}}_{i + 1}^{n - 1}} \right) \notag \\ 
&\qquad  \quad  - \sum\limits_{j = i + 1}^{n - 1} {I\left( {\left. {{\bf{Z}}\left[ j \right];{\bf{Y}}\left[ i \right]} \right|M,{{\bf{Y}}^{i - 1}},{\bf{Z}}_{j + 1}^{n - 1}} \right)},
\end{align*}
where $(a)$ follows by adding $h\left( {\left. {{\bf{Y}}\left[ i \right]} \right|M,{{\bf{Y}}^{i - 1}},{\bf{Z}}_{i + 1}^{n - 1}} \right) - h\left( {\left. {{\bf{Y}}\left[ i \right]} \right|M,{{\bf{Y}}^{i - 1}},{\bf{Z}}_{i + 1}^{n - 1}} \right) = 0$,  $(b)$ follows from the definition of the conditional mutual information,  $(c)$ and $(d)$ follow from the mutual information chain rule. 
Similarly,
\begin{align*}
I\left( {\left. {M;{\bf{Z}}\left[ i \right]} \right|{\bf{Z}}_{i + 1}^{n - 1}} \right)
&= h\left( {\left. {{\bf{Z}}\left[ i \right]} \right|{\bf{Z}}_{i + 1}^{n - 1}} \right) - h\left( {\left. {{\bf{Z}}\left[ i \right]} \right|M,{{\bf{Y}}^{i - 1}},{\bf{Z}}_{i + 1}^{n - 1}} \right) \notag \\
&\stackrel{(a)}{=} h\left( {\left. {{\bf{Y}}\left[ i \right]} \right|{{\bf{Y}}^{i - 1}}} \right) - h\left( {\left. {{\bf{Y}}\left[ i \right]} \right|M,{{\bf{Y}}^{i - 1}},{\bf{Z}}_{i + 1}^{n - 1}} \right) \notag \\
&\qquad  \quad + h\left( {\left. {{\bf{Z}}\left[ i \right]} \right|M,{{\bf{Y}}^{i - 1}},{\bf{Z}}_{i + 1}^{n - 1}} \right) - h\left( {\left. {{\bf{Z}}\left[ i \right]} \right|M,{\bf{Z}}_{i + 1}^{n - 1}} \right) \notag \\ 
&\stackrel{(b)}{=} I\left( {\left. {{\bf{V}},{{\bf{Y}}^{i - 1}};{\bf{Z}}\left[ i \right]} \right|{\bf{Z}}_{i + 1}^{n - 1}} \right) - I\left( {\left. {{{\bf{Y}}^{i - 1}};{\bf{Z}}\left[ i \right]} \right|M,{\bf{Z}}_{i + 1}^{n - 1}} \right) \notag \\
&\stackrel{(c)}{=} I\left( {\left. {{{\bf{Y}}^{i - 1}};{\bf{Z}}\left[ i \right]} \right|{\bf{Z}}_{i + 1}^{n - 1}} \right) + I\left( {\left. {M;{\bf{Z}}\left[ i \right]} \right|{{\bf{Y}}^{i - 1}},{\bf{Z}}_{i + 1}^{n - 1}} \right) - I\left( {\left. {{{\bf{Y}}^{i - 1}};{\bf{Z}}\left[ i \right]} \right|M,{\bf{Z}}_{i + 1}^{n - 1}} \right) \notag \\
&\stackrel{(d)}{=} \sum\limits_{j = 0}^{i - 1} {I\left( {\left. {{\bf{Y}}\left[ j \right];{\bf{Z}}\left[ i \right]} \right|{{\bf{Y}}^{j - 1}},{\bf{Z}}_{i + 1}^{n - 1}} \right)}  + I\left( {\left. {M;{\bf{Z}}\left[ i \right]} \right|{{\bf{Y}}^{i - 1}},{\bf{Z}}_{i + 1}^{n - 1}} \right) \notag \\ 
&\qquad  \quad - \sum\limits_{j = 0}^{i - 1} {I\left( {\left. {{\bf{Y}}\left[ j \right];{\bf{Z}}\left[ i \right]} \right|M,{{\bf{Y}}^{j - 1}},{\bf{Z}}_{i + 1}^{n - 1}} \right)},
\end{align*} 
where $(a)$ follows by adding $h\left( {\left. {{\bf{Z}}\left[ i \right]} \right|M,{{\bf{Y}}^{i - 1}},{\bf{Z}}_{i + 1}^{n - 1}} \right) - h\left( {\left. {{\bf{Z}}\left[ i \right]} \right|M,{{\bf{Y}}^{i - 1}},{\bf{Z}}_{i + 1}^{n - 1}} \right) = 0$,  $(b)$ follows from the definition of the conditional mutual information,  $(c)$ and $(d)$ follow from the mutual information chain rule. 
Note that  
\begin{equation*}
\sum\limits_{i = 0}^{n - 1} {\sum\limits_{j = i + 1}^{n - 1} {I\left( {\left. {{\bf{Z}}\left[ j \right];{\bf{Y}}\left[ i \right]} \right|{{\bf{Y}}^{i - 1}},{\bf{Z}}_{j + 1}^{n - 1}} \right)} }  = \sum\limits_{i = 0}^{n - 1} {\sum\limits_{j = 0}^{i - 1} {I\left( {\left. {{\bf{Y}}\left[ j \right];{\bf{Z}}\left[ i \right]} \right|{{\bf{Y}}^{j - 1}},{\bf{Z}}_{i + 1}^{n - 1}} \right)} },
\end{equation*}
as both represent a sum over all possible values of $I\left( {\left. {{\bf{Y}}\left[ j \right];{\bf{Z}}\left[ i \right]} \right|{{\bf{Y}}^{j - 1}},{\bf{Z}}_{i + 1}^{n - 1}} \right)$ with  $j,i \in \left\{ {0 \ldots n - 1} \right\}$ and $j > i$. Similarly, 
\begin{equation*}
\sum\limits_{i = 0}^{n - 1} {\sum\limits_{j = i + 1}^{n - 1} {I\left( {\left. {{\bf{Z}}\left[ j \right];{\bf{Y}}\left[ i \right]} \right|M,{{\bf{Y}}^{i - 1}},{\bf{Z}}_{j + 1}^{n - 1}} \right)} }  = \sum\limits_{i = 0}^{n - 1} \sum\limits_{j = 0}^{i - 1} {I\left( {\left. {{\bf{Y}}\left[ j \right];{\bf{Z}}\left[ i \right]} \right|M,{{\bf{Y}}^{j - 1}},{\bf{Z}}_{i + 1}^{n - 1}} \right)}.
\end{equation*}
It therefore follows that \eqref{eqn:ProofExtWTC5} can be written as  
\begin{align}
I\left( {M;{{\bf{Y}}^{n - 1}}} \right) - I\left( {M;{{\bf{Z}}^{n - 1}}} \right) 
&= \sum\limits_{i = 0}^{n - 1} {I\left( {\left. {M;{\bf{Y}}\left[ i \right]} \right|{{\bf{Y}}^{i - 1}}} \right)}  - \sum\limits_{i = 0}^{n - 1} {I\left( {\left. {M;{\bf{Z}}\left[ i \right]} \right|{\bf{Z}}_{i + 1}^{n - 1}} \right)} \notag \\
&= \sum\limits_{i = 0}^{n - 1} {\sum\limits_{j = i + 1}^{n - 1} {I\left( {\left. {{\bf{Z}}\left[ j \right];{\bf{Y}}\left[ i \right]} \right|{{\bf{Y}}^{i - 1}},{\bf{Z}}_{j + 1}^{n - 1}} \right)} }  + \sum\limits_{i = 0}^{n - 1} {I\left( {\left. {M;{\bf{Y}}\left[ i \right]} \right|{{\bf{Y}}^{i - 1}},{\bf{Z}}_{i + 1}^{n - 1}} \right)} \notag \\
& \qquad - \sum\limits_{i = 0}^{n - 1} {\sum\limits_{j = i + 1}^{n - 1} {I\left( {\left. {{\bf{Z}}\left[ j \right];{\bf{Y}}\left[ i \right]} \right|M,{{\bf{Y}}^{i - 1}},{\bf{Z}}_{j + 1}^{n - 1}} \right)} } \notag \\
& \qquad - \sum\limits_{i = 0}^{n - 1} {\sum\limits_{j = 0}^{i - 1} {I\left( {\left. {{\bf{Y}}\left[ j \right];{\bf{Z}}\left[ i \right]} \right|{{\bf{Y}}^{j - 1}},{\bf{Z}}_{i + 1}^{n - 1}} \right)} }  - \sum\limits_{i = 0}^{n - 1} {I\left( {\left. {M;{\bf{Z}}\left[ i \right]} \right|{{\bf{Y}}^{i - 1}},{\bf{Z}}_{i + 1}^{n - 1}} \right)} \notag \\
& \qquad  + \sum\limits_{i = 0}^{n - 1} {\sum\limits_{j = 0}^{i - 1} {I\left( {\left. {{\bf{Y}}\left[ j \right];{\bf{Z}}\left[ i \right]} \right|M,{{\bf{Y}}^{j - 1}},{\bf{Z}}_{i + 1}^{n - 1}} \right)} } \notag \\
&= \sum\limits_{i = 0}^{n - 1} {I\left( {\left. {M;{\bf{Y}}\left[ i \right]} \right|{{\bf{Y}}^{i - 1}},{\bf{Z}}_{i + 1}^{n - 1}} \right)}  - \sum\limits_{i = 0}^{n - 1} {I\left( {\left. {M;{\bf{Z}}\left[ i \right]} \right|{{\bf{Y}}^{i - 1}},{\bf{Z}}_{i + 1}^{n - 1}} \right)} \notag \\
&= \sum\limits_{i = 0}^{n - 1} {\left( {I\left( {\left. {M;{\bf{Y}}\left[ i \right]} \right|{{\bf{Y}}^{i - 1}},{\bf{Z}}_{i + 1}^{n - 1}} \right) - I\left( {\left. {M;{\bf{Z}}\left[ i \right]} \right|{{\bf{Y}}^{i - 1}},{\bf{Z}}_{i + 1}^{n - 1}} \right)} \right)}.
\label{eqn:ProofExtWTC6}
\end{align}
Define ${\bf{T}}\left[ i \right] \triangleq \left[ {{{\bf{Y}}^{i - 1}},{\bf{Z}}_{i + 1}^{n - 1}} \right]$ and ${\bf{V}}\left[ i \right] \triangleq \left[ {M,{\bf{T}}\left[ i \right]} \right]$, thus \eqref{eqn:ProofExtWTC6} results in
\begin{align}
I\left( {M;{{\bf{Y}}^{n - 1}}} \right) - I\left( {M;{{\bf{Z}}^{n - 1}}} \right) 
&= \sum\limits_{i = 0}^{n - 1} {\left( {I\left( {\left. {M;{\bf{Y}}\left[ i \right]} \right|{{\bf{Y}}^{i - 1}},{\bf{Z}}_{i + 1}^{n - 1}} \right) - I\left( {\left. {M;{\bf{Z}}\left[ i \right]} \right|{{\bf{Y}}^{i - 1}},{\bf{Z}}_{i + 1}^{n - 1}} \right)} \right)} \notag \\
&= \sum\limits_{i = 0}^{n - 1} {\left( {I\left( {\left. {M;{\bf{Y}}\left[ i \right]} \right|{\bf{T}}\left[ i \right]} \right) - I\left( {\left. {M;{\bf{Z}}\left[ i \right]} \right|{\bf{T}}\left[ i \right]} \right)} \right)}  \notag \\
&=\sum\limits_{i = 0}^{n - 1} {\left( {I\left( {\left. {M,{\bf{T}}\left[ i \right];{\bf{Y}}\left[ i \right]} \right|{\bf{T}}\left[ i \right]} \right) - I\left( {\left. {M,{\bf{T}}\left[ i \right];{\bf{Z}}\left[ i \right]} \right|{\bf{T}}\left[ i \right]} \right)} \right)}  \notag \\
&= \sum\limits_{i = 0}^{n - 1} {\left( {I\left( {\left. {{\bf{V}}\left[ i \right];{\bf{Y}}\left[ i \right]} \right|{\bf{T}}\left[ i \right]} \right) - I\left( {\left. {{\bf{V}}\left[ i \right];{\bf{Z}}\left[ i \right]} \right|{\bf{T}}\left[ i \right]} \right)} \right)}.
\label{eqn:ProofExtWTC7}
\end{align}
Note that 
\begin{align*}
&p\left( {{\bf{T}}\left[ i \right],{\bf{V}}\left[ i \right],{\bf{X}}\left[ i \right],{\bf{Y}}\left[ i \right],{\bf{Z}}\left[ i \right]} \right) \notag \\
&\qquad= p\left( {{\bf{T}}\left[ i \right]} \right)p\left( {\left. {{\bf{V}}\left[ i \right]} \right|{\bf{T}}\left[ i \right]} \right)p\left( {\left. {{\bf{X}}\left[ i \right]} \right|{\bf{V}}\left[ i \right],{\bf{T}}\left[ i \right]} \right)p\left( {\left. {{\bf{Y}}\left[ i \right],{\bf{Z}}\left[ i \right]} \right|{\bf{X}}\left[ i \right],{\bf{V}}\left[ i \right],{\bf{T}}\left[ i \right]} \right)\notag \\
&\qquad\stackrel{(a)}{=}p\left( {{\bf{T}}\left[ i \right]} \right)p\left( {\left. {{\bf{V}}\left[ i \right]} \right|{\bf{T}}\left[ i \right]} \right)p\left( {\left. {{\bf{X}}\left[ i \right]} \right|{\bf{V}}\left[ i \right]} \right)p\left( {\left. {{\bf{Y}}\left[ i \right],{\bf{Z}}\left[ i \right]} \right|{\bf{X}}\left[ i \right],M,{{\bf{Y}}^{i - 1}},{\bf{Z}}_{i + 1}^{n - 1}} \right)\notag \\
&\qquad\stackrel{(b)}{=}p\left( {{\bf{T}}\left[ i \right]} \right)p\left( {\left. {{\bf{V}}\left[ i \right]} \right|{\bf{T}}\left[ i \right]} \right)p\left( {\left. {{\bf{X}}\left[ i \right]} \right|{\bf{V}}\left[ i \right]} \right)p\left( {\left. {{\bf{Y}}\left[ i \right],{\bf{Z}}\left[ i \right]} \right|{\bf{X}}\left[ i \right]} \right),
\end{align*}
where $(a)$ follows from the definition of ${\bf{V}}\left[ i \right]$, and (b) follows  as the channel is memoryless and thus, given the channel input at time $i$ and the statistics of the output at time $i$, depends only on the statistics of the noises at time $i$,  which are mutually independent and memoryless. It thus follows that ${\bf{T}}\left[ i \right] \to {\bf{V}}\left[ i \right] \to {\bf{X}}\left[ i \right] \to {\bf{Y}}\left[ i \right],{\bf{Z}}\left[ i \right]$ form a markov chain. 
Next, plugging \eqref{eqn:ProofExtWTC7} into \eqref{eqn:ProofExtWTC4} yields
\begin{equation}
\label{eqn:ProofExtWTC8}
\left( {1 - {\epsilon _1}} \right)\left( {{R_s} - {\epsilon _3}} \right) - \frac{1}{n} - {\epsilon _2} \le \frac{1}{n}\sum\limits_{i = 0}^{n - 1} \Big( I\left( {\left. {{\bf{V}}\left[ i \right];{\bf{Y}}\left[ i \right]} \right|{\bf{T}}\left[ i \right]} \right) - I\left( {\left. {{\bf{V}}\left[ i \right];{\bf{Z}}\left[ i \right]} \right|{\bf{T}}\left[ i \right]} \right) \Big).
\end{equation}
Define an RV $Q$ uniformly distributed over $0,1,\ldots,n-1$. We can now write 
\begin{equation}
\label{eqn:ProofExtWTC8a}
\frac{1}{n}\sum\limits_{i = 0}^{n - 1} \Big( I\left( {\left. {{\bf{V}}\left[ i \right];{\bf{Y}}\left[ i \right]} \right|{\bf{T}}\left[ i \right]} \right) - I\left( {\left. {{\bf{V}}\left[ i \right];{\bf{Z}}\left[ i \right]} \right|{\bf{T}}\left[ i \right]} \right) \Big)  = I\left( {\left. {{{\bf{V}}[Q]};{{\bf{Y}}[Q]}} \right|{{\bf{T}}[Q]},Q} \right) - I\left( {\left. {{{\bf{V}}[Q]};{{\bf{Z}}[Q]}} \right|{{\bf{T}}[Q]},Q} \right).
\end{equation}
Note that for any given set of chains ${\bf{T}}\left[ i \right] \to {\bf{V}}\left[ i \right] \to {\bf{X}}\left[ i \right] \to {\bf{Y}}\left[ i \right],{\bf{Z}}\left[ i \right]$ the variable ${{\bf{V}}[Q]}$   can be defined such that for different $i$'s, the pair $\left({\bf{T}}\left[ i \right],{\bf{V}}\left[ i \right]\right)$ is mapped into a unique set,  thus we obtain the chain $Q \to {{\bf{T}}[Q]} \to {{\bf{V}}[Q]} \to {{\bf{X}}[Q]} \to {{\bf{Y}}[Q]},{{\bf{Z}}[Q]}$. 
We note that from the definition of the channel, given ${\bf X}[Q] = {\bf x}$ the joint distribution of the output does no depend on $Q$, thus  $p_{{\left. {{{\bf{Y}}[Q]},{{\bf{Z}}[Q]}} \right|{{\bf{X}}[Q]}}}\left( \left. {\bf y}, {\bf z} \right|{\bf x} \right) = p_{{\left. {{{\bf{Y}}},{{\bf{Z}}}} \right|{{\bf{X}}}}}\left( \left. {\bf y}, {\bf z} \right|{\bf x} \right)$, and that $\forall k \in \{0,1,\ldots,K-1\}$ 
\begin{equation}
\label{eqn:ProofExtWTC9}
\E\left\{ {\left\| {\bf{X}}_k[Q] \right\|}^2 \right\} = {\E_Q}\left\{ {\E\left\{ {\left. {{{\left\| {\bf{X}}_k[Q] \right\|}^2}} \right|Q} \right\}} \right\} = \frac{1}{n}\sum\limits_{i = 0}^{n - 1} \E\left\{ {{{\left\| {{{\bf{X}}_k}\left[ i \right]} \right\|}^2}} \right\}  \le P.
\end{equation}
Therefore, \eqref{eqn:ProofExtWTC8a} can be written as
\begin{align}
\frac{1}{n}\sum\limits_{i = 0}^{n - 1} \Big( I\left( {\left. {{\bf{V}}\left[ i \right];{\bf{Y}}\left[ i \right]} \right|{\bf{T}}\left[ i \right]} \right) - I\left( {\left. {{\bf{V}}\left[ i \right];{\bf{Z}}\left[ i \right]} \right|{\bf{T}}\left[ i \right]} \right) \Big) 
&= I\left( {\left. {{{\bf{V}}[Q]};{{\bf{Y}}[Q]}} \right|{{\bf{T}}[Q]}} \right) - I\left( {\left. {{{\bf{V}}[Q]};{{\bf{Z}}[Q]}} \right|{{\bf{T}}[Q]}} \right) \notag \\
&\stackrel{(a)}{=} I\left( {\left. {{\bf{V}}[Q];{{\bf{Y}}}} \right|{\bf{T}}[Q]} \right) - I\left( {\left. {{\bf{V}}[Q];{{\bf{Z}}}} \right|{\bf{T}}[Q]} \right) \notag \\
&\stackrel{(b)}{=} I\left( {\left. {{\bf{V}};{\bf{Y}}} \right|{\bf{T}}} \right) - I\left( {\left. {{\bf{V}};{\bf{Z}}} \right|{\bf{T}}} \right),
\label{eqn:ProofExtWTC10}
\end{align}
where in $(a)$ is due to the joint distribution 
\begin{align*}
p_{ {{\bf{Y}}[Q]},{{\bf{Z}}[Q]}, {{\bf{X}}[Q]},{{\bf{V}}[Q]}, {{\bf{T}}[Q]}, Q }\left(  {\bf y}, {\bf z} ,{\bf x}, {\bf v}, {\bf t}, q \right)
&=p_{{\left. {{{\bf{Y}}[Q]},{{\bf{Z}}[Q]}} \right|{{\bf{X}}[Q]}}}\left( \left. {\bf y}, {\bf z} \right|{\bf x} \right) p_{  {{\bf{X}}[Q]},{{\bf{V}}[Q]}, {{\bf{T}}[Q]}, Q }\left( {\bf x}, {\bf v}, {\bf t}, q \right) \notag \\
&= p_{\left. {\bf{Y}},{\bf{Z}} \right|{\bf{X}}}\left( \left. {\bf y}, {\bf z} \right|{\bf x} \right) p_{  {{\bf{X}}[Q]},{{\bf{V}}[Q]}, {{\bf{T}}[Q]}, Q }\left( {\bf x}, {\bf v}, {\bf t}, q \right), 
\end{align*}
and $(b)$ follows from defining ${\bf{V}} \triangleq {{\bf{V}}[Q]}$,  ${\bf{T}} \triangleq {{\bf{T}}[Q]}$, and  ${\bf{X}} \triangleq {{\bf{X}}[Q]}$ such that ${{\bf{T}}} \to {{\bf{V}}} \to {{\bf{X}}}$ form a Markov chain and  $\E\left\{ {{{\left\| {\bf{X}}_k \right\|}^2}} \right\}  \le P$ for all $0 \leq k \leq K-1$. 
This can be done since each realization of $\left({\bf V}, {\bf T}\right) = \left({\bf v}, {\bf t}\right)$ corresponds to a specific, unique value of $Q$, hence 
$\E\left\{ {{{\left\| {\bf{X}}_k \right\|}^2}} \right\} = \E_{{\bf V}, {\bf T}}\left\{\E\left\{ \left.{\left\| {\bf{X}}_k \right\|}^2\right|{\bf V}, {\bf T} \right\}\right\} = \E_{Q}\left\{\E\left\{ \left.{\left\| {\bf{X}}_k \right\|}^2\right|Q \right\}\right\}$.
Therefore, 
\begin{equation}
\label{eqn:ProofExtWTC11}
\left( {1 - {\epsilon _1}} \right)\left( {{R_s} - {\epsilon _3}} \right) - \frac{1}{n} - {\epsilon _2} \le I\left( {\left. {{\bf{V}};{\bf{Y}}} \right|{\bf{T}}} \right) - I\left( {\left. {{\bf{V}};{\bf{Z}}} \right|{\bf{T}}} \right).
\end{equation}
Next, For a given $p\left( {{\bf{T}},{\bf{V}}} \right)p\left( {\left. {\bf{X}} \right|{\bf{V}}} \right)$ where  ${{\bf{T}}} \to {{\bf{V}}} \to {{\bf{X}}}$ form a Markov chain and  $\E\left\{ {{{\left\| {\bf{X}}_k \right\|}^2}} \right\}  \le P$ for all $0 \leq k \leq K-1$, let $\mathcal{T}$ denote the set of all possible values of ${\bf T}$. We note that $\exists \tilde{\bf t} \in \mathcal{T}$ such that 
\begin{align}
I\left( {\left. {{\bf{V}};{\bf{Y}}} \right|{\bf{T}}} \right) - I\left( {\left. {{\bf{V}};{\bf{Z}}} \right|{\bf{T}}} \right) 
&= \int\limits_{{\bf{t}} \in \mathcal{T}} {I\left( {\left. {{\bf{V}};{\bf{Y}}} \right|{\bf{T}} = {\bf{t}}} \right) - I\left( {\left. {{\bf{V}};{\bf{Z}}} \right|{\bf{T}} = {\bf{t}}} \right){p_{\bf{T}}}\left( {\bf{t}} \right)d{\bf{t}}} \notag \\
&\stackrel{(a)}{\le} I\left( {\left. {{\bf{V}};{\bf{Y}}} \right|{\bf{T}} = {\bf{\tilde t}}} \right) - I\left( {\left. {{\bf{V}};{\bf{Z}}} \right|{\bf{T}} = {\bf{\tilde t}}} \right),
\label{eqn:ProofExtWTC13}
\end{align}
where $(a)$ follows is since if $\forall {\bf{t}}' \in {\mathcal{T}}$, 
\begin{equation*}
I\left( {\left. {{\bf{V}};{\bf{Y}}} \right|{\bf{T}} \! = \! {\bf{t}}'} \right) \! - \! I\left( {\left. {{\bf{V}};{\bf{Z}}} \right|{\bf{T}} \! = \! {\bf{t}}'} \right) < \int\limits_{{\bf{t}} \in \mathcal{T}} {I\left( {\left. {{\bf{V}};{\bf{Y}}} \right|{\bf{T}} \! = \! {\bf{t}}} \right) \! - \! I\left( {\left. {{\bf{V}};{\bf{Z}}} \right|{\bf{T}} \! = \! {\bf{t}}} \right){p_{\bf{T}}}\left( {\bf{t}} \right)d{\bf{t}}},
\end{equation*}
 then necessarily 
\begin{align*}
\int\limits_{{\bf{t}}' \in \mathcal{T}} {I\left( {\left. {{\bf{V}};{\bf{Y}}} \right|{\bf{T}} \! = \! {\bf{t}}'} \right) \! - \! I\left( {\left. {{\bf{V}};{\bf{Z}}} \right|{\bf{T}} \! = \! {\bf{t}}'} \right){p_{\bf{T}}}\left( {{\bf{t}}'} \right)d{\bf{t}}}' 
&< \int\limits_{{\bf{t}}' \in \mathcal{T}} {\left( {\int\limits_{{\bf{t}} \in \mathcal{T}} {I\left( {\left. {{\bf{V}};{\bf{Y}}} \right|{\bf{T}} \! = \! {\bf{t}}} \right) \! - \! I\left( {\left. {{\bf{V}};{\bf{Z}}} \right|{\bf{T}} \! = \! {\bf{t}}} \right){p_{\bf{T}}}\left( {\bf{t}} \right)d{\bf{t}}} } \right){p_{\bf{T}}}\left( {{\bf{t}}'} \right)d{\bf{t}}'} \notag \\
&\! = \! \int\limits_{{\bf{t}} \in \mathcal{T}} {I\left( {\left. {{\bf{V}};{\bf{Y}}} \right|{\bf{T}} \! = \! {\bf{t}}} \right) \! - \! I\left( {\left. {{\bf{V}};{\bf{Z}}} \right|{\bf{T}} \! = \! {\bf{t}}} \right){p_{\bf{T}}}\left( {\bf{t}} \right)d{\bf{t}}},
\end{align*}
 which is a contradiction. 
It follows from \eqref{eqn:ProofExtWTC13} that the conditional mutual information $I\left( {\left. {{\bf{V}};{\bf{Y}}} \right|{\bf{T}}} \right) - I\left( {\left. {{\bf{V}};{\bf{Z}}} \right|{\bf{T}}} \right)$ is upper bounded by maximizing the input distribution at a single value of ${\bf T}$. Since ${\bf{T}} \to {\bf{V}} \to {\bf{X}} \to {\bf{Y}},{\bf{Z}}$ form a Markov chain, it follows that the maximum is characterized by considering only the set of joint distributions $p\left( {{\bf{X}},{\bf{V}}} \right)$ which satisfy $\E\left\{ {{{\left\| {{{\bf{X}}_k}} \right\|}^2}} \right\} \le P$ for all $0 \leq k \leq K-1$.
It therefore follows from \eqref{eqn:ProofExtWTC11} that 
\begin{equation}
\label{eqn:ProofExtWTC11a}
\left( {1 - {\epsilon _1}} \right)\left( {{R_s} - {\epsilon _3}} \right) - \frac{1}{n} - {\epsilon _2} \le \mathop {\sup }\limits_{p\left( {\bf{V}}, {\bf{X}} \right):{{\bf{V}} \to {\bf{X}} \to {\bf{Y}},{\bf{Z}}},\left\{ {\E\left\{ {{{\left\| {{{\bf{X}}_k}} \right\|}^2}} \right\} \le P} \right\}_{k=0}^{K-1}} I\left( {{\bf{V}};{\bf{Y}}} \right) - I\left( {{\bf{V}};{\bf{Z}}} \right).
\end{equation}
Note that \eqref{eqn:ProofExtWTC11a} holds for all ${\epsilon _1},{\epsilon _2},{\epsilon _3} > 0$ and for all sufficiently large $n$. Thus, as the supremum is the least upper bound \cite[Def. 1.8]{Rudin:76}, it follows from taking the supremum left hand side of \eqref{eqn:ProofExtWTC12} over $n, \epsilon_1, \epsilon_2, \epsilon_3$ that
\begin{equation}
\label{eqn:ProofExtWTC12}
R_s \le \mathop {\sup }\limits_{p\left( {\bf{V}}, {\bf{X}} \right):{{\bf{V}} \to {\bf{X}} \to {\bf{Y}},{\bf{Z}}},\left\{ {\E\left\{ {{{\left\| {{{\bf{X}}_k}} \right\|}^2}} \right\} \le P} \right\}_{k=0}^{K-1}} I\left( {{\bf{V}};{\bf{Y}}} \right) - I\left( {{\bf{V}};{\bf{Z}}} \right).
\end{equation}
and therefore, 
\begin{equation*}
C_s \leq \mathop {\sup }\limits_{p\left( {\bf{V}}, {\bf{X}} \right):{{\bf{V}} \to {\bf{X}} \to {\bf{Y}},{\bf{Z}}},\left\{ {\E\left\{ {{{\left\| {{{\bf{X}}_k}} \right\|}^2}} \right\} \le P} \right\}_{k=0}^{K-1}} I\left( {{\bf{V}};{\bf{Y}}} \right) - I\left( {{\bf{V}};{\bf{Z}}} \right).
\end{equation*}
\end{proof}
\fi

%
The secrecy capacity of the $n$-MGMWTC is stated in the following proposition:
\begin{proposition}
\label{pro:Proof_Qn1}
The secrecy capacity of the $n$-MGMWTC defined in Subsection \ref{subsec:CapStep1} is given by $\CMG{n}$.
\end{proposition}
\vspace{-0.2cm}
\begin{IEEEproof} In order to obtain the secrecy capacity of the $n$-MGMWTC defined in Subsection \ref{subsec:CapStep1}, we first show that $\CMG{n}$ is the maximum achievable secrecy rate for the $n$-MGMWTC when considering {\em only codes whose blocklength is an integer multiple of $n$}, i.e, $\left[R, b \cdot n\right]$ codes, where $b \in \mathds{N}$. Then, we show that any secrecy rate achievable for the $n$-MGMWTC can be achieved by considering only codes whose blocklength is an integer multiple of $n$.

Let us consider the $n$-MGMWTC constrained to using only codes whose blocklength is an integer multiple of $n$. In this case, we can represent the channel as an equivalent $n \cdot {n_t} \times \left( {n - m} \right) \cdot {n_r} \times \left( {n - m} \right) \cdot {n_e}$ MIMO WTC (without loss of information), via the following assignments: 
Define the input of the transformed channel at time $i\in \dsN$ by the $n \cdot {n_t} \times 1$ vector ${{\bf X} _{eq}}\left[\, \tilde i \,\right] \triangleq {\bf X} _{\tilde{i} \cdot n}^{\left( \tilde i + 1 \right) \cdot n-1}$, $\tilde i \geq 0$, the output at the intended receiver at time $i\in \dsN$ by the $(n-m) \cdot {n_r} \times 1$ vector ${{\bf Y} _{eq}}\left[\, \tilde i \,\right] \triangleq {\bf Y} _{{\tilde i} \cdot n + m}^{\left( \tilde i + 1 \right) \cdot n-1}$, and the output at the eavesdropper at time $\ti \in \dsN$ by the $(n-m) \cdot {n_e} \times 1$ vector ${{\bf Z} _{eq}}\left[\, \tilde i \,\right] \triangleq {\bf Z} _{{\tilde i }  \cdot n + m}^{\left( \tilde i + 1 \right) \cdot n-1}$.
The transformation is clearly bijective, and thus, the secrecy capacity of the equivalent channel is equal to the secrecy capacity of the original $n$-MGMWTC. 
Since the $n$-MGMWTC is $n$-block memoryless, it follows from Def. \ref{def:MemorylessChannel} that the {\em equivalent transformed MIMO channel obtained above is memoryless}, with the transmitter having  $n$ times more antennas than in the $n$-MGMWTC, and both  the intended receiver and  the eavesdropper having $(n-m)$ times more antennas than in the $n$-MGMWTC.
The signals received at the intended receiver and at the eavesdropper are corrupted by the additive noise vectors ${{\bf W} _{eq}}\left[\, \tilde i \,\right] \triangleq {\bf W} _{{\tilde i } \cdot n + m}^{\left( \tilde i + 1 \right) \cdot n-1}$ and ${{\bf U} _{eq}}\left[\, \tilde i \,\right] \triangleq {\bf U} _{{\tilde i } \cdot n + m}^{\left( \tilde i + 1 \right) \cdot n-1}$, respectively. From the noise characterization in Subsection \ref{subsec:Pre_Model} and the definition of the $n$-MGMWTC, it follows that both ${{\bf W} _{eq}}\left[\, \tilde i \,\right]$ and  ${{\bf U} _{eq}}\left[\, \tilde i \,\right]$ are zero-mean Gaussian with positive-definite covariance matrices (since the elements of the random vectors are not linearly dependent, see \cite[Ch. 8.1]{Papoulis:91}), and each process  ${{\bf W} _{eq}}\left[\, \tilde i \,\right]$ and  ${{\bf U} _{eq}}\left[\, \tilde i \,\right]$ is i.i.d. in time (here we refer to the index $\tilde{i}$ as ''time").  
The secrecy capacity of the transformed channel, denoted $Q_n^{eq}$, can be expressed in the form of the result of  Csisz{\'a}r and K\"{o}rner  \cite[Eq. (11)]{Csiszar:78}\footnote{While \cite{Csiszar:78} considered discrete alphabets, it is noted that the result can be extended to incorporate continuous-valued power-constrained inputs as considered in this paper, see \cite[Sec. VI]{Csiszar:78}, \cite[Ch. 5.1]{Barros:11}, \cite[Sec. IV.A]{Wornell:10}, and \cite[Sec. I]{Shamai:09}.}
\begin{align}
Q_n^{eq} &=\!\! \mathop {\sup }\limits_{p\left( {\bf V}_{eq}, {\bf X}_{eq} \right)}\! \bigg\{ {I\left( {{{{\bf V} }_{eq}};{{{\bf Y} }_{eq}}} \right) - I\left( {{{{\bf V} }_{eq}};{{{\bf Z} }_{eq}}} \right)} \bigg\}\notag \\
&\stackrel{(a)}{=} \sup_{\substack{p\left({\bf V}^{n-1}, {\bf X}^{n-1} \right): \\ \MyBeta{n} }} \mspace{-25mu} \bigg\{ I\left( {{{{\bf V} }^{n-1}};{\bf Y} _m^{n-1}} \right)  - I\left( {{{{\bf V} }^{n-1}};{\bf Z} _m^{n-1}} \right) \bigg\}, \label{eqn:EqSecCap1}
\end{align}

\noindent where $(a)$ follows from the definition of the quantities used in the equivalent channel, and $\MyBeta{n}$ corresponds to the {\em per-symbol} power constraint of the $n$-MGMWTC.
As every channel use in the transformed MIMO channel corresponds to $n$ channel uses in the $n$-MGMWTC, it follows from \eqref{eqn:EqSecCap1} that the maximal achievable secrecy rate of the $n$-MGMWTC in bits per channel use, subject to the restriction that only codes whose blocklength is an integer multiple of $n$ are allowed, is $\frac{1}{n}Q_n^{eq} = \CMG{n}$.

Next, we show that any secrecy rate achievable for the $n$-MGMWTC can be achieved by considering only codes whose blocklength is an integer multiple of $n$: Consider a secrecy rate $R_s$ achievable for the $n$-MGMWTC and fix $\epsilon _1 > 0, \epsilon _2> 0, \epsilon _3> 0$. From Def. \ref{def:SecrecyRate}  it follows that $\exists l _0 > 0$ such that $\forall l > l_0$ there exists an $\left[R, l\right]$ code which satisfies \eqref{eqn:def_Rs1}-\eqref{eqn:def_Rs3}. Thus, by setting $b_0$ as the smallest integer for which $b_0 \cdot n \geq l_0$, it follows that for all integer $b > b_0$ there exists an $\left[R, b \cdot n\right]$ code which satisfies \eqref{eqn:def_Rs1}-\eqref{eqn:def_Rs3}. Therefore, the secrecy rate $R_s$ is also achievable when considering only codes whose blocklength is an integer multiple of $n$. We thus conclude that $\CMG{n}$ is the maximum achievable secrecy rate for the $n$-MGMWTC.
\end{IEEEproof}

\subsection{Proving that $\Cs = \mathop {\lim }\limits_{n \to \infty } \CMG{n}$}
\label{app:Proof1a}
Next, we prove that the secrecy capacity of the LGMWTC, $\Cs$, coincides with $\CMG{n}$ in the limit of $n \rightarrow \infty$.
We begin by defining 
\begin{equation}
{C_n}\left( {{{{\bf s} }_0}} \right)\! \triangleq \! \frac{1}{n} \mspace{-30mu} \sup_{\substack{p\left( {{{{\bf V} }^{n\! - \!1}},{{{\bf X} }^{n\! - \!1}}} \right): \\ \MyBeta{n} }} \mspace{-30mu} \bigg\{ I\left( {\left. {{{{\bf V} }^{n\! - \!1}};{{{\bf Y} }^{n\! - \!1}}} \right|{{{\bf S} }_0} \! = \! {{{\bf s} }_0}} \right)   - I\left( {\left. {{{{\bf V} }^{n\! - \!1}};{{{\bf Z} }^{n\! - \!1}}} \right|{{{\bf S} }_0} \! = \! {{{\bf s} }_0}} \right) \bigg\}.
\label{eqn:Sec_DefCn}
\end{equation}
The outline of the proof is as follows:
\begin{itemize}
\item First, we show in Lemma \ref{lem:Proof_EtaBound} and Lemma \ref{lem:Proof_EtaBound2} that for the LGMWTC \eqref{eqn:RxModel_2}, the mutual information between the channel inputs and any $m$ channel outputs can be upper bounded by a fixed and finite number. 
\item Next, in Lemma \ref{lem:Proof_Thm1a}, we prove that  ${\Cs} \le \mathop {\inf }\limits_{{\bf s}_0 \in \mathcal{S}_0} \left( {\mathop {\lim \inf }\limits_{n \to \infty } {C_n}\left( {{{{\bf s} }_0}} \right)} \right)$. 
\item Then, in Lemma \ref{lem:Proof_Thm1b}, we show that $\mathop {\inf }\limits_{{{{\bf s} }_0} \in \mathcal{S}_0} \left( {\mathop {\lim \inf }\limits_{n \to \infty } {C_n}\left( {{{{\bf s} }_0}} \right)} \right) {\leq} \mathop {\lim \inf }\limits_{n \to \infty } {\CMG{n}}$. 
\item Lastly, in Lemma \ref{lem:Proof_Thm1c}, we prove that $\mathop {\lim \sup }\limits_{n \rightarrow \infty} {\CMG{n}} \le {\Cs}$. 
\end{itemize}
By combining these lemmas, we conclude in Proposition \ref{pro:Proof_Qn2} that the secrecy capacity of the LGMWTC is equal to $\mathop{\lim}\limits_{n \rightarrow \infty}\CMG{n}$ and that the limit exists.

%
%


\begin{lemma}
\label{lem:Proof_EtaBound} 
There exists a finite and fixed $\eta >0$, such that for all positive integers $a, b, n, l$, satisfying $b > l$, $n > 2m$, and $n+m > a \ge m$, it holds that
\begin{subequations}
\label{eqns:ToProve1_Ron}
\begin{equation}
I\bigg( {{\bf{X}}^{b \cdot n\! + \!a\! - \!1}};{\bf{Z}}_a^{a\! + \!m\! - \!1}\Big|{\bf{Z}}_{a\! + \!m}^{n\! + \!a\! - \!1}, {\bf{Z}}_{n\! + \!a \! + \! m}^{2 n\! + \!a\! - \!1}, \ldots ,{\bf{Z}}_{\left( {b \! - \! 1} \right) \cdot n\! + \!a \! + \! m}^{b \cdot n\! + \!a\! - \!1}, {\bf U}_{a\! - \!m}^{a\! - \!1} \bigg)  \le \eta, \label{eqn:ToProve1ab}
\end{equation}
and
\begin{align}
&I\bigg( {{\bf{X}}^{b \cdot n\! + \!a\! - \!1}};{\bf{Z}}_{l \cdot n\! + \!a}^{l \cdot n\! + \!a \! + \! m\! - \!1}\Big|{\bf{Z}}_{a\! + \!m}^{n\! + \!a\! - \!1}, {\bf{Z}}_{ n\! + \!a \! + \! m}^{2 n\! + \!a\! - \!1}, \ldots ,{\bf{Z}}_{\left( {b \! - \! 1} \right) \cdot n\! + \!a \! + \! m}^{b \cdot n\! + \!a\! - \!1}, 
\notag \\
& \qquad \qquad \qquad
{{\bf{Z}}_a^{a\! + \!m\! - \!1}}, {\bf{Z}}_{n\! + \!a}^{n\! + \!a \! + \! m\! - \!1}, \ldots ,{\bf{Z}}_{\left( {l \! - \! 1} \right) \cdot n\! + \!a}^{\left( {l \! - \! 1} \right) \cdot n\! + \!a \! + \! m\! - \!1},{\bf U}_{a\! - \!m}^{a\! - \!1} \bigg)   \le \eta. \label{eqn:ToProve1b}
\end{align}
\end{subequations}
\end{lemma}
\begin{IEEEproof}
\ifextended
We begin by describing the underlying principle of the lemma, after which we present a detailed proof. 
\fi 
From the input-output relationship of the LGMWTC \eqref{eqn:RxModel_2} it follows that any sequence of $k > 0$ consecutive channel outputs corresponding to indexes $i_0, i_0+1, \ldots, i_0 +k -1$, {\em when their subsequent and preceding channel outputs are given}, depends on the channel inputs at indexes $i_0-m,i_0-m+1, \ldots, i_0 +k -1$, due to the finite length of the channel impulse response, and the dependence extends also to the channel inputs at indexes $\{i_0-2m,i_0-2m+1, \ldots, i_0 -m -1\}  \cup \{i_0+k, i_0+k + 1 \ldots, i_0 +k +m -1\}$, due to the temporal span of the noise correlation. The latter follows as, given the corresponding channel outputs, these inputs are statistically dependent on the noise at these indexes.
Therefore, similarly to the derivation in \cite[Eq. (63)-(65)]{Shamai:05}, we obtain that each of the two conditional mutual information expressions in \eqref{eqns:ToProve1_Ron}
is upper-bounded by the mean of a quadratic function of at most $4m$ channel inputs. Since the channel input ${\bf X}[i]$ is subject to a per-symbol power constraint, the lemma follows.
\ifextended

Let $n_q$ be a positive integer. In order to reduce notation clutter, for a sequence ${\bf q}[k] \in \mathds{R}^{n_q}$, $k\in \{0,1,\ldots,n-1\}$, and a function $f:\mathds{R}^{n \cdot n_q} \mapsto \mathds{R}$, we abbreviate the integration $\int_{\mathds{R}^{n_q}} \cdots \int_{\mathds{R}^{n_q}} f\left({\bf q}^{n\!-\!1}\right) d{\bf q}[0]\cdots d{\bf q}[n-1]$ as $\int f\left({\bf q}^{n\!-\!1}\right) d{\bf q}^{n\!-\!1}$. 
Define ${\bf{\check Z}} \triangleq \Big({\bf{Z}}_{a+2m}^{n+a - 1},{\bf{Z}}_{n+a + m}^{2n+a - 1},{\bf{Z}}_{2 n+a + m}^{3 n+a - 1}, \ldots ,{\bf{Z}}_{\left( {b - 1} \right) \cdot n+a + m}^{b \cdot n+a - 1}\Big)$.
We now show that $I\Big( {{\bf{X}}^{b \cdot n \! + \! a \!-\!1}}$ $;{\bf{Z}}_a^{a\!+\! m\! -\! 1}\Big|{\bf{Z}}_{a+m}^{n \! + \! a \!-\!1},{\bf{Z}}_{ n\! + \! a \! +\! m}^{2 n\! + \! a \!-\!1},$ $\ldots,$ ${\bf{Z}}_{\left( {b \!-\! 1} \right) \cdot n\! + \! a \! +\! m}^{b \cdot n\! + \! a \!-\!1}, {\bf U}_{a\! -\! m}^{a\! -\! 1}  \Big) = I\Big( \left.{\bf{X}}^{b \cdot n\! + \! a \! - \! 1};{\bf{Z}}_a^{a+m\! - \! 1}\right|{\bf{Z}}_{a+m}^{a+2m\! - \! 1}\! , {\bf{\check Z}} ,{\bf U}_{a-m}^{a-1}  \Big)$ is upper-bounded by a fixed constant. 
From the definition of the conditional mutual information of continuous random vectors \cite[Eq. 2.4.20]{Gallager:68}, it follows that 
\begin{align} 
&I\! \left( \left.{\bf{X}}^{b \cdot n\! + a - \! 1};{\bf{Z}}_a^{a\! + \! m\! - \! 1}\right|{\bf{Z}}_{a\! +\! m}^{a\! +\! 2m\! - \! 1}\! , {\bf{\check Z}} ,{\bf U}_{a\! - \! m}^{a \! -\! 1} \right) \notag \\
&\quad= \! \int \! {p_{{{\bf{Z}}_a^{a\! +\! 2m\! - \! 1}}\! ,{{\bf{X}}^{b\cdot n\! + \! a \! - \! 1}}\! ,{\bf{\check Z}}, {\bf U}_{a\! - \! m}^{a \! -\! 1}}}\! \left( {{\bf{z}}_a^{a\! +\! 2m\! - \! 1}}\! ,{{\bf{x}}^{b\cdot n\! + \! a \! - \! 1}}\! ,{\bf{\check z}}, {\bf u}_{a\! - \! m}^{a \! -\! 1}  \right) 
\notag \\
&\quad \qquad \times 
\log \frac{{{p_{{{\bf{Z}}_a^{a\! + \! m\! - \! 1}}|{{\bf{X}}^{b\cdot n\! + \! a \! - \! 1}}\! ,{\bf{Z}}_{a\! +\! m}^{a\! +\! 2m\! - \! 1}\! ,{\bf{\check Z}} ,{\bf U}_{a\! - \! m}^{a \! -\! 1} }}\! \left( {{\bf{z}}_a^{a\! + \! m\! - \! 1}}|{{\bf{x}}^{b\cdot n\! + \! a \! - \! 1}}\! ,{\bf{z}}_{a\! +\! m}^{a\! +\! 2m\! - \! 1}\! ,{\bf{\check z}} ,{\bf u}_{a\! - \! m}^{a \! -\! 1}  \right)}}{{{p_{{{\bf{Z}}_a^{a\! + \! m\! - \! 1}}|{\bf{Z}}_{a\! +\! m}^{a\! +\! 2m\! - \! 1}\! ,{\bf{\check Z}} ,{\bf U}_{a\! - \! m}^{a \! -\! 1} }}\! \left( {{{\bf{z}}_a^{a\! + \! m\! - \! 1}}|{\bf{z}}_{a\! +\! m}^{a\! +\! 2m\! - \! 1}\! ,{\bf{\check z}} ,{\bf u}_{a\! - \! m}^{a \! -\! 1} } \right)}}d{{\bf{z}}_a^{a\! +\! 2m\! - \! 1}}d{{\bf{x}}^{b\cdot n\! + \! a \! - \! 1}}d{\bf{\check z}}~d{\bf u}_{a\! - \! m}^{a \! -\! 1}  \notag \\
&\quad=\! \int \! {p_{{{\bf{Z}}_a^{a\! +\! 2m\! - \! 1}}\! ,{{\bf{X}}^{b\cdot n\! + \! a \! - \! 1}}\! ,{\bf{\check Z}}, {\bf U}_{a\! - \! m}^{a \! -\! 1}}}\! \left( {{\bf{z}}_a^{a\! +\! 2m\! - \! 1}}\! ,{{\bf{x}}^{b\cdot n\! + \! a \! - \! 1}}\! ,{\bf{\check z}}, {\bf u}_{a\! - \! m}^{a \! -\! 1}  \right)
\notag \\
&\quad \qquad \times 
\Bigg( \log \frac{p_{{{\bf{Z}}_a^{a\! + \! m\! - \! 1}}|{{\bf{X}}^{b\cdot n\! + \! a \! - \! 1}}\! ,{\bf{Z}}_{a\! +\! m}^{a\! +\! 2m\! - \! 1}\! ,{\bf{\check Z}} ,{\bf U}_{a\! - \! m}^{a \! -\! 1}}\! \left( {{\bf{z}}_a^{a\! + \! m\! - \! 1}}|{{\bf{x}}^{b\cdot n\! + \! a \! - \! 1}}\! ,{\bf{z}}_{a\! +\! m}^{a\! +\! 2m\! - \! 1}\! ,{\bf{\check z}} ,{\bf u}_{a\! - \! m}^{a \! -\! 1} \right)}{p_{{{\bf{U}}_a^{a\! + \! m\! - \! 1}}|{\bf{U}}_{a\! +\! m}^{a\! +\! 2m\! - \! 1}\! ,{\bf U}_{a\! - \! m}^{a \! -\! 1}}\! \left( {{\bf{z}}_a^{a\! + \! m\! - \! 1}}|{\bf{z}}_{a\! +\! m}^{a\! +\! 2m\! - \! 1}\! ,{\bf u}_{a\! - \! m}^{a \! -\! 1} \right)}  \notag \\ 
&\quad \qquad \qquad  \qquad 
- \log \frac{{p_{{{\bf{Z}}_a^{a\! + \! m\! - \! 1}}|{\bf{Z}}_{a\! +\! m}^{a\! +\! 2m\! - \! 1}\! ,{\bf{\check Z}} ,{\bf U}_{a\! - \! m}^{a \! -\! 1}}\! \left( {{{\bf{z}}_a^{a\! + \! m\! - \! 1}}|{\bf{z}}_{a\! +\! m}^{a\! +\! 2m\! - \! 1}\! ,{\bf{\check z}} ,{\bf u}_{a\! - \! m}^{a \! -\! 1}} \right)}}{{{p_{{{\bf{U}}_a^{a\! + \! m\! - \! 1}}|{\bf{U}}_{a\! +\! m}^{a\! +\! 2m\! - \! 1}\! ,{\bf U}_{a\! - \! m}^{a \! -\! 1}}}\! \left( {{{\bf{z}}_a^{a\! + \! m\! - \! 1}}|{\bf{z}}_{a\! +\! m}^{a\! +\! 2m\! - \! 1}\! ,{\bf u}_{a\! - \! m}^{a \! -\! 1}} \right)}} \Bigg) d{{\bf{z}}_a^{a\! +\! 2m\! - \! 1}}d{{\bf{x}}^{b\cdot n\! + \! a \! - \! 1}}d{\bf{\check z}}~d{\bf u}_{a\! - \! m}^{a \! -\! 1} 
\label{eqn:Bound1}.
\end{align}
Note that 
\begin{align}
& \int \! {p_{{{\bf{Z}}_a^{a\! +\! 2m\! - \! 1}}\! ,{{\bf{X}}^{b\cdot n\! + \! a \! - \! 1}}\! ,{\bf{\check Z}}, {\bf U}_{a\! - \! m}^{a \! -\! 1}}}\! \left( {{\bf{z}}_a^{a\! +\! 2m\! - \! 1}}\! ,{{\bf{x}}^{b\cdot n\! + \! a \! - \! 1}}\! ,{\bf{\check z}}, {\bf u}_{a\! - \! m}^{a \! -\! 1}  \right)
\notag \\
&\qquad \times 
\log \frac{p_{{{\bf{Z}}_a^{a\! + \! m\! - \! 1}}|{\bf{Z}}_{a\! +\! m}^{a\! +\! 2m\! - \! 1}\! ,{\bf{\check Z}} ,{\bf U}_{a\! - \! m}^{a \! -\! 1}}\! \left( {{\bf{z}}_a^{a\! + \! m\! - \! 1}}|{\bf{z}}_{a\! +\! m}^{a\! +\! 2m\! - \! 1}\! ,{\bf{\check z}} ,{\bf u}_{a\! - \! m}^{a \! -\! 1} \right)}{p_{{{\bf{U}}_a^{a\! + \! m\! - \! 1}}|{\bf{U}}_{a\! +\! m}^{a\! +\! 2m\! - \! 1}\! ,{\bf U}_{a\! - \! m}^{a \! -\! 1}}\! \left( {{\bf{z}}_a^{a\! + \! m\! - \! 1}}|{\bf{z}}_{a\! +\! m}^{a\! +\! 2m\! - \! 1}\! ,{\bf u}_{a\! - \! m}^{a \! -\! 1} \right)}d{{\bf{z}}_a^{a\! +\! 2m\! - \! 1}}d{{\bf{x}}^{b\cdot n\! + \! a \! - \! 1}}d{\bf{\check z}}~d{\bf u}_{a\! - \! m}^{a \! -\! 1} \notag \\
&\quad =\int \! {p_{{{\bf{Z}}_a^{a\! +\! 2m\! - \! 1}}\! ,{\bf{\check Z}}, {\bf U}_{a\! - \! m}^{a \! -\! 1}}}\! \left( {{\bf{z}}_a^{a\! +\! 2m\! - \! 1}}\! ,{\bf{\check z}}, {\bf u}_{a\! - \! m}^{a \! -\! 1}  \right)
\notag \\
&\quad \qquad \times 
\log \frac{p_{{{\bf{Z}}_a^{a\! + \! m\! - \! 1}}|{\bf{Z}}_{a\! +\! m}^{a\! +\! 2m\! - \! 1}\! ,{\bf{\check Z}} ,{\bf U}_{a\! - \! m}^{a \! -\! 1}}\! \left( {{\bf{z}}_a^{a\! + \! m\! - \! 1}}|{\bf{z}}_{a\! +\! m}^{a\! +\! 2m\! - \! 1}\! ,{\bf{\check z}} ,{\bf u}_{a\! - \! m}^{a \! -\! 1} \right)}{p_{{{\bf{U}}_a^{a\! + \! m\! - \! 1}}|{\bf{U}}_{a\! +\! m}^{a\! +\! 2m\! - \! 1}\! ,{\bf U}_{a\! - \! m}^{a \! -\! 1}}\! \left( {{\bf{z}}_a^{a\! + \! m\! - \! 1}}|{\bf{z}}_{a\! +\! m}^{a\! +\! 2m\! - \! 1}\! ,{\bf u}_{a\! - \! m}^{a \! -\! 1} \right)}d{{\bf{z}}_a^{a\! +\! 2m\! - \! 1}}d{\bf{\check z}}~d{\bf u}_{a\! - \! m}^{a \! -\! 1}. \label{eqn:Bound1a}
\end{align}
Let $D_{KL}(f||g)$ denote the Kullback-Leiber divergence between two densities $f$ and $g$ \cite[Ch. 8.5]{Cover:06}. Note that 
\begin{align*}
&\! \int \! {p_{{{\bf{Z}}_a^{a\! +\! 2m\! - \! 1}}\! ,{\bf{\check Z}}, {\bf U}_{a\! - \! m}^{a \! -\! 1}}}\! \left( {{\bf{z}}_a^{a\! +\! 2m\! - \! 1}}\! ,{\bf{\check z}}, {\bf u}_{a\! - \! m}^{a \! -\! 1}  \right)
\notag \\
&\quad \qquad \times 
\log \frac{p_{{{\bf{Z}}_a^{a\! + \! m\! - \! 1}}|{\bf{Z}}_{a\! +\! m}^{a\! +\! 2m\! - \! 1}\! ,{\bf{\check Z}} ,{\bf U}_{a\! - \! m}^{a \! -\! 1}}\! \left( {{\bf{z}}_a^{a\! + \! m\! - \! 1}}|{\bf{z}}_{a\! +\! m}^{a\! +\! 2m\! - \! 1}\! ,{\bf{\check z}} ,{\bf u}_{a\! - \! m}^{a \! -\! 1} \right)}{p_{{{\bf{U}}_a^{a\! + \! m\! - \! 1}}|{\bf{U}}_{a\! +\! m}^{a\! +\! 2m\! - \! 1}\! ,{\bf U}_{a\! - \! m}^{a \! -\! 1}}\! \left( {{\bf{z}}_a^{a\! + \! m\! - \! 1}}|{\bf{z}}_{a\! +\! m}^{a\! +\! 2m\! - \! 1}\! ,{\bf u}_{a\! - \! m}^{a \! -\! 1} \right)}d{{\bf{z}}_a^{a\! +\! 2m\! - \! 1}}d{\bf{\check z}}~d{\bf u}_{a\! - \! m}^{a \! -\! 1} \\ 
&\quad = \! \int \! \! \left(\! \int \! p_{{{\bf{Z}}_{a}^{a\! + \! m  \! -\! 1}}|{\bf{Z}}_{a\! + \! m}^{a\! + \! 2m  \! -\! 1}\! ,{\bf{\check Z}} ,{\bf U}_{a \! - \! m}^{a \! -\! 1}}\! \left( {{{\bf{z}}_{a}^{a\! + \! m  \! -\! 1}}|{\bf{z}}_{a\! + \! m}^{a\! + \! 2m  \! -\! 1}\! ,{\bf{\check z}} ,{\bf u}_{a \! - \! m}^{a \! -\! 1}} \right)\log \frac{{{p_{{{\bf{Z}}_{a}^{a\! + \! m  \! -\! 1}}|{\bf{Z}}_{a\! + \! m}^{a\! + \! 2m  \! -\! 1}\! ,{\bf{\check Z}} ,{\bf U}_{a \! - \! m}^{a \! -\! 1}}}\! \left( {{{\bf{z}}_{a}^{a\! + \! m  \! -\! 1}}|{\bf{z}}_{a\! + \! m}^{a\! + \! 2m  \! -\! 1}\! ,{\bf{\check z}} ,{\bf u}_{a \! - \! m}^{a \! -\! 1}} \right)}}{{{p_{{{\bf{U}}_{a}^{a\! + \! m  \! -\! 1}}|{\bf{U}}_{a\! + \! m}^{a\! + \! 2m  \! -\! 1}\! ,{\bf U}_{a \! - \! m}^{a \! -\! 1}}}\! \left( {{{\bf{z}}_{a}^{a\! + \! m  \! -\! 1}}|{\bf{z}}_{a\! + \! m}^{a\! + \! 2m  \! -\! 1}\! ,{\bf u}_{a \! - \! m}^{a \! -\! 1}} \right)}}d{{\bf{z}}_a^{a\! + \! m\! - \! 1}}\right) \\
&\quad  \qquad \qquad \times 
 p_{{\bf{Z}}_{a\! + \! m}^{a\! + \! 2m  \! -\! 1}\! ,{\bf{\check Z}}, {\bf U}_{a \! - \! m}^{a \! -\! 1}}\! \left( {\bf{z}}_{a\! + \! m}^{a\! + \! 2m  \! -\! 1}\! ,{\bf{\check z}}, {\bf u}_{a \! - \! m}^{a \! -\! 1} \right)  d{\bf{z}}_{a\! + \! m}^{a\! + \! 2m  \! -\! 1} d{\bf{\check z}} ~d{\bf u}_{a \! - \! m}^{a \! -\! 1}\\
&\quad = \! \int \! D_{KL}\! \left(\left.p_{{{\bf{Z}}_{a}^{a\! + \! m  \! -\! 1}}|{\bf{Z}}_{a\! + \! m}^{a\! + \! 2m  \! -\! 1} = {\bf{z}}_{a\! + \! m}^{a\! + \! 2m  \! -\! 1}\! ,{\bf{\check Z}}={\bf{\check z}} ,{\bf U}_{a \! - \! m}^{a \! -\! 1}={\bf u}_{a \! - \! m}^{a \! -\! 1}}\right\| p_{{{\bf{U}}_{a}^{a\! + \! m  \! -\! 1}}|{\bf{U}}_{a\! + \! m}^{a\! + \! 2m  \! -\! 1} = {\bf{z}}_{a\! + \! m}^{a\! + \! 2m  \! -\! 1}\! ,{\bf U}_{a \! - \! m}^{a \! -\! 1}={\bf u}_{a \! - \! m}^{a \! -\! 1}}\right) 
\notag \\
&\quad  \qquad \qquad \times 
 p_{{\bf{Z}}_{a\! + \! m}^{a\! + \! 2m  \! -\! 1}\! ,{\bf{\check Z}}, {\bf U}_{a \! - \! m}^{a \! -\! 1}}\! \left( {\bf{z}}_{a\! + \! m}^{a\! + \! 2m  \! -\! 1}\! ,{\bf{\check z}}, {\bf u}_{a \! - \! m}^{a \! -\! 1} \right)  d{\bf{z}}_{a\! + \! m}^{a\! + \! 2m  \! -\! 1}d{\bf{\check z}} ~d{\bf u}_{a \! - \! m}^{a \! -\! 1} \notag \\
&\quad\stackrel{(a)}{\ge} 0,
\end{align*}
where $(a)$ follows as $D_{KL}(f||g) \geq 0$ for all densities $f$ and $g$ \cite[Thm. 8.6.1]{Cover:06}. 
It therefore follows that \eqref{eqn:Bound1a} is non-negative, thus \eqref{eqn:Bound1} implies that 
\begin{align}
&I\! \left( \left.{\bf{X}}^{b \cdot n\! + a - \! 1};{\bf{Z}}_a^{a\! + \! m\! - \! 1}\right|{\bf{Z}}_{a\! +\! m}^{a\! +\! 2m\! - \! 1}\! , {\bf{\check Z}} ,{\bf U}_{a\! - \! m}^{a \! -\! 1} \right)  
\leq  
\! \int \! {p_{{{\bf{Z}}_a^{a\! +\! 2m\! - \! 1}}\! ,{{\bf{X}}^{b\cdot n\! + \! a \! - \! 1}}\! ,{\bf{\check Z}}, {\bf U}_{a\! - \! m}^{a \! -\! 1}}}\! \left( {{\bf{z}}_a^{a\! +\! 2m\! - \! 1}}\! ,{{\bf{x}}^{b\cdot n\! + \! a \! - \! 1}}\! ,{\bf{\check z}}, {\bf u}_{a\! - \! m}^{a \! -\! 1}  \right)
\notag \\
& \qquad \times 
\log \frac{p_{{{\bf{Z}}_a^{a\! + \! m\! - \! 1}}|{{\bf{X}}^{b\cdot n\! + \! a \! - \! 1}}\! ,{\bf{Z}}_{a\! +\! m}^{a\! +\! 2m\! - \! 1}\! ,{\bf{\check Z}} ,{\bf U}_{a\! - \! m}^{a \! -\! 1}}\! \left( {{\bf{z}}_a^{a\! + \! m\! - \! 1}}|{{\bf{x}}^{b\cdot n\! + \! a \! - \! 1}}\! ,{\bf{z}}_{a\! +\! m}^{a\! +\! 2m\! - \! 1}\! ,{\bf{\check z}} ,{\bf u}_{a\! - \! m}^{a \! -\! 1} \right)}{p_{{{\bf{U}}_a^{a\! + \! m\! - \! 1}}|{\bf{U}}_{a\! +\! m}^{a\! +\! 2m\! - \! 1}\! ,{\bf U}_{a\! - \! m}^{a \! -\! 1}}\! \left( {{\bf{z}}_a^{a\! + \! m\! - \! 1}}|{\bf{z}}_{a\! +\! m}^{a\! +\! 2m\! - \! 1}\! ,{\bf u}_{a\! - \! m}^{a \! -\! 1} \right)}d{{\bf{z}}_a^{a\! +\! 2m\! - \! 1}}d{{\bf{x}}^{b\cdot n\! + \! a \! - \! 1}}d{\bf{\check z}}~d{\bf u}_{a\! - \! m}^{a \! -\! 1}.
\label{eqn:Bound2}
\end{align}
Define ${\bf{\check U}} \triangleq \left({\bf{U}}_{a+2m}^{a+n - 1},{\bf{U}}_{n +a+ m}^{2n+a - 1}, \ldots ,{\bf{U}}_{\left( {b - 1} \right) \cdot n + a+m}^{b \cdot n + a - 1}\right)$. 
From the input-output relationship of the LGMWTC it follows that $\exists \Gm \in \mathds{R}^{\left(n_e \cdot m\right) \times \left(n_t \cdot 2m\right)}$ such that ${\bf Z}_a^{a+m-1} = \Gm {\bf X}_{a-m}^{a+m-1}  + {\bf U}_a^{a+m-1}$, and $\exists {\mathsf{\check G}} \in \mathds{R}^{\left(n_e \cdot (b \cdot (n\!-\! m)\!-\! m)\right) \times \left(n_t \cdot (b \cdot n + a)\right)}$ such that ${\bf {\check Z}} = {\mathsf{\check G}} {\bf X}^{b \cdot n+a -1} + {\bf {\check U}}$. As the LGMWTC is time-invariant it also follows that ${\bf Z}_{a+m}^{a\!+\! 2m\!-\!1}  = \Gm {\bf X}_a^{a \! + \! 2m\!-\!1}  + {\bf U}_{a+m}^{a\!+\! 2m\!-\!1}$. 
Next, we note that 
\begin{align*}
&{p_{{{\bf{Z}}_a^{a\!+ \! m\! - \! 1}}|{{\bf{X}}^{b \cdot n\! + \! a \! - \! 1}}\! ,{\bf{Z}}_{a\! + \! m}^{a\! + \! 2m\! - \! 1}\! ,{\bf{\check Z}}  ,{\bf{U}}_{a\! -\!  m}^{a\! -\! 1}}}\! \left( {\left. {{{\bf{z}}_a^{a\!+ \! m\! - \! 1}}} \right|{{\bf{x}}^{b \cdot n\! + \! a \! - \! 1}}\! ,{\bf{z}}_{a\! + \! m}^{a\! + \! 2m\! - \! 1}\! ,{\bf{\check z}}  ,{\bf{u}}_{a\! -\!  m}^{a\! -\! 1}} \right) \\
&\quad= {p_{{{\bf{U}}_a^{a\!+ \! m\! - \! 1}}|{{\bf{X}}^{b \cdot n\! + \! a \! - \! 1}}\! ,{\bf{U}}_{a\! + \! m}^{a\! + \! 2m\! - \! 1}\! ,{\bf{\check U}}\! ,{\bf{U}}_{a\! -\!  m}^{a\! -\! 1}}}\! \left( {\left. {{{\bf{z}}_a^{a\!+ \! m\! - \! 1}} \!-\! \Gm{\bf{x}}_{a\! -\!  m}^{a\! + \! m\! - \! 1}} \right|{{\bf{x}}^{b \cdot n\! + \! a \! - \! 1}}\! ,{\bf{z}}_{a\! + \! m}^{a\! + \! 2m\! - \! 1} - \Gm{{\bf{x}}_a^{a\! + \! 2m\! - \! 1}}\! ,{\bf{\check z}} \!-\! \mathsf{\check G}{{\bf{x}}^{b \cdot n\! + \! a \! - \! 1}}\!  ,{\bf{u}}_{a\! -\!  m}^{a\! -\! 1}} \right)\\
&\quad\stackrel{(a)}{=} {p_{{{\bf{U}}_a^{a\!+ \! m\! - \! 1}}|{\bf{U}}_{a\! + \! m}^{a\! + \! 2m\! - \! 1}\! ,{\bf{U}}_{a\! -\!  m}^{a\! -\! 1}}}\! \left( \left. {{{\bf{z}}_a^{a\!+ \! m\! - \! 1}} - \Gm{\bf{x}}_{a\! -\!  m}^{a\! + \! m\! - \! 1}} \right|{\bf{z}}_{a\! + \! m}^{a\! + \! 2m\! - \! 1} - \Gm{\bf{x}}_a^{a\! + \! 2m\! - \! 1}\! ,{\bf{u}}_{a\! -\!  m}^{a\! -\! 1} \right),
\end{align*} 
where $(a)$ follows since ${\bf{U}}_a^{a\!+\! m\! - \! 1}$ is statistically independent of the channel inputs ${{\bf{X}}^{b \cdot n\! + \! a \! - \! 1}}$, and since the temporal correlation of the multivariate Gaussian process ${\bf U}[i]$ is finite and shorter than $m+1$, which implies that ${\bf{U}}_a^{a\!+\! m\! - \! 1}$ is also mutually independent of ${\bf {\check U}}$. 
Let $\mathsf{C}_{{\bf U}_a^{a\!+\! m\! - \! 1}}$ be the covariance matrix of ${\bf{U}}_a^{a\!+\! m\! - \! 1}$. As the noise samples are not linearly dependent it follows that $\mathsf{C}_{{\bf U}_a^{a\!+\! m\! - \! 1}}$ is invertible. 
Define 
\begin{equation*}
\mathsf{C}_{{\bf U}_a^{a\!+\! m\! - \! 1} {\bf U}_{a\! + \!  m}^{a\! + \! 2m \! - \! 1}} \triangleq \E\left\{{{\bf{U}}_a^{a\!+\! m\! - \! 1}}\left({\bf{U}}_{a\! + \!  m}^{a\! + \! 2m \! - \! 1}\right)^T\right\}, 
\end{equation*}
\begin{equation*}
\mathsf{C}_{{\bf U}_a^{a\!+\! m\! - \! 1} {\bf U}_{a\!- \! m}^{a\! -\! 1}} \triangleq \E\left\{{{\bf{U}}_a^{a\!+\! m\! - \! 1}}\left({\bf{U}}_{a\!- \! m}^{a\! -\! 1}\right)^T\right\}, 
\end{equation*}
\begin{equation*}
\mathsf{M}_{1} \triangleq \mathsf{C}_{{\bf U}_a^{a\!+\! m\! - \! 1} {\bf U}_{a\!- \! m}^{a\! -\! 1}} \left(\mathsf{C}_{{\bf U}_a^{a\!+\! m\! - \! 1}}\right)^{-1}, 
\end{equation*}
\begin{equation*}
\mathsf{M}_{2} \triangleq \mathsf{C}_{{\bf U}_a^{a\!+\! m\! - \! 1} {\bf U}_{a\! + \!  m}^{a\! + \! 2m \! - \! 1}} \left(\mathsf{C}_{{\bf U}_a^{a\!+\! m\! - \! 1}}\right)^{-1}, 
\end{equation*}
and 
\begin{align}
\mathsf{Q} \!\triangleq \!\mathsf{C}_{{\bf U}_a^{a\!+\! m\! - \! 1}} &\! -\!  \mathsf{C}_{{\bf U}_a^{a\!+\! m\! - \! 1} {\bf U}_{a\! + \!  m}^{a\! + \! 2m \! - \! 1}}   \! \left(\mathsf{C}_{{\bf U}_a^{a\!+\! m\! - \! 1}}\!\right)^{-1}\!  \left(\mathsf{C}_{{\bf U}_a^{a\!+\! m\! - \! 1} {\bf U}_{a\! + \!  m}^{a\! + \! 2m \! - \! 1}}\!\right)^T  \notag \\
& \!-\!  \mathsf{C}_{{\bf U}_a^{a\!+\! m\! - \! 1} {\bf U}_{a\!- \! m}^{a\! -\! 1}} \left(\mathsf{C}_{{\bf U}_a^{a\!+\! m\! - \! 1}}\!\right)^{-1}\! \left(\mathsf{C}_{{\bf U}_a^{a\!+\! m\! - \! 1} {\bf U}_{a\!- \! m}^{a\! -\! 1}}\right)^T.
\label{eqn:DefQ1}
\end{align}
Note that ${{\bf{U}}_a^{a\!+\! m\! - \! 1}}$ and $\left[\left({\bf{U}}_{a\! + \!  m}^{a\! + \! 2m \! - \! 1}\right)^T, \left({\bf{U}}_{a-m}^{a -  1}\right)^T\right]^T$ are jointly Gaussian, therefore, the conditional distribution ${{\bf{U}}_a^{a\!+\! m\! - \! 1}}|\left[\left({\bf{U}}_{a\! + \!  m}^{a\! + \! 2m \! - \! 1}\right)^T,\left({\bf{U}}_{a\! -\! m}^{a\!-\! 1}\right)^T\right]^T = \left[\left({\bf{u}}_{a\! + \!  m}^{a\! + \! 2m \! - \! 1}\right)^T,\left({\bf{u}}_{ a- m}^{a-1}\right)^T\right]^T$ is a multivariate Gaussian distribution  \cite[Proposition 3.13]{Eaton:07} with mean vector 
\begin{align*}
&\E\left\{ \left. {{{\bf{U}}_a^{a\! + m\! -\!  1}}} \right|\left[ \left({\bf{U}}_{a\!+ \! m}^{a\! + \! 2m\! -\!  1}\right)^T\! ,\left({\bf{U}}_{a\! - \!  m}^{a\! -\! 1}\right)^T \right]^T = \left[ \left({\bf{u}}_{a\!+ \! m}^{a\! + \! 2m\! -\!  1}\right)^T\! ,\left({\bf{u}}_{a\! - \!  m}^{a\! -\! 1}\right)^T \right]^T \right\} \\
&\quad\stackrel{(a)}{=}\E\left\{ {{\bf{U}}_a^{a\! + m\! -\!  1}}\left[ \left({\bf{U}}_{a\!+ \! m}^{a\! + \! 2m\! -\!  1}\right)^T\! ,\left({\bf{U}}_{a\! - \!  m}^{a\! -\! 1}\right)^T \right] \right\}\left( \E\left\{ \left[ \left({\bf{U}}_{a\!+ \! m}^{a\! + \! 2m\! -\!  1}\right)^T\! ,\left({\bf{U}}_{a\! - \!  m}^{a\! -\! 1}\right)^T \right]^T\left[ \left({\bf{U}}_{a\!+ \! m}^{a\! + \! 2m\! -\!  1}\right)^T\! ,\left({\bf{U}}_{a\! - \!  m}^{a\! -\! 1}\right)^T \right] \right\} \right)^{ - 1}
\notag \\
&\quad \qquad \qquad \qquad \times 
\left[ \left({\bf{u}}_{a\!+ \! m}^{a\! + \! 2m\! -\!  1}\right)^T\! ,\left({\bf{u}}_{a\! - \!  m}^{a\! -\! 1}\right)^T \right]^T\notag \\
&\quad\stackrel{(b)}{=} \left[ {\begin{array}{*{20}{c}}
{{\mathsf{C}_{{{\bf{U}}_a^{a\! + m\! -\!  1}}{\bf{U}}_{a\!+ \! m}^{a\! + \! 2m\! -\!  1}}}}&{{\mathsf{C}_{{{\bf{U}}_a^{a\! + m\! -\!  1}}{\bf{U}}_{a\! - \!  m}^{a\! -\! 1}}}}
\end{array}} \right]{\left[ {\begin{array}{*{20}{c}}
{{\mathsf{C}_{{{\bf{U}}_a^{a\! + m\! -\!  1}}}}}&\mathsf{0}\\
\mathsf{0}&{{\mathsf{C}_{{{\bf{U}}_a^{a\! + m\! -\!  1}}}}}
\end{array}} \right]^{ - 1}}\left[ {\begin{array}{*{20}{c}}
{{\bf{u}}_{a\!+ \! m}^{a\! + \! 2m\! -\!  1}}\\
{{\bf{u}}_{a\! - \!  m}^{a\! -\! 1}}
\end{array}} \right]\notag \\ 
&\quad= \mathsf{M}_{1}{{\bf{u}}_{a\!- \! m}^{a\! -\! 1}} + \mathsf{M}_{2}{{\bf{u}}_{a\! + \!  m}^{a\! + \! 2m \! - \! 1}},
\end{align*}
where $(a)$ follows from \cite[Proposition 3.13]{Eaton:07} as both ${\bf{U}}[i]$ is zero-mean; 
$(b)$ follows from the finite temporal correlation and the stationarity of ${\bf U}[i]$ which implies that $\E\left\{ {{\bf{U}}_{a\!+ \! m}^{a\! + \! 2m\! -\!  1}{{\left( {{\bf{U}}_{a\! - \!  m}^{a\! -\! 1}} \right)}^T}} \right\} = \mathsf{0}$ and $\E\left\{ {\bf{U}}_{a\!+ \! m}^{a\! + \! 2m\! -\!  1}\left( {\bf{U}}_{a\!+ \! m}^{a\! + \! 2m\! -\!  1} \right)^T \right\} = \E\left\{ {\bf{U}}_{a\! - \!  m}^{a\! -\! 1}\left( {\bf{U}}_{a\! - \!  m}^{a\! -\! 1} \right)^T \right\} = \mathsf{C}_{{{\bf{U}}_a^{a\! + m\! -\!  1}}}$. 
Similarly, it follows from \cite[Proposition 3.13]{Eaton:07} that the covariance matrix of the conditional distribution ${{\bf{U}}_a^{a\!+\! m\! - \! 1}}|\left[\left({\bf{U}}_{a\! + \!  m}^{a\! + \! 2m \! - \! 1}\right)^T,\left({\bf{U}}_{a\! -\! m}^{a\!-\! 1}\right)^T\right]^T = \left[\left({\bf{u}}_{a\! + \!  m}^{a\! + \! 2m \! - \! 1}\right)^T,\left({\bf{u}}_{a - m}^{a-1}\right)^T\right]^T$ is $\mathsf{Q}$ defined in \eqref{eqn:DefQ1}.
It therefore follows that 
\begin{align*}
&	\log \frac{{{p_{\left. {{\bf{Z}}_a^{a \! + \! m \! - \! 1}} \right|{{\bf{X}}^{b \cdot n \! + \! a \! - \! 1}},{\bf{Z}}_{a \! + \! m}^{a \! + \! 2m \! - \! 1},{\bf{\check Z}},{\bf{U}}_{a \! - \! m}^{a \! - \! 1}}}\left( {\left. {{\bf{z}}_a^{a \! + \! m \! - \! 1}} \right|{{\bf{x}}^{b \cdot n \! + \! a \! - \! 1}},{\bf{z}}_{a \! + \! m}^{a \! + \! 2m \! - \! 1},{\bf{\check z}},{\bf{u}}_{a \! - \! m}^{a \! - \! 1}} \right)}}{{{p_{\left. {{\bf{U}}_a^{a \! + \! m \! - \! 1}} \right|{\bf{U}}_{a \! + \! m}^{a \! + \! 2m \! - \! 1},{\bf{U}}_{a \! - \! m}^{a \! - \! 1}}}\left( {\left. {{\bf{z}}_a^{a \! + \! m \! - \! 1}} \right|{\bf{z}}_{a \! + \! m}^{a \! + \! 2m \! - \! 1},{\bf{u}}_{a \! - \! m}^{a \! - \! 1}} \right)}} \notag\\
&\quad = \log \frac{{{p_{\left. {{\bf{U}}_a^{a \! + \! m \! - \! 1}} \right|{\bf{U}}_{a \! + \! m}^{a \! + \! 2m \! - \! 1},{\bf{U}}_{a \! - \! m}^{a \! - \! 1}}}\left( {\left. {{\bf{z}}_a^{a \! + \! m \! - \! 1} \! - \! \Gm {\bf{x}}_{a \! - \! m}^{a \! + \! m \! - \! 1}} \right|{\bf{z}}_{a \! + \! m}^{a \! + \! 2m \! - \! 1} \! - \! \Gm {\bf{x}}_a^{a \! + \! 2m \! - \! 1},{\bf{u}}_{a \! - \! m}^{a \! - \! 1}} \right)}}{{{p_{\left. {{\bf{U}}_a^{a \! + \! m \! - \! 1}} \right|{\bf{U}}_{a \! + \! m}^{a \! + \! 2m \! - \! 1},{\bf{U}}_{a \! - \! m}^{a \! - \! 1}}}\left( {\left. {{\bf{z}}_a^{a \! + \! m \! - \! 1}} \right|{\bf{z}}_{a \! + \! m}^{a \! + \! 2m \! - \! 1},{\bf{u}}_{a \! - \! m}^{a \! - \! 1}} \right)}} \notag \\
&\quad =  \! - \! \frac{1}{2}\log e \cdot {\left( {{\bf{z}}_a^{a \! + \! m \! - \! 1} \! - \! \Gm {\bf{x}}_{a \! - \! m}^{a \! + \! m \! - \! 1} \! - \! {\mathsf{M}_1}{\bf{u}}_{a \! - \! m}^{a \! - \! 1} \! - \! {\mathsf{M}_2}\left( {{\bf{z}}_{a \! + \! m}^{a \! + \! 2m \! - \! 1} \! - \! \Gm {\bf{x}}_a^{a \! + \! 2m \! - \! 1}} \right)} \right)^T}{\mathsf{Q}^{ \! - \! 1}} \notag \\
&\qquad \qquad \qquad \qquad \qquad \qquad \times 
\left( {{\bf{z}}_a^{a \! + \! m \! - \! 1} \! - \! \Gm {\bf{x}}_{a \! - \! m}^{a \! + \! m \! - \! 1} \! - \! {\mathsf{M}_1}{\bf{u}}_{a \! - \! m}^{a \! - \! 1} \! - \! {\mathsf{M}_2}\left( {{\bf{z}}_{a \! + \! m}^{a \! + \! 2m \! - \! 1} \! - \! \Gm {\bf{x}}_a^{a \! + \! 2m \! - \! 1}} \right)} \right) \notag \\
&\qquad  \! + \! \frac{1}{2}\log e \cdot {\left( {{\bf{z}}_a^{a \! + \! m \! - \! 1} \! - \! {\mathsf{M}_1}{\bf{u}}_{a \! - \! m}^{a \! - \! 1} \! - \! {\mathsf{M}_2}{\bf{z}}_{a \! + \! m}^{a \! + \! 2m \! - \! 1}} \right)^T}{\mathsf{Q}^{ \! - \! 1}}\left( {{\bf{z}}_a^{a \! + \! m \! - \! 1} \! - \! {\mathsf{M}_1}{\bf{u}}_{a \! - \! m}^{a \! - \! 1} \! - \! {\mathsf{M}_2}{\bf{z}}_{a \! + \! m}^{a \! + \! 2m \! - \! 1}} \right)\notag \\
&\quad = \frac{1}{2}\log e \cdot \Bigg(\left( {2{{\left( {{\bf{z}}_a^{a \! + \! m \! - \! 1} \! - \! {\mathsf{M}_1}{\bf{u}}_{a \! - \! m}^{a \! - \! 1} \! - \! {\mathsf{M}_2}{\bf{z}}_{a \! + \! m}^{a \! + \! 2m \! - \! 1}} \right)}^T}{\mathsf{Q}^{ \! - \! 1}}\left( {\Gm {\bf{x}}_{a \! - \! m}^{a \! + \! m \! - \! 1} \! - \! {\mathsf{M}_2}\Gm {\bf{x}}_a^{a \! + \! 2m \! - \! 1}} \right)} \right)  \notag \\
&\quad  \qquad \qquad \qquad  \! - \! {\left( {\Gm {\bf{x}}_{a \! - \! m}^{a \! + \! m \! - \! 1} \! - \! {\mathsf{M}_2}\Gm {\bf{x}}_a^{a \! + \! 2m \! - \! 1}} \right)^T}{\mathsf{Q}^{ \! - \! 1}}\left( {\Gm {\bf{x}}_{a \! - \! m}^{a \! + \! m \! - \! 1} \! - \! {\mathsf{M}_2}\Gm {\bf{x}}_a^{a \! + \! 2m \! - \! 1}} \right)\Bigg).
\end{align*}
Plugging this into \eqref{eqn:Bound2} yields
\begin{align}
&I\left( {\left. {{{\bf{X}}^{b \cdot n \! + \! a \! - \! 1}};{\bf{Z}}_a^{a \! + \! m \! - \! 1}} \right|{\bf{Z}}_{a \! + \! m}^{a \! + \! 2m \! - \! 1},{\bf{\check Z}},{\bf{U}}_{a \! - \! m}^{a \! - \! 1}} \right)  \notag \\
&\qquad \leq 
\log  e \cdot
\Bigg(
\! \int \! {{p_{{\bf{Z}}_a^{a \! + \! 2m \! - \! 1}{{\bf{X}}^{b \cdot n \! + \! a \! - \! 1}},{\bf{\check Z}},{\bf{U}}_{a \! - \! m}^{a \! - \! 1}}}\left( {{\bf{z}}_a^{a \! + \! 2m \! - \! 1},{{\bf{x}}^{b \cdot n \! + \! a \! - \! 1}},{\bf{\check z}},{\bf{u}}_{a \! - \! m}^{a \! - \! 1}} \right)} \cdot {\left( {{\bf{z}}_a^{a \! + \! m \! - \! 1} \! - \! {\mathsf{M}_1}{\bf{u}}_{a \! - \! m}^{a \! - \! 1} \! - \! {\mathsf{M}_2}{\bf{z}}_{a \! + \! m}^{a \! + \! 2m \! - \! 1}} \right)^T}
 \notag \\
&\qquad \qquad \qquad \qquad \qquad \qquad \times 
{\mathsf{Q}^{ \! - \! 1}}\left( {\Gm {\bf{x}}_{a \! - \! m}^{a \! + \! m \! - \! 1} \! - \! {\mathsf{M}_2}\Gm {\bf{x}}_a^{a \! + \! 2m \! - \! 1}} \right)d{\bf{z}}_a^{a \! + \! 2m \! - \! 1}d{{\bf{x}}^{b \cdot n \! + \! a \! - \! 1}}d{\bf{\check z}}~d{\bf{u}}_{a \! - \! m}^{a \! - \! 1} 
\notag \\ 
&\qquad \quad 
 \! - \! \frac{1}{2}\! \int \! {{p_{{\bf{Z}}_a^{a \! + \! 2m \! - \! 1}{{\bf{X}}^{b \cdot n \! + \! a \! - \! 1}},{\bf{\check Z}},{\bf{U}}_{a \! - \! m}^{a \! - \! 1}}}\left( {{\bf{z}}_a^{a \! + \! 2m \! - \! 1},{{\bf{x}}^{b \cdot n \! + \! a \! - \! 1}},{\bf{\check z}},{\bf{u}}_{a \! - \! m}^{a \! - \! 1}} \right)} \cdot {\left( {\Gm {\bf{x}}_{a \! - \! m}^{a \! + \! m \! - \! 1} \! - \! {\mathsf{M}_2}\Gm {\bf{x}}_a^{a \! + \! 2m \! - \! 1}} \right)^T}
 \notag \\
&\qquad \qquad \qquad \qquad \qquad \qquad \times 
{\mathsf{Q}^{ \! - \! 1}}\left( {\Gm {\bf{x}}_{a \! - \! m}^{a \! + \! m \! - \! 1} \! - \! {\mathsf{M}_2}\Gm {\bf{x}}_a^{a \! + \! 2m \! - \! 1}} \right)d{\bf{z}}_a^{a \! + \! 2m \! - \! 1}d{{\bf{x}}^{b \cdot n \! + \! a \! - \! 1}}d{\bf{\check z}}~d{\bf{u}}_{a \! - \! m}^{a \! - \! 1}\Bigg) 
\notag \\
&\qquad = 
\log e \cdot
\Bigg(\! \int \! {{p_{{\bf{Z}}_a^{a \! + \! 2m \! - \! 1},{\bf{X}}_{a \! - \! m}^{a \! + \! 2m \! - \! 1},{\bf{U}}_{a \! - \! m}^{a \! - \! 1}}}\left( {{\bf{z}}_a^{a \! + \! 2m \! - \! 1},{\bf{x}}_{a \! - \! m}^{a \! + \! 2m \! - \! 1},{\bf{u}}_{a \! - \! m}^{a \! - \! 1}} \right)} \cdot {\left( {{\bf{z}}_a^{a \! + \! m \! - \! 1} \! - \! {\mathsf{M}_1}{\bf{u}}_{a \! - \! m}^{a \! - \! 1} \! - \! {\mathsf{M}_2}{\bf{z}}_{a \! + \! m}^{a \! + \! 2m \! - \! 1}} \right)^T}
 \notag \\
&\qquad \qquad \qquad \qquad \qquad \qquad \times 
{\mathsf{Q}^{ \! - \! 1}}\left( {\Gm {\bf{x}}_{a \! - \! m}^{a \! + \! m \! - \! 1} \! - \! {\mathsf{M}_2}\Gm {\bf{x}}_a^{a \! + \! 2m \! - \! 1}} \right)d{\bf{z}}_a^{a \! + \! 2m \! - \! 1}d{\bf{x}}_{a \! - \! m}^{a \! + \! 2m \! - \! 1}d{\bf{u}}_{a \! - \! m}^{a \! - \! 1}
\notag \\ 
&\qquad \quad 
 \! - \! \frac{1}{2}\! \int \! {{p_{{\bf{X}}_{a \! - \! m}^{a \! + \! 2m \! - \! 1}}}\left( {{\bf{x}}_{a \! - \! m}^{a \! + \! 2m \! - \! 1}} \right)} \cdot {\left( {\Gm {\bf{x}}_{a \! - \! m}^{a \! + \! m \! - \! 1} \! - \! {\mathsf{M}_2}\Gm {\bf{x}}_a^{a \! + \! 2m \! - \! 1}} \right)^T}{\mathsf{Q}^{ \! - \! 1}}\left( {\Gm {\bf{x}}_{a \! - \! m}^{a \! + \! m \! - \! 1} \! - \! {\mathsf{M}_2}\Gm {\bf{x}}_a^{a \! + \! 2m \! - \! 1}} \right)d{\bf{x}}_{a \! - \! m}^{a \! + \! 2m \! - \! 1}\Bigg).
\label{eqn:Bound3}
\end{align}
Now, note that 
\begin{align}
&\! \int \! {{p_{{\bf{X}}_{a \! - \! m}^{a \! + \! 2m \! - \! 1}}}\left( {{\bf{x}}_{a \! - \! m}^{a \! + \! 2m \! - \! 1}} \right)} \cdot {\left( {\Gm {\bf{x}}_{a \! - \! m}^{a \! + \! m \! - \! 1} \! - \! {\mathsf{M}_2}\Gm {\bf{x}}_a^{a \! + \! 2m \! - \! 1}} \right)^T}{\mathsf{Q}^{ \! - \! 1}}\left( {\Gm {\bf{x}}_{a \! - \! m}^{a \! + \! m \! - \! 1} \! - \! {\mathsf{M}_2}\Gm {\bf{x}}_a^{a \! + \! 2m \! - \! 1}} \right)d{\bf{x}}_{a \! - \! m}^{a \! + \! 2m \! - \! 1} \notag  \\
& \qquad = \E\left\{ {{{\left( {\Gm {\bf{X}}_{a \! - \! m}^{a \! + \! m \! - \! 1} \! - \! {\mathsf{M}_2}\Gm {\bf{X}}_a^{a \! + \! 2m \! - \! 1}} \right)}^T}{\mathsf{Q}^{ \! - \! 1}}\left( {\Gm {\bf{X}}_{a \! - \! m}^{a \! + \! m \! - \! 1} \! - \! {\mathsf{M}_2}\Gm {\bf{X}}_a^{a \! + \! 2m \! - \! 1}} \right)} \right\}.  \label{eqn:Bound3a}
\end{align}
Also note that
\begin{align}
& \! \int \! {{p_{{\bf{Z}}_a^{a \! + \! 2m \! - \! 1},{\bf{X}}_{a \! - \! m}^{a \! + \! 2m \! - \! 1},{\bf{U}}_{a \! - \! m}^{a \! - \! 1}}}\left( {{\bf{z}}_a^{a \! + \! 2m \! - \! 1},{\bf{x}}_{a \! - \! m}^{a \! + \! 2m \! - \! 1},{\bf{u}}_{a \! - \! m}^{a \! - \! 1}} \right)} \cdot {\left( {{\bf{z}}_a^{a \! + \! m \! - \! 1} \! - \! {\mathsf{M}_1}{\bf{u}}_{a \! - \! m}^{a \! - \! 1} \! - \! {\mathsf{M}_2}{\bf{z}}_{a \! + \! m}^{a \! + \! 2m \! - \! 1}} \right)^T}{\mathsf{Q}^{ \! - \! 1}}
\notag \\
&\quad \qquad \qquad \qquad \qquad \qquad \times
\left( {\Gm {\bf{x}}_{a \! - \! m}^{a \! + \! m \! - \! 1} \! - \! {\mathsf{M}_2}\Gm {\bf{x}}_a^{a \! + \! 2m \! - \! 1}} \right)d{\bf{z}}_a^{a \! + \! 2m \! - \! 1}d{\bf{x}}_{a \! - \! m}^{a \! + \! 2m \! - \! 1}d{\bf{u}}_{a \! - \! m}^{a \! - \! 1} 
\notag \\
& \quad = \! \int \! \left( \! \int \! {p_{\left. {{\bf{Z}}_a^{a \! + \! 2m \! - \! 1},{\bf{U}}_{a \! - \! m}^{a \! - \! 1}} \right|{\bf{X}}_{a \! - \! m}^{a \! + \! 2m \! - \! 1}}}\left( {\left. {{\bf{z}}_a^{a \! + \! 2m \! - \! 1},{\bf{u}}_{a \! - \! m}^{a \! - \! 1}} \right|{\bf{x}}_{a \! - \! m}^{a \! + \! 2m \! - \! 1}} \right) \cdot {{\left( {{\bf{z}}_a^{a \! + \! m \! - \! 1} \! - \! {\mathsf{M}_1}{\bf{u}}_{a \! - \! m}^{a \! - \! 1} \! - \! {\mathsf{M}_2}{\bf{z}}_{a \! + \! m}^{a \! + \! 2m \! - \! 1}} \right)}^T}d{\bf{z}}_a^{a \! + \! 2m \! - \! 1}d{\bf{u}}_{a \! - \! m}^{a \! - \! 1}  \right) {\mathsf{Q}^{ \! - \! 1}}
\notag \\
&\quad \qquad \qquad \qquad \qquad \qquad \times
\left( {\Gm {\bf{x}}_{a \! - \! m}^{a \! + \! m \! - \! 1} \! - \! {\mathsf{M}_2}\Gm {\bf{x}}_a^{a \! + \! 2m \! - \! 1}} \right){p_{{\bf{X}}_{a \! - \! m}^{a \! + \! 2m \! - \! 1}}}\left( {{\bf{x}}_{a \! - \! m}^{a \! + \! 2m \! - \! 1}} \right)d{\bf{x}}_{a \! - \! m}^{a \! + \! 2m \! - \! 1}, 
\label{eqn:Bound4}
\end{align} 
where 
\begin{align} 
&\! \int \! {{p_{\left. {{\bf{Z}}_a^{a \! + \! 2m \! - \! 1},{\bf{U}}_{a \! - \! m}^{a \! - \! 1}} \right|{\bf{X}}_{a \! - \! m}^{a \! + \! 2m \! - \! 1}}}\left( {\left. {{\bf{z}}_a^{a \! + \! 2m \! - \! 1},{\bf{u}}_{a \! - \! m}^{a \! - \! 1}} \right|{\bf{x}}_{a \! - \! m}^{a \! + \! 2m \! - \! 1}} \right)\left( {{\bf{z}}_a^{a \! + \! m \! - \! 1} \! - \! {\mathsf{M}_1}{\bf{u}}_{a \! - \! m}^{a \! - \! 1} \! - \! {\mathsf{M}_2}{\bf{z}}_{a \! + \! m}^{a \! + \! 2m \! - \! 1}} \right)d{\bf{z}}_a^{a \! + \! 2m \! - \! 1}d{\bf{u}}_{a \! - \! m}^{a \! - \! 1}} 
 \notag \\
& \quad = \E\left\{ {\left. {{\bf{Z}}_a^{a \! + \! m \! - \! 1}} \right|{\bf{X}}_{a \! - \! m}^{a \! + \! 2m \! - \! 1} \! = \! {\bf{x}}_{a \! - \! m}^{a \! + \! 2m \! - \! 1}} \right\} \! - \! {\mathsf{M}_1}\E\left\{ {\left. {{\bf{U}}_{a \! - \! m}^{a \! - \! 1}} \right|{\bf{X}}_{a \! - \! m}^{a \! + \! 2m \! - \! 1} \! = \! {\bf{x}}_{a \! - \! m}^{a \! + \! 2m \! - \! 1}} \right\} \! - \! {\mathsf{M}_2}\E\left\{ {\left. {{\bf{Z}}_{a \! + \! m}^{a \! + \! 2m \! - \! 1}} \right|{\bf{X}}_{a \! - \! m}^{a \! + \! 2m \! - \! 1} \! = \! {\bf{x}}_{a \! - \! m}^{a \! + \! 2m \! - \! 1}} \right\} \notag \\
& \quad \stackrel{(a)}{=} \Gm {\bf{x}}_{a \! - \! m}^{a \! + \! m \! - \! 1} \! - \! {\mathsf{M}_2}\Gm {\bf{x}}_a^{a \! + \! 2m \! - \! 1}, 
\label{eqn:Bound5}
\end{align} 
where $(a)$ follows since ${\bf{Z}}_a^{a \! + \! m \! - \! 1} = \Gm {\bf{X}}_{a \! - \! m}^{a \! + \! m \! - \! 1} \! + \!{\bf{U}}_a^{a \! + \! m \! - \! 1}$ and ${\bf{Z}}_{a \! + \! m}^{a \! + \! 2m \! - \! 1} = \Gm {\bf{X}}_a^{a \! + \! 2m \! - \! 1} \! + \!{\bf{U}}_{a \! + \! m}^{a \! + \! 2m \! - \! 1}$, and since the mean of the noise ${\bf{U}}[i]$ is zero and ${\bf{U}}[i]$ is independent of the channel input ${\bf{X}}[i]$. 
Plugging \eqref{eqn:Bound5} into \eqref{eqn:Bound4} yields 
\begin{align}
& \! \int \! {{p_{{\bf{Z}}_a^{a \! + \! 2m \! - \! 1},{\bf{X}}_{a \! - \! m}^{a \! + \! 2m \! - \! 1},{\bf{U}}_{a \! - \! m}^{a \! - \! 1}}}\left( {{\bf{z}}_a^{a \! + \! 2m \! - \! 1},{\bf{x}}_{a \! - \! m}^{a \! + \! 2m \! - \! 1},{\bf{u}}_{a \! - \! m}^{a \! - \! 1}} \right)} {\left( {{\bf{z}}_a^{a \! + \! m \! - \! 1} \! - \! {\mathsf{M}_1}{\bf{u}}_{a \! - \! m}^{a \! - \! 1} \! - \! {\mathsf{M}_2}{\bf{z}}_{a \! + \! m}^{a \! + \! 2m \! - \! 1}} \right)^T}{\mathsf{Q}^{ \! - \! 1}}
\notag \\
&\qquad \qquad \qquad \qquad \qquad \qquad \times
\left( {\Gm {\bf{x}}_{a \! - \! m}^{a \! + \! m \! - \! 1} \! - \! {\mathsf{M}_2}\Gm {\bf{x}}_a^{a \! + \! 2m \! - \! 1}} \right)d{\bf{z}}_a^{a \! + \! 2m \! - \! 1}d{\bf{x}}_{a \! - \! m}^{a \! + \! 2m \! - \! 1}d{\bf{u}}_{a \! - \! m}^{a \! - \! 1}
\notag \\
& \qquad = \! \int \! {{{\left( {\Gm {\bf{x}}_{a \! - \! m}^{a \! + \! m \! - \! 1} \! - \! {\mathsf{M}_2}\Gm {\bf{x}}_a^{a \! + \! 2m \! - \! 1}} \right)}^T}} {\mathsf{Q}^{ \! - \! 1}}\left( {\Gm {\bf{x}}_{a \! - \! m}^{a \! + \! m \! - \! 1} \! - \! {\mathsf{M}_2}\Gm {\bf{x}}_a^{a \! + \! 2m \! - \! 1}} \right){p_{{\bf{X}}_{a \! - \! m}^{a \! + \! 2m \! - \! 1}}}\left( {{\bf{x}}_{a \! - \! m}^{a \! + \! 2m \! - \! 1}} \right)d{\bf{x}}_{a \! - \! m}^{a \! + \! 2m \! - \! 1} \notag \\
& \qquad = \E\left\{ {{{\left( {\Gm {\bf{X}}_{a \! - \! m}^{a \! + \! m \! - \! 1} \! - \! {\mathsf{M}_2}\Gm {\bf{X}}_a^{a \! + \! 2m \! - \! 1}} \right)}^T}{\mathsf{Q}^{ \! - \! 1}}\left( {\Gm {\bf{X}}_{a \! - \! m}^{a \! + \! m \! - \! 1} \! - \! {\mathsf{M}_2}\Gm {\bf{X}}_a^{a \! + \! 2m \! - \! 1}} \right)} \right\}. 
\label{eqn:Bound6}
\end{align}
Plugging \eqref{eqn:Bound3a} and \eqref{eqn:Bound6} into \eqref{eqn:Bound3} yields
\begin{align}
&I\left( {\left. {{{\bf{X}}^{b \cdot n \! + \! a \! - \! 1}};{\bf{Z}}_a^{a \! + \! m \! - \! 1}} \right|{\bf{Z}}_{a \! + \! m}^{a \! + \! 2m \! - \! 1},{\bf{\check Z}},{\bf{U}}_{a \! - \! m}^{a \! - \! 1}} \right)
\notag \\
&\qquad \qquad \qquad \leq 
\frac{1}{2} \log e \cdot \E\left\{ {{{\left( {\Gm {\bf{X}}_{a \! - \! m}^{a \! + \! m \! - \! 1} \! - \! {\mathsf{M}_2}\Gm {\bf{X}}_a^{a \! + \! 2m \! - \! 1}} \right)}^T}{\mathsf{Q}^{ \! - \! 1}}\left( {\Gm {\bf{X}}_{a \! - \! m}^{a \! + \! m \! - \! 1} \! - \! {\mathsf{M}_2}\Gm {\bf{X}}_a^{a \! + \! 2m \! - \! 1}} \right)} \right\},
\label{eqn:Bound7}
\end{align}
with $\mathsf{Q}$ defined in \eqref{eqn:DefQ1}.
Note that the expression in \eqref{eqn:Bound7} is the mean of a quadratic function of $3m$ channel inputs, which depends only on the noise correlation function through $\mathsf{Q}$, the channel transfer function through $\mathsf{G}$, and the power of the input signal. Consequently, $\exists \left\{\alpha_{i_1}\right\}_{i_1=a\! - \! m}^{a\! +\! 2m\! - \! 1} \in \mathds{R}$, $\left|\alpha_{i_1}\right| < \infty$, such that \eqref{eqn:Bound7} can be written as 
\begin{align}
I\left( {\left. {{{\bf{X}}^{b \cdot n \! + \! a \! - \! 1}};{\bf{Z}}_a^{a \! + \! m \! - \! 1}} \right|{\bf{Z}}_{a \! + \! m}^{a \! + \! 2m \! - \! 1},{\bf{\check Z}},{\bf{U}}_{a \! - \! m}^{a \! - \! 1}} \right) 
&\leq  \sum\limits_{{i_1} =  a\! - \! m}^{a\! +\! 2m\! - \! 1}  \left|\alpha _{{i_1}}\right|  \E\left\{\left\|{\bf{X}}\left[ {{i_1}} \right]\right\|^2\right\}  \notag \\
&\stackrel{(a)}{\leq} \sum\limits_{{i_1} =  a\! - \! m}^{a\! +\! 2m\! - \! 1}  \left|\alpha _{{i_1}}\right| P,
\label{eqn:Bound8}
\end{align}
where  $(b)$ follows since the input signal ${\bf X}[i]$ is subject to a per-symbol power constraint. 
It follows from \eqref{eqn:Bound8} that $\exists \eta$ finite such that $I\left( {\left. {{{\bf{X}}^{b \cdot n \! + \! a \! - \! 1}};{\bf{Z}}_a^{a \! + \! m \! - \! 1}} \right|{\bf{Z}}_{a \! + \! m}^{a \! + \! 2m \! - \! 1},{\bf{\check Z}},{\bf{U}}_{a \! - \! m}^{a \! - \! 1}} \right) \leq \eta$.

Next, we prove \eqref{eqn:ToProve1b} using a similar derivation:
Note that
\begin{align*}
&I\left( {\left. {{{\bf{X}}^{b \cdot n \! + \! a \! - \! 1}};{\bf{Z}}_{l \cdot n \! + \! a}^{l \cdot n \! + \! a \! + \! m \! - \! 1}} \right|{\bf{Z}}_{a \! + \! m}^{n \! + \! a \! - \! 1},{\bf{Z}}_{n \! + \! a \! + \! m}^{2n \! + \! a \! - \! 1}, \ldots ,{\bf{Z}}_{\left( {b \! - \! 1} \right) \cdot n \! + \! a \! + \! m}^{b \cdot n \! + \! a \! - \! 1},{\bf{Z}}_a^{a \! + \! m \! - \! 1},{\bf{Z}}_{n \! + \! a}^{n \! + \! a \! + \! m \! - \! 1}, \ldots ,{\bf{Z}}_{\left( {l \! - \! 1} \right) \cdot n \! + \! a}^{\left( {l \! - \! 1} \right) \cdot n \! + \! a \! + \! m \! - \! 1},{\bf{U}}_{a \! - \! m}^{a \! - \! 1}} \right) \notag \\
&=I\left( {\left. {{{\bf{X}}^{b \cdot n \! + \! a \! - \! 1}};{\bf{Z}}_{l \cdot n \! + \! a}^{l \cdot n \! + \! a \! + \! m \! - \! 1}} \right|{\bf{Z}}_a^{l \cdot n \! + \! a \! - \! 1},{\bf{Z}}_{l \cdot n \! + \! a \! + \! m}^{\left( {l \! + \! 1} \right) \cdot n \! + \! a \! - \! 1},{\bf{Z}}_{\left( {l \! + \! 1} \right) \cdot n \! + \! a \! + \! m}^{\left( {l \! + \! 2} \right) \cdot n \! + \! a \! - \! 1}, \ldots ,{\bf{Z}}_{\left( {b \! - \! 1} \right) \cdot n \! + \! a \! + \! m}^{b \cdot n \! + \! a \! - \! 1},{\bf{U}}_{a \! - \! m}^{a \! - \! 1}} \right).
\end{align*}
Define ${\bf{Z}}' \triangleq \Big( {\bf{Z}}_{l \cdot n \! + \! a \! - \! m}^{l \cdot n \! + \! a \! - \! 1},{\bf{Z}}_{l \cdot n \! + \! a \! + \! m}^{l \cdot n \! + \! a \! + \! 2m \! - \! 1} \Big)$ and $\tilde{\bf{Z}} \triangleq \Big( {\bf{Z}}_a^{l \cdot n \! + \! a \! - \! m},{\bf{Z}}_{l \cdot n \! + \! a \! + \! 2m \! - \! 1}^{\left( {l \! + \! 1} \right) \cdot n \! + \! a \! - \! 1},{\bf{Z}}_{\left( {l \! + \! 1} \right) \cdot n \! + \! a \! + \! m}^{\left( {l \! + \! 2} \right) \cdot n \! + \! a \! - \! 1}, \ldots ,{\bf{Z}}_{\left( {b \! - \! 1} \right) \cdot n \! + \! a \! + \! m}^{b \cdot n \! + \! a \! - \! 1} \Big)$, thus
\begin{align} 
&I\left( {\left. {{{\bf{X}}^{b \cdot n \! + \! a \! - \! 1}};{\bf{Z}}_{l \cdot n \! + \! a}^{l \cdot n \! + \! a \! + \! m \! - \! 1}} \right|{\bf{Z}}_a^{l \cdot n \! + \! a \! - \! 1},{\bf{Z}}_{l \cdot n \! + \! a \! + \! m}^{\left( {l \! + \! 1} \right) \cdot n \! + \! a \! - \! 1}, \ldots ,{\bf{Z}}_{\left( {b \! - \! 1} \right) \cdot n \! + \! a \! + \! m}^{b \cdot n \! + \! a \! - \! 1},{\bf{U}}_{a \! - \! m}^{a \! - \! 1}} \right) 
\notag \\
&= I\left( {\left. {{{\bf{X}}^{b \cdot n \! + \! a \! - \! 1}};{\bf{Z}}_{l \cdot n \! + \! a}^{l \cdot n \! + \! a \! + \! m \! - \! 1}} \right|{\bf{Z}}',{\bf{\tilde Z}},{\bf{U}}_{a \! - \! m}^{a \! - \! 1}} \right)
\notag \\
&= \int {{p_{{\bf{Z}}_{l \cdot n \! + \! a}^{l \cdot n \! + \! a \! + \! m \! - \! 1},{{\bf{X}}^{b \cdot n \! + \! a \! - \! 1}},{\bf{Z}}',{\bf{\tilde Z}},{\bf{U}}_{a \! - \! m}^{a \! - \! 1}}}\left( {{\bf{z}}_{l \cdot n \! + \! a}^{l \cdot n \! + \! a \! + \! m \! - \! 1},{{\bf{x}}^{b \cdot n \! + \! a \! - \! 1}},{\bf{z}}',{\bf{\tilde z}},{\bf{u}}_{a \! - \! m}^{a \! - \! 1}} \right)} 
\notag \\
&\quad  \times 
\log \frac{{{p_{{\bf{Z}}_{l \cdot n \! + \! a}^{l \cdot n \! + \! a \! + \! m \! - \! 1}|{{\bf{X}}^{b \cdot n \! + \! a \! - \! 1}},{\bf{Z}}',{\bf{\tilde Z}},{\bf{U}}_{a \! - \! m}^{a \! - \! 1}}}\left( {{\bf{z}}_{l \cdot n \! + \! a}^{l \cdot n \! + \! a \! + \! m \! - \! 1}|{{\bf{x}}^{b \cdot n \! + \! a \! - \! 1}},{\bf{z}}',{\bf{\tilde z}},{\bf{u}}_{a \! - \! m}^{a \! - \! 1}} \right)}}{{{p_{{\bf{Z}}_{l \cdot n \! + \! a}^{l \cdot n \! + \! a \! + \! m \! - \! 1}|{\bf{Z}}',{\bf{\tilde Z}},{\bf{U}}_{a \! - \! m}^{a \! - \! 1}}}\left( {{\bf{z}}_{l \cdot n \! + \! a}^{l \cdot n \! + \! a \! + \! m \! - \! 1}|{\bf{z}}',{\bf{\tilde z}},{\bf{u}}_{a \! - \! m}^{a \! - \! 1}} \right)}}d{\bf{z}}_{l \cdot n \! + \! a}^{l \cdot n \! + \! a \! + \! m \! - \! 1}d{{\bf{x}}^{b \cdot n \! + \! a \! - \! 1}}d{\bf{z}}'d{\bf{\tilde z}}d{\bf{u}}_{a \! - \! m}^{a \! - \! 1} 
\notag \\
&= \int {{p_{{\bf{Z}}_{l \cdot n \! + \! a}^{l \cdot n \! + \! a \! + \! m \! - \! 1},{{\bf{X}}^{b \cdot n \! + \! a \! - \! 1}},{\bf{Z}}',{\bf{\tilde Z}},{\bf{U}}_{a \! - \! m}^{a \! - \! 1}}}\left( {{\bf{z}}_{l \cdot n \! + \! a}^{l \cdot n \! + \! a \! + \! m \! - \! 1},{{\bf{x}}^{b \cdot n \! + \! a \! - \! 1}},{\bf{z}}',{\bf{\tilde z}},{\bf{u}}_{a \! - \! m}^{a \! - \! 1}} \right)} 
\notag \\
&\quad  \times
\log \frac{{{p_{{\bf{Z}}_{l \cdot n \! + \! a}^{l \cdot n \! + \! a \! + \! m \! - \! 1}|{{\bf{X}}^{b \cdot n \! + \! a \! - \! 1}},{\bf{Z}}',{\bf{\tilde Z}},{\bf{U}}_{a \! - \! m}^{a \! - \! 1}}}\left( {{\bf{z}}_{l \cdot n \! + \! a}^{l \cdot n \! + \! a \! + \! m \! - \! 1}|{{\bf{x}}^{b \cdot n \! + \! a \! - \! 1}},{\bf{z}}',{\bf{\tilde z}},{\bf{u}}_{a \! - \! m}^{a \! - \! 1}} \right)}}{{{p_{{\bf{U}}_{l \cdot n \! + \! a}^{l \cdot n \! + \! a \! + \! m \! - \! 1}|{\bf{U}}_{l \cdot n \! + \! a \! - \! m}^{l \cdot n \! + \! a \! - \! 1},{\bf{U}}_{l \cdot n \! + \! a \! + \! m}^{l \cdot n \! + \! a \! + \! 2m \! - \! 1}}}\left( {{\bf{z}}_{l \cdot n \! + \! a}^{l \cdot n \! + \! a \! + \! m \! - \! 1}|{\bf{z}}_{l \cdot n \! + \! a \! - \! m}^{l \cdot n \! + \! a \! - \! 1},{\bf{z}}_{l \cdot n \! + \! a \! + \! m}^{l \cdot n \! + \! a \! + \! 2m \! - \! 1}} \right)}}d{\bf{z}}_{l \cdot n \! + \! a}^{l \cdot n \! + \! a \! + \! m \! - \! 1}d{{\bf{x}}^{b \cdot n \! + \! a \! - \! 1}}d{\bf{z}}'d{\bf{\tilde z}}d{\bf{u}}_{a \! - \! m}^{a \! - \! 1}
\notag \\
&\! - \! \int {{p_{{\bf{Z}}_{l \cdot n \! + \! a}^{l \cdot n \! + \! a \! + \! m \! - \! 1},{{\bf{X}}^{b \cdot n \! + \! a \! - \! 1}},{\bf{Z}}',{\bf{\tilde Z}},{\bf{U}}_{a \! - \! m}^{a \! - \! 1}}}\left( {{\bf{z}}_{l \cdot n \! + \! a}^{l \cdot n \! + \! a \! + \! m \! - \! 1},{{\bf{x}}^{b \cdot n \! + \! a \! - \! 1}},{\bf{z}}',{\bf{\tilde z}},{\bf{u}}_{a \! - \! m}^{a \! - \! 1}} \right)} 
\notag \\
&\quad  \times
\log \frac{{{p_{{\bf{Z}}_{l \cdot n \! + \! a}^{l \cdot n \! + \! a \! + \! m \! - \! 1}|{\bf{Z}}',{\bf{\tilde Z}},{\bf{U}}_{a \! - \! m}^{a \! - \! 1}}}\left( {{\bf{z}}_{l \cdot n \! + \! a}^{l \cdot n \! + \! a \! + \! m \! - \! 1}|{\bf{z}}',{\bf{\tilde z}},{\bf{u}}_{a \! - \! m}^{a \! - \! 1}} \right)}}{{{p_{{\bf{U}}_{l \cdot n \! + \! a}^{l \cdot n \! + \! a \! + \! m \! - \! 1}|{\bf{U}}_{l \cdot n \! + \! a \! - \! m}^{l \cdot n \! + \! a \! - \! 1},{\bf{U}}_{l \cdot n \! + \! a \! + \! m}^{l \cdot n \! + \! a \! + \! 2m \! - \! 1}}}\left( {{\bf{z}}_{l \cdot n \! + \! a}^{l \cdot n \! + \! a \! + \! m \! - \! 1}|{\bf{z}}_{l \cdot n \! + \! a \! - \! m}^{l \cdot n \! + \! a \! - \! 1},{\bf{z}}_{l \cdot n \! + \! a \! + \! m}^{l \cdot n \! + \! a \! + \! 2m \! - \! 1}} \right)}}d{\bf{z}}_{l \cdot n \! + \! a}^{l \cdot n \! + \! a \! + \! m \! - \! 1}d{{\bf{x}}^{b \cdot n \! + \! a \! - \! 1}}d{\bf{z}}'d{\bf{\tilde z}}d{\bf{u}}_{a \! - \! m}^{a \! - \! 1}. 
\label{eqn:ExtBound1}
\end{align}
Note that 
\begin{align} 
&\int {{p_{{\bf{Z}}_{l \cdot n \! + \! a}^{l \cdot n \! + \! a \! + \! m \! - \! 1},{{\bf{X}}^{b \cdot n \! + \! a \! - \! 1}},{\bf{Z}}',{\bf{\tilde Z}},{\bf{U}}_{a \! - \! m}^{a \! - \! 1}}}\left( {{\bf{z}}_{l \cdot n \! + \! a}^{l \cdot n \! + \! a \! + \! m \! - \! 1},{{\bf{x}}^{b \cdot n \! + \! a \! - \! 1}},{\bf{z}}',{\bf{\tilde z}},{\bf{u}}_{a \! - \! m}^{a \! - \! 1}} \right)} 
\notag \\
&\quad  \times 
\log \frac{{{p_{{\bf{Z}}_{l \cdot n \! + \! a}^{l \cdot n \! + \! a \! + \! m \! - \! 1}|{\bf{Z}}',{\bf{\tilde Z}},{\bf{U}}_{a \! - \! m}^{a \! - \! 1}}}\left( {{\bf{z}}_{l \cdot n \! + \! a}^{l \cdot n \! + \! a \! + \! m \! - \! 1}|{\bf{z}}',{\bf{\tilde z}},{\bf{u}}_{a \! - \! m}^{a \! - \! 1}} \right)}}{{{p_{{\bf{U}}_{l \cdot n \! + \! a}^{l \cdot n \! + \! a \! + \! m \! - \! 1}|{\bf{U}}_{l \cdot n \! + \! a \! - \! m}^{l \cdot n \! + \! a \! - \! 1},{\bf{U}}_{l \cdot n \! + \! a \! + \! m}^{l \cdot n \! + \! a \! + \! 2m \! - \! 1}}}\left( {{\bf{z}}_{l \cdot n \! + \! a}^{l \cdot n \! + \! a \! + \! m \! - \! 1}|{\bf{z}}_{l \cdot n \! + \! a \! - \! m}^{l \cdot n \! + \! a \! - \! 1},{\bf{z}}_{l \cdot n \! + \! a \! + \! m}^{l \cdot n \! + \! a \! + \! 2m \! - \! 1}} \right)}}d{\bf{z}}_{l \cdot n \! + \! a}^{l \cdot n \! + \! a \! + \! m \! - \! 1}d{{\bf{x}}^{b \cdot n \! + \! a \! - \! 1}}d{\bf{z}}'d{\bf{\tilde z}}d{\bf{u}}_{a \! - \! m}^{a \! - \! 1}
\notag \\
&= \int {{p_{{\bf{Z}}_{l \cdot n \! + \! a}^{l \cdot n \! + \! a \! + \! m \! - \! 1},{\bf{Z}}',{\bf{\tilde Z}},{\bf{U}}_{a \! - \! m}^{a \! - \! 1}}}\left( {{\bf{z}}_{l \cdot n \! + \! a}^{l \cdot n \! + \! a \! + \! m \! - \! 1},{\bf{z}}',{\bf{\tilde z}},{\bf{u}}_{a \! - \! m}^{a \! - \! 1}} \right)} 
\notag \\
&\quad  \times 
\log \frac{{{p_{{\bf{Z}}_{l \cdot n \! + \! a}^{l \cdot n \! + \! a \! + \! m \! - \! 1}|{\bf{Z}}',{\bf{\tilde Z}},{\bf{U}}_{a \! - \! m}^{a \! - \! 1}}}\left( {{\bf{z}}_{l \cdot n \! + \! a}^{l \cdot n \! + \! a \! + \! m \! - \! 1}|{\bf{z}}',{\bf{\tilde z}},{\bf{u}}_{a \! - \! m}^{a \! - \! 1}} \right)}}{{{p_{{\bf{U}}_{l \cdot n \! + \! a}^{l \cdot n \! + \! a \! + \! m \! - \! 1}|{\bf{U}}_{l \cdot n \! + \! a \! - \! m}^{l \cdot n \! + \! a \! - \! 1},{\bf{U}}_{l \cdot n \! + \! a \! + \! m}^{l \cdot n \! + \! a \! + \! 2m \! - \! 1}}}\left( {{\bf{z}}_{l \cdot n \! + \! a}^{l \cdot n \! + \! a \! + \! m \! - \! 1}|{\bf{z}}_{l \cdot n \! + \! a \! - \! m}^{l \cdot n \! + \! a \! - \! 1},{\bf{z}}_{l \cdot n \! + \! a \! + \! m}^{l \cdot n \! + \! a \! + \! 2m \! - \! 1}} \right)}}d{\bf{z}}_{l \cdot n \! + \! a}^{l \cdot n \! + \! a \! + \! m \! - \! 1}d{\bf{z}}'d{\bf{\tilde z}}d{\bf{u}}_{a \! - \! m}^{a \! - \! 1}
\notag \\
&= \int \bigg( \int {p_{{\bf{Z}}_{l \cdot n \! + \! a}^{l \cdot n \! + \! a \! + \! m \! - \! 1}|{\bf{Z}}',{\bf{\tilde Z}},{\bf{U}}_{a \! - \! m}^{a \! - \! 1}}}\left( {{\bf{z}}_{l \cdot n \! + \! a}^{l \cdot n \! + \! a \! + \! m \! - \! 1}|{\bf{z}}',{\bf{\tilde z}},{\bf{u}}_{a \! - \! m}^{a \! - \! 1}} \right)
\notag \\
&\qquad  \times 
\log \frac{{{p_{{\bf{Z}}_{l \cdot n \! + \! a}^{l \cdot n \! + \! a \! + \! m \! - \! 1}|{\bf{Z}}',{\bf{\tilde Z}},{\bf{U}}_{a \! - \! m}^{a \! - \! 1}}}\left( {{\bf{z}}_{l \cdot n \! + \! a}^{l \cdot n \! + \! a \! + \! m \! - \! 1}|{\bf{z}}',{\bf{\tilde z}},{\bf{u}}_{a \! - \! m}^{a \! - \! 1}} \right)}}{{{p_{{\bf{U}}_{l \cdot n \! + \! a}^{l \cdot n \! + \! a \! + \! m \! - \! 1}|{\bf{U}}_{l \cdot n \! + \! a \! - \! m}^{l \cdot n \! + \! a \! - \! 1},{\bf{U}}_{l \cdot n \! + \! a \! + \! m}^{l \cdot n \! + \! a \! + \! 2m \! - \! 1}}}\left( {{\bf{z}}_{l \cdot n \! + \! a}^{l \cdot n \! + \! a \! + \! m \! - \! 1}|{\bf{z}}_{l \cdot n \! + \! a \! - \! m}^{l \cdot n \! + \! a \! - \! 1},{\bf{z}}_{l \cdot n \! + \! a \! + \! m}^{l \cdot n \! + \! a \! + \! 2m \! - \! 1}} \right)}}d{\bf{z}}_{l \cdot n \! + \! a}^{l \cdot n \! + \! a \! + \! m \! - \! 1}  \bigg) {p_{{\bf{Z}}',{\bf{\tilde Z}},{\bf{U}}_{a \! - \! m}^{a \! - \! 1}}}\left( {{\bf{z}}',{\bf{\tilde z}},{\bf{u}}_{a \! - \! m}^{a \! - \! 1}} \right)d{\bf{z}}'d{\bf{\tilde z}}d{\bf{u}}_{a \! - \! m}^{a \! - \! 1}
\notag \\
&= \int {{D_{KL}}\left( {\left. {{p_{{\bf{Z}}_{l \cdot n \! + \! a}^{l \cdot n \! + \! a \! + \! m \! - \! 1}|{\bf{Z}}' = {\bf{z}}',{\bf{\tilde Z}} = {\bf{\tilde z}},{\bf{U}}_{a \! - \! m}^{a \! - \! 1} = {\bf{u}}_{a \! - \! m}^{a \! - \! 1}}}} \right\|{p_{{\bf{U}}_{l \cdot n \! + \! a}^{l \cdot n \! + \! a \! + \! m \! - \! 1}|{\bf{U}}_{l \cdot n \! + \! a \! - \! m}^{l \cdot n \! + \! a \! - \! 1} = {\bf{z}}_{l \cdot n \! + \! a \! - \! m}^{l \cdot n \! + \! a \! - \! 1},{\bf{U}}_{l \cdot n \! + \! a \! + \! m}^{l \cdot n \! + \! a \! + \! 2m \! - \! 1} = {\bf{z}}_{l \cdot n \! + \! a \! + \! m}^{l \cdot n \! + \! a \! + \! 2m \! - \! 1}}}} \right)} 
\notag \\
&\quad  \times 
{p_{{\bf{Z}}',{\bf{\tilde Z}},{\bf{U}}_{a \! - \! m}^{a \! - \! 1}}}\left( {{\bf{z}}',{\bf{\tilde z}},{\bf{u}}_{a \! - \! m}^{a \! - \! 1}} \right)d{\bf{z}}'d{\bf{\tilde z}}d{\bf{u}}_{a \! - \! m}^{a \! - \! 1}
\notag \\
&\stackrel{(a)}{\geq} 0,
\label{eqn:ExtBound2}
\end{align}
where $(a)$ follows as $D_{KL}(f||g) \geq 0$ for all densities $f$ and $g$ \cite[Thm. 8.6.1]{Cover:06}. 
It therefore follows that \eqref{eqn:ExtBound2} is non\! - \!negative, thus \eqref{eqn:ExtBound1} implies that 
\begin{align} 
&I\left( {\left. {{{\bf{X}}^{b \cdot n \! + \! a \! - \! 1}};{\bf{Z}}_{l \cdot n \! + \! a}^{l \cdot n \! + \! a \! + \! m \! - \! 1}} \right|{\bf{Z}}_a^{l \cdot n \! + \! a \! - \! 1},{\bf{Z}}_{l \cdot n \! + \! a \! + \! m}^{\left( {l \! + \! 1} \right) \cdot n \! + \! a \! - \! 1}, \ldots ,{\bf{Z}}_{\left( {b \! - \! 1} \right) \cdot n \! + \! a \! + \! m}^{b \cdot n \! + \! a \! - \! 1},{\bf{U}}_{a \! - \! m}^{a \! - \! 1}} \right) 
\notag \\
& \le \int {{p_{{\bf{Z}}_{l \cdot n \! + \! a}^{l \cdot n \! + \! a \! + \! m \! - \! 1},{{\bf{X}}^{b \cdot n \! + \! a \! - \! 1}},{\bf{Z}}',{\bf{\tilde Z}},{\bf{U}}_{a \! - \! m}^{a \! - \! 1}}}\left( {{\bf{z}}_{l \cdot n \! + \! a}^{l \cdot n \! + \! a \! + \! m \! - \! 1},{{\bf{x}}^{b \cdot n \! + \! a \! - \! 1}},{\bf{z}}',{\bf{\tilde z}},{\bf{u}}_{a \! - \! m}^{a \! - \! 1}} \right)} 
\notag \\
&\quad  \times 
\log \frac{{{p_{{\bf{Z}}_{l \cdot n \! + \! a}^{l \cdot n \! + \! a \! + \! m \! - \! 1}|{{\bf{X}}^{b \cdot n \! + \! a \! - \! 1}},{\bf{Z}}',{\bf{\tilde Z}},{\bf{U}}_{a \! - \! m}^{a \! - \! 1}}}\left( {{\bf{z}}_{l \cdot n \! + \! a}^{l \cdot n \! + \! a \! + \! m \! - \! 1}|{{\bf{x}}^{b \cdot n \! + \! a \! - \! 1}},{\bf{z}}',{\bf{\tilde z}},{\bf{u}}_{a \! - \! m}^{a \! - \! 1}} \right)}}{{{p_{{\bf{U}}_{l \cdot n \! + \! a}^{l \cdot n \! + \! a \! + \! m \! - \! 1}|{\bf{U}}_{l \cdot n \! + \! a \! - \! m}^{l \cdot n \! + \! a \! - \! 1},{\bf{U}}_{l \cdot n \! + \! a \! + \! m}^{l \cdot n \! + \! a \! + \! 2m \! - \! 1}}}\left( {{\bf{z}}_{l \cdot n \! + \! a}^{l \cdot n \! + \! a \! + \! m \! - \! 1}|{\bf{z}}_{l \cdot n \! + \! a \! - \! m}^{l \cdot n \! + \! a \! - \! 1},{\bf{z}}_{l \cdot n \! + \! a \! + \! m}^{l \cdot n \! + \! a \! + \! 2m \! - \! 1}} \right)}}d{\bf{z}}_{l \cdot n \! + \! a}^{l \cdot n \! + \! a \! + \! m \! - \! 1}d{{\bf{x}}^{b \cdot n \! + \! a \! - \! 1}}d{\bf{z}}'d{\bf{\tilde z}}d{\bf{u}}_{a \! - \! m}^{a \! - \! 1}.
\label{eqn:ExtBound3}
\end{align}
Define ${\bf{U}}' \triangleq \Big( {\bf{U}}_{l \cdot n \! + \! a \! - \! m}^{l \cdot n \! + \! a \! - \! 1},{\bf{U}}_{l \cdot n \! + \! a \! + \! m}^{l \cdot n \! + \! a \! + \! 2m \! - \! 1} \Big)$ and $\tilde{\bf{U}} \triangleq \Big( {\bf{U}}_a^{l \cdot n \! + \! a \! - \! m},{\bf{U}}_{l \cdot n \! + \! a \! + \! 2m \! - \! 1}^{\left( {l \! + \! 1} \right) \cdot n \! + \! a \! - \! 1},{\bf{U}}_{\left( {l \! + \! 1} \right) \cdot n \! + \! a \! + \! m}^{\left( {l \! + \! 2} \right) \cdot n \! + \! a \! - \! 1}, \ldots ,{\bf{U}}_{\left( {b \! - \! 1} \right) \cdot n \! + \! a \! + \! m}^{b \cdot n \! + \! a \! - \! 1} \Big)$. Note that since the channel is LTI, then $\exists \mathsf{G}, \mathsf{G}_1, \mathsf{G}_2$ such that ${\bf{\tilde Z}} = {\mathsf{G}_1}{{\bf{X}}^{b \cdot n \! + \! a \! - \! 1}} \! + \! {\bf{\tilde U}}$, ${\bf{Z}}' = {\mathsf{G}_2}{\bf{X}}_{l \cdot n \! + \! a \! - \! 2m}^{l \cdot n \! + \! a \! + \! 2m \! - \! 1} \! + \! {\bf{U}}'$, ${\bf{Z}}_{l \cdot n \! + \! a \! - \! m}^{l \cdot n \! + \! a \! - \! 1} = \mathsf{G}{\bf{X}}_{l \cdot n \! + \! a \! - \! 2m}^{l \cdot n \! + \! a \! - \! 1} \! + \! {\bf{U}}_{l \cdot n \! + \! a \! - \! m}^{l \cdot n \! + \! a \! - \! 1}$, ${\bf{Z}}_{l \cdot n \! + \! a}^{l \cdot n \! + \! a \! + \! m \! - \! 1} = \mathsf{G}{\bf{X}}_{l \cdot n \! + \! a \! - \! m}^{l \cdot n \! + \! a \! + \! m \! - \! 1} \! + \! {\bf{U}}_{l \cdot n \! + \! a}^{l \cdot n \! + \! a \! + \! m \! - \! 1}$, and ${\bf{Z}}_{l \cdot n \! + \! a \! + \! m}^{l \cdot n \! + \! a \! + \! 2m \! - \! 1} = \mathsf{G}{\bf{X}}_{l \cdot n \! + \! a}^{l \cdot n \! + \! a \! + \! 2m \! - \! 1} \! + \! {\bf{U}}_{l \cdot n \! + \! a \! + \! m}^{l \cdot n \! + \! a \! + \! 2m \! - \! 1}$. 
Now, 
\begin{align}
&{p_{{\bf{Z}}_{l \cdot n \! + \! a}^{l \cdot n \! + \! a \! + \! m \! - \! 1}|{{\bf{X}}^{b \cdot n \! + \! a \! - \! 1}},{\bf{Z}}',{\bf{\tilde Z}},{\bf{U}}_{a \! - \! m}^{a \! - \! 1}}}\left( {{\bf{z}}_{l \cdot n \! + \! a}^{l \cdot n \! + \! a \! + \! m \! - \! 1}|{{\bf{x}}^{b \cdot n \! + \! a \! - \! 1}},{\bf{z}}',{\bf{\tilde z}},{\bf{u}}_{a \! - \! m}^{a \! - \! 1}} \right)
\notag \\
&= {p_{{\bf{U}}_{l \cdot n \! + \! a}^{l \cdot n \! + \! a \! + \! m \! - \! 1}|{{\bf{X}}^{b \cdot n \! + \! a \! - \! 1}},{\bf{U}}',{\bf{\tilde U}},{\bf{U}}_{a \! - \! m}^{a \! - \! 1}}}\left( {{\bf{z}}_{l \cdot n \! + \! a}^{l \cdot n \! + \! a \! + \! m \! - \! 1} \! - \! \mathsf{G}{\bf{x}}_{l \cdot n \! + \! a \! - \! m}^{l \cdot n \! + \! a \! + \! m \! - \! 1}|{{\bf{x}}^{b \cdot n \! + \! a \! - \! 1}},{\bf{z}}' \! - \! {\mathsf{G}_2}{\bf{x}}_{l \cdot n \! + \! a \! - \! 2m}^{l \cdot n \! + \! a \! + \! 2m \! - \! 1},{\bf{\tilde z}} \! - \! {\mathsf{G}_1}{{\bf{x}}^{b \cdot n \! + \! a \! - \! 1}},{\bf{u}}_{a \! - \! m}^{a \! - \! 1}} \right)
\notag \\ 
&= {p_{{\bf{U}}_{l \cdot n \! + \! a}^{l \cdot n \! + \! a \! + \! m \! - \! 1}|{\bf{U}}'}}\left( {{\bf{z}}_{l \cdot n \! + \! a}^{l \cdot n \! + \! a \! + \! m \! - \! 1} \! - \! \mathsf{G}{\bf{x}}_{l \cdot n \! + \! a \! - \! m}^{l \cdot n \! + \! a \! + \! m \! - \! 1}|{\bf{z}}' \! - \! {\mathsf{G}_2}{\bf{x}}_{l \cdot n \! + \! a \! - \! 2m}^{l \cdot n \! + \! a \! + \! 2m \! - \! 1}} \right) 
\notag \\ 
&= {p_{{\bf{U}}_{l \cdot n \! + \! a}^{l \cdot n \! + \! a \! + \! m \! - \! 1}|{\bf{U}}_{l \cdot n \! + \! a \! - \! m}^{l \cdot n \! + \! a \! - \! 1},{\bf{U}}_{l \cdot n \! + \! a \! + \! m}^{l \cdot n \! + \! a \! + \! 2m \! - \! 1}}}\left( {{\bf{z}}_{l \cdot n \! + \! a}^{l \cdot n \! + \! a \! + \! m \! - \! 1} \! - \! \mathsf{G}{\bf{x}}_{l \cdot n \! + \! a \! - \! m}^{l \cdot n \! + \! a \! + \! m \! - \! 1}|{\bf{z}}_{l \cdot n \! + \! a \! - \! m}^{l \cdot n \! + \! a \! - \! 1} \! - \! \mathsf{G}{\bf{x}}_{l \cdot n \! + \! a \! - \! 2m}^{l \cdot n \! + \! a \! - \! 1},{\bf{z}}_{l \cdot n \! + \! a \! + \! m}^{l \cdot n \! + \! a \! + \! 2m \! - \! 1} \! - \! \mathsf{G}{\bf{x}}_{l \cdot n \! + \! a}^{l \cdot n \! + \! a \! + \! 2m \! - \! 1}} \right).
\label{eqn:ExtBound4}
\end{align}
From the stationarity of ${\bf U}[i]$ it follows that ${\bf{U}}_{l \cdot n \! + \! a}^{l \cdot n \! + \! a \! + \! m \! - \! 1}|{\bf{U}}_{l \cdot n \! + \! a \! - \! m}^{l \cdot n \! + \! a \! - \! 1},{\bf{U}}_{l \cdot n \! + \! a \! + \! m}^{l \cdot n \! + \! a \! + \! 2m \! - \! 1} \stackrel{d}{=} {\bf{U}}_a^{a \! + \! m \! - \! 1}|{\bf{U}}_{a \! - \! m}^{a \! - \! 1},{\bf{U}}_{a \! + \! m}^{a \! + \! 2m \! - \! 1}$. As was shown in the proof of \eqref{eqn:ToProve1ab}, Eqn. \eqref{eqn:ExtBound4} implies that
\begin{align}
&\log \frac{{{p_{{\bf{Z}}_{l \cdot n \! + \! a}^{l \cdot n \! + \! a \! + \! m \! - \! 1}|{{\bf{X}}^{b \cdot n \! + \! a \! - \! 1}},{\bf{Z}}',{\bf{\tilde Z}},{\bf{U}}_{a \! - \! m}^{a \! - \! 1}}}\left( {{\bf{z}}_{l \cdot n \! + \! a}^{l \cdot n \! + \! a \! + \! m \! - \! 1}|{{\bf{x}}^{b \cdot n \! + \! a \! - \! 1}},{\bf{z}}',{\bf{\tilde z}},{\bf{u}}_{a \! - \! m}^{a \! - \! 1}} \right)}}{{{p_{{\bf{U}}_{l \cdot n \! + \! a}^{l \cdot n \! + \! a \! + \! m \! - \! 1}|{\bf{U}}_{l \cdot n \! + \! a \! - \! m}^{l \cdot n \! + \! a \! - \! 1},{\bf{U}}_{l \cdot n \! + \! a \! + \! m}^{l \cdot n \! + \! a \! + \! 2m \! - \! 1}}}\left( {{\bf{z}}_{l \cdot n \! + \! a}^{l \cdot n \! + \! a \! + \! m \! - \! 1}|{\bf{z}}_{l \cdot n \! + \! a \! - \! m}^{l \cdot n \! + \! a \! - \! 1},{\bf{z}}_{l \cdot n \! + \! a \! + \! m}^{l \cdot n \! + \! a \! + \! 2m \! - \! 1}} \right)}}  
\notag \\
&= \log \frac{{{p_{{\bf{U}}_{l \cdot n \! + \! a}^{l \cdot n \! + \! a \! + \! m \! - \! 1}|{\bf{U}}_{l \cdot n \! + \! a \! - \! m}^{l \cdot n \! + \! a \! - \! 1},{\bf{U}}_{l \cdot n \! + \! a \! + \! m}^{l \cdot n \! + \! a \! + \! 2m \! - \! 1}}}\left( {{\bf{z}}_{l \cdot n \! + \! a}^{l \cdot n \! + \! a \! + \! m \! - \! 1} \! - \! \mathsf{G}{\bf{x}}_{l \cdot n \! + \! a \! - \! m}^{l \cdot n \! + \! a \! + \! m \! - \! 1}|{\bf{z}}_{l \cdot n \! + \! a \! - \! m}^{l \cdot n \! + \! a \! - \! 1} \! - \! \mathsf{G}{\bf{x}}_{l \cdot n \! + \! a \! - \! 2m}^{l \cdot n \! + \! a \! - \! 1},{\bf{z}}_{l \cdot n \! + \! a \! + \! m}^{l \cdot n \! + \! a \! + \! 2m \! - \! 1} \! - \! \mathsf{G}{\bf{x}}_{l \cdot n \! + \! a}^{l \cdot n \! + \! a \! + \! 2m \! - \! 1}} \right)}}{{{p_{{\bf{U}}_{l \cdot n \! + \! a}^{l \cdot n \! + \! a \! + \! m \! - \! 1}|{\bf{U}}_{l \cdot n \! + \! a \! - \! m}^{l \cdot n \! + \! a \! - \! 1},{\bf{U}}_{l \cdot n \! + \! a \! + \! m}^{l \cdot n \! + \! a \! + \! 2m \! - \! 1}}}\left( {{\bf{z}}_{l \cdot n \! + \! a}^{l \cdot n \! + \! a \! + \! m \! - \! 1}|{\bf{z}}_{l \cdot n \! + \! a \! - \! m}^{l \cdot n \! + \! a \! - \! 1},{\bf{z}}_{l \cdot n \! + \! a \! + \! m}^{l \cdot n \! + \! a \! + \! 2m \! - \! 1}} \right)}}
\notag \\
&=  \! - \! \frac{1}{2}\log e \cdot {\left( {{\bf{z}}_{l \cdot n \! + \! a}^{l \cdot n \! + \! a \! + \! m \! - \! 1} \! - \! \mathsf{G}{\bf{x}}_{l \cdot n \! + \! a \! - \! m}^{l \cdot n \! + \! a \! + \! m \! - \! 1} \! - \! {\mathsf{M}_1}\left( {{\bf{z}}_{l \cdot n \! + \! a \! - \! m}^{l \cdot n \! + \! a \! - \! 1} \! - \! \mathsf{G}{\bf{x}}_{l \cdot n \! + \! a \! - \! 2m}^{l \cdot n \! + \! a \! - \! 1}} \right) \! - \! {\mathsf{M}_2}\left( {{\bf{z}}_{l \cdot n \! + \! a \! + \! m}^{l \cdot n \! + \! a \! + \! 2m \! - \! 1} \! - \! \mathsf{G}{\bf{x}}_{l \cdot n \! + \! a}^{l \cdot n \! + \! a \! + \! 2m \! - \! 1}} \right)} \right)^T}{\mathsf{Q}^{ \! - \! 1}}
\notag \\
&\quad  \times 
\left( {{\bf{z}}_{l \cdot n \! + \! a}^{l \cdot n \! + \! a \! + \! m \! - \! 1} \! - \! \mathsf{G}{\bf{x}}_{l \cdot n \! + \! a \! - \! m}^{l \cdot n \! + \! a \! + \! m \! - \! 1} \! - \! {\mathsf{M}_1}\left( {{\bf{z}}_{l \cdot n \! + \! a \! - \! m}^{l \cdot n \! + \! a \! - \! 1} \! - \! \mathsf{G}{\bf{x}}_{l \cdot n \! + \! a \! - \! 2m}^{l \cdot n \! + \! a \! - \! 1}} \right) \! - \! {\mathsf{M}_2}\left( {{\bf{z}}_{l \cdot n \! + \! a \! + \! m}^{l \cdot n \! + \! a \! + \! 2m \! - \! 1} \! - \! \mathsf{G}{\bf{x}}_{l \cdot n \! + \! a}^{l \cdot n \! + \! a \! + \! 2m \! - \! 1}} \right)} \right)
\notag \\
& \! + \! \frac{1}{2}\log e \cdot {\left( {{\bf{z}}_{l \cdot n \! + \! a}^{l \cdot n \! + \! a \! + \! m \! - \! 1} \! - \! {\mathsf{M}_1}{\bf{z}}_{l \cdot n \! + \! a \! - \! m}^{l \cdot n \! + \! a \! - \! 1} \! - \! {\mathsf{M}_2}{\bf{z}}_{l \cdot n \! + \! a \! + \! m}^{l \cdot n \! + \! a \! + \! 2m \! - \! 1}} \right)^T}{\mathsf{Q}^{ \! - \! 1}}\left( {{\bf{z}}_{l \cdot n \! + \! a}^{l \cdot n \! + \! a \! + \! m \! - \! 1} \! - \! {\mathsf{M}_1}{\bf{z}}_{l \cdot n \! + \! a \! - \! m}^{l \cdot n \! + \! a \! - \! 1} \! - \! {\mathsf{M}_2}{\bf{z}}_{l \cdot n \! + \! a \! + \! m}^{l \cdot n \! + \! a \! + \! 2m \! - \! 1}} \right)
\notag \\
&= \log e \cdot {\left( {{\bf{z}}_{l \cdot n \! + \! a}^{l \cdot n \! + \! a \! + \! m \! - \! 1} \! - \! {\mathsf{M}_1}{\bf{z}}_{l \cdot n \! + \! a \! - \! m}^{l \cdot n \! + \! a \! - \! 1} \! - \! {\mathsf{M}_2}{\bf{z}}_{l \cdot n \! + \! a \! + \! m}^{l \cdot n \! + \! a \! + \! 2m \! - \! 1}} \right)^T}{\mathsf{Q}^{ \! - \! 1}}\left( {\mathsf{G}{\bf{x}}_{l \cdot n \! + \! a \! - \! m}^{l \cdot n \! + \! a \! + \! m \! - \! 1} \! - \! {\mathsf{M}_1}\mathsf{G}{\bf{x}}_{l \cdot n \! + \! a \! - \! 2m}^{l \cdot n \! + \! a \! - \! 1} \! - \! {\mathsf{M}_2}\mathsf{G}{\bf{x}}_{l \cdot n \! + \! a}^{l \cdot n \! + \! a \! + \! 2m \! - \! 1}} \right)
\notag \\
& \! - \! \frac{1}{2}\log e \cdot {\left( {\mathsf{G}{\bf{x}}_{l \cdot n \! + \! a \! - \! m}^{l \cdot n \! + \! a \! + \! m \! - \! 1} \! - \! {\mathsf{M}_1}\mathsf{G}{\bf{x}}_{l \cdot n \! + \! a \! - \! 2m}^{l \cdot n \! + \! a \! - \! 1} \! - \! {\mathsf{M}_2}\mathsf{G}{\bf{x}}_{l \cdot n \! + \! a}^{l \cdot n \! + \! a \! + \! 2m \! - \! 1}} \right)^T}{\mathsf{Q}^{ \! - \! 1}}\left( {\mathsf{G}{\bf{x}}_{l \cdot n \! + \! a \! - \! m}^{l \cdot n \! + \! a \! + \! m \! - \! 1} \! - \! {\mathsf{M}_1}\mathsf{G}{\bf{x}}_{l \cdot n \! + \! a \! - \! 2m}^{l \cdot n \! + \! a \! - \! 1} \! - \! {\mathsf{M}_2}\mathsf{G}{\bf{x}}_{l \cdot n \! + \! a}^{l \cdot n \! + \! a \! + \! 2m \! - \! 1}} \right).
\label{eqn:ExtBound5}
\end{align}
Plugging \eqref{eqn:ExtBound5} into \eqref{eqn:ExtBound3} yields
\begin{align}
&I\left( {\left. {{{\bf{X}}^{b \cdot n \! + \! a \! - \! 1}};{\bf{Z}}_{l \cdot n \! + \! a}^{l \cdot n \! + \! a \! + \! m \! - \! 1}} \right|{\bf{Z}}_a^{l \cdot n \! + \! a \! - \! 1},{\bf{Z}}_{l \cdot n \! + \! a \! + \! m}^{\left( {l \! + \! 1} \right) \cdot n \! + \! a \! - \! 1}, \ldots ,{\bf{Z}}_{\left( {b \! - \! 1} \right) \cdot n \! + \! a \! + \! m}^{b \cdot n \! + \! a \! - \! 1},{\bf{U}}_{a \! - \! m}^{a \! - \! 1}} \right) 
\notag \\
& \le \log e \cdot \int {{p_{{\bf{Z}}_{l \cdot n \! + \! a}^{l \cdot n \! + \! a \! + \! m \! - \! 1},{{\bf{X}}^{b \cdot n \! + \! a \! - \! 1}},{\bf{Z}}',{\bf{\tilde Z}},{\bf{U}}_{a \! - \! m}^{a \! - \! 1}}}\left( {{\bf{z}}_{l \cdot n \! + \! a}^{l \cdot n \! + \! a \! + \! m \! - \! 1},{{\bf{x}}^{b \cdot n \! + \! a \! - \! 1}},{\bf{z}}',{\bf{\tilde z}},{\bf{u}}_{a \! - \! m}^{a \! - \! 1}} \right)} {\left( {{\bf{z}}_{l \cdot n \! + \! a}^{l \cdot n \! + \! a \! + \! m \! - \! 1} \! - \! {\mathsf{M}_1}{\bf{z}}_{l \cdot n \! + \! a \! - \! m}^{l \cdot n \! + \! a \! - \! 1} \! - \! {\mathsf{M}_2}{\bf{z}}_{l \cdot n \! + \! a \! + \! m}^{l \cdot n \! + \! a \! + \! 2m \! - \! 1}} \right)^T}{\mathsf{Q}^{ \! - \! 1}}
\notag \\
&\quad  \times 
\left( {\mathsf{G}{\bf{x}}_{l \cdot n \! + \! a \! - \! m}^{l \cdot n \! + \! a \! + \! m \! - \! 1} \! - \! {\mathsf{M}_1}\mathsf{G}{\bf{x}}_{l \cdot n \! + \! a \! - \! 2m}^{l \cdot n \! + \! a \! - \! 1} \! - \! {\mathsf{M}_2}\mathsf{G}{\bf{x}}_{l \cdot n \! + \! a}^{l \cdot n \! + \! a \! + \! 2m \! - \! 1}} \right)d{\bf{z}}_{l \cdot n \! + \! a}^{l \cdot n \! + \! a \! + \! m \! - \! 1}d{{\bf{x}}^{b \cdot n \! + \! a \! - \! 1}}d{\bf{z}}'d{\bf{\tilde z}}d{\bf{u}}_{a \! - \! m}^{a \! - \! 1}
\notag \\
& \! - \! \frac{1}{2}\log e \cdot \int {{p_{{\bf{Z}}_{l \cdot n \! + \! a}^{l \cdot n \! + \! a \! + \! m \! - \! 1},{{\bf{X}}^{b \cdot n \! + \! a \! - \! 1}},{\bf{Z}}',{\bf{\tilde Z}},{\bf{U}}_{a \! - \! m}^{a \! - \! 1}}}\left( {{\bf{z}}_{l \cdot n \! + \! a}^{l \cdot n \! + \! a \! + \! m \! - \! 1},{{\bf{x}}^{b \cdot n \! + \! a \! - \! 1}},{\bf{z}}',{\bf{\tilde z}},{\bf{u}}_{a \! - \! m}^{a \! - \! 1}} \right)} {\left( {\mathsf{G}{\bf{x}}_{l \cdot n \! + \! a \! - \! m}^{l \cdot n \! + \! a \! + \! m \! - \! 1} \! - \! {\mathsf{M}_1}\mathsf{G}{\bf{x}}_{l \cdot n \! + \! a \! - \! 2m}^{l \cdot n \! + \! a \! - \! 1} \! - \! {\mathsf{M}_2}\mathsf{G}{\bf{x}}_{l \cdot n \! + \! a}^{l \cdot n \! + \! a \! + \! 2m \! - \! 1}} \right)^T}
\notag \\
&\qquad \quad  \times 
{\mathsf{Q}^{ \! - \! 1}} \left( {\mathsf{G}{\bf{x}}_{l \cdot n \! + \! a \! - \! m}^{l \cdot n \! + \! a \! + \! m \! - \! 1} \! - \! {\mathsf{M}_1}\mathsf{G}{\bf{x}}_{l \cdot n \! + \! a \! - \! 2m}^{l \cdot n \! + \! a \! - \! 1} \! - \! {\mathsf{M}_2}\mathsf{G}{\bf{x}}_{l \cdot n \! + \! a}^{l \cdot n \! + \! a \! + \! 2m \! - \! 1}} \right)d{\bf{z}}_{l \cdot n \! + \! a}^{l \cdot n \! + \! a \! + \! m \! - \! 1}d{{\bf{x}}^{b \cdot n \! + \! a \! - \! 1}}d{\bf{z}}'d{\bf{\tilde z}}d{\bf{u}}_{a \! - \! m}^{a \! - \! 1}
\notag \\
& = \log e \cdot \int {{p_{{\bf{Z}}_{l \cdot n \! + \! a \! - \! m}^{l \cdot n \! + \! a \! + \! 2m \! - \! 1},{\bf{X}}_{l \cdot n \! + \! a \! - \! 2m}^{l \cdot n \! + \! a \! + \! 2m \! - \! 1}}}\left( {{\bf{z}}_{l \cdot n \! + \! a \! - \! m}^{l \cdot n \! + \! a \! + \! 2m \! - \! 1},{\bf{x}}_{l \cdot n \! + \! a \! - \! 2m}^{l \cdot n \! + \! a \! + \! 2m \! - \! 1}} \right)} {\left( {{\bf{z}}_{l \cdot n \! + \! a}^{l \cdot n \! + \! a \! + \! m \! - \! 1} \! - \! {\mathsf{M}_1}{\bf{z}}_{l \cdot n \! + \! a \! - \! m}^{l \cdot n \! + \! a \! - \! 1} \! - \! {\mathsf{M}_2}{\bf{z}}_{l \cdot n \! + \! a \! + \! m}^{l \cdot n \! + \! a \! + \! 2m \! - \! 1}} \right)^T}{\mathsf{Q}^{ \! - \! 1}} 
\notag \\
&\qquad \quad  \times 
\left( {\mathsf{G}{\bf{x}}_{l \cdot n \! + \! a \! - \! m}^{l \cdot n \! + \! a \! + \! m \! - \! 1} \! - \! {\mathsf{M}_1}\mathsf{G}{\bf{x}}_{l \cdot n \! + \! a \! - \! 2m}^{l \cdot n \! + \! a \! - \! 1} \! - \! {\mathsf{M}_2}\mathsf{G}{\bf{x}}_{l \cdot n \! + \! a}^{l \cdot n \! + \! a \! + \! 2m \! - \! 1}} \right)d{\bf{z}}_{l \cdot n \! + \! a \! - \! m}^{l \cdot n \! + \! a \! + \! 2m \! - \! 1}d{\bf{x}}_{l \cdot n \! + \! a \! - \! 2m}^{l \cdot n \! + \! a \! + \! 2m \! - \! 1}
\notag \\
& \! - \! \frac{1}{2}\log e \cdot \int {{p_{{\bf{X}}_{l \cdot n \! + \! a \! - \! 2m}^{l \cdot n \! + \! a \! + \! 2m \! - \! 1}}}\left( {{\bf{x}}_{l \cdot n \! + \! a \! - \! 2m}^{l \cdot n \! + \! a \! + \! 2m \! - \! 1}} \right)} {\left( {\mathsf{G}{\bf{x}}_{l \cdot n \! + \! a \! - \! m}^{l \cdot n \! + \! a \! + \! m \! - \! 1} \! - \! {\mathsf{M}_1}\mathsf{G}{\bf{x}}_{l \cdot n \! + \! a \! - \! 2m}^{l \cdot n \! + \! a \! - \! 1} \! - \! {\mathsf{M}_2}\mathsf{G}{\bf{x}}_{l \cdot n \! + \! a}^{l \cdot n \! + \! a \! + \! 2m \! - \! 1}} \right)^T}{\mathsf{Q}^{ \! - \! 1}}  
\notag \\
&\qquad \quad  \times 
\left( {\mathsf{G}{\bf{x}}_{l \cdot n \! + \! a \! - \! m}^{l \cdot n \! + \! a \! + \! m \! - \! 1} \! - \! {\mathsf{M}_1}\mathsf{G}{\bf{x}}_{l \cdot n \! + \! a \! - \! 2m}^{l \cdot n \! + \! a \! - \! 1} \! - \! {\mathsf{M}_2}\mathsf{G}{\bf{x}}_{l \cdot n \! + \! a}^{l \cdot n \! + \! a \! + \! 2m \! - \! 1}} \right)d{\bf{x}}_{l \cdot n \! + \! a \! - \! 2m}^{l \cdot n \! + \! a \! + \! 2m \! - \! 1}.
\label{eqn:ExtBound6}
\end{align}
Next, note that 
\begin{align}
&\int {{p_{{\bf{Z}}_{l \cdot n \! + \! a \! - \! m}^{l \cdot n \! + \! a \! + \! 2m \! - \! 1},{\bf{X}}_{l \cdot n \! + \! a \! - \! 2m}^{l \cdot n \! + \! a \! + \! 2m \! - \! 1}}}\left( {{\bf{z}}_{l \cdot n \! + \! a \! - \! m}^{l \cdot n \! + \! a \! + \! 2m \! - \! 1},{\bf{x}}_{l \cdot n \! + \! a \! - \! 2m}^{l \cdot n \! + \! a \! + \! 2m \! - \! 1}} \right)} {\left( {{\bf{z}}_{l \cdot n \! + \! a}^{l \cdot n \! + \! a \! + \! m \! - \! 1} \! - \! {\mathsf{M}_1}{\bf{z}}_{l \cdot n \! + \! a \! - \! m}^{l \cdot n \! + \! a \! - \! 1} \! - \! {\mathsf{M}_2}{\bf{z}}_{l \cdot n \! + \! a \! + \! m}^{l \cdot n \! + \! a \! + \! 2m \! - \! 1}} \right)^T}{\mathsf{Q}^{ \! - \! 1}}
\notag \\
&\quad  \times 
\left( {\mathsf{G}{\bf{x}}_{l \cdot n \! + \! a \! - \! m}^{l \cdot n \! + \! a \! + \! m \! - \! 1} \! - \! {\mathsf{M}_1}\mathsf{G}{\bf{x}}_{l \cdot n \! + \! a \! - \! 2m}^{l \cdot n \! + \! a \! - \! 1} \! - \! {\mathsf{M}_2}\mathsf{G}{\bf{x}}_{l \cdot n \! + \! a}^{l \cdot n \! + \! a \! + \! 2m \! - \! 1}} \right)d{\bf{z}}_{l \cdot n \! + \! a \! - \! m}^{l \cdot n \! + \! a \! + \! 2m \! - \! 1}d{\bf{x}}_{l \cdot n \! + \! a \! - \! 2m}^{l \cdot n \! + \! a \! + \! 2m \! - \! 1}
\notag \\
& = \int \left( {\int {{p_{{\bf{Z}}_{l \cdot n \! + \! a \! - \! m}^{l \cdot n \! + \! a \! + \! 2m \! - \! 1}|{\bf{X}}_{l \cdot n \! + \! a \! - \! 2m}^{l \cdot n \! + \! a \! + \! 2m \! - \! 1}}}\left( {{\bf{z}}_{l \cdot n \! + \! a \! - \! m}^{l \cdot n \! + \! a \! + \! 2m \! - \! 1}|{\bf{x}}_{l \cdot n \! + \! a \! - \! 2m}^{l \cdot n \! + \! a \! + \! 2m \! - \! 1}} \right)} {{\left( {{\bf{z}}_{l \cdot n \! + \! a}^{l \cdot n \! + \! a \! + \! m \! - \! 1} \! - \! {\mathsf{M}_1}{\bf{z}}_{l \cdot n \! + \! a \! - \! m}^{l \cdot n \! + \! a \! - \! 1} \! - \! {\mathsf{M}_2}{\bf{z}}_{l \cdot n \! + \! a \! + \! m}^{l \cdot n \! + \! a \! + \! 2m \! - \! 1}} \right)}^T}d{\bf{z}}_{l \cdot n \! + \! a \! - \! m}^{l \cdot n \! + \! a \! + \! 2m \! - \! 1}} \right){\mathsf{Q}^{ \! - \! 1}}
\notag \\
&\qquad \quad  \times 
\left( {\mathsf{G}{\bf{x}}_{l \cdot n \! + \! a \! - \! m}^{l \cdot n \! + \! a \! + \! m \! - \! 1} \! - \! {\mathsf{M}_1}\mathsf{G}{\bf{x}}_{l \cdot n \! + \! a \! - \! 2m}^{l \cdot n \! + \! a \! - \! 1} \! - \! {\mathsf{M}_2}\mathsf{G}{\bf{x}}_{l \cdot n \! + \! a}^{l \cdot n \! + \! a \! + \! 2m \! - \! 1}} \right) {p_{{\bf{X}}_{l \cdot n \! + \! a \! - \! 2m}^{l \cdot n \! + \! a \! + \! 2m \! - \! 1}}}\left( {{\bf{x}}_{l \cdot n \! + \! a \! - \! 2m}^{l \cdot n \! + \! a \! + \! 2m \! - \! 1}} \right)d{\bf{x}}_{l \cdot n \! + \! a \! - \! 2m}^{l \cdot n \! + \! a \! + \! 2m \! - \! 1}
\notag \\
& = \int \bigg( \E\left\{ {{\bf{Z}}_{l \cdot n \! + \! a}^{l \cdot n \! + \! a \! + \! m \! - \! 1}|{\bf{X}}_{l \cdot n \! + \! a \! - \! 2m}^{l \cdot n \! + \! a \! + \! 2m \! - \! 1} \!=\! {\bf{x}}_{l \cdot n \! + \! a \! - \! 2m}^{l \cdot n \! + \! a \! + \! 2m \! - \! 1}} \right\} \! - \! {\mathsf{M}_1}\E\left\{ {{\bf{Z}}_{l \cdot n \! + \! a \! - \! m}^{l \cdot n \! + \! a \! - \! 1}|{\bf{X}}_{l \cdot n \! + \! a \! - \! 2m}^{l \cdot n \! + \! a \! + \! 2m \! - \! 1} \!=\! {\bf{x}}_{l \cdot n \! + \! a \! - \! 2m}^{l \cdot n \! + \! a \! + \! 2m \! - \! 1}} \right\} 
\notag \\
&\qquad  - \! {\mathsf{M}_2}\E\left\{ {{\bf{Z}}_{l \cdot n \! + \! a \! + \! m}^{l \cdot n \! + \! a \! + \! 2m \! - \! 1}|{\bf{X}}_{l \cdot n \! + \! a \! - \! 2m}^{l \cdot n \! + \! a \! + \! 2m \! - \! 1} \!=\! {\bf{x}}_{l \cdot n \! + \! a \! - \! 2m}^{l \cdot n \! + \! a \! + \! 2m \! - \! 1}} \right\} \bigg)^T{\mathsf{Q}^{ \! - \! 1}}\left( {\mathsf{G}{\bf{x}}_{l \cdot n \! + \! a \! - \! m}^{l \cdot n \! + \! a \! + \! m \! - \! 1} \! - \! {\mathsf{M}_1}\mathsf{G}{\bf{x}}_{l \cdot n \! + \! a \! - \! 2m}^{l \cdot n \! + \! a \! - \! 1} \! - \! {\mathsf{M}_2}\mathsf{G}{\bf{x}}_{l \cdot n \! + \! a}^{l \cdot n \! + \! a \! + \! 2m \! - \! 1}} \right)
\notag \\
&\qquad \quad  \times
{p_{{\bf{X}}_{l \cdot n \! + \! a \! - \! 2m}^{l \cdot n \! + \! a \! + \! 2m \! - \! 1}}}\left( {{\bf{x}}_{l \cdot n \! + \! a \! - \! 2m}^{l \cdot n \! + \! a \! + \! 2m \! - \! 1}} \right)d{\bf{x}}_{l \cdot n \! + \! a \! - \! 2m}^{l \cdot n \! + \! a \! + \! 2m \! - \! 1} 
\notag \\
& = \int {\left( {\mathsf{G}{\bf{x}}_{l \cdot n \! + \! a \! - \! m}^{l \cdot n \! + \! a \! + \! m \! - \! 1} \! - \! {\mathsf{M}_1}\mathsf{G}{\bf{x}}_{l \cdot n \! + \! a \! - \! 2m}^{l \cdot n \! + \! a \! - \! 1} \! - \! {\mathsf{M}_2}\mathsf{G}{\bf{x}}_{l \cdot n \! + \! a}^{l \cdot n \! + \! a \! + \! 2m \! - \! 1}} \right){\mathsf{Q}^{ \! - \! 1}}{{\left( {\mathsf{G}{\bf{x}}_{l \cdot n \! + \! a \! - \! m}^{l \cdot n \! + \! a \! + \! m \! - \! 1} \! - \! {\mathsf{M}_1}\mathsf{G}{\bf{x}}_{l \cdot n \! + \! a \! - \! 2m}^{l \cdot n \! + \! a \! - \! 1} \! - \! {\mathsf{M}_2}\mathsf{G}{\bf{x}}_{l \cdot n \! + \! a}^{l \cdot n \! + \! a \! + \! 2m \! - \! 1}} \right)}^T}} 
\notag \\
&\qquad \quad  \times 
{p_{{\bf{X}}_{l \cdot n \! + \! a \! - \! 2m}^{l \cdot n \! + \! a \! + \! 2m \! - \! 1}}}\left( {{\bf{x}}_{l \cdot n \! + \! a \! - \! 2m}^{l \cdot n \! + \! a \! + \! 2m \! - \! 1}} \right)d{\bf{x}}_{l \cdot n \! + \! a \! - \! 2m}^{l \cdot n \! + \! a \! + \! 2m \! - \! 1}.
\label{eqn:ExtBound7}
\end{align}
Thus, plugging \eqref{eqn:ExtBound7} into \eqref{eqn:ExtBound6} results in
\begin{align}
&I\left( {\left. {{{\bf{X}}^{b \cdot n \! + \! a \! - \! 1}};{\bf{Z}}_{l \cdot n \! + \! a}^{l \cdot n \! + \! a \! + \! m \! - \! 1}} \right|{\bf{Z}}_a^{l \cdot n \! + \! a \! - \! 1},{\bf{Z}}_{l \cdot n \! + \! a \! + \! m}^{\left( {l \! + \! 1} \right) \cdot n \! + \! a \! - \! 1}, \ldots ,{\bf{Z}}_{\left( {b \! - \! 1} \right) \cdot n \! + \! a \! + \! m}^{b \cdot n \! + \! a \! - \! 1},{\bf{U}}_{a \! - \! m}^{a \! - \! 1}} \right) 
\notag \\
&\le \frac{1}{2}\log e \cdot \int {{p_{{\bf{X}}_{l \cdot n \! + \! a \! - \! 2m}^{l \cdot n \! + \! a \! + \! 2m \! - \! 1}}}\left( {{\bf{x}}_{l \cdot n \! + \! a \! - \! 2m}^{l \cdot n \! + \! a \! + \! 2m \! - \! 1}} \right)} {\left( {\mathsf{G}{\bf{x}}_{l \cdot n \! + \! a \! - \! m}^{l \cdot n \! + \! a \! + \! m \! - \! 1} \! - \! {\mathsf{M}_1}\mathsf{G}{\bf{x}}_{l \cdot n \! + \! a \! - \! 2m}^{l \cdot n \! + \! a \! - \! 1} \! - \! {\mathsf{M}_2}\mathsf{G}{\bf{x}}_{l \cdot n \! + \! a}^{l \cdot n \! + \! a \! + \! 2m \! - \! 1}} \right)^T}{\mathsf{Q}^{ \! - \! 1}}
\notag \\
&\qquad \quad  \times 
\left( {\mathsf{G}{\bf{x}}_{l \cdot n \! + \! a \! - \! m}^{l \cdot n \! + \! a \! + \! m \! - \! 1} \! - \! {\mathsf{M}_1}\mathsf{G}{\bf{x}}_{l \cdot n \! + \! a \! - \! 2m}^{l \cdot n \! + \! a \! - \! 1} \! - \! {\mathsf{M}_2}\mathsf{G}{\bf{x}}_{l \cdot n \! + \! a}^{l \cdot n \! + \! a \! + \! 2m \! - \! 1}} \right)d{\bf{x}}_{l \cdot n \! + \! a \! - \! 2m}^{l \cdot n \! + \! a \! + \! 2m \! - \! 1}
\notag \\
&= \frac{1}{2}\log e \cdot \E\bigg\{ {{\left( {\mathsf{G}{\bf{X}}_{l \cdot n \! + \! a \! - \! m}^{l \cdot n \! + \! a \! + \! m \! - \! 1} \! - \! {\mathsf{M}_1}\mathsf{G}{\bf{X}}_{l \cdot n \! + \! a \! - \! 2m}^{l \cdot n \! + \! a \! - \! 1} \! - \! {\mathsf{M}_2}\mathsf{G}{\bf{X}}_{l \cdot n \! + \! a}^{l \cdot n \! + \! a \! + \! 2m \! - \! 1}} \right)}^T}{\mathsf{Q}^{ \! - \! 1}}
\notag \\
&\qquad \quad  \times 
\left( {\mathsf{G}{\bf{X}}_{l \cdot n \! + \! a \! - \! m}^{l \cdot n \! + \! a \! + \! m \! - \! 1} \! - \! {\mathsf{M}_1}\mathsf{G}{\bf{X}}_{l \cdot n \! + \! a \! - \! 2m}^{l \cdot n \! + \! a \! - \! 1} \! - \! {\mathsf{M}_2}\mathsf{G}{\bf{X}}_{l \cdot n \! + \! a}^{l \cdot n \! + \! a \! + \! 2m \! - \! 1}} \right) \bigg\}.
\label{eqn:ExtBound8}
\end{align}
Note that the expression in \eqref{eqn:ExtBound8} is the mean of a quadratic function of $4m$ channel inputs, which depends only on the noise correlation function through $\mathsf{Q}$, the channel transfer function through $\mathsf{G}$, and the power of the input signal. Consequently, as was shown in \eqref{eqn:Bound8}, since the input signal ${\bf X}[i]$ is subject to a per-symbol power constraint, 
it follows that $\exists \eta$ finite such that $I\Big( \left. {{\bf{X}}^{b \cdot n \! + \! a \! - \! 1}};{\bf{Z}}_{l \cdot n \! + \! a}^{l \cdot n \! + \! a \! + \! m \! - \! 1} \right|{\bf{Z}}_a^{l \cdot n \! + \! a \! - \! 1},{\bf{Z}}_{l \cdot n \! + \! a \! + \! m}^{\left( {l \! + \! 1} \right) \cdot n \! + \! a \! - \! 1}, \ldots ,{\bf{Z}}_{\left( {b \! - \! 1} \right) \cdot n \! + \! a \! + \! m}^{b \cdot n \! + \! a \! - \! 1},{\bf{U}}_{a \! - \! m}^{a \! - \! 1} \Big)  \leq \eta$. 
Hence, the lemma follows.
\fi
\end{IEEEproof}
\smallskip

\begin{lemma}
\label{lem:Proof_EtaBound2} 
There exists a finite and fixed $\tilde{\eta} >0$, such that for any positive integer $n > 2m$, and for all initial states ${\bf s}_0 \in \mathcal{S}_0$, it holds that
\begin{equation}
I\left( \left.{{\bf{X}}^{n-1}};{\bf{Y}}^{m-1}\right|{{\bf{Y}}_m^{n-1}}, {\bf S}_0 = {\bf s}_0 \right) \le \tilde{\eta}. \label{eqn:ToProve1a}
\end{equation}
\end{lemma}
\begin{IEEEproof}
Note that 
\vspace{-0.15cm}
\begin{align}
\!\!\!I\!\left( {\left. {{{\bf{X}}^{n \! - \!  1}};{{\bf{Y}}^{m \! - \!  1}}} \right|{\bf{Y}}_m^{n \! - \!  1},{{\bf{S}}_0} \! = \!  {{\bf{s}}_0}} \right) 
&\! = \!  h \mspace{-3mu} \left(\! {\left. {{{\bf{Y}}^{m \! - \!  1}}} \right|{\bf{Y}}_m^{n \! - \!  1},{{\bf{S}}_0} \! = \!  {{\bf{s}}_0}} \right) \mspace{-3mu} - \mspace{-3mu}  h \mspace{-3mu}\left(\! {\left. {{{\bf{Y}}^{m \! - \!  1}}} \right|{{\bf{X}}^{n \! - \!  1}},{\bf{Y}}_m^{n \! - \!  1},{{\bf{S}}_0} \! = \!  {{\bf{s}}_0}} \right) \notag \\
&\!\stackrel{(a)}{\leq} \mspace{-3mu} h \mspace{-3mu} \left(\! {\left. {{{\bf{Y}}^{m \! - \!  1}}} \mspace{-2mu} \right|{{\bf{S}}_0} \! = \!  {{\bf{s}}_0}} \right) \mspace{-3mu} - \mspace{-4mu}  h \mspace{-4mu}\left(\! {\left. {{{\bf{Y}}^{m \! - \!  1}}} \mspace{-2mu} \right| \mspace{-2mu}{{\bf{X}}^{n \! - \!  1}}, \mspace{-2mu} {\bf{Y}}_m^{n \! - \!  1},{{\bf{S}}_0} \mspace{-3mu} = \mspace{-3mu}  {{\bf{s}}_0}} \right)\mspace{-3mu},
\label{eqn:SecBound1}
\end{align}
where $(a)$ follows since conditioning reduces entropy \cite[Ch. 8.6]{Cover:06}. 
From the input-output relationship of the LGMWTC it follows that 
$\exists \mathsf{H}_1,\mathsf{H}_0 \in \mathds{R}^{\left(n_r \cdot m\right) \times \left(n_t \cdot m\right)}$ such that ${\bf{Y}}^{m - 1} = \mathsf{H}_1 {\bf X}^{m-1} + \mathsf{H}_0 {\bf X}_{-m}^{-1}  + {\bf W}^{m-1}$ and $\exists {\mathsf{\check H}} \in \mathds{R}^{\left(n_r \cdot (n\!-\! m)\right) \times \left(n_t \cdot n\right)}$ such that ${\bf{Y}}_m^{n - 1} = {\mathsf{\check H}} {\bf X}^{n -1} + {\bf{W}}_m^{n - 1}$. 
Therefore,
\vspace{-0.15cm}
\begin{align}
h\left( {\left. {{{\bf{Y}}^{m - 1}}} \right|{{\bf{X}}^{n - 1}},{\bf{Y}}_m^{n - 1},{{\bf{S}}_0} = {{\bf{s}}_0}} \right) 
&= h\left( {\left. {\mathsf{H}_1{{\bf{X}}^{m - 1}}\! +\! \mathsf{H}_0{\bf{X}}_{ - m}^{ - 1} + {{\bf{W}}^{m - 1}}} \right|{{\bf{X}}^{n - 1}},{\bf{Y}}_m^{n - 1},{{\bf{S}}_0} \!=\! {{\bf{s}}_0}} \right)
\notag \\
&\stackrel{(a)}{=} h\left( {\left. {{{\bf{W}}^{m - 1}}} \right|{{\bf{X}}^{n - 1}}, {\bf{W}}_m^{n - 1},{\bf{W}}_{ - m}^{ - 1} = {\bf{w}}_{ - m}^{ - 1}} \right) \notag \\
&\stackrel{(b)}{=} h\left( {\left. {{{\bf{W}}^{m - 1}}} \right|{\bf{W}}_m^{n - 1},{\bf{W}}_{ - m}^{ - 1} = {\bf{w}}_{ - m}^{ - 1}} \right)
\notag \\
&\stackrel{(c)}{=} h\left( \left. {{{\bf{W}}^{m - 1}}} \right|{\bf{W}}_m^{2m - 1},{\bf{W}}_{ - m}^{ - 1} = {\bf{w}}_{ - m}^{ - 1} \right) \notag \\
&= \int\limits_{{\bf{w}}_m^{2m \! - \! 1} \in \dsR^{n_r \cdot m}}  h\left( \left. {{{\bf{W}}^{m \! - \! 1}}} \right|{\bf{W}}_m^{2m \! - \! 1} \!= \! {\bf{w}}_m^{2m \! - \! 1},{\bf{W}}_{ \! - \! m}^{ \! - \! 1} \!= \! {\bf{w}}_{ \! - \! m}^{ \! - \! 1} \right) \notag \\
&\qquad\qquad\qquad\qquad\qquad \times p_{{\bf{W}}_m^{2m \! - \! 1}}\left({\bf{w}}_m^{2m \! - \! 1}\right)d{\bf{w}}_m^{2m \! - \! 1},
\label{eqn:SecBound2}
\end{align}
where $(a)$  follows as ${{\bf S} _0} \mspace{-3mu} = \mspace{-3mu} \Big[ \left({\bf X} _{ - m}^{-1}\right)^T,\left({\bf W} _{ - m}^{-1}\right)^T,$ $ \left({\bf U} _{ - m}^{-1}\right)^T \Big]^T$;
$(b)$ follows since the noise ${\bf W}[i]$ is independent of the channel input ${\bf X}[i]$;
$(c)$ follows since the temporal correlation of the multivariate Gaussian process ${\bf W}[i]$ is finite and shorter than $m+1$, and therefore ${\bf{W}}^{m - 1}$ is independent of ${\bf{W}}_{2m}^{n - 1}$. 
Since ${\bf{W}}^{m - 1}$ and $\left[\left({\bf{W}}_m^{2m \!-\! 1}\right)^T,\left({\bf{W}}_{\!-\! m}^{\!-\! 1}\right)^T\right]^T$ are jointly Gaussian, the conditional distribution $\left. {\bf{W}}^{m - 1} \right|{\bf{W}}_m^{2m - 1} = {\bf{w}}_m^{2m - 1},{\bf{W}}_{ - m}^{ - 1} = {\bf{w}}_{ - m}^{ - 1}$  is a multivariate Gaussian distribution \cite[Proposition 3.13]{Eaton:07}, with covariance matrix $\mathsf{\tilde Q} \in \dsR^{(n_r m)\times (n_r m)}$ given by
\vspace{-0.15cm}
\begin{align}
&\mathsf{\tilde Q}\!
\triangleq \! \E\left\{\!{\bf{W}}^{m \! - \! 1}\!\left({\bf{W}}^{m \! - \! 1}\right)^T\right\}\! - \! \E\left\{\!{\bf{W}}^{m \! - \! 1}\!\left[\!\left({\bf{W}}_m^{2m \!-\! 1}\right)^T\!\!\!,\!\!\left({\bf{W}}_{\!-\! m}^{\!-\! 1}\right)^T\right]\!\right\} \notag \\
&\quad\times\!
\Bigg(\!\E\bigg\{\!\left[\left({\bf{W}}_m^{2m \!-\! 1}\right)^T\!\!,\left({\bf{W}}_{\!-\! m}^{\!-\! 1}\right)^T\right]^T\!
\left[\left({\bf{W}}_m^{2m \!-\! 1}\right)^T\!\!,\left({\bf{W}}_{\!-\! m}^{\!-\! 1}\right)^T\right]\!\bigg\}\!\Bigg)^{\! - \!1} \notag \\
&\quad\times \!\E\left\{\left[\left({\bf{W}}_m^{2m \!-\! 1}\right)^T\!\!,\left({\bf{W}}_{\!-\! m}^{\!-\! 1}\right)^T\right]^T\!\left({\bf{W}}^{m \! - \! 1}\right)^T\right\}.
\label{eqn:SecBound3}
\end{align}
We note that as the noise samples are not linearly dependent\footnote{Note that for any pair of jointly-Gaussian real-valued random vectors {\bf A} and {\bf B}, such that the entries of $\left[{\bf A}^T,{\bf B}^T\right]^T$ are not linearly dependent, it follows from \cite[Ch. 3.5]{Gallager:13} that the entries of ${\bf{A}}$ conditioned on  ${\bf{B}} = {\bf{b}}$ are also not linearly dependent.}, it follows that $\big|\mathsf{\tilde Q}\big| > 0$ \cite[Ch. 8.1]{Papoulis:91}.
Then, from the differential entropy of a multivariate Gaussian RV \cite[Thm. 8.4.1]{Cover:06} we conclude that \eqref{eqn:SecBound2} can be written as
\begin{equation}
\mspace{-3mu} h \mspace{-3mu} \left(\! {\left. {{{\bf{Y}}^{m\! -\! 1}}} \right|{{\bf{X}}^{n\! -\! 1}},{\bf{Y}}_m^{n\! -\! 1},{{\bf{S}}_0}\! =\! {{\bf{s}}_0}} \right)\! =\!
\frac{1}{2}\log\left(\!\left(2 \pi e\right)^{n_r m} \big|\mathsf{\tilde Q} \big|\right) \mspace{-2mu},
\label{eqn:SecBound4}
\end{equation}
where $\big|\mathsf{\tilde Q} \big|$ is positive, finite, and independent of $n$.
Next, note that 
\vspace{-0.15cm}
\begin{align}
 h \mspace{-3mu}\left( {\left. {{{\bf{Y}}^{m \! - \! 1}}} \right|{{\bf{S}}_0} \!=\! {{\bf{s}}_0}} \right) 
& = \mspace{-3mu} h \mspace{-3mu}\left( {\left. {\mathsf{H}_1{{\bf{X}}^{m \! - \!  1}} \! + \! \mathsf{H}_0{\bf{X}}_{  - \!  m}^{  - \!  1} \! + \! {{\bf{W}}^{m \! - \!  1}}} \right|{{\bf{S}}_0} \!=\! {{\bf{s}}_0}} \right)
\notag \\
& = \mspace{-3mu} h \mspace{-3mu}\left( {\left. {\mathsf{H}_1{{\bf{X}}^{m - 1}} \! + \! {{\bf{W}}^{m \! - \! 1}}} \right|{\bf{X}}_{ \! - \! m}^{ \! - \! 1} \!=\! {\bf{x}}_{  - \! m}^{  - \! 1},{\bf{W}}_{  - \! m}^{  - \! 1} \!=\! {\bf{w}}_{  - \! m}^{  - \! 1}} \right).
\label{eqn:SecBound4a}
\end{align}
Let 
$\mathsf{K}_{\bf Y}$ be the covariance matrix of the conditional distribution $\left. \mathsf{H}_1{{\bf{X}}^{m - 1}} + {{\bf{W}}^{m - 1}} \right|{\bf{X}}_{ - m}^{ - 1} = {\bf{x}}_{ - m}^{ - 1},{\bf{W}}_{ - m}^{ - 1} = {\bf{w}}_{ - m}^{ - 1}$,
$\mathsf{K}_{\bf X}$ be the covariance matrix of $ {\bf{X}}^{m - 1}$,
and $\mathsf{K}_{\bf W}$ be the covariance matrix of the conditional distribution $\left. {{\bf{W}}^{m - 1}} \right|{\bf{W}}_{ - m}^{ - 1} = {\bf{w}}_{ - m}^{ - 1}$. 
Since the channel input ${\bf X}[i]$ is subject to a per-symbol power constraint $P$ for $i \geq 0$, it follows that the entries of $\mathsf{K}_{\bf X}$ are all not larger than $P$ for any initial state ${\bf{x}}_{ - m}^{ - 1}$. 
As ${{\bf{W}}^{m - 1}}$ and ${\bf{W}}_{ - m}^{ - 1}$ are jointly Gaussian, it follows from \cite[Proposition 3.13]{Eaton:07} that $\mathsf{K}_{\bf W}$ is independent of the realization of ${\bf{W}}_{ - m}^{ - 1}$, ${\bf{w}}_{ - m}^{ - 1}$.
Since ${\bf X}[i]\big|{\bf{X}}_{ - m}^{ - 1}, {\bf{W}}_{ - m}^{ - 1} \stackrel{d}{=} {\bf X}[i]\big|{\bf{X}}_{ - m}^{ - 1}$ and ${\bf W}[i]\big|{\bf{X}}_{ - m}^{ - 1}, {\bf{W}}_{ - m}^{ - 1} \stackrel{d}{=} {\bf W}[i]\big|{\bf{W}}_{ - m}^{ - 1}$ are mutually independent, and the encoder is independent of the initial channel state, it follows that $\mathsf{K}_{\bf Y} = \mathsf{H}_1 \mathsf{K}_{\bf X} \mathsf{H}_1^T + \mathsf{K}_{\bf W}$. As the noise samples are not linearly dependent, we obtain $\left|\mathsf{K}_{\bf Y}\right| > 0$ \cite[Ch. 8.1]{Papoulis:91}. 
Defining $\gamma_k$ as $\gamma_k \triangleq \sum\limits_{{k_1} = 0}^{  m \cdot {n_t} - 1} \sum\limits_{{k_2} = 0}^{ m \cdot {n_t} - 1} \left|\left( {{\mathsf{H}_1}} \right)_{k,{k_2}}\left( {{\mathsf{H}_1}} \right)_{k,{k_1}}\right|$,
it follows  from Hadamard's inequality \cite[Thm. 17.9.2]{Cover:06} that 
\vspace{-0.15cm}
\begin{align*}
\left|\mathsf{K}_{\bf Y} \right| 
&\leq \prod\limits_{k = 0}^{m \cdot {n_r} - 1} {{{\left( \mathsf{K}_{\bf Y} \right)}_{k,k}}} \notag \\
&= \prod\limits_{k = 0}^{m \cdot {n_r} - 1} {\left( {{{\left( {{\mathsf{H}_1}\mathsf{K}_{\bf X}\mathsf{H}_1^T} \right)}_{k,k}} + {{\left(\mathsf{K}_{\bf W} \right)}_{k,k}}} \right)} \notag \\
&= \prod\limits_{k = 0}^{m \cdot {n_r} - 1} \Bigg( \sum\limits_{{k_1} = 0}^{m \cdot {n_t} - 1} {\sum\limits_{{k_2} = 0}^{m \cdot {n_t} - 1} {{{\left( {{\mathsf{H}_1}} \right)}_{k,{k_2}}}{{\left( \mathsf{K}_{\bf X} \right)}_{{k_2},{k_1}}}} {{\left( {{\mathsf{H}_1}} \right)}_{k,{k_1}}}}   + {{\left( \mathsf{K}_{\bf W} \right)}_{k,k}} \Bigg) \notag \\
&\le \prod\limits_{k = 0}^{m \cdot {n_r} - 1} {\left( {{\gamma _k}P + {{\left( \mathsf{K}_{\bf W} \right)}_{k,k}}} \right)}.
\end{align*}

\vspace{-0.1cm}
It follows that $\left|\mathsf{K}_{\bf Y} \right|$ is positive, finite, and independent of $n$. 
Plugging \eqref{eqn:SecBound4} and \eqref{eqn:SecBound4a} into \eqref{eqn:SecBound1} leads to 
\vspace{-0.15cm}
\begin{align}
\!I\left( {\left. {{{\bf{X}}^{n - 1}};{{\bf{Y}}^{m - 1}}} \right|{\bf{Y}}_m^{n - 1},{{\bf{S}}_0} = {{\bf{s}}_0}} \right) 
&\leq h\left( {\left. {\mathsf{H}_1{{\bf{X}}^{m - 1}} + {{\bf{W}}^{m - 1}}} \right|{\bf{X}}_{ - m}^{ - 1} = {\bf{x}}_{ - m}^{ - 1},{\bf{W}}_{ - m}^{ - 1} = {\bf{w}}_{ - m}^{ - 1}} \right)- \frac{1}{2}\log\left(\left(2 \pi e\right)^{n_r m} \big|\mathsf{\tilde Q} \big|\right) \notag \\
&\stackrel{(a)}{\leq} \frac{1}{2}\log\left(\left(2 \pi e\right)^{n_r m} \left|\mathsf{K}_{\bf Y} \right|\right) - \frac{1}{2}\log\left(\left(2 \pi e\right)^{n_r m} \big|\mathsf{\tilde Q} \big|\right),
\label{eqn:SecBound5}
\end{align}
where $(a)$ follows since $ h\big( \left. {\mathsf{H}_1{{\bf{X}}^{m - 1}} + {{\bf{W}}^{m - 1}}} \right|{\bf{X}}_{ - m}^{ - 1} = {\bf{x}}_{ - m}^{ - 1},{\bf{W}}_{ - m}^{ - 1} = {\bf{w}}_{ - m}^{ - 1} \big) $ is upper-bounded by the differential entropy of an $n_r \cdot m \times 1$  multivariate Gaussian RV with the same covariance matrix \cite[Thm. 8.6.5]{Cover:06}.
It therefore follows from \eqref{eqn:SecBound5} that $\exists \tilde{\eta}$ independent of $n$ such that $I\left( {\left. {{{\bf{X}}^{n - 1}};{{\bf{Y}}^{m - 1}}} \right|{\bf{Y}}_m^{n - 1},{{\bf{S}}_0} = {{\bf{s}}_0}} \right) \leq \tilde{\eta}$.
\end{IEEEproof}

\begin{remark}
	\label{rem:Constraint}
{\em	Note that the per-symbol power constraint \eqref{eqn:Constraint1} is required in the proofs of Lemmas \ref{lem:Proof_EtaBound}  and \ref{lem:Proof_EtaBound2}
in order to upper bound the mutual information between a transmitted block of $b \cdot n + a $  channel inputs (for Lemma \ref{lem:Proof_EtaBound})
or $n$ channel inputs (for Lemma \ref{lem:Proof_EtaBound2}) and a received
block of $m$ channel outputs, by a finite quantity independent of $n$. Consequently, the per-symbol constraint is essential for proving the asymptotic secrecy capacity equivalence stated in Proposition \ref{Pro:MainThm1}.}
\end{remark}

\begin{lemma}
\label{lem:Proof_Thm1a}
The secrecy capacity of the LGMWTC satisfies 
${\Cs} \le \mathop {\inf }\limits_{{\bf s}_0 \in \mathcal{S}_0} \left( {\mathop {\lim \inf }\limits_{n \to \infty } {C_n}\left( {{{{\bf s} }_0}} \right)} \right)$.
\end{lemma}
\begin{IEEEproof}
We prove the lemma by showing that every secrecy rate $R_s$ achievable for the LGMWTC satisfies ${R_s} \le \mathop {\lim \inf }\limits_{n \to \infty } {C_n}\left( {{{{\bf s} }_0}} \right)$ for any initial state ${{{{\bf s} }_0}}$. By definition, if $R_s$ is achievable for the LGMWTC, then for every non-negative triplet $\epsilon _1, \epsilon _2, \epsilon _3 > 0$ and for all sufficiently large $n$ there exists an $[R, n]$ code, such that \eqref{eqn:def_Rs1}-\eqref{eqn:def_Rs3} are satisfied. Fix ${{\bf S}}_0 = \tilde{{\bf s}}_0$, and recall that from Fano's inequality \cite[Sec. 2.10]{Cover:06} it follows that
\begin{align}
H\left( {\left. M \right|{{{\bf Y} }^{n-1} },{{{\bf S} }_0} = {{{\bf{\tilde s}} }_0}} \right)
&\leq 1 + \Pr \left( \left.{M \ne \hat M}\right|{{{\bf S} }_0} = {{{\bf{\tilde s}} }_0} \right)\cdot nR \notag \\
&\stackrel{(a)}{\leq} 1 + {\epsilon _1}\cdot nR
, \label{eqn:app_proof1}
\end{align}
where $(a)$ follows from \eqref{eqn:def_Rs1} since $\Pr\left(\left. {M \ne \hat M}\right|{{{\bf S} }_0} = {{{\bf{\tilde s}} }_0} \right) \le \mathop {\sup }\limits_{{\bf s}_0 \in \mathcal{S}_0} \Pr\left( {\left. {M \ne \hat M} \right|{{{\bf S} }_0} = {{{\bf s} }_0}} \right) \le {\epsilon _1}$. Therefore, 
\begin{align}
\!\!\!\!\!I\left( {\left. {M;{{{\bf Y} }^{n\!- \!1} }} \right|{{{\bf S} }_0}  \! = \! {{{\bf{\tilde s}} }_0}} \right) \!- \! I\left( {\left. {M;{{{\bf Z} }^{n\!- \!1} }} \right|{{{\bf S} }_0} \! = \! {{{\bf{\tilde s}} }_0}} \right) 
&  =  H\left( {\left. M \right|{{{\bf S} }_0} \! = \! {{{\bf{\tilde s}} }_0}} \right) \!- \! H\left( {\left. M \right|{{{\bf Y} }^{n\!- \!1} },{{{\bf S} }_0} \! = \! {{{\bf{\tilde s}} }_0}} \right)  \!- \! I\left( {\left. {M;{{{\bf Z} }^{n\!- \!1} }} \right|{{{\bf S} }_0} \! = \! {{{\bf{\tilde s}} }_0}} \right)\notag \\
&\stackrel{(a)}{\geq} H\left( {\left. M \right|{{{\bf S} }_0} = {{{\bf{\tilde s}} }_0}} \right)  - 1 - {\epsilon _1}\cdot nR- {\epsilon _2} \cdot n \notag \\
&\stackrel{(b)}{=} nR  - 1 - {\epsilon _1}\cdot nR - {\epsilon _2} \cdot n, \label{eqn:app_proof2}
\end{align}
where $(a)$ follows from \eqref{eqn:app_proof1} and from \eqref{eqn:def_Rs2}, as $I\!\left( {\left. {M;{{{\bf Z} }^{n-1} }} \right|{{{\bf S} }_0} = {{{\bf{\tilde s}} }_0}} \right)\! \le\! \mathop {\sup }\limits_{{\bf s}_0 \in \mathcal{S}_0} I\!\left( {\left. {M;{{{\bf Z} }^{n-1} }} \right|{{{\bf S} }_0} = {{{\bf s} }_0}} \right) < {\epsilon _2} \cdot n$; and $(b)$ follows since $M$ is uniformly distributed and is independent of ${{\bf S}}_0$.
Combining \eqref{eqn:def_Rs3} and \eqref{eqn:app_proof2} leads to
\begin{align}
\left( {1 - {\epsilon _1}} \right)\left( {{R_s} - {\epsilon _3}} \right) - \frac{1}{n } - {\epsilon _2} 
& \le \frac{1}{n}\bigg( I\left( {\left. {M;{{{\bf Y} }^{n-1} }} \right|{{{\bf S} }_0} = {{{\bf{\tilde s}} }_0}} \right)
- I\left( {\left. {M;{{{\bf Z} }^{n-1} }} \right|{{{\bf S} }_0} = {{{\bf{\tilde s}} }_0}} \right) \bigg)
\notag \\
& \stackrel{(a)}{\leq}
\frac{1}{n }\!\!\!\! \sup_{\substack{p\left( {{{{\bf V} }^{n-1} },{{{\bf X} }^{n-1} }} \right): \\ \MyBeta{n}} } \mspace{-25mu} \bigg\{ I\left( {\left. {{{{\bf V} }^{n-1} };{{{\bf Y} }^{n-1} }} \right|{{{\bf S} }_0} = {{{\bf{\tilde s}} }_0}} \right)  - I\left( {\left. {{{\bf V} }^{n-1} };{{{\bf Z} }^{n-1} } \right|{{{\bf S} }_0} = {{{\bf{\tilde s}} }_0}} \right) \bigg\} \notag \\
&\quad\equiv
{C_n }\left( {{{{\bf{\tilde s}} }_0}} \right), \label{eqn:app_proof3}
\end{align}
where $(a)$ follows since we can define a pair $\left({{\bf V} }^{n-1},{{\bf X} }^{n-1}\right)$  such that ${{{\bf V} }^{n-1} }$ is a random variable representing the uniformly distributed message $M$ and $p\left({{{{\bf X} }^{n-1} }|{{{\bf V} }^{n-1} }}\right)$ is defined by the encoder of the $[R, n]$ code (either deterministic or stochastic), as done in the proof of \cite[Lemma 4]{Bloch:08}. 
Since \eqref{eqn:app_proof3} holds for all sufficiently large $n$, it follows from \cite[Thm. 3.19]{Rudin:76} that  $\mathop {\lim \inf }\limits_{n  \to \infty }\left(\left( {1 - {\epsilon _1}} \right)\left( {{R_s} - {\epsilon _3}} \right) -\frac{1}{n} - {\epsilon _2}\right) \le \mathop {\lim \inf }\limits_{n  \to \infty } {C_n }\left( {{{{\bf{\tilde s}} }_0}} \right)$, thus
\begin{equation}
\label{eqn:app_proof4}
\left( {1 - {\epsilon _1}} \right)\left( {{R_s} - {\epsilon _3}} \right) - {\epsilon _2} \le \mathop {\lim \inf }\limits_{n  \to \infty } {C_n }\left( {{{{\bf{\tilde s}} }_0}} \right).
\end{equation}
Since  $\epsilon _1$, $\epsilon _2$, and $\epsilon _3$ can also be made arbitrarily small, \eqref{eqn:app_proof4} implies that
\begin{equation}
\label{eqn:app_proof5}
R_s \le \mathop {\lim \inf }\limits_{n  \to \infty } {C_n }\left( {{{{\bf{\tilde s}} }_0}} \right),
\end{equation}
and as \eqref{eqn:app_proof5} is true for any achievable secrecy rate $R_s$, we conclude that for all ${{{{\bf{\tilde s}} }_0}}\in\mSzro$, ${\Cs} \le \mathop {\lim \inf }\limits_{n  \to \infty } {C_n }\left( {{{{\bf{\tilde s}} }_0}} \right)$\footnote{As the supremum is defined as the least upper bound \cite[Def. 1.8]{Rudin:76}, it follows that if every achievable secrecy rate $R_s$ is not larger than a given real number $\gamma \in \mathds{R}$, then the supremum of all achievable secrecy rates, $C_s$, is also not larger than $\gamma$.}, thus ${\Cs} \le \mathop {\inf }\limits_{{\bf s}_0 \in \mathcal{S}_0}  \left( {\mathop {\lim \inf }\limits_{n \to \infty } {C_n}\left( {{{{\bf s} }_0}} \right)} \right)$.
\end{IEEEproof}
\smallskip

\begin{lemma}
\label{lem:Proof_Thm1b} 
$\CMG{n}$, defined in \eqref{eqn:Sec_DefQn},
satisfies 
$\mathop {\inf }\limits_{{{{\bf s} }_0} \in \mathcal{S}_0} \left( {\mathop {\lim \inf }\limits_{n \to \infty } {C_n}\left( {{{{\bf s} }_0}} \right)} \right) {\leq} \mathop {\lim \inf }\limits_{n \to \infty } {\CMG{n}}$.
\end{lemma}
\begin{IEEEproof}
First, we show that for all ${{{{\bf s} }_0}} \in \mathcal{S}_0$, ${C_n}\left( {{{{\bf s} }_0}} \right) \leq \CMG{n} + \frac{\tilde{\eta}}{n}$.
Note that 
\begin{align}
I\left( {\left. {{{{\bf V} }^{n\!-\!1}};{{{\bf Y} }^{n\!-\!1}}} \right|{{{\bf S} }_0} \!=\! {{{\bf s} }_0}} \right) 
&  
\stackrel{(a)}{=} I \left( {\left. {{{{\bf V} }^{n\!-\!1}};{\bf Y} _m^{n\!-\!1}} \right|{{{\bf S} }_0} \!=\! {{{\bf s} }_0}} \right)
\! + \! I \left( {\left. {{{{\bf V} }^{n\!-\!1}};{{{\bf Y} }^{m\!-\!1}}} \right|{\bf Y} _m^{n\!-\!1},{{{\bf S} }_0} \!=\! {{{\bf s} }_0}} \right)
\notag \\
&\stackrel{(b)}{\leq}I\left( {\left. {{{{\bf V} }^{n\!-\!1}};{\bf Y} _m^{n\!-\!1}} \right|{{{\bf S} }_0} \!=\! {{{\bf s} }_0}} \right) \! + \! I\left( {\left. {{{{\bf X} }^{n\!-\!1}};{{{\bf Y} }^{m\!-\!1}}} \right|{\bf Y} _m^{n\!-\!1},{{{\bf S} }_0} \!=\! {{{\bf s} }_0}} \right) \notag \\
&
\stackrel{(c)}{\leq}I\left( {\left. {{{{\bf V} }^{n\!-\!1}};{\bf Y} _m^{n\!-\!1}} \right|{{{\bf S} }_0} \!=\! {{{\bf s} }_0}} \right) \! + \! \tilde{\eta}  \notag \\
&
\stackrel{(d)}{=}I\left( {{{{\bf V} }^{n\!-\!1}};{\bf Y} _m^{n\!-\!1}}  \right) + \tilde{\eta}
, \label{eqn:app2_proof1}
\end{align}
where $(a)$ follows from the mutual information chain rule \cite[Sec. 2.5]{Cover:06}; 
$(b)$ follows from the data processing inequality and the Markov chain\footnote{The Markov chain ${\bf V}^{n-1}|{\bf S}_0 \rightarrow {\bf X}^{n-1}|{\bf S}_0 \rightarrow {\bf Y}^{n-1}|{\bf S}_0$ is a short notation for the relationship $p\left( \left.{{{\bf{V}}^{n - 1}}}, {{{\bf{X}}^{n - 1}}}, {{{\bf{Y}}^{n - 1}}}\right|{{\bf{S}}_0} = {{\bf{s}}_0} \right) = p\left( \left. {{{\bf{V}}^{n - 1}}} \right|{{\bf{S}}_0}  = {{\bf{s}}_0}\right)p\left( \left. {{{\bf{X}}^{n - 1}}} \right|{{{\bf{V}}^{n - 1}}}, {{\bf{S}}_0}  = {{\bf{s}}_0}\right) p\big( \left. {{{\bf{Y}}^{n - 1}}} \right|{{{\bf{X}}^{n - 1}}}, $ ${{\bf{S}}_0} = {{\bf{s}}_0} \big)$, for all ${\bf s}_0 \in \mathcal{S}_0$, see, e.g., \cite{Dabora:10a}.} ${\bf V}^{n-1}|{\bf S}_0 \rightarrow {\bf X}^{n-1}|{\bf S}_0 \rightarrow {\bf Y}^{n-1}|{\bf S}_0$;
$(c)$ follows from Lemma \ref{lem:Proof_EtaBound2}; 
and $(d)$ follows since ${\bf Y}[i]$ is independent of the initial state $\forall i \geq m$.
Using \eqref{eqn:app2_proof1} in the definition of $C_n\left({\bf s}_0\right)$ in \eqref{eqn:Sec_DefCn} we obtain
\begin{align*}
{C_n}\left( {{{{\bf s} }_0}} \right) 
&\stackrel{(a)}{\leq} \frac{1}{n} \!\!\!\!\sup_{\substack{ p\left( {{{{\bf V} }^{n-1}},{{{\bf X} }^{n-1}}} \right): \\ \MyBeta{n}} } \mspace{-25mu} \bigg\{I\left( {{{{\bf V} }^{n-1}};{\bf Y} _m^{n-1}} \right) + \tilde{\eta}   - I\left( {\left. {{{{\bf V} }^{n-1}};{{{\bf Z} }^{n-1}}} \right|{{{\bf S} }_0} = {{{\bf s} }_0}} \right) \bigg\}
\notag \\
&
\stackrel{(b)}{\leq} 
\frac{1}{n} \mspace{-25mu} \sup_{ \substack{p\left( {{{{\bf V} }^{n-1}},{{{\bf X} }^{n-1}}} \right): \\ \MyBeta{n} } } \mspace{-25mu} \bigg\{I\left(  {{{{\bf V} }^{n-1}};{\bf Y} _m^{n-1}}  \right)    - I\left( {\left. {{{{\bf V} }^{n-1}};{\bf Z} _m^{n-1}} \right|{{{\bf S} }_0} = {{{\bf s} }_0}} \right) \bigg\} + \frac{1}{n} \tilde{\eta}
\notag\\
& \stackrel{(c)}{=}
\frac{1}{n} \mspace{-25mu} \sup_{ \substack{p\left( {{{{\bf V} }^{n-1}},{{{\bf X} }^{n-1}}} \right): \\ \MyBeta{n} } } \mspace{-25mu} \bigg\{ I\left( {{{{\bf V} }^{n-1}};{\bf Y} _m^{n-1}}  \right)  - I\left(  {{{{\bf V} }^{n-1}};{\bf Z} _m^{n-1}}  \right) \bigg\} + \frac{1}{n} \tilde{\eta} \notag \\
&\equiv
\CMG{n} + \frac{ \tilde{\eta}}{n},
\end{align*}
where $(a)$ follows from \eqref{eqn:app2_proof1}; 
$(b)$ follows from the non-negativity of the mutual information which implies that $I\left( {\left. {{{{\bf V} }^{n-1}};{\bf Z} _m^{n-1}} \right|{{{\bf S} }_0} = {{{\bf s} }_0}} \right) \leq I\left( {\left. {{{{\bf V} }^{n-1}};{{{\bf Z} }^{n-1}}} \right|{{{\bf S} }_0} = {{{\bf s} }_0}} \right)$; 
and $(c)$ follows since ${\bf Z}[i]$ is independent of the initial state $\forall i \geq m$.

Now, since for all ${{{{\bf s} }_0}} \in \mathcal{S}_0$, ${C_n}\left( {{{{\bf s} }_0}} \right) \leq \CMG{n} + \frac{\tilde{\eta}}{n}$, then $ \mathop {\lim \inf }\limits_{n \to \infty } {C_n}\left( {{{{\bf s} }_0}} \right) {\leq} \mathop {\lim \inf }\limits_{n \to \infty } {\CMG{n}}$, therefore, 
\begin{equation*}
\mathop {\inf }\limits_{{{{\bf s} }_0} \in \mathcal{S}_0} \left( {\mathop {\lim \inf }\limits_{n \to \infty } {C_n}\left( {{{{\bf s} }_0}} \right)} \right) {\leq} \mathop {\lim \inf }\limits_{n \to \infty } {\CMG{n}}.
\end{equation*}
This proves the lemma. 
\end{IEEEproof}
\smallskip

\begin{lemma}
\label{lem:Proof_Thm1c} $\CMG{n}$, defined in \eqref{eqn:Sec_DefQn},
satisfies 
$\mathop {\lim \sup }\limits_{n \rightarrow \infty} {\CMG{n}} \le {\Cs}$.
\end{lemma}
\begin{IEEEproof}
In order to prove the lemma we show that every non-negative $R_s <  \mathop {\lim \sup }\limits_{n \rightarrow \infty} {\CMG{n}}$ is an achievable secrecy rate for the LGMWTC\footnote{Note that if every non-negative $R_s <  \mathop {\lim \sup }\limits_{n \rightarrow \infty} {\CMG{n}}$ satisfies $R_s \leq \Cs$, then necessarily, $\mathop {\lim \sup }\limits_{n \rightarrow \infty} {\CMG{n}} \le {\Cs}$. This follows since if $\mathop {\lim \sup }\limits_{n \rightarrow \infty} {\CMG{n}} > {\Cs}$ then $\exists \tilde{R}_s$, such that $\Cs < \tilde{R}_s < \mathop {\lim \sup }\limits_{n \rightarrow \infty} {\CMG{n}}$, i.e., $\tR_s$ does not satisfy out initial assumption.}
To that aim, consider such $R_s <  \mathop {\lim \sup }\limits_{n \rightarrow \infty} {\CMG{n}}$:
From \cite[Thm. 5.5]{Amann:05} it follows that if $R_s <  \mathop {\lim \sup }\limits_{n \rightarrow \infty} {\CMG{n}}$, then {\em there are infinitely many values of $n \in \mathds{N}$ such that $R_s \le \CMG{n}$}, hence, $R_s$ is an achievable secrecy rate  for the $n$-MGMWTC for these values of $n$. Consequently, it follows that  for a given real number $\eta > 0$ and for any arbitrarily fixed non-negative triplet
$\epsilon _1, \epsilon _2, \epsilon _3 > 0$,  $\exists n > n_1 \triangleq \left\lceil \frac{2\eta}{\epsilon_2}\right\rceil$ such that $R_s$ is an achievable secrecy rate for the $n$-MGMWTC.
Note that since $n > n_1$ it follows that ${\epsilon _2} - \frac{2\eta}{n} > 0$. 
By Def. \ref{def:SecrecyRate}, the achievability of $R_s$ implies that 	we can find a sufficiently large $b_0 \in \mathds{N}$, such that for all integer $b > b_0$ there exists an $\left[R_1, b\cdot n\right]$ code for the  $n$-MGMWTC which satisfies 
\begin{subequations}
\label{eqn:app_Rs}
\begin{equation}
\mathop {\sup }\limits_{{{{\bf s} }_0} \in \mathcal{S}_0} P_e^{b\cdot n} \left( {{{{\bf s} }_0}} \right) 
\stackrel{(a)}{=}P_e^{b\cdot n}  \le {\epsilon _1},
\label{eqn:app_Rs1}
\end{equation}
\begin{align}
\mathop {\sup }\limits_{{{{\bf s} }_0} \in \mathcal{S}_0}
\frac{1}{{b\cdot n}}I\left( {\left. M;{\bf Z}_m^{n-1}, {\bf Z}_{ n + m}^{2 n-1}, \ldots,{\bf Z}_{(b-1)\cdot n + m}^{b\cdot n-1} \right|{{{\bf S} }_0}= {{{\bf s} }_0}} \right) 
&\stackrel{(b)}{=} \frac{1}{{b\cdot n}}I\left(  M;{\bf Z}_m^{n-1},{\bf Z}_{n + m}^{2 n-1},\ldots,{\bf Z}_{(b-1)\cdot n + m}^{b\cdot n-1} \right) \notag \\
& \le {\epsilon _2} - \frac{2\eta}{n},
\label{eqn:app_Rs2} 
\end{align}
and
\begin{equation}
\label{eqn:app_Rs3}
R_1 \geq R_s - \frac{\epsilon_3}{2},
\end{equation}
\end{subequations}
where $(a)$ and $(b)$ follow since the $n$-MGMWTC is $n$-block memoryless, hence, the channel outputs and the probability of error are independent of the initial state, when the length of the codeword is an integer multiple of $n$.
Denote this code by $\mathcal{C} _{b \cdot n}^{MG}$, and recall that from the definition of the $n$-MGMWTC,  it follows that the decoders at the intended receiver and at the eavesdropper use only the last $n-m$ channel outputs out of each block of $n$ consecutive channel outputs of the LGMWTC. Let ${\bf X}_{MG}^{b \cdot n-1}$ denote the codeword of length $b \cdot n$ used for transmitting a message  $\zeta\in \mathcal{M}$ via the code $\mathcal{C} _{b \cdot n}^{MG}$ for the $n$-MGMWTC.
 
Next, based on the code $\mathcal{C} _{b \cdot n}^{MG}$, we construct a code for the LGMWTC with codeword length $l = b\cdot n + a$, where $a$ can be selected arbitrarily from $a \in \{m,m+1,\ldots,n+m-1\}$. We denote this code by $\mathcal{C} _{l}^{LG}$, and in the following we analyze the performance of $\mathcal{C} _{l}^{LG}$.
In the analysis we use $\bar{\bf Y}[i]$ and $\bar{\bf Z}[i]$ to denote the channel outputs of the LGMWTC at the intended receiver and at the eavesdropper, respectively, when the code $\mathcal{C} _{l}^{LG}$ is employed.
The encoder of the $\mathcal{C} _{l}^{LG}$ code encodes the message $\zeta \in \mathcal{M}$ into the codeword ${\bf X}_{LG}^{l-1}$ by setting ${\bf X}_{LG}^{a-1} = \mathsf{0}_{a\cdot n_t \times 1}$ and
setting ${\bf X}_{LG,a}^{l-1}$ to be equal to the codeword used for transmitting $\zeta$ using  the $\mathcal{C} _{b \cdot n}^{MG}$ code, i.e., ${\bf X}_{LG,a}^{l-1}={\bf X}_{MG}^{b \cdot n-1}$ for the
same message $\zeta$.	
The decoder of the $\mathcal{C} _{l}^{LG}$ code  discards the first $a$ channel outputs of the codeword, and then discards the first $m$ channel outputs of each block of $n$ channel outputs.
The remaining channel outputs, namely $\bar{\bf Y}_{LG} \triangleq \bigg(\bar{\bf Y}_{a+m}^{n+a-1},\bar{\bf Y}_{n +a+ m}^{2n+a-1},\ldots,\bar{\bf Y}_{(b-1)\cdot n +a+ m}^{b\cdot n+a-1}\bigg)$,
are then used for decoding the message using the decoder for the $\mathcal{C} _{b \cdot n}^{MG}$ code.

Define
\begin{equation*}
{\bf Y}_{MG} \triangleq \bigg({\bf Y}_{m}^{n-1},{\bf Y}_{n + m}^{2 n-1},\ldots,{\bf Y}_{(b-1)\cdot n + m}^{b\cdot n-1}\bigg),
\end{equation*}
 and
 \begin{equation*}
 {\bf W}_{a} \triangleq \bigg({\bf W}_{a+m}^{n+a-1},{\bf W}_{n +a+ m}^{2 n+a-1},\ldots,{\bf W}_{(b-1)\cdot n +a+ m}^{b\cdot n+a-1}\bigg).
 \end{equation*}
It follows from the definition of the LGMWTC that $\exists \bar{\mathsf{H}} \in \mathds{R}^{n_r\cdot b \cdot (n-m) \times n_t \cdot b \cdot n}$ such that $\bar{\bf Y}_{LG} = \bar{\mathsf{H}}{\bf X}_{LG,a}^{l-1} +{\bf W}_{a}$ and also  ${\bf Y}_{MG} = \bar{\mathsf{H}}{\bf X}_{MG}^{b \cdot n-1} +{\bf W}_{0}$. 
It now follows from the stationarity of ${\bf W}[i]$ and the relationship between $\mathcal{C} _{l}^{LG}$ and  $\mathcal{C} _{b \cdot n}^{MG}$ that the decoder for the $\mathcal{C} _{l}^{LG}$ code operates on channel outputs
which have the same statistical characterization as the channel outputs ${\bf Y}_{MG}$,
which result from transmitting codewords using
the $\mathcal{C} _{b \cdot n}^{MG}$ code. {\em Hence, the probability of error for the code $\mathcal{C} _{l}^{LG}$ is identical to that for the code $\mathcal{C} _{b \cdot n}^{MG}$}.
Similarly, by defining $\bar{\bf Z}_{LG} \! \triangleq \! \bigg(\bar{\bf Z}_{a\! + \!m}^{n\! + \!a\! - \!1},\bar{\bf Z}_{n \! + \!a\! + \! m}^{2 n\! + \!a\! - \!1},\ldots,\bar{\bf Z}_{(b\! - \!1)\cdot n \! + \!a\! + \! m}^{b\cdot n\! + \!a\! - \!1}\bigg)$, ${\bf Z}_{MG} \! \triangleq \! \bigg({\bf Z}_{m}^{n\! - \!1},{\bf Z}_{n \! + \! m}^{2n\! - \!1},\ldots,{\bf Z}_{(b\! - \!1)\cdot n \! + \! m}^{b\cdot n\! - \!1}\bigg)$ and ${\bf U}_{a} \! \triangleq \! \bigg({\bf U}_{a\! + \!m}^{n\! + \!a\! - \!1},{\bf U}_{n \! + \!a\! + \! m}^{2 n\! + \!a\! - \!1}, \ldots,{\bf U}_{(b\! - \!1)\cdot n \! + \!a\! + \! m}^{b\cdot n\! + \!a\! - \!1}\bigg)$, it follows that  $\exists \bar{\mathsf{G}} \in \mathds{R}^{n_e\cdot b \cdot (n-m) \times n_t \cdot b \cdot n}$ such that $\bar{\bf Z}_{LG} = \bar{\mathsf{G}}{\bf X}_{LG,a}^{l-1} +{\bf U}_{a}$ and also  ${\bf Z}_{MG} = \bar{\mathsf{G}}{\bf X}_{MG}^{b \cdot n-1} +{\bf U}_{0}$, which implies that $\bar{\bf Z}_{LG}$ and ${\bf Z}_{MG}$ both have the same statistical characterization. Consequently,
\begin{align}
I\left( {M;{{\bf{Z}}_{MG}}} \right) 
&\stackrel{(a)}{=} I\left( {M,{{\bf{X}}_{MG}^{b \cdot n - 1}};{{\bf{Z}}_{MG}}} \right) - I\left( {\left. {{{\bf{X}}_{MG}^{b \cdot n - 1}};{{\bf{Z}}_{MG}}} \right|M} \right)\notag \\
&\stackrel{(b)}{=} I\left( {\bf{X}}_{MG}^{b \cdot n - 1};{\bf{Z}}_{MG} \right) - I\left( {\left. {{{\bf{X}}_{MG}^{b \cdot n - 1}};{{\bf{Z}}_{MG}}} \right|M} \right)\notag \\
&\stackrel{(c)}{=} I\left( {{\bf{X}}_{LG,a}^{l - 1};{{{\bf{\bar Z}}}_{LG}}} \right) - I\left( {\left. {{\bf{X}}_{LG,a}^{l - 1};{{{\bf{\bar Z}}}_{LG}}} \right|M} \right)\notag \\
&\stackrel{(d)}{=} I\left( {M;{{{\bf{\bar Z}}}_{LG}}} \right),  \label{eqn:Leakage_Trans0a}
\end{align}
where $(a)$ follows from the chain rule for mutual information \cite[Ch. 2.5]{Cover:06}; 
$(b)$ follows since $M \rightarrow {\bf{X}}_{MG}^{b \cdot n - 1} \rightarrow {\bf{Z}}_{MG}$ form a Markov chain; 
$(c)$ follows from the combination of the following three properties: $(1)$ the stationarity of ${\bf U}[i]$; $(2)$ the definition of the encoder of the $\mathcal{C} _{l}^{LG}$ code; $(3)$ the fact that the channel matrix $\bar{\mathsf{G}}$  is identical for both the LGMWTC and the $n$-MGMWTC, which imply that the joint distribution of $\left( {\bf{X}}_{MG}^{b \cdot n - 1},{\bf{Z}}_{MG}\right) $ is identical to the joint distribution of $\left( {\bf{X}}_{LG,a}^{l - 1},{{{\bf{\bar Z}}}_{LG}}\right) $, and also implies that the joint distribution of $\left( {\bf{X}}_{MG}^{b \cdot n - 1},{\bf{Z}}_{MG}\right) $ given $M$ is identical to the joint distribution of $\left( {\bf{X}}_{LG,a}^{l - 1},{{{\bf{\bar Z}}}_{LG}}\right) $ given $M$;
and $(d)$ follows from the construction of the $\mathcal{C} _{l}^{LG}$ code, which sets ${\bf X}_{LG}^{a-1}$ to be the all zero vector, and by applying the reverse of the transition from (a) to (b).

Next, let $\bar{\bf Z}^{l-1}$ denote the entire set of $l$ vector channel outputs obtained when transmitting using the code $\mathcal{C} _{l}^{LG}$. When this transmission is applied,
the information leakage rate for the LGMWTC satisfies
\begin{align}
\mathop {\sup }\limits_{{{{\bf s} }_0} \in \mathcal{S}_0}
\frac{1}{l}I\left( \left. M;\bar{\bf Z}^{l-1}  \right|{{{\bf S} }_0} ={{{\bf s} }_0}\right) 
& =\mathop {\sup }\limits_{{{{\bf s} }_0} \in \mathcal{S}_0} 
\frac{1}{{b\cdot n}+a }I\left( \left. M;\bar{\bf Z}^{b\cdot n+a-1}  \right|{{{\bf S} }_0} ={{{\bf s} }_0}\right)
\notag \\
& \stackrel{(a)}{=} \mathop {\sup }\limits_{{{{\bf s} }_0} \in \mathcal{S}_0} 
\frac{1}{{b\cdot n}+a }\bigg( I\left( {\left. {M;{{\bar{\bf{ Z}}}^{a - 1}}} \right|{{\bf{S}}_0} = {{\bf{s}}_0}} \right)  \notag \\
& \qquad \qquad \qquad \qquad \quad + I\left( {\left. {M;\bar{\bf{ Z}}_a^{b\cdot n +a - 1}} \right|{{\bar{\bf{ Z}}}^{a - 1}},{{\bf{S}}_0} = {{\bf{s}}_0}} \right) \bigg) \notag \\
& \stackrel{(b)}{=} \mathop {\sup }\limits_{{{{\bf s} }_0} \in \mathcal{S}_0} \frac{1}{{b\cdot n + a} }I\left( {\left. {M;\bar{\bf{ Z}}_a^{b \cdot n + a - 1}} \right|{{\bar{\bf{ Z}}}^{a - 1}},{{\bf{S}}_0} = {{\bf{s}}_0}} \right) \notag \\
& \stackrel{(c)}{=} \mathop {\sup }\limits_{{{{\bf s} }_0} \in \mathcal{S}_0} \frac{1}{{b\cdot n + a} }I\left( {\left. {M;\bar{\bf{ Z}}_a^{b \cdot n + a - 1}} \right|{{{\bf{U}}}^{a - 1}},{{\bf{S}}_0} = {{\bf{s}}_0}} \right) \notag \\
& \stackrel{(d)}{=} \frac{1}{{b\cdot n + a}} I\left( {\left. {M;\bar{\bf{ Z}}_a^{b \cdot n + a - 1}} \right|{\bf{U}}_{a - m}^{a - 1}} \right),
\label{eqn:Leakage_Trans1}
\end{align}
where $(a)$ follows from the chain rule for mutual information \cite[Ch. 2.5]{Cover:06}; 
$(b)$ follows since when using the code $\mathcal{C} _{l}^{LG}$, the first $a$ channel outputs depend only on the initial state and the noise, hence, $M$ and $\bar{\bf{ Z}}^{a - 1}$ are mutually independent;
$(c)$ follows since ${\bf{ X}}_{LG}^{a - 1}$ is all zeros, thus $\exists \check{\mathsf{G}} \in \mathds{R}^{n_e\cdot a \times n_t \cdot m}$ such that $\bar{\bf{ Z}}^{a - 1} = \check{\mathsf{G}}{\bf X}_{-m}^{-1} +{\bf{ U}}^{a - 1}$; 
$(d)$ follows since the finite memory of the channel implies that $\bar{\bf{ Z}}_a^{b \cdot n + a - 1}$ is independent of the initial state and of ${{\bf{U}}}^{a-m- 1}$, regardless of the code. 
This can be shown by noting that we can define a matrix  $\check{\check{\mathsf{G}}} \in \mathds{R}^{n_e\cdot b \cdot n \times n_t (b \cdot n + m)}$ such that $\bar{\bf{ Z}}_a^{b \cdot n + a - 1} = \check{\check{\mathsf{G}}}{\bf X}_{LG,a-m}^{b\cdot n+a-1} +{\bf{ U}}_{a}^{b\cdot n+a-1}$, and noting that ${\bf X}_{LG}[i]$ is independent of both ${{\bf{U}}}^{a-m- 1}$ and ${{\bf{S}}_0}$ for all $a \le i \le b\cdot n +a -1$, and
that, due to the finite memory of the noise, then for all $i\ge a \ge m$   ${\bf U}[i]$ is independent of both ${{\bf{S}}_0}$ as well as ${{\bf{U}}}^{a-m- 1}$.
Next, we note that $\frac{1}{{b\cdot n} }I\left( \left. {M;{{\bar{\bf Z} }_a^{b\cdot n + a\! - \!1} }} \right|{\bf{U}}_{a \! - \! m}^{a \! - \! 1}\right)$ can be upper bounded as stated in \eqref{eqn:Epsilon2_a},
\begin{figure*}[!t]
\normalsize
\begin{align}
&\frac{1}{{b\cdot n} }I\left( \left. {M;{{\bar{\bf Z} }_a^{b\cdot n + a\! - \!1} }} \right|{\bf{U}}_{a \! - \! m}^{a \! - \! 1}\right)
\notag \\
&\quad \stackrel{(a)}{=} \frac{1}{{b\cdot n} }I\left( \left. M;\bar{\bf Z} _{a\! + \!m}^{n\! + \!a\! - \!1},\bar{\bf Z}_{n \! + \!a\! + \! m}^{2 n\! + \!a\! - \!1},\ldots,\bar{\bf Z}_{(b\! - \!1)\cdot n \! + \!a\! + \! m}^{b\cdot n\! + \!a\! - \!1}  \right|{\bf{U}}_{a \! - \! m}^{a \! - \! 1}\right)  
\notag \\
&\quad \qquad \quad
\! + \! \frac{1}{b\cdot n}\bigg(I\Big(\left. M;\bar{\bf{Z}}_a^{a\! + \!m\! - \!1}\right|\bar{\bf{Z}}_{a\! + \!m}^{n\! + \!a\! - \!1}, \bar{\bf Z}_{n \! + \!a\! + \! m}^{2 n\! + \!a\! - \!1},\ldots ,\bar{\bf{Z}}_{\left( {b \! - \! 1} \right) \cdot n \! + \! a\! + \!m}^{b \cdot n\! + \!a\! - \!1}, {\bf{U}}_{a \! - \! m}^{a \! - \! 1} \Big) 
\notag \\
&\quad \qquad \qquad \qquad
\! + \!  \sum\limits_{k=1}^{b\! - \!1}I\Big(\left. M;\bar{\bf{Z}}_{k \cdot n \! + \! a}^{k \cdot n \! + \!a\! + \! m\! - \!1}\right|\bar{\bf{Z}}_{a \! + \! m}^{n\!+ \! a - \!1}, \bar{\bf{Z}}_{ n \! + \!a\! + \! m}^{2 n\! + \!a\! - \!1}, \ldots ,\bar{\bf{Z}}_{\left( {b \! - \! 1} \right) \cdot n \! + \!a\! + \! m}^{b \cdot n\! + \!a\! - \!1},  
\notag \\
&\qquad \qquad \qquad \qquad \qquad \qquad \qquad \qquad \qquad
{\bar{\bf{Z}}_a^{a\! + \!m\! - \!1}}, \bar{\bf{Z}}_{n\! + \!a}^{n \! + \!a\! + \! m\! - \!1}, \ldots ,\bar{\bf{Z}}_{\left( {k \! - \! 1} \right) \cdot n\! + \!a}^{\left( {k \! - \! 1} \right) \cdot n \! + \!a\! + \! m\! - \!1},{\bf{U}}_{a \! - \! m}^{a \! - \! 1} \Big)  \bigg)\notag \\
&\quad \stackrel{(b)}{\leq} \frac{1}{{b\cdot n} }I\left( \left. M;\bar{\bf Z} _{a\! + \!m}^{n\! + \!a\! - \!1},\bar{\bf Z}_{ n\! + \!a \! + \! m}^{2 n\! + \!a\! - \!1}, \ldots,\bar{\bf Z}_{(b\! - \!1)\cdot n\! + \!a \! + \! m}^{b\cdot n\! + \!a\! - \!1}  \right|{\bf{U}}_{a \! - \! m}^{a \! - \! 1}\right)  
\notag \\
&\quad \qquad \quad
\! + \! \frac{1}{b\cdot n}\bigg(I\Big(\left. {\bf{X}}_{LG}^{b\cdot n\! + \!a\! - \!1};\bar{\bf{Z}}_a^{a\! + \!m\! - \!1}\right|\bar{\bf{Z}}_{a\! + \!m}^{n\! + \!a\! - \!1}, \bar{\bf{Z}}_{n\! + \!a \! + \! m}^{2 n\! + \!a\! - \!1}, \ldots ,\bar{\bf{Z}}_{\left( {b \! - \! 1} \right) \cdot n\! + \!a \! + \! m}^{b \cdot n\! + \!a\! - \!1}, {\bf{U}}_{a \! - \! m}^{a \! - \! 1} \Big)
\notag \\
&\quad \qquad \qquad \qquad
\! + \!  \sum\limits_{k=1}^{b\! - \!1}I\Big(\left. {\bf{X}}_{LG}^{b\cdot n\! + \!a\! - \!1};\bar{\bf{Z}}_{k \cdot n\! + \!a}^{k \cdot n\! + \!a \! + \! m\! - \!1}\right|\bar{\bf{Z}}_{a\! + \!m}^{n\! + \!a\! - \!1}, \bar{\bf{Z}}_{ n \! + \!a\! + \! m}^{2 n\! + \!a\! - \!1}, \ldots ,\bar{\bf{Z}}_{\left( {b \! - \! 1} \right) \cdot n \! + \!a\! + \! m}^{b \cdot n\! + \!a\! - \!1},
\notag \\
&\qquad \qquad \qquad \qquad \qquad \qquad \qquad \qquad \qquad \qquad
\bar{\bf{Z}}_a^{a\! + \!m\! - \!1}, \bar{\bf{Z}}_{n\! + \!a}^{n\! + \!a \! + \! m\! - \!1}, \ldots ,\bar{\bf{Z}}_{\left( {k \! - \! 1} \right) \cdot n\! + \!a}^{\left( {k \! - \! 1} \right) \cdot n\! + \!a \! + \! m\! - \!1}, {\bf{U}}_{a \! - \! m}^{a \! - \! 1} \Big)  \bigg) \notag \\
&\quad \stackrel{(c)}{\leq} \frac{1}{{b\cdot n} }I\left( M;\bar{\bf Z} _{a\! + \!m}^{n\! + \!a\! - \!1}, \bar{\bf Z}_{ n \! + \!a\! + \! m}^{2 n\! + \!a\! - \!1}, \ldots,\bar{\bf Z}_{(b\! - \!1)\cdot n \! + \!a\! + \! m}^{b\cdot n\! + \!a\! - \!1} \right) \! + \! \frac{\eta}{n} \notag \\
&\quad \stackrel{(d)}{=} \frac{1}{{b\cdot n} }I\left( M;{\bf Z} _{m}^{n\! - \!1}, {\bf Z}_{n \! + \! m}^{2 n\! - \!1}, \ldots,{\bf Z}_{(b\! - \!1)\cdot n \! + \! m}^{b\cdot n\! - \!1} \right) \! + \! \frac{\eta}{n} \notag \\
&\quad \stackrel{(e)}{\le} {\epsilon _2} \! - \! \frac{\eta}{n}, \label{eqn:Epsilon2_a}
\end{align}
\hrulefill
\vspace*{4pt}
\end{figure*}
where $(a)$ follows from the chain rule for mutual information \cite[Ch. 2.5]{Cover:06}; $(b)$ follows from the data-processing inequality \cite[Ch. 2.8]{Cover:06}; $(c)$ follows from Lemma \ref{lem:Proof_EtaBound}, and from the finite memory of the channel which implies that for $i \geq a+m$, $\bar{\bf Z}[i]$ is independent of ${\bf{U}}_{a - m}^{a - 1}$;
$(d)$ follows from \eqref{eqn:Leakage_Trans0a}; 
$(e)$ follows from \eqref{eqn:app_Rs2}.
Plugging \eqref{eqn:Epsilon2_a} into \eqref{eqn:Leakage_Trans1} yields 
\begin{equation*}
\mathop {\sup }\limits_{{{{\bf s} }_0} \in \mathcal{S}_0} 
\frac{1}{l}I\left( \left. M;\bar{\bf Z}^{l-1}  \right|{{{\bf S} }_0} ={{{\bf s} }_0}\right) 
\leq \frac{b\cdot n}{{b\cdot n + a}} \left({\epsilon _2} - \frac{\eta}{n}\right) \leq {\epsilon _2}.
\end{equation*}
The code rate for $\mathcal{C} _{l}^{LG}$ is obtained from
\begin{align*}
R_{LG}
&= R_1 \cdot \frac{b \cdot n }{b \cdot n + a} \\
&\stackrel{(a)}{\geq}\left( R_s - \frac{\epsilon_3}{2} \right)\frac{b \cdot n }{b \cdot n + a},
\end{align*}
where $(a)$ follows from \eqref{eqn:app_Rs3}. Thus, for sufficiently large $b$, namely, $b > \frac{2a(R_s - \epsilon_3)}{n \cdot \epsilon_3}$, it follows that $R_{LG} \geq R_s - \epsilon_3$.
It therefore follows that for all  sufficiently large $b$ and $a \in \{m,m+1,\ldots,n+m-1\}$, there exists a code for the LGMWTC with blocklength $l = b \cdot n + a$ which satisfies \eqref{eqn:def_Rs1}-\eqref{eqn:def_Rs3}. 
Consequently, for any secrecy rate $R_s \leq \mathop {\lim \sup }\limits_{n \rightarrow \infty} {\CMG{n}}$, $\exists l_0 \in \mathds{N}$ large enough such that reliable secure communications is achievable for the LGMWTC at any rate arbitrarily close to $R_s$,  for all blocklengths larger than $l_0$.
Thus, $R_s \leq \Cs$, from which it follows that $\mathop {\lim \sup }\limits_{n \rightarrow \infty} {\CMG{n}} \le {\Cs}$.
\end{IEEEproof}
\smallskip
\begin{remark}
\label{cmt:Proof1a}
{\em 
Note that without an eavesdropper, the $n$-MGMWTC becomes an instance to the $n$-block memoryless Gaussian multiterminal channel
($n$-MGMC), defined in \cite[Appendix A]{Goldsmith:01}, and the LGMWTC  becomes an instance to the linear Gaussian multiterminal channel (LGMC), defined in \cite[Appendix A]{Goldsmith:01}.
In  \cite[Lemma 2]{Goldsmith:01} it is shown that the capacity of the $n$-MGMC  
is not greater than the capacity of the LGMC {\em for all} $n > 2m$. However, when the eavesdropper is present, the {\em secrecy capacity} of the $n$-MGMWTC can be shown to be upper-bounded by that of the LGMWTC {\em only for $n \rightarrow \infty$}, as the information leakage due to the first $m$ channel outputs of each $n$-block received at the eavesdropper, which are not accounted for in the leakage model of the $n$-MGMWTC, is negligible only for asymptotic blocklengths with $n \rightarrow \infty$.
}
\end{remark}
%
\begin{proposition}
\label{pro:Proof_Qn2}
The secrecy capacity of the LGMWTC defined in \eqref{eqn:RxModel_2} subject to the power constraint in \eqref{eqn:Constraint1} satisfies 
\begin{equation*}
{\Cs} = \mathop {\lim }\limits_{n \to \infty } {\CMG{n}},
\end{equation*}
where $\CMG{n}$ is the secrecy capacity of the $n$-MGMWTC, which is stated in \eqref{eqn:Sec_DefQn}, and the limit exists.
\end{proposition}
\begin{IEEEproof}
By combining the above lemmas it follows that 
\begin{align*}
\mathop {\lim \sup }\limits_{n \to \infty } {\CMG{n}} 
&\stackrel{(a)}{\leq} \Cs \\
&\stackrel{(b)}{\leq} \mathop {\inf }\limits_{{{{\bf s} }_0} \in \mathcal{S}_0} \left( {\mathop {\lim \inf }\limits_{n \to \infty } {C_n}\left( {{{{\bf s} }_0}} \right)} \right) \\
&\stackrel{(c)}{\leq} \mathop {\lim \inf }\limits_{n \to \infty } {\CMG{n}},
\end{align*}
where $(a)$ follows from Lemma \ref{lem:Proof_Thm1c}, $(b)$ follows from Lemma \ref{lem:Proof_Thm1a}, and $(c)$ follows from Lemma \ref{lem:Proof_Thm1b}. 
Since $\mathop {\lim \inf }\limits_{n \to \infty } {\CMG{n}} \leq \mathop {\lim \sup }\limits_{n \to \infty } {\CMG{n}}$, 
it follows from  \cite[Sec. 3.18]{Rudin:76} that
\begin{equation*}
\Cs =  \mathop {\lim }\limits_{n \to \infty } {\CMG{n}},
\end{equation*}
and the limit exists. This proves the proposition.
\end{IEEEproof}
\smallskip

\subsection{Proving that $\mathop {\lim }\limits_{n \to \infty } \CMG{n}$ is Equal to $\mathop {\lim }\limits_{n \to \infty } \CCG{n}$}
\label{app:Proof1b}
We next prove that in the limit of $n \rightarrow \infty$, the $n$-CGMWTC and the $n$-MGMWTC have the same secrecy capacity.
This is done in the following steps:
\begin{itemize}
	\item First, we obtain in Lemma \ref{lem:Proof_Rn2} an expression for the secrecy capacity of the $n$-CGMWTC, $\CCG{n}$, by proving that it can be transformed into an equivalent memoryless MIMO WTC.
	\item Next, in Lemma \ref{lem:Proof_Rn3} we prove that for a single $n$-block, the mutual information between the channel input and the last $n-m$ channel outputs is the same for both the $n$-CGMWTC and the $n$-MGMWTC. 	
	\item Then, we show in Lemma \ref{lem:Proof_EtaBound3} that the mutual information between the channel inputs and any $m$ channel outputs of the $n$-CGMWTC can be upper bounded by a fixed and finite number.
	\item Lastly, in Proposition \ref{pro:Proof_Rn2} we use Lemma \ref{lem:Proof_Rn2}, Lemma \ref{lem:Proof_Rn3}, and Lemma \ref{lem:Proof_EtaBound3} to prove  that there exists a finite $\utilde{\eta}$, such that $\forall n > 2m$,  $\CMG{n} - \frac{\utilde{\eta}}{n} \leq \CCG{n} \leq \CMG{n} + \frac{\utilde{\eta}}{n}$, thus in the limit of $n \rightarrow \infty$, $\CMG{n}$ is equal to $\CCG{n}$.
\end{itemize}

\begin{lemma}
\label{lem:Proof_Rn2}
The secrecy capacity of the $n$-CGMWTC subject to the power constraint in \eqref{eqn:Constraint1} is given by
\begin{equation}
\CCG{n} = \frac{1}{n} \mspace{-25mu} \sup_{ \substack{ p\left( {\bf V}^{n-1},{\bf X}^{n-1} \right): \\ \MyBeta{n} }} \mspace{-25mu} \bigg\{ I\left( {{\bf V} }^{n-1};\utilde{\bf{Y}}^{n-1} \right) - I\left( {{{{\bf V} }^{n-1}};\utilde{\bf{Z}} ^{n-1}} \right) \bigg\}.
\label{eqn:Proof_Rn2}
\end{equation}
\end{lemma}
\begin{IEEEproof}
The proof of this lemma follows the same outline as in the proof of Proposition \ref{pro:Proof_Qn1}.
We first show that \eqref{eqn:Proof_Rn2} characterizes the maximum achievable secrecy rate when considering only codes whose blocklength is an integer multiple of $n$, i.e, $\left[R, b \cdot n\right]$ codes where $b$ is a positive integer. This is proved by transforming the $n$-CGMWTC into an equivalent memoryless MIMO WTC using a bijective transformation, and then characterizing the capacity of the transformed channel.
Then, we show that every secrecy rate achievable for the $n$-CGMWTC can be achieved by considering only codes whose blocklength is an integer multiple of $n$.

Let us consider the $n$-CGMWTC subject to the constraint that only codes with blocklengths that are integer multiples of $n$ are allowed. In this case we can transform the channel into an equivalent $n \cdot {n_t} \times n \cdot {n_r} \times n \cdot {n_e}$ memoryless MIMO wiretap channel without loss of information, via the following assignment: Define the input of the transformed channel by the $n \cdot {n_t} \times 1$ vector ${{\bf X} _{eq}}\left[\, \tilde i \,\right] \triangleq {\bf X} _{ {\tilde i}\cdot n}^{\left( {\tilde i + 1} \right) \cdot n-1}$, $\tilde i \geq 0$, the output at the intended receiver by the $n \cdot {n_r} \times 1$ vector ${\utilde{\bf Y} _{eq}}\left[\, \tilde i \,\right] \triangleq \utilde{\bf Y} _{ {\tilde i } \cdot n }^{\left( {\tilde i + 1} \right) \cdot n-1}$, and the output at the eavesdropper by the $n \cdot {n_e} \times 1$ vector ${\utilde{\bf Z} _{eq}}\left[\, \tilde i \,\right] \triangleq \utilde{\bf Z} _{ {\tilde i } \cdot n }^{\left( {\tilde i + 1} \right) \cdot n-1}$. The transformation is clearly bijective thus, the secrecy capacity of the transformed channel is equal to the secrecy capacity of the original channel.
Since the $n$-CGMWTC is $n$-block memoryless, it follows from Def. \ref{def:MemorylessChannel} and Def. \ref{def:BlockMemorylessChannel} that the transformed MIMO channel is memoryless. The outputs at the intended receiver and at the eavesdropper are corrupted by the additive noise vectors ${\utilde{\bf W} _{eq}}\left[\, \tilde i \,\right] \triangleq \utilde{\bf W} _{ {\tilde i } \cdot n }^{\left( {\tilde i + 1} \right) \cdot n-1}$ and ${\utilde{\bf U} _{eq}}\left[\, \tilde i \,\right] \triangleq \utilde{\bf U} _{ {\tilde i } \cdot n }^{\left( {\tilde i + 1} \right) \cdot n-1}$, respectively. From the definition of $\utilde{\bf W}[i]$ and $\utilde{\bf U}[i]$ in Section
 \ref{sec:SecCap} it follows that both ${\utilde{\bf W} _{eq}}\left[\, \tilde i \,\right]$ and  ${\utilde{\bf U} _{eq}}\left[\, \tilde i \,\right]$ are zero-mean real Gaussian vectors with positive-definite covariance matrices (this follows since the elements of the random vectors are not linearly dependent, see \cite[Ch. 8.1]{Papoulis:91}). From the construction of the transformed equivalent channel, the definition of the noises for the $n$-CGMWTC in the proof outline of Thm. \ref{Thm:MainThm2}, and of block-memorylessness  in Def. \ref{def:BlockMemorylessChannel}, it follows that both  ${\utilde{\bf W} _{eq}}\left[\, \tilde i \,\right]$ and  ${\utilde{\bf U} _{eq}}\left[\, \tilde i \,\right]$ are i.i.d. and mutually independent.
The secrecy capacity of the transformed channel, denoted $C_n^{eq}$,  can be written in the form of the result of  Csisz{\'a}r and K\"{o}rner \cite[Eq. (11)]{Csiszar:78}:
\begin{equation*}
C_n^{eq}= \mspace{-25mu} \sup_{\substack{ p\left( {\bf V}^{n-1},{\bf X}^{n-1} \right): \\ \MyBeta{n} } } \mspace{-25mu} \Big\{I\left( {{{{\bf V} }^{n-1}};\utilde{\bf Y}^{n-1}} \right)  - I\left( {{{{\bf V} }^{n-1}};\utilde{\bf Z} ^{n-1}} \right) \Big\},
\end{equation*}
where the constraint $\MyBeta{n}$ follows from the {\em per-symbol} power constraint of the $n$-CGMWTC.
As each MIMO channel use corresponds to $n$ channel uses in the original channel, it follows that the achievable secrecy rate of the $n$-CGMWTC, subject to the constraint that only codes with blocklengths that are integer multiples of $n$ are allowed, in bits per channel use, is $\frac{1}{n}C_n^{eq}$, which coincides with \eqref{eqn:Proof_Rn2}.

Next, we show that any secrecy rate achievable for the $n$-CGMWTC can be achieved by considering only codes with blocklengths that are integer multiples of $n$: Consider a secrecy rate $R_s$ achievable for the $n$-CGMWTC and fix $\epsilon _1$, $\epsilon _2$, and $\epsilon _3$ to arbitrary positive real numbers. From Def. \ref{def:SecrecyRate}  it follows that $\exists n _0 > 0$ such that $\forall l > n _0$ there exists an $\left[R, l\right]$ code which satisfies \eqref{eqn:def_Rs1}-\eqref{eqn:def_Rs3}.
Thus, by setting $b_0$ as the smallest integer such that $b_0 \cdot n \geq n _0$ it follows that for all integer $b > b_0$ there exists a $\left[R, b \cdot n\right]$ code which satisfies \eqref{eqn:def_Rs1}-\eqref{eqn:def_Rs3}. Therefore, the secrecy rate $R_s$ is also achievable when considering only codes whose blocklength is an integer multiple of $n$. We therefore conclude that \eqref{eqn:Proof_Rn2} denotes the maximum achievable secrecy rate for the $n$-CGMWTC, which completes the proof of the lemma.
\end{IEEEproof}
\smallskip

\begin{lemma}
\label{lem:Proof_Rn3}
For any joint distribution $p\left( {\bf V}^{n-1},{\bf X}^{n-1} \right)$ such that ${\bf V}^{n-1} \rightarrow {\bf X}^{n-1}\rightarrow {\bf Y}_m^{n-1},{\bf Z}_m^{n-1}$ and  ${\bf V}^{n-1} \rightarrow {\bf X}^{n-1}\rightarrow \utilde{\bf Y}_m^{n-1},\utilde{\bf Z}_m^{n-1}$ form a Markov chain, the channel outputs of the $n$-MGMWTC and of the $n$-CGMWTC satisfy 
\begin{subequations}
\label{eqn:Proof_Rn3}
\begin{equation}
\label{eqn:Proof_Rn3a}
I\left({\bf V} ^{n-1};{\bf{Y}}_m^{n-1} \right) = I\left({\bf V} ^{n-1};{\bf{\utilde Y}}_m^{n-1} \right),
\end{equation}
and
\begin{equation}
\label{eqn:Proof_Rn3b}
I\left({\bf V} ^{n-1};{\bf{Z}}_m^{n-1} \right) = I\left({\bf V} ^{n-1};{\bf{\utilde Z}}_m^{n-1} \right).
\end{equation}
\end{subequations}
\end{lemma}
\begin{IEEEproof}
It follows from \eqref{eqn:RxModel_2} and \eqref{eqn:CRxModel_2} that $\exists \utilde{\mathsf{H}} \in \mathds{R}^{\left(n_r\cdot (n\! - \!m)\right) \times \left(n_t \cdot n\right)}$ such that ${\bf{Y}}_m^{n\! - \!1} = \utilde{\mathsf{H}}{\bf{X}}^{n\! - \!1} + {\bf{W}}_m^{n\! - \!1}$ and $\utilde{\bf{Y}}_m^{n\! - \!1} = \utilde{\mathsf{H}}{\bf{X}}^{n\! - \!1} + \utilde{\bf{W}}_m^{n\! - \!1}$. Hence, as the channel input ${\bf{X}}^{n\! - \!1}$ is independent of the noise in both channels it follows that
\begin{align}
\!\!\!p_{\left.{\bf{Y}}_m^{n\! - \! 1}\right|{\bf X} ^{n\! - \! 1}}\left(\left.{\bf{y}}_m^{n\! - \! 1}\right|{\bf x} ^{n\! - \! 1}\right)
&= \!p_{{\bf{W}}_m^{n\! - \! 1}}\left({\bf{y}}_m^{n\! - \! 1} - \utilde{\mathsf{H}}{\bf x} ^{n\! - \! 1}\right)
\notag \\
&\stackrel{(a)}{=}\! p_{\utilde{\bf{W}}_m^{n\! - \! 1}}\left({\bf{y}}_m^{n\! - \! 1} - \utilde{\mathsf{H}}{\bf x} ^{n\! - \! 1}\right)
\notag \\
&= \!p_{\left.\utilde{\bf{Y}}_m^{n\! - \! 1}\right|{\bf X} ^{n\! - \! 1}}\left(\left.{\bf{y}}_m^{n\! - \! 1}\right|{\bf x} ^{n\! - \! 1}\right),  \label{eqn:Proof_Rn3c}
\end{align}
where $(a)$ follows as, by definition, the random vectors ${\bf{W}}_m^{n\! - \!1}$ and $\utilde{\bf{W}}_m^{n\! - \!1}$ are identically distributed, as both are zero-mean real Gaussian random vectors with the same correlation matrix: To see this, the $\forall i_1, i_2 \in \{m, m+1, \ldots , n-1\}$ we write
\begin{align*}
\E\left\{\utilde{\bf{W}}[i_1]\utilde{\bf{W}}^H[i_2]\right\}
&= \CUw{i_1 - i_2} \notag \\
&=  \Cw{i_1 - i_2} + \Cw{i_1 - i_2 + n} + \Cw{i_1 - i_2 - n} \notag \\
&\stackrel{(b)}{=} \Cw{i_1 - i_2} \notag \\
&= \E\left\{{\bf{W}}[i_1]{\bf{W}}^H[i_2]\right\},
\end{align*}
where $(b)$ follows from \eqref{eqn:CorrModel1a} as $|i_1 - i_2| < n-m$. 
It therefore follows that 
\begin{align}
{p_{{{\bf{V}}^{n \! - \! 1}},{\bf{Y}}_m^{n \! - \! 1}}}\left( {{{\bf{v}}^{n \! - \! 1}},{\bf{y}}_m^{n \! - \! 1}} \right) 
&= \int\limits_{\xvec^{n-1}\in\dsR^{n_t \cdot n}} {{p_{{{\bf{V}}^{n \! - \! 1}},{{\bf{X}}^{n \! - \! 1}},{\bf{Y}}_m^{n \! - \! 1}}}\left( {{{\bf{v}}^{n \! - \! 1}},{{\bf{x}}^{n \! - \! 1}},{\bf{y}}_m^{n \! - \! 1}} \right)} d{{\bf{x}}^{n \! - \! 1}} \notag \\
&\stackrel{(a)}{=} \int\limits_{\xvec^{n-1}\in\dsR^n} {p_{\left. {{\bf{Y}}_m^{n \! - \! 1}} \right|{{\bf{X}}^{n \! - \! 1}}}}\left( {\left. {{\bf{y}}_m^{n \! - \! 1}} \right|{{\bf{x}}^{n \! - \! 1}}} \right) {p_{{{\bf{V}}^{n \! - \! 1}},{{\bf{X}}^{n \! - \! 1}}}}\left( {{{\bf{v}}^{n \! - \! 1}},{{\bf{x}}^{n \! - \! 1}}} \right) d{{\bf{x}}^{n \! - \! 1}} \notag \\
&\stackrel{(b)}{=} \int\limits_{\xvec^{n-1}\in\dsR^{n_t \cdot n}} {p_{\left. {\utilde{\bf{Y}}_m^{n \! - \! 1}} \right|{{\bf{X}}^{n \! - \! 1}}}}\left( {\left. {{\bf{y}}_m^{n \! - \! 1}} \right|{{\bf{x}}^{n \! - \! 1}}} \right){p_{{{\bf{V}}^{n \! - \! 1}},{{\bf{X}}^{n \! - \! 1}}}}\left( {{{\bf{v}}^{n \! - \! 1}},{{\bf{x}}^{n \! - \! 1}}} \right) d{{\bf{x}}^{n \! - \! 1}} \notag \\
&\stackrel{(c)}{=} \int\limits_{\xvec^{n-1}\in\dsR^{n_t \cdot n}} {{p_{{{\bf{V}}^{n \! - \! 1}},{{\bf{X}}^{n \! - \! 1}},\utilde{\bf{Y}}_m^{n \! - \! 1}}}\left( {{{\bf{v}}^{n \! - \! 1}},{{\bf{x}}^{n \! - \! 1}},{\bf{y}}_m^{n \! - \! 1}} \right)} d{{\bf{x}}^{n \! - \! 1}} \notag \\
&= {p_{{{\bf{V}}^{n \! - \! 1}},\utilde{\bf{Y}}_m^{n \! - \! 1}}}\left( {{{\bf{v}}^{n \! - \! 1}},{\bf{y}}_m^{n \! - \! 1}} \right),
\label{eqn:Proof_Rn3d}
\end{align}
where $(a)$ follows since ${\bf V}^{n-1} \rightarrow {\bf X}^{n-1}\rightarrow {\bf Y}_m^{n-1}$ form a Markov chain; 
$(b)$ follows from \eqref{eqn:Proof_Rn3c}; 
and $(c)$ follows since ${\bf V}^{n-1} \rightarrow {\bf X}^{n-1}\rightarrow \utilde{\bf Y}_m^{n-1}$ form a Markov chain. Equality
\eqref{eqn:Proof_Rn3d} directly leads to  \eqref{eqn:Proof_Rn3a}. The proof of \eqref{eqn:Proof_Rn3b} is obtained using similar steps with the letters $Y$ and  $W$ in the derivations of \eqref{eqn:Proof_Rn3c} and \eqref{eqn:Proof_Rn3d} replaced by $Z$ and $U$, respectively. This completes the proof of the lemma.
\end{IEEEproof}
\smallskip

\begin{lemma}
\label{lem:Proof_EtaBound3} 
There exists a finite and fixed $\utilde{\eta} >0$, such that the channel outputs of the $n$-CGMWTC satisfy
\begin{subequations}
\label{eqn:Proven1}
\begin{equation}
I \left(\left. {{{{\bf X} }^{n-1}};\utilde{\bf{Y}}^{m- 1}} \right|\utilde{\bf{Y}}_m^{n-1} \right)\le \utilde{\eta}, \label{eqn:Proven1a}
\end{equation}
and
\begin{equation}
I \left(\left. {{{{\bf X} }^{n-1}};\utilde{\bf{Z}}^{m- 1}} \right|\utilde{\bf{Z}}_m^{n-1} \right)\le \utilde{\eta}. \label{eqn:Proven1b}
\end{equation}
\end{subequations}
\end{lemma}
\begin{IEEEproof}
\ifextended
Note that  
\begin{equation}
I \left(\left. {{{{\bf X} }^{n-1}};\utilde{\bf{Y}}^{m- 1}} \right|\utilde{\bf{Y}}_m^{n-1} \right)
= h\left( {\left. \utilde{\bf{Y}}^{m - 1} \right|\utilde{\bf{Y}}_m^{n - 1}} \right) - h\left( {\left. {{\utilde{\bf{Y}}^{m - 1}}} \right|{\bf{X}}^{n - 1},\utilde{\bf{Y}}_m^{n - 1}} \right),
\label{eqn:Sec2Bound1}
\end{equation}
where
\begin{align}
h\left( {\left. \utilde{\bf{Y}}^{m - 1}  \right|{\bf{X}}^{n - 1}, \utilde{\bf{Y}}_m^{n - 1}} \right)
&\stackrel{(a)}{=} h\left( {\left. {{\utilde{\bf{W}}^{m - 1}}} \right|{\bf{X}}^{n - 1},\utilde{\bf{W}}_m^{n - 1}} \right)
\notag \\
&\stackrel{(b)}{=}h\left( {\left. {{\utilde{\bf{W}}^{m - 1}}} \right|\utilde{\bf{W}}_m^{n - 1}} \right) \notag \\
&\stackrel{(c)}{=} h\left( {\left. \utilde{\bf{W}}^{m - 1} \right|\utilde{\bf{W}}_m^{2m - 1}, \utilde{\bf{W}}_{n-m}^{n - 1}} \right) \notag \\
&= \int\limits_{\left( {\bf{w}}_m^{2m \! - \! 1},{\bf{w}}_{n \! - \! m}^{n \! - \! 1}\right)  \in \dsR^{n_r \cdot 2m}}   h\left( \left. {{\utilde{\bf{W}}^{m \! - \! 1}}} \right|\utilde{\bf{W}}_m^{2m \! - \! 1} \!= \! \utilde{\bf{w}}_m^{2m \! - \! 1},\utilde{\bf{W}}_{n \! - \! m}^{n \! - \! 1} \!= \! \utilde{\bf{w}}_{n \! - \! m}^{n \! - \! 1} \right) \notag \\
& \qquad \qquad \qquad \qquad \quad \times p_{\utilde{\bf{W}}_m^{2m \! - \! 1},\utilde{\bf{W}}_{n \! - \! m}^{n \! - \! 1}}\left(\utilde{\bf{w}}_m^{2m \! - \! 1},\utilde{\bf{w}}_{n \! - \! m}^{n \! - \! 1}\right)d\utilde{\bf{w}}_m^{2m \! - \! 1}d\utilde{\bf{w}}_{n \! - \! m}^{n \! - \! 1},
\label{eqn:Sec2Bound2}
\end{align}
where $(a)$ follows since 
$\exists \utilde{\mathsf{\check H}} \in \mathds{R}^{\left(n_r \cdot m\right) \times \left(n_t \cdot n\right)}$ such that $\utilde{\bf{Y}}^{m - 1} = \utilde{\mathsf{\check H}} {\bf X}^{n-1}  + \utilde{\bf W}^{m-1}$, and $\exists \utilde{\mathsf{\ddot H}} \in \mathds{R}^{\left(n_r \cdot (n\!-\! m)\right) \times \left(n_t \cdot n\right)}$ such that $\utilde{\bf{Y}}_m^{n - 1} = \utilde{\mathsf{\ddot H}} {\bf X}^{n -1} + \utilde{\bf{W}}_m^{n - 1}$; 
$(b)$ follows since the noise $\utilde{\bf W}[i]$ is independent of the channel input ${\bf X}^{n-1}$;
$(c)$ follows since the finite temporal correlation of the multivariate Gaussian process $\utilde{\bf W}[i]$ is circular, finite, and smaller than $m+1$, which implies that $\utilde{\bf{W}}^{m - 1}$ is  mutually independent of $\utilde{\bf{W}}_{2m}^{n - m - 1}$. 

Since $\utilde{\bf{W}}^{m - 1}$ and $\left[\left(\utilde{\bf{W}}_m^{2m - 1}\right)^T,\left(\utilde{\bf{W}}_{n-m}^{n - 1}\right)^T\right]^T$ are jointly Gaussian, the conditional distribution $\left. {{\utilde{\bf{W}}^{m - 1}}} \right|\utilde{\bf{W}}_m^{2m - 1} = \utilde{\bf{w}}_m^{2m - 1}, \utilde{\bf{W}}_{n-m}^{n - 1} = \utilde{\bf{w}}_{n-m}^{n - 1}$  is a multivariate Gaussian distribution \cite[Proposition 3.13]{Eaton:07} with covariance matrix 
\begin{align}
&\utilde{\mathsf{Q}} \triangleq \E\left\{\utilde{\bf{W}}^{m \! - \! 1}\!\left(\utilde{\bf{W}}^{m \! - \! 1}\right)^T\right\} \! - \! \E\left\{\utilde{\bf{W}}^{m \! - \! 1}\!\left[\left(\utilde{\bf{W}}_m^{2m - 1}\right)^T,\left(\utilde{\bf{W}}_{n-m}^{n - 1}\right)^T\right]\right\} \Bigg(\E\bigg\{\left[\left(\utilde{\bf{W}}_m^{2m - 1}\right)^T,\left(\utilde{\bf{W}}_{n-m}^{n - 1}\right)^T\right]^T\!
 \notag \\
& \qquad\qquad \qquad  \times 
\left[\left(\utilde{\bf{W}}_m^{2m - 1}\right)^T,\left(\utilde{\bf{W}}_{n-m}^{n - 1}\right)^T\right]\bigg\}\Bigg)^{\! - \!1} 
 \E\left\{\left[\left(\utilde{\bf{W}}_m^{2m - 1}\right)^T,\left(\utilde{\bf{W}}_{n-m}^{n - 1}\right)^T\right]^T\!\left(\utilde{\bf{W}}^{m \! - \! 1}\right)^T\right\}.
\label{eqn:Sec2Bound3}
\end{align}
Since the noise samples are not linearly dependent, it follows that $\left|\utilde{\mathsf{Q}}\right| > 0$ \cite[Ch. 8.1]{Papoulis:91}.  
Using the expression for the differential entropy of a multivariate Gaussian RV \cite[Thm. 8.4.1]{Cover:06} we obtain that \eqref{eqn:Sec2Bound2} can be written as
\begin{equation}
h\left( \left. \utilde{\bf{Y}}^{m - 1} \right|{\bf{X}}^{n - 1},\utilde{\bf{Y}}_m^{n - 1} \right)  = 
\frac{1}{2}\log\left(\left(2 \pi e\right)^m \left|\utilde{\mathsf{Q}} \right|\right).
\label{eqn:Sec2Bound4}
\end{equation}
Plugging \eqref{eqn:Sec2Bound4} into \eqref{eqn:Sec2Bound1} leads to 
\begin{align}
I\left( {\left. {{{\bf{X}}^{n - 1}};{\utilde{\bf{Y}}^{m - 1}}} \right|\utilde{\bf{Y}}_m^{n - 1}} \right) 
&= h\left( {\left. {{\utilde{\bf{Y}}^{m - 1}}} \right|\utilde{\bf{Y}}_m^{n - 1}} \right) - \frac{1}{2}\log\left(\left(2 \pi e\right)^m \left|\utilde{\mathsf{Q}}\right|\right) \notag \\
&\stackrel{(a)}{\leq} h\left(  {{\utilde{\bf{Y}}^{m - 1}}} \right) - \frac{1}{2}\log\left(\left(2 \pi e\right)^m \left|\utilde{\mathsf{Q}} \right|\right), 
\label{eqn:Sec2Bound5}
\end{align}
where $(a)$ follows since conditioning reduces entropy \cite[Ch. 8.6]{Cover:06}. 
Define $\utilde{\bf X}_{n-m}^{m-1} \triangleq \left[\left({\bf X}_{n-m}^{n-1}\right)^T, \left({\bf X}^{m-1}\right)^T\right]^T$. From the input-output relationship of the $n$-CGMWTC it follows that $\exists \utilde{\mathsf{H}}_1 \in \mathds{R}^{\left(n_r \cdot m\right) \times \left(n_t \cdot 2\cdot m\right)}$ such that $\utilde{\bf{Y}}^{m - 1} = \utilde{\mathsf{H}}_1 \utilde{\bf X}_{n-m}^{m-1}  + \utilde{\bf W}^{m-1}$.
Let 
$\mathsf{K}_{\utilde{\bf Y}}$ be the covariance matrix of $\utilde{\bf{Y}}^{m - 1}$,
$\mathsf{K}_{\utilde{\bf X}}$ be the covariance matrix of $\utilde{\bf X}_{n-m}^{m-1}$,
and $\mathsf{K}_{\utilde{\bf W}}$ be the covariance matrix of $\utilde{\bf{W}}^{m - 1}$. 
Since the channel input ${\bf X}[i]$, $i \geq 0$, is subject to a per-symbol power constraint $P$, it follows that the entries of $\mathsf{K}_{\utilde{\bf X}}$ are all not larger than $P$. 
Since ${\bf X}[i]$ and ${\bf W}[i]$ are mutually independent, it follows that $\mathsf{K}_{\utilde{\bf Y}} = \utilde{\mathsf{H}}_1 \mathsf{K}_{\utilde{\bf X}} \utilde{\mathsf{H}}_1^T + \mathsf{K}_{\utilde{\bf W}}$, and since the noise samples are not linearly dependent, we obtain $\left|\mathsf{K}_{\utilde{\bf Y}}\right| > 0$ \cite[Ch. 8.1]{Papoulis:91}.  
Define $\utilde{\gamma}_k \triangleq \sum\limits_{{k_1} = 0}^{ 2\cdot m \cdot {n_t} - 1} \sum\limits_{{k_2} = 0}^{ 2\cdot m \cdot {n_t} - 1} \left|\left( {{\utilde{\mathsf{H}}_1}} \right)_{k,{k_2}}\left( {{\utilde{\mathsf{H}}_1}} \right)_{k,{k_1}}\right|$. 
It follows from Hadamard's inequality \cite[Thm. 17.9.2]{Cover:06} that 
\begin{align*}
\left|\mathsf{K}_{\utilde{\bf Y}} \right| 
&\leq \prod\limits_{k = 0}^{m \cdot {n_r} - 1} {{{\left( \mathsf{K}_{\utilde{\bf Y}} \right)}_{k,k}}} \notag \\
&= \prod\limits_{k = 0}^{m \cdot {n_r} - 1} {\left( {{{\left( \utilde{\mathsf{H}}_1\mathsf{K}_{\utilde{\bf X}}\utilde{\mathsf{H}}_1^T \right)}_{k,k}} + {{\left(\mathsf{K}_{\utilde{\bf W}} \right)}_{k,k}}} \right)} \notag \\
&= \prod\limits_{k = 0}^{m \cdot {n_r} - 1} {\left( {\sum\limits_{{k_1} = 0}^{ 2\cdot m \cdot {n_t} - 1} {\sum\limits_{{k_2} = 0}^{ 2\cdot m \cdot {n_t} - 1} {{{\left( \utilde{\mathsf{H}}_1 \right)}_{k,{k_2}}}{{\left( \mathsf{K}_{\utilde{\bf X}} \right)}_{{k_2},{k_1}}}} {{\left( \utilde{\mathsf{H}}_1 \right)}_{k,{k_1}}}}  + {{\left( \mathsf{K}_{\utilde{\bf W}} \right)}_{k,k}}} \right)} \notag \\
&\le \prod\limits_{k = 0}^{m \cdot {n_r} - 1} {\left( {{\utilde{\gamma} _k}P + {{\left( \mathsf{K}_{\utilde{\bf W}} \right)}_{k,k}}} \right)}.
\end{align*}
It follows that $\left|\mathsf{K}_{\utilde{\bf Y}} \right|$ is positive, finite, and independent of $n$. 
Since $ h\left(  {{\utilde{\bf{Y}}^{m - 1}}} \right)$ is upper-bounded by the differential entropy of an $n_r \cdot m \times 1$  multivariate Gaussian RV with the same covariance matrix \cite[Thm. 8.6.5]{Cover:06}, it follows from \eqref{eqn:Sec2Bound5} that 
\begin{equation}
I\left( {\left. {{{\bf{X}}^{n - 1}};{\utilde{\bf{Y}}^{m - 1}}} \right|\utilde{\bf{Y}}_m^{n - 1}} \right) 
\leq  \frac{1}{2}\log\left(\left(2 \pi e\right)^m \left|\mathsf{K}_{\utilde{\bf Y}} \right|\right) - \frac{1}{2}\log\left(\left(2 \pi e\right)^m \left|\utilde{\mathsf{Q}} \right|\right), 
\label{eqn:Sec2Bound6}
\end{equation}
From \eqref{eqn:Sec2Bound6} we conclude $\exists \utilde{\eta}$ independent of $n$ such that $I\left( {\left. {{{\bf{X}}^{n - 1}};{\utilde{\bf{Y}}^{m - 1}}} \right|\utilde{\bf{Y}}_m^{n - 1}} \right) \leq \utilde{\eta}$.
 The proof of \eqref{eqn:Proven1b} is obtained using similar steps with the letters $Y$ and $W$ in the derivation leading to \eqref{eqn:Sec2Bound6} are replaced by $Z$ and $U$, respectively. This completes the proof of the lemma. 
\fi
\end{IEEEproof}

\begin{proposition}
\label{pro:Proof_Rn2}
The secrecy capacity of the $n$-MGMWTC and the secrecy capacity of the $n$-CGMWTC satisfy
\begin{equation}
\label{eqn:pro_Rn2}
\mathop {\lim }\limits_{n \to \infty } {\CMG{n}} =  \mathop {\lim }\limits_{n \to \infty } \CCG{n},
\end{equation}
and the limits exist.
\end{proposition}
\begin{IEEEproof}
It follows from the mutual information chain rule \cite[Sec. 2.5]{Cover:06} that 
\begin{subequations}
\begin{align}
 I\left( {{{{\bf V} }^{n-1}};\utilde{\bf{Y}}^{n-1}} \right)  
&  = I\left( {{{{\bf V} }^{n-1}};\utilde{\bf{Y}}_m^{n-1}} \right) +I \left(\left. {{{{\bf V} }^{n-1}};\utilde{\bf{Y}}^{m- 1}} \right|\utilde{\bf{Y}}_m^{n-1} \right) \notag \\
&  \stackrel{(a)}{\leq} I\left( {{{{\bf V} }^{n-1}};\utilde{\bf{Y}}_m^{n-1}} \right) +I \left(\left. {{{{\bf X} }^{n-1}};\utilde{\bf{Y}}^{m- 1}} \right|\utilde{\bf{Y}}_m^{n-1} \right) \notag \\
&  \stackrel{(b)}{\leq} I\left( {{{{\bf V} }^{n-1}};\utilde{\bf{Y}}_m^{n-1}} \right) + {\utilde{\eta}},
\label{eqn:pro_Rn2a}
\end{align}
where $(a)$ follows from the data processing inequality \cite[Thm. 2.8.1]{Cover:06} and the Markov chain $\mathbf{V}^{n-1} \rightarrow \mathbf{X}^{n-1} \rightarrow \utilde{\mathbf{Y}}^{n-1},\utilde{\mathbf{Z}}^{n-1}$, and $(b)$ follows from Lemma \ref{lem:Proof_EtaBound3}.
Similarly,  
\begin{equation}
\label{eqn:pro_Rn2c}
I\left( {{{{\bf V} }^{n-1}};\utilde{\bf{Z}}^{n-1}} \right) \leq I\left( {{{{\bf V} }^{n-1}};\utilde{\bf{Z}}_m^{n-1}} \right) + {\utilde{\eta}}.
\end{equation}
\end{subequations}
Now, for any given joint distribution  $p\left( {\bf V}^{n-1} ,{\bf X}^{n-1} \right)$, it follows from \eqref{eqn:pro_Rn2a} that
\begin{align}
 I\left( {{{{\bf V} }^{n-1}};\utilde{\bf{Y}}^{n-1}} \right) - I\left( {{{{\bf V} }^{n-1}};\utilde{\bf{Z}} ^{n-1}} \right)
&\leq I \left( {{{{\bf V} }^{n-1}};  \utilde{\bf{Y}}_m^{n-1}} \right) + {\utilde{\eta}} - I\left( {{{{\bf V} }^{n-1}};\utilde{\bf{Z}} ^{n-1}} \right) \notag \\
&\stackrel{(a)}{\leq} I\left( {{{{\bf V} }^{n-1}};\utilde{\bf{Y}}_m^{n-1}} \right)  - I\left( {{{{\bf V} }^{n-1}};\utilde{\bf{Z}}_m^{n-1}} \right) + {\utilde{\eta}} \notag \\
&\stackrel{(b)}{=} I \left( {{{{\bf V} }^{n-1}};{\bf{ Y}}_m^{n-1}} \right)  -  I \left( {{{{\bf V} }^{n-1}};{\bf{ Z}}_m ^{n-1}} \right) + {\utilde{\eta}},
\label{eqn:pro_Rn2bNew}
\end{align}
where $(a)$ follows from the chain rule for mutual information \cite[Sec. 2.5]{Cover:06} and the fact that mutual information is non-negative; $(b)$ follows  from Lemma \ref{lem:Proof_Rn3}.
From Lemma \ref{lem:Proof_Rn2} it follows that
\begin{align}
\!\!\CCG{n}
&= \frac{1}{n} \mspace{-25mu} \sup_{ \substack{ p\left( {\bf V}^{n-1} ,{\bf X}^{n-1} \right): \\ \MyBeta{n}}} \mspace{-25mu} \bigg\{I\left( {{{{\bf V} }^{n-1}};\utilde{\bf{Y}}^{n-1}} \right)  - I\left( {{{{\bf V} }^{n-1}};\utilde{\bf{Z}} ^{n-1}} \right) \bigg\} \notag \\
&\stackrel{(a)}{\leq} \frac{1}{n} \mspace{-25mu} \sup_{ \substack{ p\left( {\bf V}^{n-1} ,{\bf X}^{n-1} \right): \\ \MyBeta{n} }} \mspace{-25mu} \bigg\{ I \left( {{{{\bf V} }^{n-1}};{\bf{ Y}}_m^{n-1}} \right) -  I \left( {{{{\bf V} }^{n-1}};{\bf{ Z}}_m ^{n-1}} \right) \bigg\} + \frac{\utilde{\eta}}{n} \notag \\
&= \CMG{n} + \frac{\utilde{\eta}}{n},\label{eqn:pro_Rn2b}
\end{align}
where $(a)$ follows from  \eqref{eqn:pro_Rn2bNew}.

Next, for any given $p\left( {\bf V}^{n-1} ,{\bf X}^{n-1} \right)$, we also have the following relationship
\begin{align}
\!\!\!\! {I\left( {{{{\bf V} }^{n-1}};{\bf{Y}}_m^{n-1}} \right) - I\left( {{{{\bf V} }^{n-1}};{\bf{ Z}}_m ^{n-1}} \right)} 
&\stackrel{(a)}{=}  I\left( {{{{\bf V} }^{n-1}};\utilde{\bf{Y}}_m^{n-1}} \right) - I\left( {{{{\bf V} }^{n-1}};\utilde{\bf{ Z}}_m ^{n-1}} \right) \notag \\
&\stackrel{(b)}{\leq}  I\left( {{{{\bf V} }^{n-1}};\utilde{\bf{Y}}_m^{n-1}} \right) - I\left( {{{{\bf V} }^{n-1}};\utilde{\bf{ Z}}^{n-1}} \right)+ \utilde{\eta}  \notag \\
&\stackrel{(c)}{\leq}  I\left( {{{{\bf V} }^{n-1}};\utilde{\bf{Y}}^{n-1}} \right) - I\left( {{{{\bf V} }^{n-1}};\utilde{\bf{ Z}}^{n-1}}  \right) + {\utilde{\eta}},
\label{eqn:pro_Rn2dNew}
\end{align}
where $(a)$ follows from  Lemma \ref{lem:Proof_Rn3}; 
$(b)$ follows from \eqref{eqn:pro_Rn2c};
$(c)$ follows from the chain rule for mutual information \cite[Sec. 2.5]{Cover:06} and as mutual information is non-negative which implies that $I\left( {{{{\bf V} }^{n-1}};\utilde{\bf{Y}}^{n-1}} \right) \geq I\left( {{{{\bf V} }^{n-1}};\utilde{\bf{Y}}_m^{n-1}} \right)$.
From the definition of $\CMG{n}$ in \eqref{eqn:Sec_DefQn} it follows that
\begin{align}
{\CMG{n}} 
&= \frac{1}{n} \mspace{-25mu} \sup_{ \substack{ p\left( {\bf V}^{n-1} ,{\bf X}^{n-1} \right): \\ \MyBeta{n} } } \mspace{-25mu} \bigg\{I\left( {{{{\bf V} }^{n-1}};{\bf{Y}}_m^{n-1}} \right)  - I\left( {{{{\bf V} }^{n-1}};{\bf{ Z}}_m ^{n-1}} \right) \bigg\} \notag \\
&\stackrel{(a)}{\leq} \frac{1}{n}\mspace{-25mu} \sup_{ \substack{ p\left( {\bf V}^{n-1} ,{\bf X}^{n-1} \right): \\ \MyBeta{n} }} \mspace{-25mu} \bigg\{ I\left( {{{{\bf V} }^{n-1}};\utilde{\bf{Y}}^{n-1}} \right)  - I\left( {{{{\bf V} }^{n-1}};\utilde{\bf{ Z}}^{n-1}}  \right) \bigg\} + \frac{\utilde{\eta}}{n} \notag \\
&= \CCG{n} + \frac{\utilde{\eta}}{n},\label{eqn:pro_Rn2d}
\end{align}
where $(a)$ follows from \eqref{eqn:pro_Rn2dNew}. 
 Combining \eqref{eqn:pro_Rn2b} and \eqref{eqn:pro_Rn2d} yields
\begin{equation}
\label{eqn:pro_Rn2e}
\CMG{n} - \frac{\utilde{\eta}}{n} \leq \CCG{n} \leq \CMG{n} + \frac{\utilde{\eta}}{n}.
\end{equation}
Since $\mathop {\lim }\limits_{n  \to \infty }\frac{\utilde{\eta}}{n} = 0$, and since $\mathop {\lim }\limits_{n \to \infty } {\CMG{n}} = {\Cs}$ exists, letting $n \rightarrow \infty$ in \eqref{eqn:pro_Rn2e}  proves the proposition.
\end{IEEEproof}
\smallskip

Combining Propositions \ref{pro:Proof_Qn2} and \ref{pro:Proof_Rn2} proves Proposition \ref{Pro:MainThm1}.

\section{Proof of Proposition \ref{pro:Expres_Rn}}
\label{app:Proof2}
\vspace{-0.1cm}
Recall that the secrecy capacity of the $n$-CGMWTC subject to the {\em per-symbol} power constraint \eqref{eqn:Constraint1}, is denoted by $\CCG{n}$.
In order to derive the expression in \eqref{eqn:pro_main2}   for $\CCG{n}$,
we  first derive the secrecy capacity of the $n$-CGMWTC subject to the {\em time-averaged} power constraint \cite[Sec. II]{Wornell:10}, \cite[Eq. (7)]{Massey:88}, \cite[Eq. (7)]{Goldsmith:01}:
\begin{subequations} 
\label{eqn:Constraint1eq}
\begin{equation}
\label{eqn:Constraint1eqa}
\E\left\{ {\frac{1}{{{l}}}\sum\limits_{i = 0}^{l - 1} {{{\left\| {{\bf X} \left[ i \right]} \right\|}^2}} } \right\} \le P, 
\end{equation}
for all blocklengths $l$, and specifically, {\em for each $n$-block of the $n$-CGMWTC} we require
\begin{equation}
\label{eqn:Constraint1eqb}
\E\left\{ {\frac{1}{{{n}}}\sum\limits_{i = 0}^{n - 1} {{{\left\| {{\bf X} \left[ i \right]} \right\|}^2}} } \right\} \le P. 
\end{equation}
\end{subequations} 
We denote the secrecy capacity of the $n$-CGMWTC subject to \eqref{eqn:Constraint1eq} with $\CCGTA{n}$.
The derivation consists of the following steps:
\begin{itemize}
	\item First, in Lemma \ref{lem:Proof_Noise1} we show that applying the DFT transforms the $n$-CGMWTC into a set of independent parallel MIMO WTCs.
We then explain that any achievable rate for the $n$-CGMWTC  can be obtained by considering only codewords whose length is an integer multiple of $n$.

	\item  Next,  in Lemma \ref{lem:ParallelChannels} we derive the maximal achievable secrecy rate given a fixed power allocation, when the blocklength is an integer multiple of $n$, by a simple extension of the results of \cite[Thm. 1]{Li:10}. We conclude that $\CCGTA{n}$ can be written as a maximization of the sum of the per-subchannel secrecy capacities over all power allocations satisfying a specified sum-power constraint.

	\item  Then, in Lemma \ref{lem:Proof_Noise3} we characterize symmetry conditions on the maximal achievable secrecy rate expression and on the optimal input distribution for
 the $n$-CGMWTC, subject to \eqref{eqn:Constraint1eq}. This results in an explicit expression for $\CCGTA{n}$ stated in \eqref{eqn:ParallelChannels3}.

	\item Lastly,  in Corollary \ref{cor:PerSymbol} we prove that $\CCG{n} = \CCGTA{n}$.
\end{itemize}
The approach of characterizing the capacity of a channel subject to a per-symbol power constraint by considering a time-averaged power constraint was also used in \cite{Massey:88} for the point-to-point LTI channel (without an eavesdropper).

We begin with some preliminary properties of the multivariate DFT defined in  \eqref{eqn:MultiDFT}.
Recall that in  \eqref{eqn:MultiDFT}, each entry $l \in \{0,1,\ldots, n_q - 1\}$ of the multivariate DFT $\left\{{\bf \hat{q}}[k]\right\}_{k=0}^{n-1}$ is obtained as the scalar DFT of the $l$-th entries of the sequence of vectors $\left\{{\bf q}[i]\right\}_{i=0}^{n-1}$. 
Consequently, the following properties of the  multivariate DFT of real-valued multivariate sequences can be obtained as straightforward extensions of the corresponding properties of  scalar DFTs, see, e.g., \cite[Ch. 8.5-8.6]{Oppenheim:89}:
\begin{enumerate}[label={\em P\arabic*}]
	\item 	The multivariate DFT defined in \eqref{eqn:MultiDFT} is invertible, and the  $l$-th entry of the inverse DFT is obtained as the scalar inverse DFT of the set of the $l$-th entries of the sequence of vectors $\left\{{\bf \hat{q}}[k]\right\}_{k=0}^{n-1}$. Hence, we can write
	\begin{equation}
	\label{eqn:InvDFT}
	{\bf q}[i] = \frac{1}{n}\sum\limits_{k=0}^{n-1}{\bf \hat{q}}[k]e^{j2\pi \frac{ik}{n}}.
	\end{equation}
	\item Since ${\bf q}[i]$ is real, then ${\bf \hat{q}}[k] = \left({\bf \hat{q}}[n-k]\right)^*$ for all $1 \leq k \leq n-1$. Consequently, $\left\{{\bf \hat{q}}[k]\right\}_{k=0}^{n-1}$ can be obtained from $\left\{{\bf \hat{q}}[k]\right\}_{k=0}^{\Lm}$. Note that ${\bf \hat{q}}[0]$ is real for any $n$ and that ${\bf \hat{q}}\left[\Lm\right]$ is real for even $n$. 
	\item \label{itm:Parseval} Parseval's relationship for the multivariate DFT is given by 
	$\sum\limits_{i=0}^{n-1}\left\|{\bf{q}}[i]\right\|^2=\frac{1}{n}\sum\limits_{k=0}^{n-1}\left\|{\bf{\hat q}}[k]\right\|^2$.
	\item \label{itm:Conv} The DFT of a multivariate circular convolution is the product of the corresponding DFT sequences: Let $n_{q_1}, n_{q_2} \in \mathds{N}$,
	and consider the pair of sequences of length $n$, ${\bf p}[i] \in \mathds{R}^{n_{q_1}}$ and ${\mathsf{R}}[i] \in \mathds{R}^{n_{q_2} \times n_{q_1}}$, $i \in \mathcal{N}$. 
	Let  $\left\{{\bf \hat{p}}[k]\right\}_{k=0}^{n-1}$ be the $n$-point DFT of $\left\{{\bf p}[i]\right\}_{i=0}^{n-1}$, and define  ${\mathsf{\hat R}}[k] \triangleq \sum\limits_{i=0}^{n-1}{\mathsf{R}}[i]e^{-j2\pi \frac{ik}{n}}$, $k \in \mathcal{N}$. Consider the sequence $\left\{{\bf q}[i]\right\}_{i=0}^{n-1}$ given by ${\bf{q}}[i] = \sum\limits_{\tau = 0}^{n-1} {{{\mathsf{R}}}[\tau]{{\bf{p}}}\left[\left(\left(i - \tau\right)\right)_n\right]}$. The $n$-point DFT of $\left\{{\bf q}[i]\right\}_{i=0}^{n-1}$ is given by
	${\bf{\hat q}}[k] = {\mathsf{\hat R}}[k]{\bf \hat{p}}[k]$,
	$k \in \mathcal{N}$.
\end{enumerate}

\subsection{\underline{Step 1}: Transforming the $n$-CGMWTC into a Set of Independent Parallel MIMO WTCs}
\label{app:Proof2a}
Focusing on the  $n$-CGMWTC, consider the input sequence transmitted during one $n$-block, ${\bf X}^{n-1}$, and the corresponding channel outputs observed at the intended receiver and at the eavesdropper, denoted $\utilde{\bf Y}^{n-1}$ and $\utilde{\bf Z}^{n-1}$, respectively.
Recall that by definition of the $n$-CGMWTC,
the outputs are independent of the initial channel state ${\bf S}_0$.
For $\tau \in \mathcal{N}$ define the zero-padded extensions of the Tx--Rx and of the Tx--Ev channel impulse responses by $\utilde{\mathsf{H}}[\tau]$ and $\utilde{\mathsf{G}}[\tau]$, respectively, where $\utilde{\mathsf{H}}[\tau] = {\mathsf{H}}[\tau]$ and $\utilde{\mathsf{G}}[\tau] = {\mathsf{G}}[\tau]$ for $0 \leq \tau \leq  m$, while $\utilde{\mathsf{H}}[\tau] = {\mathsf{0}}_{n_r \times n_t}$ and $\utilde{\mathsf{G}}[\tau] = {\mathsf{0}}_{n_e \times n_t}$ for $m < \tau < n$.
Using these definitions, Eqn. \eqref{eqn:CRxModel_2} can be written as
\begin{subequations}
\label{eqn:CRxModel_4}
\begin{equation}
\utilde{\bf{Y}}[i] = \sum\limits_{\tau = 0}^{n-1} {{\utilde{\mathsf{H}}}[\tau]{{\bf{X}}}\left[\left(\left(i - \tau\right)\right)_n\right]}  + \utilde{\bf{W}}[i] \label{eqn:CRxModel_4a}
\end{equation}
\begin{equation}
\utilde{\bf{Z}}[i] = \sum\limits_{\tau = 0}^{n-1} {{\utilde{\mathsf{G}}}[\tau]{{\bf{X}}}\left[\left(\left(i - \tau\right)\right)_n\right]}  + \utilde{\bf{U}}[i], \label{eqn:CRxModel_4b}
\end{equation}
\end{subequations}
$i \in \mathcal{N}$. 
Let $\left\{{\bf{\hat{X}}}[k]\right\}_{k=0}^{n-1}$, $\left\{\utilde{\bf{\hat{Y}}}[k]\right\}_{k=0}^{n-1}$, and $\big\{\utilde{\bf{\hat{Z}}}[k]\big\}_{k=0}^{n-1}$ be the $n$-point DFTs of $\big\{{\bf{X}}[i]\big\}_{i=0}^{n-1}$, $\big\{\utilde{\bf{Y}}[i]\big\}_{i=0}^{n-1}$, and $\big\{\utilde{\bf{Z}}[i]\big\}_{i=0}^{n-1}$, respectively.
Note that ${\hat{\mathsf{H}}}[k]$ and ${\hat{\mathsf{G}}}[k]$, defined in Section \ref{subsec:CapStep2}
 in terms of $\left\{{\mathsf{H}}[\tau]\right\}_{\tau=0}^{m}$ and $\left\{{\mathsf{G}}[\tau]\right\}_{\tau=0}^{m}$, can be equivalently stated in terms of $\big\{\utilde{\mathsf{H}}[\tau]\big\}_{\tau=0}^{n-1}$ and $\big\{\utilde{\mathsf{G}}[\tau]\big\}_{\tau=0}^{n-1}$ via ${\hat{\mathsf{H}}}[k] = \sum\limits_{\tau=0}^{n-1}\utilde{\mathsf{H}}[\tau]e^{-j2\pi \frac{\tau k}{n}}$ and ${\hat{\mathsf{G}}}[k] = \sum\limits_{\tau=0}^{n-1}\utilde{\mathsf{G}}[\tau]e^{-j2\pi \frac{\tau k}{n}}$.
Using the sequences $\left\{\utilde{\bf{\hat{W}}}[k]\right\}_{k=0}^{n-1}$ and $\left\{\utilde{\bf{\hat{U}}}[k]\right\}_{k=0}^{n-1}$, which correspond to the DFTs of  $\left\{\utilde{\bf{W}}[i]\right\}_{i=0}^{n-1}$ and  $\left\{\utilde{\bf{U}}[i]\right\}_{i=0}^{n-1}$, respectively (see Subsection \ref{subsec:CapStep2}) and
property \ref{itm:Conv} for the DFT of a multivariate circular convolution, we obtain the following relationships:
\vspace{-0.15cm}
\begin{subequations}
\label{eqn:CRxModel_5}
\begin{equation}
\utilde{\bf{\hat Y}}[k] = {\hat{\mathsf{H}}}[k]{\bf{\hat X}}[k] + \utilde{\bf{\hat W}}[k] \label{eqn:CRxModel_5a}
\end{equation}
\begin{equation}
\utilde{\bf{\hat Z}}[k] = {\hat{\mathsf{G}}}[k]{\bf{\hat X}}[k] + \utilde{\bf{\hat U}}[k], \label{eqn:CRxModel_5b}
\end{equation}
\end{subequations}

\vspace{-0.2cm}
\noindent $k \in \mathcal{N}$.
Since the DFT is an invertible transformation and the channel outputs are real, it follows that 
the channel outputs $\left\{\utilde{\bf{Y}}[i]\right\}_{i \in \mathcal{N}}$ and  $\left\{\utilde{\bf{Z}}[i]\right\}_{i \in \mathcal{N}}$ can be obtained from \eqref{eqn:CRxModel_5} for $k\in\{0,1,\ldots,\Lm\} \triangleq \LmSet$. Therefore, it is sufficient to consider $\left\{\utilde{\bf{\hat Y}}[k]\right\}_{k\in \LmSet}$ and  $\left\{\utilde{\bf{\hat Z}}[k]\right\}_{k\in \LmSet}$ for deriving the secrecy capacity of the $n$-CGMWTC. 
Define next
\vspace{-0.15cm}
\begin{align}
\label{eqn:Pk_Def}
P_k \triangleq \E\left\{\left\|{\bf{\hat X}}[k]\right\|^2\right\}.
\end{align}

\vspace{-0.2cm}
\noindent The average power constraint \eqref{eqn:Constraint1eq} yields a per $n$-block power constraint
\vspace{-0.15cm}
\begin{align}
n^2P 
&\geq n\E\left\{ \sum\limits_{i = 0}^{n - 1} \left\| {\bf X} \left[ i \right] \right\|^2  \right\} \notag \\
&\stackrel{(a)}{=}  \E\left\{ \sum\limits_{k = 0}^{n - 1} \left\| {\bf{\hat X}}[k] \right\|^2  \right\} \nonumber \\
&  = \sum\limits_{k = 0}^{n - 1}P_k,
\label{eqn:constraint_DFT1}
\end{align}
where $(a)$ follows from Parseval's relationship (property \ref{itm:Parseval}). As ${\bf \hat{X}}[k] = \left({\bf \hat{X}}[n-k]\right)^*$, it follows that $P_{n-k} = P_k$ must hold.
In conclusion, when the codeword length is an integer multiple of $n$, then the secrecy capacity of the $n$-CGMWTC \eqref{eqn:CRxModel_2} subject to the time-averaged power constraint \eqref{eqn:Constraint1eq} is equal to the secrecy capacity of the memoryless WTC \eqref{eqn:CRxModel_5} subject to the power constraint \eqref{eqn:constraint_DFT1}. 

Finally, as explained in the last paragraph in the proof of Lemma \ref{lem:Proof_Rn2}, the capacity of the $n$-CGMWTC can be completely characterized by considering only codewords whose length is an integer multiple of $n$.

%
\begin{lemma}
\label{lem:Proof_Noise1}
For $k \in \LmSet$, $\utilde{\bf{\hat W}}[k]$ and $\utilde{\bf{\hat U}}[k]$ are zero mean Gaussian random vectors statistically independent over $k$, i.e., for all $k_1 \neq k_2$, $\utilde{\bf{\hat W}}\left[k_1\right]$ and $\utilde{\bf{\hat W}}\left[k_2\right]$ are independent, and $\utilde{\bf{\hat U}}\left[k_1\right]$ and $\utilde{\bf{\hat U}}\left[k_2\right]$ are independent.
For $1 \leq k < \frac{n}{2}$, $\utilde{\bf{\hat W}}[k]$ and $\utilde{\bf{\hat U}}[k]$ are circularly symmetric {\em complex} random vectors, and for $k=0$, and also for $k = \frac{n}{2}$ when $n$  is even, $\utilde{\bf{\hat W}}[k]$ and $\utilde{\bf{\hat U}}[k]$ are zero-mean {\em real} Gaussian random vectors.
The covariance matrices are given by
\begin{subequations}
\label{eqn:Lemma_NoiseCov}
\begin{equation}
\label{eqn:Lemma_NoiseCovW}
\CUHw{k} \triangleq \E\left\{\utilde{\bf{\hat W}}[k]\left(\utilde{\bf{\hat W}}[k]\right)^H\right\} = n\sum\limits_{\tau  =  - m }^{m } \Cw{\tau}e^{ - j2\pi \frac{k\tau }{n}} 
\end{equation}
and
\begin{equation}
\label{eqn:Lemma_NoiseCovU}
\CUHu{k} \triangleq \E\left\{\utilde{\bf{\hat U}}[k]\left(\utilde{\bf{\hat U}}[k]\right)^H\right\} = n\sum\limits_{\tau  =  - m }^{m } \Cu{\tau}e^{ - j2\pi \frac{k\tau }{n}}.
\end{equation}
\end{subequations}
Furthermore, for each fixed $k_1$, $\utilde{\bf{\hat W}}[k_1]$ obtained from different $n$-blocks are i.i.d., and also $\utilde{\bf{\hat U}}[k_1]$ obtained from different $n$-blocks are i.i.d.
Finally, $\utilde{\bf{\hat W}}[k_1]$ and $\utilde{\bf{\hat U}}[k_2]$ are mutually independent for any $(k_1, k_2) \in \LmSet \times \LmSet$
\end{lemma}
\begin{IEEEproof}
The proof follows similar arguments to those used in the proof in \cite[Appendix B]{Goldsmith:01} 
\ifextended
for scalar noises. 
We first note that since both $\utilde{\bf{\hat W}}\left[k\right]$ and ${\utilde{\bf{\hat U}}}\left[k\right]$ are defined as linear combinations of random Gaussian vectors, each has a zero mean, it follows that $\left\{\utilde{\bf{\hat W}}\left[k\right]\right\}_{k \in \LmSet}$ are zero-mean, jointly Gaussian, and $\left\{\utilde{\bf{\hat U}}\left[k\right]\right\}_{k \in \LmSet}$ are zero-mean, jointly Gaussian.
As $\utilde{\bf{W}}\left[i_1\right]$ and $\utilde{\bf{U}}\left[i_2\right]$ are mutually independent for all $i_1$ and $i_2$, it follows that $\utilde{\bf{\hat W}}\left[k_1\right]$ and $\utilde{\bf{\hat U}}\left[k_2\right]$ are mutually independent for all $k_1, k_2$. 
%
Next, writing explicitly the DFTs we have
\begin{align}
\E\left\{ {\utilde{\bf{\hat W}}\left[ {{k_1}} \right]{{\left( {\utilde{\bf{\hat W}}\left[ {{k_2}} \right]} \right)}^H}} \right\} &= \E\left\{ {\left( {\sum\limits_{{i_1} = 0}^{n - 1} {{\bf{\utilde W}}\left[ {{i_1}} \right]{e^{ - j2\pi \frac{{{k_1}{i_1}}}{n}}}} } \right){{\left( {\sum\limits_{{i_2} = 0}^{n - 1} {{\bf{\utilde W}}\left[ {{i_2}} \right]{e^{ - j2\pi \frac{{{k_2}{i_2}}}{n}}}} } \right)}^H}} \right\} \notag \\
&\quad= \sum\limits_{i_1 = 0}^{n - 1} \sum\limits_{i_2 = 0}^{n - 1} \E\left\{ {\bf{\utilde W}}\left[ i_1 \right]{\bf{\utilde W}}^T\left[ i_2 \right] \right\}e^{ - j2\pi \frac{{{k_1}{i_1}}}{n}}e^{j2\pi \frac{{{k_2}{i_2}}}{n}}  \notag \\
&\quad\stackrel{(a)}{=} \sum\limits_{{i_1} = 0}^{n - 1} {\sum\limits_{{i_2} = 0}^{n - 1} {\left(\sum\limits_{l=-1}^1 \Cw{{i_1} - {i_2} + l \cdot n} \right){e^{ - j2\pi \frac{k_1 i_1 - k_2 i_2}{n}}}} },  
\label{eqn:Proof_Noise1a}
\end{align}
where $(a)$ is obtained by plugging the expression for $\CUw{ \tau  }$ from \eqref{eqn:CorrModel1a}. 
Note that \eqref{eqn:Proof_Noise1a} can be written as 
\begin{align}
&\sum\limits_{{i_1} = 0}^{n \! - \! 1} \sum\limits_{{i_2} = 0}^{n \! - \! 1} \!\Cw{{i_1} \! - \! {i_2}} {e^{ \! - \! j2\pi \frac{{{k_1}{i_1} \! - \! {k_2}{i_2}}}{n}}}   \! + \! \sum\limits_{{i_1} = 0}^{n \! - \! 1} \sum\limits_{{i_2} = 0}^{n \! - \! 1} \!\Cw{{i_1} \! - \! {i_2} \! - \! n}{e^{ \! - \! j2\pi \frac{{{k_1}{i_1} \! - \! {k_2}{i_2}}}{n}}}   \! + \! \sum\limits_{{i_1} = 0}^{n \! - \! 1} \sum\limits_{{i_2} = 0}^{n \! - \! 1} \!\Cw{{i_1} \! - \! {i_2} \! + \! n}{e^{ \! - \! j2\pi \frac{{{k_1}{i_1} \! - \! {k_2}{i_2}}}{n}}}  \notag \\
&\stackrel{(a)}{=} 
\sum\limits_{{i_1} = 0}^{n \! - \! 1} \sum\limits_{\tau  = {i_1} \! - \! n \! + \! 1}^{{i_1}} \!\Cw{\tau}{e^{ \! - \! j2\pi \frac{{{i_1}\left( {{k_1} \! - \! {k_2}} \right) \! + \! \tau {k_2}}}{n}}}   \! + \! \sum\limits_{{i_1} = 0}^{n \! - \! 1} \sum\limits_{\tau  = {i_1} \! - \! 2n \! + \! 1}^{{i_1} \! - \! n} \!\Cw{\tau}{e^{ \! - \! j2\pi \frac{{{i_1}\left( {{k_1} \! - \! {k_2}} \right) \! + \! \tau {k_2}}}{n}}}   \! + \! \sum\limits_{{i_1} = 0}^{n \! - \! 1} \sum\limits_{\tau  = {i_1} \! + \! 1}^{{i_1} \! + \! n} \!\Cw{\tau}{e^{ \! - \! j2\pi \frac{{{i_1}\left( {{k_1} \! - \! {k_2}} \right) \! + \! \tau {k_2}}}{n}}}  \notag \\
&= \sum\limits_{{i_1} = 0}^{n \! - \! 1} \sum\limits_{\tau  = {i_1} \! - \! 2n \! + \! 1}^{{i_1} \! + \! n} \!\Cw{\tau}{e^{ \! - \! j2\pi \frac{{{i_1}\left( {{k_1} \! - \! {k_2}} \right) \! + \! \tau {k_2}}}{n}}}  \notag \\
&\stackrel{(b)}{=} 
\sum\limits_{{i_1} = 0}^{n \! - \! 1} \sum\limits_{\tau  =  \! - \! m \! }^{m \! } \!\Cw{\tau}{e^{ \! - \! j2\pi \frac{{{i_1}\left( {{k_1} \! - \! {k_2}} \right) \! + \! \tau {k_2}}}{n}}}, \label{eqn:Proof_NoiseAid1}
\end{align}
where $(a)$ follows from setting $\tau = i_1 - i_2$, $\tau = i_1 - i_2 -n$, and $\tau = i_1 - i_2 +n$ in the first sum, second sum, and third sum, respectively, 
and $(b)$ follows 
since $\Cw{\tau} = \mathsf{0}_{n_r \times n_r}$ for $|\tau| > m$, and since $n > 2m$, thus $\forall 0 \leq i_1 < n$, $\left[-m,m\right] \subset \left[i_1 -2n +1, i_1 +n\right]$. 
%
%
It follows that \eqref{eqn:Proof_Noise1a} yields
\begin{align*}
\E\left\{ {\utilde{\bf{\hat W}}\left[ {{k_1}} \right]{{\left( {\utilde{\bf{\hat W}}\left[ {{k_2}} \right]} \right)}^H}} \right\} 
&= \sum\limits_{\tau  =  - m }^{m } \Cw{\tau}{\sum\limits_{{i_1} = 0}^{n - 1} {{e^{ - j2\pi \frac{{{i_1}\left( {{k_1} - {k_2}} \right) + \tau {k_2}}}{n}}}} }  \notag \\
&= \sum\limits_{\tau  =  - m }^{m } \Cw{\tau}{{e^{ - j2\pi \frac{{{k_2}\tau }}{n}}}}\sum\limits_{{i_1} = 0}^{n - 1} {{e^{ - j2\pi \frac{{{i_1}\left( {{k_1} - {k_2}} \right)}}{n}}}}.  
\end{align*}
Let $\delta[n]$ denote the Kronecker delta function. Note that since $k_1, k_2 \in \mathcal{N}$, it follows that $\left|k_1 - k_2\right| < n$, thus for $k_1 \neq k_2$, $\sum\limits_{{i_1} = 0}^{n - 1} {{e^{ - j2\pi \frac{{{i_1}\left( {{k_1} - {k_2}} \right)}}{n}}}} =0$, while for $k_1=k_2$, $\sum\limits_{{i_1} = 0}^{n - 1} {{e^{ - j2\pi \frac{{{i_1}\left( {{k_1} - {k_2}} \right)}}{n}}}}=n$. Therefore, $\sum\limits_{{i_1} = 0}^{n - 1} {{e^{ - j2\pi \frac{{{i_1}\left( {{k_1} - {k_2}} \right)}}{n}}}} = n\delta\left[k_1-k_2\right]$ and 
\begin{equation}
\label{eqn:MultNoiseCorr1}
\E\left\{ {\utilde{\bf{\hat W}}\left[ {{k_1}} \right]{{\left( {\utilde{\bf{\hat W}}\left[ {{k_2}} \right]} \right)}^H}} \right\} 
= n\delta\left[k_1-k_2\right]\sum\limits_{\tau  =  - m }^{m } {\Cw{\tau}{e^{ - j2\pi \frac{{{k_2}\tau }}{n}}}}.
\end{equation}

Note that for $k=0$ and for $n$ even then also for $k= \frac{n}{2}$, $\utilde{\bf{\hat W}}[k]$ and $\utilde{\bf{\hat U}}[k]$ are real-valued, while for $1 \leq k < \frac{n}{2}$, $\utilde{\bf{\hat W}}[k]$ and $\utilde{\bf{\hat U}}[k]$ are complex-valued. We therefore observe the pseudo-covariance:
\begin{align}
\E\left\{ {\utilde{\bf{\hat W}}[k_1]{{\left( {\utilde{\bf{\hat W}}[k_2]} \right)}^T}} \right\}
&= \E\left\{ \left( {\sum\limits_{{i_1} = 0}^{n - 1} {{\bf{\utilde W}}\left[ {{i_1}} \right]{e^{ - j2\pi \frac{{k_1{i_1}}}{n}}}} } \right)\left( {\sum\limits_{{i_2} = 0}^{n - 1} {{\bf{\utilde W}}\left[ {{i_2}} \right]{e^{ - j2\pi \frac{{k_2{i_2}}}{n}}}} } \right)^T \right\}\notag \\
&\quad= \sum\limits_{{i_1} = 0}^{n - 1} {\sum\limits_{{i_2} = 0}^{n - 1} {\CUw{ i_1 - i_2}{e^{ - j2\pi \frac{{k{i_1}}}{n}}}{e^{ - j2\pi \frac{{k{i_2}}}{n}}}} } \notag \\
&\quad\stackrel{(a)}{=} \sum\limits_{{i_1} = 0}^{n - 1} {\sum\limits_{{i_2} = 0}^{n - 1} {\left(\sum\limits_{l=-1}^{l=1} \Cw{ {{i_1} - {i_2}} +l \cdot n } \right){e^{ - j2\pi \frac{{\left( {k_1{i_1} + k_2{i_2}} \right)}}{n}}}} },  
\label{eqn:Proof_Noise2a}
\end{align}
where $(a)$ is obtained by plugging the expression for $\CUw{ \tau}$ stated in \eqref{eqn:CorrModel1a}. 
Repeating the derivation leading to \eqref{eqn:Proof_NoiseAid1} with $-k_2$ instead of $k_2$, we can write 
%
\eqref{eqn:Proof_Noise2a} as
\begin{align}
\E\left\{ {\utilde{\bf{\hat W}}[k_1]{{\left( {\utilde{\bf{\hat W}}[k_2]} \right)}^T}} \right\} 
&= \sum\limits_{\tau  =  - m}^m \Cw{\tau}  \sum\limits_{{i_1} = 0}^{n - 1} {{e^{ - j2\pi \frac{{{i_1}\left( {{k_1} + {k_2}} \right) - \tau {k_2}}}{n}}}} \notag \\
&= \sum\limits_{\tau  =  - m}^m \Cw{\tau}{e^{j2\pi \frac{{\tau {k_2}}}{n}}} \sum\limits_{{i_1} = 0}^{n - 1} {{e^{ - j2\pi \frac{{{i_1}\left( {{k_1} + {k_2}} \right)}}{n}}}}.
\label{eqn:Proof_Noise2b}
\end{align}
Note that if $1 \leq k_1 < \frac{n}{2}$ or if $1 \leq k_2 < \frac{n}{2}$, then $0 < k_1 + k_2 < n$, thus $\sum\limits_{{i_1} = 0}^{n - 1} {{e^{ - j2\pi \frac{{{i_1}\left( {{k_1} + {k_2}} \right)}}{n}}}}=0$. 
It therefore follows that the pseudo-covariance is $\mathsf{0}_{n_r \times n_r}$, except when $k_1 = k_2 = 0$ and $k_1 = k_2 = \frac{n}{2}$, which corresponds to real-valued $\utilde{\bf{\hat W}}[k_1]$, i.e., the pseudo-covariance matrix is equal to the covariance matrix.

It also follows from \eqref{eqn:MultNoiseCorr1} and \eqref{eqn:Proof_Noise2b} that for $1 \leq k  < \frac{n}{2}$, $\utilde{\bf{\hat W}}\left[k\right]$ is jointly proper complex \cite[Def. 1]{Massey:93}. 
For $k = 0$ and, when $n$ is even then also for $k = \frac{n}{2}$, $\utilde{\bf{\hat W}}\left[k\right]$ is real.

Since $\left\{\utilde{\bf{\hat W}}\left[k\right]\right\}_{k \in \LmSet}$ are jointly Gaussian, it follows from \eqref{eqn:MultNoiseCorr1} and from \eqref{eqn:Proof_Noise2b} that $\left\{\utilde{\bf{\hat W}}\left[k\right]\right\}_{k \in \LmSet}$ are mutually independent.
Since for $1 \leq k < \frac{n}{2}$, $\utilde{\bf{\hat W}}[k]$ is zero-mean proper complex Gaussian random vector, it follows that $\utilde{\bf{\hat W}}[k]$ is circularly symmetric \cite[Thm. 4]{Gallager:08}.  
The fact that for a fixed $k_1$, $\utilde{\bf{\hat W}}[k_1]$ obtained from different $n$-blocks are i.i.d. follows as $\utilde{\bf{\hat W}}[k_1]$ is a function of only $\left\{\utilde{\bf{W}}[i]\right\}_{i=0}^{n-1}$ which are i.i.d. over different $n$-blocks due to the $n$-block memorylessness of the $n$-CGMWTC.
The proof for $\utilde{\bf{\hat U}}[k]$ is similar to the proof for $\utilde{\bf{\hat W}}[k]$. This concludes the proof of Lemma \ref{lem:Proof_Noise1}.
\fi 
\end{IEEEproof}
\smallskip
Since the noises  $\left\{\utilde{\bf{\hat W}}\left[k\right]\right\}_{k \in \LmSet}$,  $\left\{\utilde{\bf{\hat U}}\left[k\right]\right\}_{k \in \LmSet}$ are mutually independent it follows that the channels \eqref{eqn:CRxModel_5} are {\em parallel} Gaussian channels.

\subsection{\underline{Step 2}: The Maximal Achievable Secrecy Rate $\CCGTA{n}$}
\label{app:Proof2b}
Define for $k\in \LmSet$
\begin{align}
R_n^k\left(P_k\right) &\triangleq \mspace{-25mu} \sup_{ \substack{{p}\left({\bf{V}}[k],{\bf{\hat X}}[k]\right), \\ \E\left\{\left\|{\bf{\hat X}}[k]\right\|^2\right\} \leq P_k }} \mspace{-15mu}  \bigg\{ I\left( {\bf{ V}}[k];\utilde{\bf{\hat Y}}[k] \right)  - I\left( {\bf{ V}}[k];\utilde{\bf{\hat Z}}[k] \right) \bigg\}.
\label{eqn:DefIndRate1}
\end{align}
Note that $R_n^k\left(P_k\right)$ represents the secrecy capacity of an $n_t \times n_r \times n_e$ {\em memoryless} MIMO WTC with additive Gaussian noise i.i.d. in time, subject to input power constraint $P _k$ \cite[Corollary 1]{Wornell:10}.
Let $R{'_n}\left( {\left\{ {{P_k}} \right\}_{k = 0}^{{\Lm}}} \right)$ denote the maximal achievable secrecy rate for the WTC \eqref{eqn:CRxModel_5} subject to a  given a set of per-channel power constraints $\left\{ \E\left\{\left\|{\bf{\hat X}}[k]\right\|^2\right\} \leq P_k\right\}_{k=0}^{\Lm}$.
\begin{lemma}
\label{lem:ParallelChannels}
When the codeword length is restricted to be an integer multiple of $n$, $R{'_n}\left( {\left\{ {{P_k}} \right\}_{k = 0}^{{\Lm}}} \right)$ satisfies:
\begin{equation}
R'_n\left(\left\{P_k\right\}_{k=0}^{\Lm}\right) = \sum\limits_{k=0}^{\Lm}R_n^k\left(P_k\right).
\label{eqn:ParallelChannels}
\end{equation} 
\end{lemma}
\begin{IEEEproof}
 From Lemma \ref{lem:Proof_Noise1} we have that the noises at each subchannel $k$ are each i.i.d. over different $n$-blocks,
 and that the noises at subchannel $k$ are independent of the noises at all other subchannels.
 It thus  follows that \eqref{eqn:CRxModel_5} can be considered as $\Lm +1$ {\em parallel memoryless MIMO WTCs}
 (e.g., by extending the definition in \cite[Sec. 1.3]{Li:10} for scalar channels to the MIMO case).
 In \cite[Thm. 1]{Li:10} it was shown that the secrecy capacity of independent memoryless parallel {\em scalar} WTCs is given by the sum of the secrecy capacities of each subchannel. Although  differently from \cite[Sec. 1.3]{Li:10}, which considered the secrecy capacity of parallel scalar WTCs, in the current analysis we consider the maximization of the achievable secrecy rate of parallel {\em MIMO} WTCs {\em subject to a fixed per-subchannel power allocation}, the proof for our case follows identical steps to the proof of \cite[Thm. 1]{Li:10}, and thus it is not repeated here.
\end{IEEEproof}
Note that \eqref{eqn:ParallelChannels} is the {\em maximum} achievable secrecy rate for the WTC \eqref{eqn:CRxModel_5} {\em for a given assignment of $\left\{P_k\right\}_{k=0}^{\Lm}$} when the codeword length is an integer multiple of $n$; The secrecy capacity of the $n$-CGMWTC subject to the power constraint \eqref{eqn:Constraint1eq} is therefore obtained by finding the assignment of $\left\{P_k\right\}_{k=0}^{\Lm}$ which maximizes \eqref{eqn:ParallelChannels} subject to \eqref{eqn:constraint_DFT1} while the set
$\big\{ P_k\big\}_{k=0}^{n-1}$ is constrained to satisfy $P_{n-k} = P_k$ for every $1 \leq k < \frac{n}{2}$.
As each of the $\lfloor \frac{n}{2} \rfloor$ channel uses - one for each subchannel in the set of $\Lm$  parallel subchannels, corresponds to $n$ channel uses of the $n$-CGMWTC,
we can summarize the above discussion in the following result
\begin{align} 
   \CCGTA{n} 
& = \frac{1}{n} \mathop{\max }\limits_{ \substack{ \left\{P_k\right\}_{k=0}^{\Lm}: \\ \sum\limits_{k=0}^{n-1}P_k \leq n^2 P, \, P_{n-k} = P_k > 0} } \mspace{-30mu}  R'_n\left(\left\{P_k\right\}_{k=0}^{\Lm}\right)
\notag \\
&=  \mathop{\max }\limits_{ \substack{ \left\{P_k\right\}_{k=0}^{\Lm}: \\ \sum\limits_{k=0}^{n-1}P_k \leq n^2 P,\, P_{n-k} = P_k > 0}} \mspace{-30mu}  \frac{1}{n}\sum\limits_{k=0}^{\Lm}R_n^k\left(P_k\right).
\label{eqn:ParallelChannels2} 
\end{align}

\subsection{\underline{Step 3}: Deriving an Explicit Expression for the Maximization \eqref{eqn:ParallelChannels2}}
\label{app:Proof2c}
Define $\tilde{\mathcal{L}}^n$ as $\tilde{\mathcal{L}}^n=\left\{0\right\}$ for $n$ odd and $\tilde{\mathcal{L}}^n=\left\{0, \frac{n}{2}\right\}$ for $n$ even.
From Lemma \ref{lem:Proof_Noise1}, it follows that for $1 \leq k < \frac{n}{2}$, the $k$-th subchannel is a {\em complex} memoryless MIMO WTC with circularly symmetric complex normal additive white Gaussian noise. For the remaining values of $k$, i.e., for $k \in \tilde{\mathcal{L}}^n$,  it follows from Lemma \ref{lem:Proof_Noise1}
 that the $k$-th subchannel is a {\em real} memoryless Gaussian MIMO WTC.
For a fixed $\rho \geq 0$, let $\mathcal{Q}_{\rho}$ be the set of $n_t \times n_t$ Hermitian positive semi-definite matrices $\mathsf{Q}$ such that ${\rm {Tr}}\left( {\mathsf{Q}} \right) \leq \rho$. 
We define\footnote{Following \cite[Eqn. (20)]{Wornell:10} and \cite[Thm. 1]{Hassibi:11}, $\tilde{R}_n^k(\rho)$ can be written as a maximization over $\mathcal{Q}_\rho$, instead of a supremum.} for $k \in\mathcal{N}$,
\begin{equation}
\label{eqn:Def_R_nk}
\tilde{R}_n^k\left(\rho \right) \triangleq \mathop {\max }\limits_{\mathsf{Q} \in \mathcal{Q}_{\rho}}\frac{1}{2} \log \frac{{\left| {{\mathsf{I}}_{{n_r}}} + {\hat{\mathsf{H}}}[k]\mathsf{Q}\left({\hat{\mathsf{H}}}[k]\right)^H \left(\CUHw{k}\right)^{-1} \right|}}{{\left| {{\mathsf{I}}_{{n_e}}} + {\hat{\mathsf{G}}}[k]\mathsf{Q}\left({\hat{\mathsf{G}}}[k]\right)^H \left(\CUHu{k}\right)^{-1} \right|}}.
\end{equation}
\begin{lemma}
\label{lem:Proof_Noise3}
For $\Lm < k < n$, $\tilde{R}_n^k\left(\rho \right) = \tilde{R}_n^{n-k}\left(\rho \right)$. When $\tilde{R}_n^k\left(\rho \right)$ is obtained with $\mathsf{Q}_{opt}$, then $\tilde{R}_n^{n-k}\left(\rho \right)$ is obtained with $\mathsf{Q}_{opt}^*$.
\end{lemma}
\begin{IEEEproof}
Define 
\begin{equation*}
\mathsf{F}_r^k\!\left(\mathsf{Q}\right)\! \triangleq\! {{\mathsf{I}}_{{n_r}}}\! + \!\left(\CUHw{k}\right)^{\!-\frac{1}{2}}{\hat{\mathsf{H}}}[k]\mathsf{Q}\!\left({\hat{\mathsf{H}}}[k]\right)^H \!\!\! \left(\CUHw{k}\right)^{\!-\frac{1}{2}}, 
\end{equation*}
and  
\begin{equation*}
\mathsf{F}_e^k\!\left(\mathsf{Q}\right)\! \triangleq \!{{\mathsf{I}}_{{n_e}}} \!+ \!\left(\CUHu{k}\right)^{\!-\frac{1}{2}} {\hat{\mathsf{G}}}[k]\mathsf{Q}\!\left({\hat{\mathsf{G}}}[k]\right)^H\!\!\! \left(\CUHu{k}\right)^{\!-\frac{1}{2}}. 
\end{equation*}
Since $\CUHw{k}$ and $\CUHu{k}$ are positive-definite Hermitian matrices $\forall k \in\mathcal{N}$, it follows from \cite[Thm. 7.2.6]{Horn:85} that $\left(\CUHw{k}\right)^{\!-\frac{1}{2}}$ and $\left(\CUHu{k}\right)^{\!-\frac{1}{2}}$ are also positive-definite Hermitian matrices. Thus, $\forall \mathsf{Q} \in \mathcal{Q}_{\rho}$, $\mathsf{F}_r^k\left(\mathsf{Q}\right)$ and  $\mathsf{F}_e^k\left(\mathsf{Q}\right)$ are Hermitian matrices.

As ${{\mathsf{H}}}[\tau]$ and  ${{\mathsf{G}}}[\tau]$ are real matrices, it follows that ${\hat{\mathsf{H}}}[n-k] = \left({\hat{\mathsf{H}}}[k]\right)^*$ and ${\hat{\mathsf{G}}}[n-k] = \left({\hat{\mathsf{G}}}[k]\right)^*$. 
Note that 
\ifextended
$\mathsf{I}_{n_r} = \left(\CUHw{k}\right)^{-\frac{1}{2}}\left(\CUHw{k}\right)^{-\frac{1}{2}}\CUHw{k}$, and since $\left(\mathsf{I}_{n_r}\right)^* = \mathsf{I}_{n_r}$ it follows that
\begin{align}
\mathsf{I}_{n_r} 
&= \left(\left(\CUHw{k}\right)^{-\frac{1}{2}}\left(\CUHw{k}\right)^{-\frac{1}{2}}\CUHw{k}\right)^* \notag \\
&\stackrel{(a)}{=} \left(\left(\CUHw{k}\right)^{-\frac{1}{2}}\right)^*\left(\left(\CUHw{k}\right)^{-\frac{1}{2}}\right)^*\left(\CUHw{k}\right)^* \notag \\ 
&\stackrel{(b)}{=} \left(\left(\CUHw{k}\right)^{-\frac{1}{2}}\right)^*\left(\left(\CUHw{k}\right)^{-\frac{1}{2}}\right)^*\CUHw{n-k},
\label{eqn:CUW_nk}
\end{align}
where $(a)$ follows as for all matrices $\mathsf{A}_1$, $\mathsf{A}_2$ of compatible dimensions, $\left(\mathsf{A}_1\mathsf{A}_2\right)^* = \mathsf{A}_1^*\mathsf{A}_2^*$~\cite[Ch. 3.6]{Meyer:00}; and $(b)$ follows from \eqref{eqn:Lemma_NoiseCov} as $\CUHw{n-k} = \left(\CUHw{k}\right)^*$. 
It follows from \eqref{eqn:CUW_nk} that 
$\left(\CUHw{n-k}\right)^{-\frac{1}{2}} = \left(\left(\CUHw{k}\right)^{-\frac{1}{2}}\right)^*$. 
Similarly, $\left(\CUHu{n-k}\right)^{-\frac{1}{2}} = \left(\left(\CUHu{k}\right)^{-\frac{1}{2}}\right)^*$.
\fi
%
It therefore follows that
\ifextended 
\begin{align}
 \left(\mathsf{F}_r^{k}\left(\mathsf{Q}\right)\right)^* 
&= \left(\mathsf{I}_{n_r}\! + \!\left(\CUHw{k}\right)^{\!-\frac{1}{2}}{\hat{\mathsf{H}}}[k]\mathsf{Q}\!\left({\hat{\mathsf{H}}}[k]\right)^H \! \left(\CUHw{k}\right)^{\!-\frac{1}{2}}\right)^* \notag \\
&\stackrel{(a)}{=} 
\mathsf{I}_{n_r}\! + \!\left(\left(\CUHw{k}\right)^{\!-\frac{1}{2}}\right)^*\left({\hat{\mathsf{H}}}[k]\right)^*\mathsf{Q}^*\!\left({\hat{\mathsf{H}}}[k]\right)^T \! \left(\left(\CUHw{k}\right)^{\!-\frac{1}{2}}\right)^* \notag \\
&\stackrel{(b)}{=} 
\mathsf{I}_{n_r}\! + \!\left(\CUHw{n-k}\right)^{\!-\frac{1}{2}}{\hat{\mathsf{H}}}[n-k]\mathsf{Q}^*\!\left({\hat{\mathsf{H}}}[n-k]\right)^H \! \left(\CUHw{n-k}\right)^{\!-\frac{1}{2}} \notag \\
&= \mathsf{F}_r^{n-k}\left(\mathsf{Q}^*\right),
\label{eqn:CUW_nk2}
\end{align}
where $(a)$ follows from \cite[Ch. 3.6]{Meyer:00}, and
$(b)$ follows 
\fi
from plugging $\left(\CUHw{n-k}\right)^{-\frac{1}{2}} = \left(\left(\CUHw{k}\right)^{-\frac{1}{2}}\right)^*$ and ${\hat{\mathsf{H}}}[n-k] = \left({\hat{\mathsf{H}}}[k]\right)^*$. 
Similarly, $\mathsf{F}_e^k\left(\mathsf{Q}\right) = \left(\mathsf{F}_e^{n-k}\left(\mathsf{Q}^*\right)\right)^*$. 
Therefore
\vspace{-0.15cm}
\begin{align*}
\tilde{R}_n^{n-k}\left(\rho \right) 
&\stackrel{(a)}{=} \mathop {\max }\limits_{\mathsf{Q} \in \mathcal{Q}_{\rho}}\frac{1}{2} \log \frac{{\left| \mathsf{F}_r^{n-k}\left(\mathsf{Q}\right) \right|}}{{\left| \mathsf{F}_e^{n-k}\left(\mathsf{Q}\right)\right|}} \notag \\
&\stackrel{(b)}{=} \mathop {\max }\limits_{\mathsf{Q} \in \mathcal{Q}_{\rho}}\frac{1}{2} \log \frac{{\left| \mathsf{F}_r^{k}\left(\mathsf{Q}^*\right) \right|}}{{\left| \mathsf{F}_e^{k}\left(\mathsf{Q}^*\right)\right|}} \notag \\
&\stackrel{(c)}{=} \mathop {\max }\limits_{\mathsf{Q} \in \mathcal{Q}_{\rho}}\frac{1}{2} \log \frac{{\left| \mathsf{F}_r^{k}\left(\mathsf{Q}\right) \right|}}{{\left| \mathsf{F}_e^{k}\left(\mathsf{Q}\right)\right|}} \notag \\
&= \tilde{R}_n^{k}\left(\rho \right), 
\end{align*}
where $(a)$ follows from applying Sylvester's determinant theorem \cite[Ch. 6.2]{Meyer:00}
to \eqref{eqn:Def_R_nk}; 
$(b)$ follows since the determinant of a Hermitian matrix is the same as the determinant of its conjugate \cite[Ch. 7.5]{Meyer:00}, thus 
$\left| \mathsf{F}_r^{n-k}\left(\mathsf{Q}\right) \right| = \left| \left(\mathsf{F}_r^{n-k}\left(\mathsf{Q}\right)\right)^* \right| = \left| \left(\mathsf{F}_r^{k}\left(\mathsf{Q}^*\right)\right) \right|$ and $\left| \mathsf{F}_e^{n-k}\left(\mathsf{Q}\right) \right| = \left| \left(\mathsf{F}_e^{n-k}\left(\mathsf{Q}\right)\right)^* \right| = \left| \left(\mathsf{F}_e^{k}\left(\mathsf{Q}^*\right)\right) \right|$;
$(c)$ follows since the definition of $\mathcal{Q}_{\rho}$ implies that if $\mathsf{Q} \in \mathcal{Q}_{\rho}$ then also $\mathsf{Q}^* \in \mathcal{Q}_{\rho}$. 
Let $\mathsf{Q}_{opt} \triangleq \mathop {\argmax}\limits_{\mathsf{Q} \in \mathcal{Q}_{\rho}}\frac{1}{2} \log \frac{{\left| \mathsf{F}_r^{k}\left(\mathsf{Q}\right) \right|}}{{\left| \mathsf{F}_e^{k}\left(\mathsf{Q}\right)\right|}}$. 
Since $\frac{1}{2} \log \frac{{\left| \mathsf{F}_r^{k}\left(\mathsf{Q}_{opt}\right) \right|}}{{\left| \mathsf{F}_e^{k}\left(\mathsf{Q}_{opt}\right)\right|}}  = \frac{1}{2} \log \frac{{\left| \mathsf{F}_r^{n-k}\left(\mathsf{Q}_{opt}^*\right) \right|}}{{\left| \mathsf{F}_e^{n-k}\left(\mathsf{Q}_{opt}^*\right)\right|}}$,
it follows that $\mathsf{Q}_{opt}^*$ maximizes $\frac{1}{2} \log \frac{{\left| \mathsf{F}_r^{n-k}\left(\mathsf{Q}\right) \right|}}{{\left| \mathsf{F}_e^{n-k}\left(\mathsf{Q}\right)\right|}}$. This proves the lemma. 
\end{IEEEproof}
\smallskip

It follows from \cite[Thm. 1]{Wornell:10}, \cite[Thm. 1]{Hassibi:11},  and \cite[Corollary 1]{Shamai:09}
that $R_n^k\left(P_k\right)$ defined in \eqref{eqn:DefIndRate1} is given by
\vspace{-0.15cm}
\begin{align}
\label{eqn:MIMO_cap_Rn}
R_n^k\left(P_k\right) = \left\{\begin{array}{*{20}{c}}
{2\tilde{R}_n^k\left(P_k\right) \qquad 1 \leq k <  \frac{n}{2}}\\
{\tilde{R}_n^k\left(P_k\right) \qquad \quad k \in \tilde{\mathcal{L}}^n}
\end{array}\right. ,
\end{align}
and that the maximizing channel input for the $k$-th subchannel, ${\hat{\bf X}}[k]$, is circularly symmetric complex normal for  $1 \leq k <  \frac{n}{2}$ and zero-mean normal for $k \in \tilde{\mathcal{L}}^n$, with the covariance matrix of ${\hat{\bf X}}[k]$, denoted $\CUHx{k}$, satisfying ${\rm Tr}\left(\CUHx{k}\right) \leq P_k$.
Let ${\mathcal{B}}^n_P$ be the set which contains all sets of non-negative scalars $\left\{P_k\right\}_{k=0}^{n-1}$ such that $\sum\limits_{k=0}^{n-1}P_k \leq n^2P$ and $P_k = P_{n-k}$.
Plugging \eqref{eqn:MIMO_cap_Rn} into \eqref{eqn:ParallelChannels2} yields
\vspace{-0.15cm}
\begin{align}
\vspace{-0.1cm}
& \CCGTA{n}
= \mathop {\max }\limits_{\left\{P_k\right\}_{k=0}^{n-1} \in {\mathcal{B}}^n_P} \frac{1}{n}\left(\sum\limits_{k=0}^{\Lm}\tilde{R}_n^k\left(P_k\right) + \sum\limits_{k=1}^{\Ltm}\tilde{R}_n^k\left(P_{k}\right)\right) \notag \\
&\quad\qquad= \mathop {\max }\limits_{\left\{P_k\right\}_{k=0}^{n-1} \in {\mathcal{B}}^n_P} \frac{1}{n}\left(\sum\limits_{k=0}^{\Lm}\tilde{R}_n^k\left(P_k\right) + \sum\limits_{k=\Lm + 1}^{n-1}\tilde{R}_n^k\left(P_{k}\right)\right) \notag \\
&\quad\qquad\stackrel{(a)}{=} \mathop {\max }\limits_{\left\{P_k\right\}_{k=0}^{n-1} \in {\mathcal{B}}^n_P} \frac{1}{n}\sum\limits_{k=0}^{n-1}\tilde{R}_n^k\left(P_k\right) \nonumber\\
&\quad\qquad\stackrel{(b)}{=} \mathop {\max }\limits_{\left\{\CUHx{k}\right\}_{k=0}^{n-1} \in {\mathcal{\hat{C}}}^n_P}  \frac{1}{2n}
\sum\limits_{k=0}^{n-1} \log \hat{\psi}[k], \label{eqn:ParallelChannels3}
\end{align}
where $(a)$ follows since for $\Lm < k < n$ we conclude from Lemma \ref{lem:Proof_Noise3} that  $\tilde{R}_n^k\left(P_k\right) = \tilde{R}_n^k\left(P_{n-k}\right) = \tilde{R}_n^{n-k}\left(P_{n-k}\right)$; and $(b)$ follows from plugging the definition of $\tilde{R}_n^k\left(P_k\right)$ from \eqref{eqn:Def_R_nk}, and from Lemma \ref{lem:Proof_Noise3}
combined with \eqref{eqn:constraint_DFT1} which restrict the sets of matrices to ${\mathcal{\hat{C}}}^n_P$, and from the definition of $\hat{\psi}[k]$ in \eqref{eqn:psihatkDef}.
Note that \eqref{eqn:ParallelChannels3} coincides with \eqref{eqn:pro_main2}. 
%

\subsection{\underline{Step 4}: Proving that $\CCG{n} = \CCGTA{n}$}
\label{app:Proof2d}
So far we have derived $\CCGTA{n}$, the secrecy capacity of the $n$-CGMWTC subject to the {\em time-averaged} power constraint \eqref{eqn:Constraint1eq} when the blocklength is an integer multiple of $n$. However, we are interested in $\CCG{n}$, the secrecy capacity of the $n$-CGMWTC subject to a {\em per-symbol} power constraint \eqref{eqn:Constraint1}. As the per-symbol constraint \eqref{eqn:Constraint1} is more restrictive than the time-averaged constraint \eqref{eqn:Constraint1eq}, it follows that $\CCG{n} \leq \CCGTA{n}$. In this subsection we prove that the secrecy capacities are, in fact, equal.
Let $\left\{\CUHxOpt{k}\right\}_{k=0}^{n-1}$ be the set of $n$ matrices which maximize \eqref{eqn:ParallelChannels3}. 
From the derivation of the secrecy capacity of the $n$-CGMWTC subject to a {\em time-averaged} power constraint \eqref{eqn:Constraint1eq} derived in subsections \ref{app:Proof2a}--\ref{app:Proof2c}, we note the following characteristics of the optimal channel input:
\begin{itemize}
	\item As ${\bf X}[i]$ is a real multivariate sequence it follows from the properties of the DFT that 
    for $\Lm < k < n$, $\hat{\bf X}[k]$ is obtained from $\hat{\bf X}[k] = \left(\hat{\bf X}[n-k]\right)^*$.
	\item From the characteristics of the channel inputs which achieve the secrecy capacity for real-valued memoryless Gaussian MIMO WTCs \cite[Corollary 1]{Ekrem:12}\footnote{Note that the real-valued memoryless Gaussian MIMO WTC is a special case of the real-valued memoryless Gaussian MIMO BC with common and confidential messages studied in \cite{Ekrem:12}, when there is no common message and only a single confidential message.}
	it follows that for $k \in \tilde{\mathcal{L}}^n$, $\hat{\bf X}[k]$ is a zero-mean real-valued Gaussian random vecore with covariance matrix $\CUHxOpt{k}$.
	\item From the characteristics of the channel inputs which achieve the secrecy capacity for complex memoryless Gaussian MIMO WTCs with 
        circularly symmetric complex normal noise \cite[Thm. 1]{Wornell:10}\footnote{
	Note that \cite{Wornell:10} showed that circularly symmetric complex normal inputs are optimal for complex memoryless Gaussian MIMO WTCs with additive circularly-symmetric complex normal (ACSCN) noise, subject to the more general time-averaged power constraint, which subsumes the per-symbol power constraint. It directly follows
from \cite{Wornell:10} and \cite[Lemma 1]{Wornell:10a} that the optimal codebook which achieves the secrecy capacity in \cite{Wornell:10} also satisfies the per-symbol power constraint. Thus, we conclude that circularly symmetric complex normal inputs are also optimal for complex memoryless Gaussian MIMO WTCs with ACSCN noise subject to the per-symbol power constraint.}
	it follows that for $1 \leq k  < \frac{n}{2}$, $\hat{\bf X}[k]$ is a circularly symmetric complex normal random vector with covariance matrix $\CUHxOpt{k}$.
	\item As the subchannels are independent, the optimal input which achieves \eqref{eqn:ParallelChannels3} satisfies $p\left({\bf{\hat X}}^{\Lm}\right) = \prod\limits_{k=0}^{\Lm}p\left({\bf{\hat X}}[k]\right)$, i.e., $\left\{{\bf{\hat X}}\left[ k \right]\right\}_{k=0}^{\Lm}$ are mutually independent RVs.
\end{itemize}
The above characteristics give rise to the following corollary:
\begin{corollary}
\label{cor:PerSymbol}
The secrecy capacity of the $n$-CGMWTC with a time-averaged power constraint, $\CCGTA{n}$, is obtained with an equal per-symbol power allocation.
\end{corollary}
\begin{IEEEproof}
Let $\mathcal{L}_-^n$ be set of indexes $k \in \mathcal{N}$ such that $k \notin \tilde{\mathcal{L}}^n$. Note that for $k \in \tilde{\mathcal{L}}_-^n$, $\hat{\bf X}[k] = \left(\hat{\bf X}[n-k]\right)^*$. 
We consider the autocorrelation of the time-domain optimal channel input which obtains $\CCGTA{n}$. As the time-domain channel input of the $n$-CGMWTC is real-valued we can write:
\begin{align}
\E\left\{ {{\bf{X}}\left[ {{i_1}} \right]\big({{\bf{X}}}\left[ {{i_2}} \right]\big)^T} \right\}
&  = \E\left\{ {{\bf{X}}\left[ {{i_1}} \right]\big({{\bf{X}}}\left[ {{i_2}} \right]\big)^H} \right\} \notag \\
&\stackrel{(a)}{=}
\frac{1}{{{n^2}}}\sum\limits_{{k_1} = 0}^{n \! - \! 1} {\sum\limits_{{k_2} = 0}^{n - 1} {\E\left\{ {{\bf{\hat X}}\left[ {{k_1}} \right]\big({{{\bf{\hat X}}}}\left[ {{k_2}} \right]\big)^H} \right\}} } {e^{j2\pi \frac{{{i_1}{k_1} \! - \! {i_2}{k_2}}}{n}}} \notag \\
&\stackrel{(b)}{=}
\frac{1}{{{n^2}}}\bigg( \sum\limits_{{k_1} = 0}^{n \! - \! 1} {\E\left\{ {{\bf{\hat X}}\left[ {{k_1}} \right]\big({{{\bf{\hat X}}}}\left[ {{k_1}} \right]\big)^H} \right\}} {e^{j2\pi {k_1}\frac{{{i_1} \! - \! {i_2}}}{n}}}  + \! \sum\limits_{{k_1} \in \mathcal{L}_-^n} {\E\left\{ {{\bf{\hat X}}\left[ {{k_1}} \right]\big({{{\bf{\hat X}}}}\left[ {{k_1}} \right]\big)^T} \right\}} {e^{j2\pi {k_1}\frac{{{i_1} \! + \! {i_2}}}{n}}} \bigg) \notag \\
&\stackrel{(c)}{=}
\frac{1}{{{n^2}}} \sum\limits_{{k_1} = 0}^{n \! - \! 1} {\E\left\{ {{\bf{\hat X}}\left[ {{k_1}} \right]\big({{{\bf{\hat X}}}}\left[ {{k_1}} \right]\big)^H} \right\}} e^{j2\pi {k_1}\frac{{{i_1} \! - \! {i_2}}}{n}},
\label{eqn:PerSymbol1}
\end{align}
where $(a)$ follows by plugging the inverse DFT \eqref{eqn:InvDFT};
$(b)$ follows since   $\left\{{\bf{\hat X}}\left[ k \right]\right\}_{k=0}^{\Lm}$ are zero-mean and mutually independent, thus $\E\left\{ {{\bf{\hat X}}\left[ {{k_1}} \right]\big({{{\bf{\hat X}}}}\left[ {{k_2}} \right]\big)^H} \right\}$ is non zero only when $k_2 = k_1$ and when $k_2 = n-k_1$,
and since $\hat{\bf X}[k] = \left(\hat{\bf X}[n-k]\right)^*$; 
$(c)$ follows since for ${k} \in \mathcal{L}_-^n$, the optimal $\hat{\bf X}[k]$ is circularly symmetric complex normal, thus
$\E\left\{ {{\bf{\hat X}}\left[ k \right]\big({{{\bf{\hat X}}}}\left[ k \right]\big)^T} \right\} = \mathsf{0}_{n_t\times n_t}$ \cite[Sec. III-A.]{Massey:93}.
It follows from \eqref{eqn:PerSymbol1} that $\E\left\{ {{\bf{X}}\left[ i \right]\big({{\bf{X}}}\left[i \right]\big)^T} \right\} = \frac{1}{{{n^2}}} \sum\limits_{{k_1} = 0}^{n - 1} {\E\left\{ {{\bf{\hat X}}\left[ {{k_1}} \right]\big({{{\bf{\hat X}}}}\left[ {{k_1}} \right]\big)^H} \right\}}$, thus, the covariance matrix of the time-domain optimal channel input ${\bf X}[i]$ which achieves $\CCGTA{n}$ is independent of the time index $i$, $i\in \mN$. As the power constraint is the same for all $n$-blocks, we conclude that  $\CCGTA{n}$  is obtained with an equal per-symbol power allocation, namely satisfies \eqref{eqn:Constraint1} with equality.
\end{IEEEproof}
\smallskip
%
Since $\CCG{n} \leq \CCGTA{n}$, then Corollary \ref{cor:PerSymbol} implies that $\CCG{n} = \CCGTA{n}$.
Combining this with \eqref{eqn:ParallelChannels3}, and noting that it is enough to consider blocklengths which are integer multiples of $n$ to characterize the secrecy capacity of the $n$-CGMWTC, proves that $\CCG{n}$ is obtained by \eqref{eqn:pro_main2}.

\ifextended
\section{Proof of Proposition \ref{Pro:MainThm2}}
\label{app:Proof3}
\vspace{-0.1cm}
We now show that in the limit of $n \rightarrow \infty$, $\CCG{n}$ stated in \eqref{eqn:pro_main2b} coincides with \eqref{eqn:thm_res}. 
Following the approach of \cite[Sec. V]{Goldsmith:01}, this is done by showing that $\CCG{n}$ can be written as averaging over $n$ uniformly distributed samples of a Riemann integrable function, thus  
in the limit of 
$n \rightarrow \infty$, $\CCG{n}$ can be written as an integral \cite[Def. 6.2]{Rudin:76}.
For $k = 0,1,\ldots,n-1$,  
let $\omega_k = \frac{2\pi k}{n}$. By definition, $\mathsf{\hat H}[k] = \mathsf {H}'\left(\omega_k\right)$ and $\mathsf{\hat G}[k] = \mathsf {G}'\left(\omega_k\right)$. From \eqref{eqn:Lemma_NoiseCov} it follows that $\CUHw{k} = n\CCw{\omega_k}$ and $\CUHu{k} = n\CCu{\omega_k}$. 
For a fixed $n > 2m$, by setting $\CCx{\omega_k}= \frac{1}{n}\CUHx{k}$, we can write $\CCG{n}$ of \eqref{eqn:pro_main2b} as
\begin{equation}
\! \CCG{n}
= \mathop {\max }
\limits_{\left\{\CCx{\omega_k}\right\}_{k=0}^{n-1} \in {\mathcal{\hat{C}}}^n_{\frac{P}{n}}}  \frac{1}{2n} 
\sum\limits_{k=0}^{n-1} \log \frac{\left| {{\mathsf{I}}_{{n_r}}} + \mathsf{H}'\left(\omega_k\right)\CCx{\omega_k}\left(\mathsf {H}'\left(\omega_k\right)\right)^H  \left(\CCw{\omega_k}\right)^{-1} \right|}{\left| {{\mathsf{I}}_{{n_e}}} + \mathsf{G}'\left(\omega_k\right)\CCx{\omega_k}\left(\mathsf {G}'\left(\omega_k\right)\right)^H  \left(\CCu{\omega_k}\right)^{-1} \right|}. 
\label{eqn:Rn_Fixedn_1}
\end{equation}

Define $f(\omega) \triangleq \frac{1}{2} \log \frac{\left| {{\mathsf{I}}_{{n_r}}} + \mathsf{H}'\left(\omega\right)\CCx{\omega}\left(\mathsf {H}\left(\omega\right)\right)^H \left(\CCw{\omega}\right)^{-1} \right|}{\left| {{\mathsf{I}}_{{n_e}}} + \mathsf{G}'\left(\omega\right)\CCx{\omega}\left(\mathsf {G}\left(\omega\right)\right)^H \left(\CCu{\omega}\right)^{-1} \right|}$.
Eq. \eqref{eqn:Rn_Fixedn_1} can be written as
\begin{equation}
\label{eqn:Rn_Fixedn_2}
\CCG{n} =  \mathop {\max }\limits_{\left\{\CCx{\omega_k}\right\}_{k=0}^{n-1} \in {\mathcal{\hat{C}}}^n_{\frac{P}{n}}} \frac{1}{n} \sum\limits_{k=0}^{n-1} f\left(\omega_k\right).
\end{equation}
Next, we show that $f(\omega)$ is Riemann integrable. 
From \cite[Thm. 11.33]{Rudin:76},
if a function is bounded and continuous almost everywhere on $\left[0,2\pi\right)$, then it is Riemann integrable. 
Define $f_N(\omega) \triangleq \log {\left| {{\mathsf{I}}_{{n_r}}} \!+\! \mathsf{H}'\left(\omega\right)\CCx{\omega}\left(\mathsf {H}'\left(\omega\right)\right)^H \left(\CCw{\omega}\right)^{-1} \right|}$ and $f_D(\omega) \triangleq \log {\left| {{\mathsf{I}}_{{n_e}}} \!+\! \mathsf{G}'\left(\omega\right)\CCx{\omega}\left(\mathsf {G}'\left(\omega\right)\right)^H \left(\CCu{\omega}\right)^{-1} \right|}$.
From \cite[Thm. 17.9.7]{Cover:06} it follows that both $f_N(\omega)$ and $f_D(\omega)$ are non-negative. 
Following \cite[Eqns. (29)-(31)]{Goldsmith:01}, we assume that the entries of $\CCx{\omega}$ are Riemann integrable functions over $\left[0,2\pi\right)$. 
From the definition in Thm. \ref{Thm:MainThm2}, the entries of the matrices $\mathsf{H}'\left(\omega\right)$, $\mathsf{G}'\left(\omega\right)$, $\CCw{\omega}$, and $\CCu{\omega}$ are  each a finite linear combination of exponential functions, and are therefore continuous with respect to $\omega$, with a finite number of finite magnitude extremums in any given finite interval.  
Thus, from \cite[Thm. 11.33]{Rudin:76} we conclude that the entries of $\mathsf{H}'\left(\omega\right)$, $\mathsf{G}'\left(\omega\right)$, $\CCw{\omega}$, and $\CCu{\omega}$ are all Riemann integrable. 
From Cramer's rule \cite[Ch. 0.8]{Horn:85} and since the determinant of a matrix is obtained as a sum of products of its elements \cite[Ch. 6.1]{Meyer:00}, 
 it follows that each entry in the inverse matrix is obtained as a finite sum of products of entries of the original matrix. 
From \cite[Thm. 6.12, Thm. 6.13]{Rudin:76} we obtain that sum and products of Riemann integrable functions are Riemann integrable.
Hence, when the entries of a matrix are Riemann integrable functions, so are the entries of its inverse, and therefore, the elements of $\left(\CCw{\omega}\right)^{-1}$ and $\left(\CCu{\omega}\right)^{-1}$ are also Riemann integrable functions. It this follows that the entries of ${{\mathsf{I}}_{{n_r}}} + \mathsf{H}'\left(\omega\right)\CCx{\omega}\left(\mathsf {H}\left(\omega\right)\right)^H \left(\CCw{\omega}\right)^{-1}$ and of ${{\mathsf{I}}_{{n_e}}} + \mathsf{G}'\left(\omega\right)\CCx{\omega}\left(\mathsf {G}\left(\omega\right)\right)^H \left(\CCu{\omega}\right)^{-1}$ are Riemann integrable. 
Consequently, as 
$\log\left(\cdot\right)$ is continuous over the set of positive real numbers, it follows from  \cite[Thm. 6.11, Thm. 6.12, Thm. 6.13]{Rudin:76} that $f_N(\omega)$ and $f_D(\omega)$ are Riemann integrable functions on $\left[0,2\pi\right)$. Since the sum of Riemann integrable functions is also  Riemann integrable \cite[Thm. 6.12]{Rudin:76}, it follows that $f(\omega)$ is also Riemann integrable. 
Thus, setting $\Delta\omega^{(n)} = \frac{2\pi}{n}$ yields $\mathop {\lim }\limits_{{n} \to \infty } \frac{1}{{{n}}}\sum\limits_{k = 0}^{n - 1} f\left( {{\omega _k}} \right)  = \mathop {\lim }\limits_{{n} \to \infty } \frac{1}{{2\pi }}\sum\limits_{k = 0}^{n - 1} f\left( {{\omega _k}} \right)\Delta \omega^{(n)} = \frac{1}{{2\pi }}\int\limits_{ 0 }^{2\pi}  {f\left( \omega  \right)} d\omega$. 

Note that as we set $\CCx{\omega_k}$ to be $\frac{1}{n}\CUHx{k}$ we have $\CCx{\omega_k} =\CCx{\omega_{n-k}}^*$, $\forall k \in \{1,2,\ldots,n-1\}$. Therefore, in the limiting case of $n \rightarrow \infty$ it follows that  $\CCx{\omega} =\CCx{2\pi-\omega}^*$.
From the definitions in Thm. \ref{Thm:MainThm2}, it follows that also $\mathsf{H}'\left(\omega\right)=\mathsf{H}'\left(2\pi-\omega\right)^*$, $\mathsf{G}'\left(\omega\right)=\mathsf{G}'\left(2\pi-\omega\right)^*$, $\CCw{\omega}=\CCw{2\pi-\omega}^*$, and $\CCu{\omega}=\CCu{2\pi-\omega}^*$, therefore $f(\omega) = f(2\pi-\omega)$, i.e., $\int\limits_{ 0 }^{2\pi}  {f\left( \omega  \right)} d\omega = 2\int\limits_{ 0 }^{\pi}  {f\left( \omega  \right)} d\omega$.

Using similar arguments, we now show that the power constraint in the limit of $n \rightarrow \infty$ implies that $\CCx{\omega} \in \mathcal{C}_P$. 
Define $t(\omega) \triangleq {\rm {Tr}}\left( \CCx{\omega} \right)$ and note that as $\CCx{\omega} =\CCx{2\pi-\omega}^*$, it follows that $t(\omega)=t(2\pi-\omega)$. 
As the elements of $\CCx{\omega}$ are Riemann integrable, it follows that $t(\omega)$ is also Riemann integrable, thus 
\begin{align}
\mathop {\lim }\limits_{{n} \to \infty } \frac{1}{n}\sum\limits_{k=0}^{n-1}{\rm {Tr}}\left( \CCx{\omega_k} \right) 
&= \mathop {\lim }\limits_{{n} \to \infty } \frac{1}{n}\sum\limits_{k=0}^{n-1}t\left(\omega_k\right) \notag \\
&= \frac{1}{{2\pi }}\int\limits_{\omega = 0}^{2\pi}  {t\left( \omega  \right)} d\omega \notag \\
&\stackrel{(a)}{=} \frac{1}{{\pi }}\int\limits_{\omega =  0 }^\pi  {\rm {Tr}}\left( \CCx{\omega} \right)d\omega. \label{eqn:Const_Fixedn_1}
\end{align} 
where $(a)$ follows as $t(\omega)=t(2\pi-\omega)$. 
It follows from \eqref{eqn:Const_Fixedn_1} that if $\left\{\CCx{\omega_k}\right\}_{k=0}^{n-1} \in {\mathcal{\hat{C}}}^n_{\frac{P}{n}}$, then, in the limit of $n \rightarrow \infty$, 
$\CCx{\omega} \in \mathcal{C}_P$. 
Thus, by letting $n \rightarrow \infty$ in \eqref{eqn:Rn_Fixedn_2}, 
it follows that $\mathop {\lim }\limits_{{n} \to \infty } \CCG{n}$ is given by \eqref{eqn:thm_res}. 
Since it was shown in Proposition \ref{Pro:MainThm1} that $\Cs = \mathop {\lim }\limits_{{n} \to \infty }\CCG{n}$, the proposition follows.

\fi

\section{Proof of Proposition \ref{Pro:Condition_LGMWTC}}
\label{app:Proof4} 
In order to prove Proposition \ref{Pro:Condition_LGMWTC}, we first show that if \eqref{Pro:SuffCond} is satisfied, then the secrecy capacity of the LGMWTC, $\Cs$, is strictly positive; then, we prove that if $\Cs$ is strictly positive, it follows that \eqref{Pro:SuffCond} must be satisfied. The proof follows a similar outline to that of \cite[Corollary 2]{Wornell:10}.
Before we begin, we note that defining $\HCw{\omega} = \left( \CCw{\omega} \right)^{ - \frac{1}{2}}\mathsf{H}'\left( \omega  \right)$ and  $\GCw{\omega} = \left( \CCu{\omega} \right)^{ - \frac{1}{2}}\mathsf{G}'\left( \omega  \right)$, and applying Sylvester's determinant theorem \cite[Ch. 6.2]{Meyer:00}, \eqref{eqn:thm_res} can be written as
\begin{equation}
\Cs \!=\!  \mathop {\max }\limits_{\CCx{\omega} \in \mathcal{C}_P } \frac{1}{{2\pi }}\!\int\limits_{ \omega = 0 }^\pi\!  {\log \frac{\left| {{\mathsf{I}}_{{n_r}}} \!+\! \HCw{\omega}\CCx{\omega}\left(\HCw{\omega}\right)^H \right|}{\left| {{\mathsf{I}}_{{n_e}}} \!+\! \GCw{\omega}\CCx{\omega}\left(\GCw{\omega}\right)^H  \right|}} d\omega.
\label{eqn:thm_res2}
\end{equation}

Assume that \eqref{Pro:SuffCond} is satisfied, then, $\forall \omega \in \Omega$, there exists a vector 
${\bf{v}}\left( \omega  \right)$ such that 
\begin{equation}
\left\| {\GCw{\omega}{\bf{v}}\left( \omega  \right)} \right\| < \left\| \HCw{\omega}{\bf{v}}\left( \omega  \right) \right\|. 
\label{eqn:proof4_a}
\end{equation}
Note that if ${\bf{v}}\left( \omega  \right)$ satisfies \eqref{eqn:proof4_a} then $\frac{{\bf{v}}\left( \omega  \right)}{\left\| {{\bf{v}}\left( \omega  \right)} \right\|}$ also satisfies \eqref{eqn:proof4_a}, hence we can consider only vectors ${\bf{v}}\left( \omega  \right)$ such that $\left\| {{\bf{v}}\left( \omega  \right)} \right\| = 1$.
Now, let $|\Omega|$ denote the Lebesgue measure of $\Omega$ and set $\CCx{ \omega} = \mathsf{0}_{n_t \times n_t}$ for  $\omega \notin \Omega$ and $\CCx{ \omega} = \frac{\pi \cdot P}{|\Omega|}  {\bf{v}}\left( \omega  \right){{\left( {{\bf{v}}\left( \omega  \right)} \right)}^H} $ for  $\omega \in \Omega$. Note that $\CCx{ \omega} $ is a positive-definite Hermitian matrix
which satisfies \eqref{eqn:thm_const}, hence $\CCx{ \omega} \in \mathcal{C}_P$.
It follows that 
\begin{align} 
\!\!\!\!\Cs\!
&\stackrel{(a)}{\geq}\! \frac{1}{{2\pi }}\!\!\int\limits_{\omega \in \Omega}\!\!\!  \log\! \frac{\left| {{{\bf{I}}_{{n_r}}}  \!+\! \frac{\pi \cdot P}{|\Omega|}\HCw{\omega}{\bf{v}}\!\left( \omega  \right)\!{{\left( {{\bf{v}}\!\left( \omega  \right)} \right)}^H}\!{{\left( \HCw{\omega} \right)}^H}} \right|}{\left| {{{\bf{I}}_{{n_e}}} \!+\! \frac{\pi \cdot P}{|\Omega|}\GCw{\omega}{\bf{v}}\!\left( \omega  \right)\!{{\left( {{\bf{v}}\!\left( \omega  \right)} \right)}^H}\!{{\left( \GCw{\omega} \right)}^H}} \right|} d\omega \notag \\
&\stackrel{(b)}{=} \frac{1}{{2\pi }}\int\limits_{\omega \in \Omega}  \Bigg( \log \left( {1 + \frac{\pi \cdot P}{|\Omega|}\cdot{{\left\| {\HCw{\omega}{\bf{v}}\left( \omega  \right)} \right\|}^2}} \right)   - \log \left( {1 + \frac{\pi \cdot P}{|\Omega|}\cdot{{\left\| {\GCw{\omega}{\bf{v}}\left( \omega  \right)} \right\|}^2}} \right) \Bigg) d\omega, \label{eqn:app5_proof1}
\end{align}
where $(a)$ follows from plugging $\CCx{ \omega }$ defined above into \eqref{eqn:thm_res2}, and $(b)$ follows from Sylvester's determinant theorem \cite[Ch. 6.2]{Meyer:00}. 
Note that $\forall \omega \in \Omega$, $\log \left( {1 + \frac{\pi \cdot P}{|\Omega|}\cdot{{\left\| {\HCw{\omega}{\bf{v}}\left( \omega  \right)} \right\|}^2}} \right) > \log \left( {1 + \frac{\pi \cdot P}{|\Omega|}\cdot{{\left\| {\GCw{\omega}{\bf{v}}\left( \omega  \right)} \right\|}^2}} \right)$. As the Lebesgue measure of $\Omega$ is non-zero, it follows that \eqref{eqn:app5_proof1} is strictly positive, thus $\Cs$ is strictly positive, i.e., \eqref{Pro:SuffCond} is a sufficient condition for a strictly positive secrecy capacity.

Next, we show that if \eqref{Pro:SuffCond} is not satisfied, then $\Cs = 0$.  
Let $\CCxopt{\omega} \in \mathcal{C}_P$ be the maximizing covariance matrix for \eqref{eqn:thm_res2}. Thus, we have
\begin{equation}
\label{eqn:app5_proof2}
{\Cs} \!=\! \frac{1}{{2\pi }}\!\int\limits_{\omega = 0}^\pi\!  {\log \frac{{\left| {{{\mathsf{I}}_{{n_r}}}\! +\! \HCw{\omega}\CCxopt{\omega}{{\left( \HCw{\omega} \right)}^H}} \right|}}{{\left| {{{\mathsf{I}}_{{n_e}}} \!+\! \GCw{\omega}\CCxopt{\omega}{{\left( \GCw{\omega} \right)}^H}} \right|}}} d\omega.
\end{equation}
Since $\CCxopt{\omega} \in \mathcal{C}_P$, it follows that $\forall \omega  \in \left[ { 0 ,\pi } \right)$, $\CCxopt{\omega}$ is a positive semi-definite Hermitian matrix, thus, from \cite[Ch. 7.5-7.6] {Meyer:00} it can be written as $\CCxopt{\omega} = {\mathsf{L}}\left( \omega  \right){\left( {{\mathsf{L}}\left( \omega  \right)} \right)^H}$. Plugging this decomposition into \eqref{eqn:app5_proof2} we write
\begin{align}
{\Cs} 
&= \frac{1}{{2\pi }}\!\!\int\limits_{\omega = 0}^\pi \!\! {\log \frac{{\left| {{{\mathsf{I}}_{{n_r}}} + \HCw{\omega}{\mathsf{L}}\left( \omega  \right){{\left( {{\mathsf{L}}\left( \omega  \right)} \right)}^H}{{\left( \HCw{\omega} \right)}^H}} \right|}}{{\left| {{{\mathsf{I}}_{{n_e}}} + \GCw{\omega}{\mathsf{L}}\left( \omega  \right){{\left( {{\mathsf{L}}\left( \omega  \right)} \right)}^H}{{\left( \GCw{\omega}\right)}^H}} \right|}}} d\omega \notag \\
&\stackrel{(a)}{=}\frac{1}{{2\pi }}\!\!\int\limits_{\omega = 0}^\pi \!\! {\log \frac{{\left| {{{\mathsf{I}}_{{n_t}}} + {{\left( {{\mathsf{L}}\left( \omega  \right)} \right)}^H}{{\left( \HCw{\omega} \right)}^H}\HCw{\omega}{\mathsf{L}}\left( \omega  \right)} \right|}}{{\left| {{{\mathsf{I}}_{{n_t}}} + {{\left( {{\mathsf{L}}\left( \omega  \right)} \right)}^H}{{\left( \GCw{\omega} \right)}^H}\GCw{\omega}{\mathsf{L}}\left( \omega  \right)} \right|}}} d\omega, \label{eqn:app5_proof3}
\end{align}
where $(a)$ follows from  Sylvester's determinant theorem \cite[Ch. 6.2]{Meyer:00}. 
Now, define 
\begin{equation*}
{\mathsf{B}}\left( \omega  \right) \triangleq {\left( {{\mathsf{L}}\left( \omega  \right)} \right)^H}{\left( \GCw{\omega}\right)^H}\GCw{\omega}{\mathsf{L}}\left( \omega  \right) - {\left( {{\mathsf{L}}\left( \omega  \right)} \right)^H}{\left( \HCw{\omega} \right)^H}\HCw{\omega}{\mathsf{L}}\left( \omega  \right).
\end{equation*}
${\mathsf{B}}\left( \omega  \right)$ is clearly Hermitian. If \eqref{Pro:SuffCond} is not satisfied, then 
for all $\omega  \in \left[ { 0 ,\pi } \right)$, possibly except for a zero-measure subset of $\left[ { 0 ,\pi } \right)$, ${\mathsf{B}}\left( \omega  \right)$ is positive semi-definite, since $\forall {\bf a} \in \mathds{C}^{n_t}$ 
\begin{align*}
{\bf a}^H {\mathsf{B}}\left( \omega  \right) {\bf a} 
&= {\left\| {\GCw{\omega}{\mathsf{L}}\left( \omega  \right){\bf{a}}} \right\|^2} - {\left\| {\HCw{\omega}{\mathsf{L}}\left( \omega  \right){\bf{a}}} \right\|^2}\notag \\
&\stackrel{(a)}{=} {\left\| {\GCw{\omega}{\bf{\tilde a}}} \right\|^2} - {\left\| {\HCw{\omega}{\bf{\tilde a}}} \right\|^2}\notag \\
&\stackrel{(b)}{\geq} 0,
\end{align*}
where $(a)$ follows by setting ${\bf{\tilde a}} \triangleq {\mathsf{L}}\left( \omega  \right){\bf{a}}$, and $(b)$ follows since 
\begin{equation*}
\frac{{\left\| {\HCw{\omega}{\bf{\tilde a}}} \right\|}}{{\left\| {\GCw{\omega}{\bf{\tilde a}}} \right\|}} \le \mathop {\sup }\limits_{\bf{v} \in \mathds{C}^{n_t}} \frac{{\left\| {\HCw{\omega}{\bf{v}}} \right\|}}{{\left\| {\GCw{\omega}{\bf{v}}} \right\|}} \le 1.
\end{equation*}
Let $\lambda_{\mathsf{G},k}(\omega)$ and $\lambda_{\mathsf{H},k}(\omega)$ be the $k$-th largest eigenvalue of ${\left( {{\mathsf{L}}\left( \omega  \right)} \right)^H}{\left( \GCw{\omega} \right)^H}\GCw{\omega}{\mathsf{L}}\left( \omega  \right)$ and the $k$-th largest eigenvalue of ${\left( {{\mathsf{L}}\left( \omega  \right)} \right)^H}{\left( \HCw{\omega} \right)^H}\HCw{\omega}{\mathsf{L}}\left( \omega  \right)$, respectively, $k \in \{1,2,\ldots, n_t\}$. 
As ${\mathsf{B}}\left( \omega  \right)$ is a positive semi-definite Hermitian matrix, it follows from the min-max theorem \cite[Ch. 7.5]{Meyer:00} \cite[Ch. 7.7]{Horn:85} that $\forall k \in \{1,2,\ldots, n_t\}$,
\begin{equation}
\lambda_{\mathsf{G},k}(\omega)\geq \lambda_{\mathsf{H},k}(\omega) \geq 0, 
\label{eqn:app5_proof3a}
\end{equation}
where the non-negativity of the eigenvalues $\lambda_{\mathsf{G},k}(\omega)$ and $\lambda_{\mathsf{H},k}(\omega)$  follows  since ${{\left( {{\mathsf{L}}\left( \omega  \right)} \right)}^H}{{\left( \GCw{\omega}\right)}^H}\GCw{\omega}{\mathsf{L}}\left( \omega  \right)$ and ${{\left( {{\mathsf{L}}\left( \omega  \right)} \right)}^H}{{\left( \HCw{\omega}\right)}^H}\HCw{\omega}{\mathsf{L}}\left( \omega  \right)$ are positive semi-definite.
Therefore, for all $\omega  \in \left[ { 0 ,\pi } \right)$, except for maybe a zero-measure subset of $\left[ { 0 ,\pi } \right)$, it follows that 
\begin{align}
\left| {{\mathsf{I}}_{{n_t}}} + {{\left( {{\mathsf{L}}\left( \omega  \right)} \right)}^H}{{\left( \GCw{\omega}\right)}^H}\GCw{\omega}{\mathsf{L}}\left( \omega  \right) \right| 
&  \stackrel{(a)}{=} \prod\limits_{k=1}^{n_t} \left(1 + \lambda_{\mathsf{G},k}(\omega)\right) \notag \\
&\stackrel{(b)}{\ge} \prod\limits_{k=1}^{n_t} \left(1 + \lambda_{\mathsf{H},k}(\omega)\right) \notag \\
&\stackrel{(c)}{=} \left| {{{\mathsf{I}}_{{n_t}}} + {{\left( {{\mathsf{L}}\left( \omega  \right)} \right)}^H}{{\left( \HCw{\omega}\right)}^H}\HCw{\omega}{\mathsf{L}}\left( \omega  \right)} \right|,
\label{eqn:app5_proof4}
\end{align}
where $(a)$ and $(c)$ follow from \cite[Ch. 7.5]{Meyer:00} since ${{\left( {{\mathsf{L}}\left( \omega  \right)} \right)}^H}{{\left( \GCw{\omega}\right)}^H}\GCw{\omega}{\mathsf{L}}\left( \omega  \right)$ and ${{\left( {{\mathsf{L}}\left( \omega  \right)} \right)}^H}{{\left( \HCw{\omega}\right)}^H}\HCw{\omega}{\mathsf{L}}\left( \omega  \right)$ are Hermitian; $(b)$ follows from \eqref{eqn:app5_proof3a}.
Applying the relationship \eqref{eqn:app5_proof4} to \eqref{eqn:app5_proof3} yields $\Cs \leq 0$, therefore \eqref{Pro:SuffCond} is a necessary condition for a strictly positive secrecy capacity.
This completes our proof.

\section{Proof of Corollary \ref{Cor:NBPLC}}
\label{app:Proof5} 
We first consider only blocklengths which are an integer multiple of $\ScPLCnt$.
Define the $\ScPLCnt \times 1$ multivariate processes $\VecPLCX\left[\tilde{i}\right]$, $\VecPLCY\left[\tilde{i}\right]$, and $\VecPLCZ\left[\tilde{i}\right]$, using the following assignments: $\left(\VecPLCX\left[\, \tilde{i} \,\right]\right)_k = X\left[\,\tilde{i}\cdot \ScPLCnt + k\right]$, $\left(\VecPLCY\left[\, \tilde{i} \,\right]\right)_k = \ScPLCY\left[\,\tilde{i}\cdot \ScPLCnt + k\right]$, and $\left(\VecPLCZ\left[\, \tilde{i} \,\right]\right)_k = \ScPLCZ\left[\,\tilde{i}\cdot \ScPLCnt + k\right]$, respectively, $k \in \ScPLCnSet$.
It follows from \eqref{eqn:cor_PLCAvgConstb} that 
\begin{align}
\E\left\{{\left\| {\VecPLCX\left[\, \tilde i \, \right]} \right\|}^2 \right\}
&= \sum\limits_{k = 0}^{\ScPLCnt - 1}\E \left\{{\left| {X\left[\, {\tilde{i} \cdot \ScPLCnt + k} \right]} \right|}^2\right\} \notag \\
&\leq  {P}\cdot {\ScPLCnt},
\label{eqn:cor_PLCConst2}
\end{align} 
thus, $\VecPLCX\left[\, \tilde{i} \,\right]$ is subject to a per-symbol power constraint $ {P}\cdot {\ScPLCnt}$.
From \cite[Appendix B]{Shlezinger:15}, it follows that the {\em scalar} NB-PLC WTC \eqref{eqn:RxModel_1} can be transformed into the following equivalent MIMO Gaussian channel with finite memory $m=1$:  
\begin{subequations}
\label{eqn:RxModel_12}
\begin{equation}
\VecPLCY\left[\, \tilde{i} \,\right] \! = \! \sum\limits_{\tilde{\tau}  =  0}^{1} {\VecPLCH\left[\, \tilde{\tau} \,\right]\VecPLCX\left[\, \tilde{i} \! - \! \tilde{\tau}\,\right]}  \!+\! \VecPLCW\left[\, \tilde{i} \,\right] \label{eqn:RxModel_12a}
\end{equation}
\begin{equation}
\VecPLCZ\left[\, \tilde{i} \,\right] \! = \! \sum\limits_{\tilde{\tau}  =  0}^{1} {\VecPLCG\left[\, \tilde{\tau} \,\right]\VecPLCX\left[\, \tilde{i} \! - \! \tilde{\tau}\,\right]}  \!+\! \VecPLCU\left[\, \tilde{i} \,\right], \label{eqn:RxModel_12b}
\end{equation}
\end{subequations}
where the transformation is bijective as we consider only codes with blocklength which is an integer multiple of $\ScPLCnt$.
It follows from Thm. \ref{Thm:MainThm2} that the secrecy capacity of the equivalent MIMO WTC \eqref{eqn:RxModel_12} is given by $\ScPLCnt \cdot C_{s,PLC}$, where $C_{s,PLC}$ is given by \eqref{eqn:cor_PLC}, subject to \eqref{eqn:cor_PLCConst}.
Since each channel input in the equivalent MIMO channel corresponds to $\ScPLCnt$ channel inputs in the original NB-PLC WTC, the corollary follows for codes whose blocklength which is an integer multiple of $\ScPLCnt$. 
Lastly, we note that as $\ScPLCnt$ is fixed and finite, then any achievable secrecy rate can be achieved using codes whose blocklength is an integer multiple of $\ScPLCnt$, where the proof is similar to that of Proposition \ref{pro:Proof_Qn1} and of Lemma \ref{lem:Proof_Rn2}. 

\end{appendices}

%

\end{document}